НАЦІОНАЛЬНА АКАДЕМІЯ НАУК УКРАЇНИ

Інститут ядерних досліджень

На правах рукопису

Даневич Федір Анатолійович

# Експериментальні дослідження подвійного бета–розпаду атомних ядер

Спеціальність 01.04.16

фізика ядра, елементарних частинок і високих енергій

Дисертація на здобуття наукового ступеня

доктора фізико–математичних наук

Науковий консультант

Здесенко Юрій Георгійович

д.ф.–м.н., професор,

чл. кор. НАН України

Київ–2005

NATIONAL ACADEMY OF SCIENCES OF UKRAINE
Institute for Nuclear Research

Manuscript (in Ukrainian)

Fedor A. Danevich

# Experimental research of double beta decay of atomic nuclei

Science area specification (according to the Higher Attestation Commission of Ukraine)
01.04.16
(nuclear physics, particle physics, high energy physics)

A thesis for the degree Doctor of Science
(physics and mathematics)

Scientific advisor
Prof. Yuri G. Zdesenko
corresponding member of
the National Academy of Sciences of Ukraine[1]

Kyiv-2005

[1] Deceased September 1, 2004

# Abstract

Results of several 2β decay experiments performed with the help of low-background crystal scintillators are presented.

A low-background set-up based on large volume (100 – 200 cm$^3$) $CdWO_4$ crystal scintillators was build in the Solotvina Underground Laboratory (Ukraine) to check radioactive contamination of different materials. A sensitivity of the set-up to $^{40}$K and $^{228}$Th contamination in ~1 kg sample was estimated as 0.04 Bq/kg and 0.007 Bq/kg, respectively (for a measuring time 24 h and with a statistical accuracy of 30%). The facility was also used to measure the radio-purity of scintillators on the level of 0.01 – 0.01 mBq/kg (by using pulse-shape discrimination and time-amplitude analyses).

An ultra-low background scintillation spectrometer with enriched $^{116}CdWO_4$ crystal scintillators with improved passive and active shielding was developed. A background counting rate of 0.04 counts / (yr × keV × kg) in 2.5 – 3.2 MeV energy interval was reached. It has allowed to carry out high sensitivity experiment to search for 2β processes in cadmium and tungsten isotopes. In particular, the half-life of the two neutrino 2β decay of $^{116}$Cd has been measured as $T_{1/2} = 2.9^{+0.4}_{-0.3} \times 10^{19}$ yr. The most stringent half-life limit on the neutrinoless 2β decay of $^{116}$Cd has been established as $T_{1/2} \geq 1.7\ (2.6) \times 10^{23}$ yr at 90% (68%) CL. The restrictions on the Majorana neutrino mass and right-handed admixtures in the weak interaction: $\langle m_\nu \rangle \leq 1.9$ eV, $\langle \eta \rangle \leq 2.5 \times 10^{-8}$, $\langle \lambda \rangle \leq <2.2 \times 10^{-6}$ at 90% CL were obtained. Neglecting right-handed contribution a limit on the effective Majorana mass of neutrino was set as $\langle m_\nu \rangle \leq 1.7\ (1.4)$ eV at 90% (68%) CL. The value of the R-parity violating parameter of minimal SUSY standard model was restricted by the $T_{1/2}$ limit to the level of $\varepsilon \leq 7.0\ (6.3) \times 10^{-4}$ at 90% (68%) CL. Moreover, using the experimental bound on the 0ν2β decay with one Majoron emission: $T_{1/2} \geq 0.8\ (1.8) \times 10^{22}$ yr at 90% (68%) CL, and theoretical calculations, the effective Majoron-neutrino coupling constant was restricted to $\langle g_M \rangle \leq 4.6\ (3.1) \times 10^{-5}$. All these results are among the strongest constraints obtained up to date in the direct 2β decay experiments with $^{76}$Ge, $^{82}$Se, $^{100}$Mo, $^{130}$Te, and $^{136}$Xe.

New half-life bounds at the level of $10^{17} - 10^{21}$ yr were set for 2β processes in $^{64}$Zn, $^{70}$Zn, $^{106}$Cd, $^{108}$Cd, $^{114}$Cd, $^{136}$Ce, $^{138}$Ce, $^{142}$Ce, $^{160}$Gd, $^{180}$W, and $^{186}$W by using low-background $CdWO_4$, GSO, and $ZnWO_4$ crystal scintillators.

The claim of discovery of the 0ν2β decay of $^{76}$Ge [Mod. Phys. Lett. A 16 (2001) 2409] was considered critically. A firm conclusion about prematurity of such a claim was derived on the basis of a simple statistical analysis of the measured spectra. The conclusion was also proved by analyzing the cumulative data sets of the Heidelberg–Moscow and IGEX experiments. It allows to establish the highest worldwide half-life limit on the 0ν2β decay of $^{76}$Ge: $T_{1/2} \geq 2.5\ (4.2) \times 10^{25}$ yr at 90% (68%) CL, which corresponds to the most stringent constraint on the Majorana neutrino mass: $\langle m_\nu \rangle \leq 0.31\ (0.24)$ eV.

Energy resolution, $\alpha/\beta$ ratio, pulse-shape discrimination for $\gamma$ rays and $\beta$ particles, radioactive contamination were studied with $CaWO_4$, $ZnWO_4$, $CdWO_4$, $PbWO_4$, GSO, $CeF_3$, yttrium–aluminum garnet doped with neodymium (YAG:Nd) crystal scintillators. Applicability of these scintillators to search for $2\beta$ decay was discussed.

The demands of the future high sensitivity $2\beta$ decay experiments, aiming to observe the neutrinoless $2\beta$ decay or to advance restrictions on the neutrino mass to $\langle m_\nu \rangle \leq 0.01$ eV, were considered. Requirements for their sensitivity and discovery potential were formulated. Two projects of high sensitivity $2\beta$ experiments were proposed. The unique features of the CTF and BOREXINO set-ups were proposed to use for a high sensitivity study of $^{116}Cd$ neutrinoless $2\beta$ decay (CAMEO project). The sensitivity of the CAMEO experiment (in terms of the $T_{1/2}$ limit for $0\nu2\beta$ decay) was estimated as $10^{26}$ yr with 65 kg of $^{116}CdWO_4$ crystals. The last value corresponds to a limit on the neutrino mass of $\langle m_\nu \rangle \leq 0.06$ eV. Moreover, with 1000 kg of $^{116}CdWO_4$ crystals located in the BOREXINO apparatus the neutrino mass limit can be pushed down to $\langle m_\nu \rangle \sim 0.02$ eV. The CARVEL experiment to search for the $0\nu2\beta$ decay of $^{48}Ca$ with the help of enriched $^{48}CaWO_4$ crystal scintillators has been considered. Despite rather high radioactive contaminations of a $CaWO_4$ crystal used in the pilot experiment, the background rate of the $CaWO_4$ detector in the energy region 3.6 – 5.4 MeV (energy window of the $^{48}Ca$ neutrinoless $2\beta$ decay) was reduced down to 0.07 counts / (yr $\times$ keV $\times$ kg). With ~100 kg array of the $^{48}CaWO_4$ crystals the sensitivity of the CARVEL experiment was estimated as $T_{1/2} > 10^{27}$ yr. This value corresponds to the neutrino mass constraint $\langle m_\nu \rangle < 0.04 – 0.09$ eV.

ЗМІСТ

## ПЕРЕЛІК УМОВНИХ ПОЗНАЧЕНЬ, СИМВОЛІВ, ОДИНИЦЬ, СКОРОЧЕНЬ І ТЕРМІНІВ

**$2\nu2\beta$** двохнейтринний подвійний бета–розпад;

**$0\nu2\beta$** безнейтринний подвійний бета–розпад;

**$2\nu2\varepsilon$** двохнейтринний подвійний електронний захват;

**$0\nu2\varepsilon$** безнейтринний подвійний електронний захват;

**$2\nu\varepsilon\beta^+$** двохнейтринний електронний захват з вильотом позитрона;

**$0\nu\varepsilon\beta^+$** безнейтринний електронний захват з вильотом позитрона;

**$2\nu\beta^+\beta^+$** двохнейтринний подвійний бета–розпад з вильотом позитронів;

**$0\nu\beta^+\beta^+$** безнейтринний подвійний бета–розпад з вильотом позитронів;

**$Q_{\beta\beta}$** енергія $2\beta$–переходу;

**$\nu_e$, $\nu_\mu$, $\nu_\tau$** електронне, мюонне, тау–лептонне нейтрино;

**$\langle m_\nu \rangle$** ефективна маса нейтрино, яка проявляється в $0\nu2\beta$–розпаді;

**СМ** стандартна модель елементарних частинок;

**ТВО** теорії великого об‘єднання. Теорії, що мають на меті об‘єднати всі відомі взаємодії: електромагнітну, слабку (які вже об’єднані в рамках теорії Глешоу, Вайнберга, Салама), сильну і, можливо, гравітаційну;

**CL** довірча ймовірність (confidence level).

**ФЕП** фотоелектронний помножувач;

**м.в.е.** метри водяного еквіваленту. Незалежна від складу і густини порід оцінка глибини підземних лабораторії.

**FWHM** енергетична роздільна здатність детектора, визначена як ширина піка на половині його висоти (Full Width at Half Maximum);

**ч.с.с.** число ступенів свободи при підгонці даних методом найменших квадратів;

**о.с.** основний стан ядра

**C.L.** (confidence level) довірча імовірність;

**Радіочистота детектора** кількість домішок радіонуклідів (виражена в одиницях питомої активності, вагових частин, тощо) будь–якого походження, що містяться всередині або на поверхні детектора.

**Космогенні радіонукліди** радіонукліди, що виникають в речовині під дією опромінення космічними променями.

**Пасивний захист** конструкція з матеріалів, що оточують детектор і використовуються для зменшення впливу зовнішніх випромінювань шляхом їх поглинання.

**Активний захист** комплекс детекторів іонізуючого випромінювання, що оточують детектор і використовуються для зменшення впливу зовнішніх та/або внутрішніх фонових випромінювань шляхом їх реєстрації.

**ppm** (part per million) одиниця виміру концентрації домішок ($10^{-6}$ г/г).

# ВСТУП

Результати експериментів, в яких вимірювались потоки нейтрино від Сонця, з верхніх шарів атмосфери, від реакторів та прискорювача, можуть бути пояснені у припущенні про наявність нейтринних осциляцій, процесу, можливого лише за умови, що нейтрино має ненульову масу. Ці експерименти однозначно свідчать про існування нових фізичних явищ за межами сучасної теорії елементарних частинок, так званої Стандартної Моделі (СМ), в якій розглядаються безмасові нейтрино. Спостереження осциляцій нейтрино різко підвищило інтерес до пошуків безнейтринного подвійного бета–розпаду атомних ядер (0ν2β), оскільки ці експерименти є чутливими до природи цієї частинки (нейтрино Майорани чи Дірака) і дозволяють перевірити закон збереження лептонного заряду та виміряти ефективну масу нейтрино Майорани („осциляційні” експерименти чутливі лише до різниці квадратів мас масових станів нейтрино) величиною ≤0.1–0.01 еВ, яка слідує з аналізу даних осциляційних експериментів. Крім того, експерименти по пошуку 0ν2β–розпаду дають найбільш жорсткі обмеження на присутність домішок правих струмів в слабкій взаємодії, константу зв’язку нейтрино з майороном та інші параметри сучасних теорій елементарних частинок – розширень СМ. Ці результати важливі також для астрофізики та космології, оскільки припускається, що масивні нейтрино можуть давати внесок у релятивістську компоненту так званої "темної матерії" Всесвіту. Слід підкреслити, що навіть *відсутність спостереження* процесу 0ν2β–розпаду (а цей процес, не дивлячись на майже півстолітні зусилля, все ще не спостерігався) на все вищому рівні чутливості дозволяє зробити важливі висновки про властивості нейтрино та слабкої взаємодії. Пошуки безнейтринного подвійного бета–розпаду розглядаються зараз як одна з найбільш важливих задач нейтринної фізики [1,2,3,4,5]. Реєстрація і вимірювання дозволеної в рамках СМ двохнейтринної моди 2β–розпаду (2ν2β), самого рідкісного процесу розпаду атомних ядер, дозволяє уточнювати методи теоретичних розрахунків матричних ядерних елементів для безнейтринної моди, розвивати техніку наднизькофонових вимірювань. Таким чином, експериментальні дослідження процесів подвійного бета–розпаду атомних ядер є однією з найбільш актуальних задач фізики ядра, елементарних частинок та астрофізики.

До дисертації увійшли результати виконання держбюджетних тем, що виконувались у відділі фізики лептонів ІЯД НАНУ у 1993 – 1997 рр.: “Дослідження рідкісних процесів за участю лептонів (в тому числі подвійного бета–розпаду атомних ядер), пов‘язаних з фундаментальними властивостями елементарних частинок” (ДР № 0193V028265); у 1998–2000 рр.: “Дослідження властивостей атомного ядра та елементарних частинок в подвійному бета–розпаді та інших рідкісних процесах” (ДР №0198V003595); у 2001–2004 рр. “Дослідження властивостей нейтрино та пошуки ефектів за межами Стандартної Моделі елементарних частинок в експериментах по вивченню подвійного бета–розпаду атомних ядер та інших рідкісних або заборонених процесів” (ДР №0101V000409). Автор був співвиконавцем цих тем.

**Мета і задачі дослідження.** Метою роботи були експериментальні дослідження процесів подвійного бета–розпаду атомних ядер з якомога вищою чутливістю для вивчення властивостей нейтрино та слабкої взаємодії, зокрема, встановлення природи нейтрино (нейтрино Майорани або Дірака), перевірки закону збереження лептонного заряду, пошуків маси нейтрино майоранівської природи, встановлення обмежень на параметри домішок правих токів в слабкій взаємодії, константу зв’язку нейтрино з майороном, параметр мінімальної суперсиметричної моделі з порушенням R–парності. Для досягнення цієї мети необхідно було розробити детектори з якомога нижчим рівнем фону, високими енергетичною роздільною здатністю та ефективністю реєстрації процесів $2\beta$–розпаду, обладнані електронними системами реєстрації даних, які б могли стабільно функціонувати протягом місяців і років під землею. Необхідно було розробити методи аналізу даних з використанням інформації про амплітуду, час та форму сцинтиляційних сигналів. Задачами досліджень були також пошуки шляхів подальшого підвищення чутливості експериментів та розробка нових детекторів, що дали б змогу розширити перелік досліджуваних ядер. Для цього були проаналізовані джерела фону детекторів для пошуку процесів 2β–розпаду та вивчені сцинтиляційні характеристики, форма сцинтиляційних сигналів, рівень радіоактивних домішок в сцинтиляційних кристалах вольфраматів кальцію, цинку, кадмію, свинцю, ортосилікату гадолінію, фториду церію та алюмо-ітрієвого гранату, активованого неодимом.

**Наукова новизна одержаних результатів.** В роботі представлено ряд нових результатів, основними з них є такі. Виміряний з точністю близько 10% період напіврозпаду ядра $^{116}$Cd відносно двохнейтринного подвійного бета–розпаду. Одержане найвище обмеження на період напіврозпаду відносно безнейтринного 2β–розпаду $^{116}$Cd та одне з найбільш жорстких обмежень на ефективну майоранівську масу нейтрино, параметри домішок правих токів в слабкій взаємодії, константу зв'язку нейтрино з майороном та параметр мінімальної суперсиметричної моделі з порушенням R–парності. Нові обмеження одержані також для безнейтринних мод 2β–розпаду ядра $^{116}$Cd на збуджені рівні $2^+_1$, $0^+_1$ та $0^+_2$ $^{116}$Sn. Здійснено пошуки і одержані нові обмеження на періоди напіврозпаду відносно різних мод і каналів 2β–розпаду ядер $^{64}$Zn, $^{70}$Zn, $^{76}$Ge, $^{106}$Cd, $^{108}$Cd, $^{114}$Cd, $^{136}$Ce, $^{138}$Ce, $^{142}$Ce, $^{160}$Gd, $^{180}$W, $^{186}$W. Досліджені сцинтиляційні властивості, форму сцинтиляційних сигналів та радіоактивну чистоту сцинтиляційних кристалів вольфраматів кадмію, кальцію, цинку, свинцю, ортосилікату гадолінію, алюмо-ітрієвого гранату, активованого неодимом (останній кристал вперше досліджено як сцинтилятор).

**Практичне значення одержаних результатів.** Одержані в ході виконання дисертаційної роботи результати, а саме, одні з кращих обмежень на майоранівську масу нейтрино, параметри домішок правих токів в слабкій взаємодії, константу зв'язку нейтрино з майороном та параметр мінімальної суперсиметричної моделі з порушенням R–парності, важливі для розвитку теорії елементарних частинок, зокрема, теорії нейтрино та слабкої взаємодії. Виміряний з точністю близько ±10% період напіврозпаду ядра $^{116}$Cd відносно двохнейтринної моди 2$\beta$-розпаду допоможе точніше обчислити матричні елементи для безнейтринної моди процесу, що дуже важливо для отримання більш точних значень обмежень (чи абсолютних значень у випадку спостереження цього процесу) на масу нейтрино та інші параметри. В результаті проведених досліджень набула подальшого розвитку техніка наднизькофонової сцинтиляційної спектрометрії, розроблені методи аналізу сигналів за формою та часово-амплітудного аналізу подій. Поставлені експерименти показали можливості подальшого підвищення чутливості експериментів по пошуку 2β-розпаду з чутливістю до маси нейтрино на рівні 0.1 – 0.01 еВ. Виконаний аналіз даних двох експериментів по пошуку 2β-розпаду ядра $^{76}$Ge важливий як тому, що було отримане краще в світі обмеження на масу нейтрино, так і тому, що були розвинуті методи статистичного аналізу даних експериментів з малою статистикою. Це важливо для коректної інтерпретації даних таких експериментів. Досліджені властивості сцинтиляційних кристалів $CaWO_4$, $ZnWO_4$, $CdWO_4$, $PbWO_4$, GSO, $CeF_3$, YAG:Nd важливі для їх застосування як у фундаментальних дослідженнях так і для вирішення прикладних задач.

**Особистий внесок здобувача** полягає в участі в розробці, монтажі, настройці експериментальних установок, проведенні вимірювань, аналізі та інтерпретації даних всіх описаних експериментів, підготовці публікацій. Автором вперше було запропоновано використати аналіз форми сцинтиляційних сигналів для ідентифікації подій та зменшення фону сцинтиляційних детекторів вольфрамату кадмію. За участю здобувача цей метод був розроблений і використаний для всіх подальших експериментів по пошуку подвійного бета–розпаду та альфа–розпаду природного вольфраму. Вперше було запропоновано використати сцинтилятори $ZnWO_4$ та YAG:Nd для пошуків подвійного бета–розпаду ядер цинку та неодиму.

Було оцінено ступінь радіочистоти сцинтиляційних кристалів вольфраматів кальцію, цинку, кадмію, свинцю, ортосилікату гадолінію та фториду церію. Було показано, що більша частина радіоактивних домішок в кристалах $^{116}CdWO_4$ концентрується в тонкому поверхневому шарі.

Значна частина описаних в дисертації досліджень є логічним продовженням і розвитком досліджень, що були виконані і увійшли до кандидатської дисертації автора «Поиск 2β–распада $^{116}$Cd с помощью сцинтилляторов вольфрамата кадмия” (Київ, 1994 р). В той же час, всі описані в даній роботі результати були одержані після захисту кандидатської дисертації здобувача.

Результати дисертації доповідалися на міжнародних конференціях:

1 4th Int. Symp. on Weak and Electromagn. Interactions in Nuclei WEIN–95, Osaka, Japan, June 12–16, 1995.

2 4th Int. Workshop on Theor. and Phenomen. Aspects of Underground Phys. TAUP–95, Toledo, Spain, Sept. 17–21, 1995.

3 46 Сов. по яд. спектроскопии и структуре ат. ядра. Москва, 18–21 июня 1996 г. – Санкт–Петербург, 1996

4 “Topics in Astroparticle and Underground Physiscs” (TAUP 97), Gran Sasso, Italy, September 1997.

5 Int. Workshop on Tungstate Crystals, Roma, Italy, Oct. 12–14, 1998. Univ. ‘La Sapienza’, 1999

6 17th Int. Workshop on Weak Interactions and Neutrinos “WIN’99”, Cape Town, South Africa, 24–30 Jan. 1999

7 Міжн. конф. студентів і мол. науковців з теор. та експ. фізики ″Евріка – 2001″, Львів, 16–18.05.2001. Львів, ЛНУ, 2001

8 International Conference “Non–Accelerator New Physics” (NANP–2001), Dubna, Russia, June 19—23, 2001

9 51 Сов. по яд. спектроскопии и структуре ат. ядра. Саров, 3–8 сентября 2001 г. – Саров, 2001

10 Int. Conf. “Topics in Astroparticle and Underground Physics” (TAUP–2001) Gran Sasso, Italy, September 8–12, 2001

11 52 Сов. по яд. спектроскопии и структуре ат. ядра. Москва, 18–22 июня 2002 г. – Москва.

12 Neutrino Physics and Astrophysics. May 25-30, 2002. Munich, Germany.

13 4th Int. Workshop on Neutrino Oscillations and Their Origin, Kanazawa, Japan, 10–14.02.2003

14 Int. Conf. “Non–Accelerator New Physics” (NANP–2003) Dubna, Russia, June 23—28, 2003

15 Int. Conf. “Topics in Astroparticle and Underground Physics” (TAUP–2003) September 5 – 9, 2003 University of Washington, Seattle

16 II Int.School on Neutrino Physics in memory of Bruno Pontecorvo, 7 – 18 September, 2003, Alushta, Ukraine

17 3th Int. Workshop on Phys. Aspects Luminisc. Complex Oxide Dielectrics, 14–17.09.2004, Kharkiv, Ukraine

18 Low Radioactivity Technique (LRT–2004), Sudburry, Canada, December 2004.

19 11 нац. конф. по росту кристаллов НКРК–2004, Москва, 13–17.12.2004 Москва, ИК РАН, 2004.

20 Annual International Conference "Relativistic Astrophysics, Gravitation and Cosmology" May 22–26, 2005 Kyiv, Ukraine.

21 Int. Conf. “Non–Accelerator New Physics” (NANP–2005) Dubna, Russia, June 20–25, 2005.

22 Int. Conf. on Inorganic Scintillatrors and their Industrial Applications, (SCINT–2005) Alushta, Ukraine, 19–23 September 2005.

Результати дисертації доповідались на щорічних конференціях Інституту ядерних досліджень НАНУ в 1997, 1998, 1999, 2001, 2002, 2003, 2004, 2005 рр., на семінарах в Інституті ядерних проблем Мінського університету (4 листопада 2004 р, Мінськ, Білорусія), Національній Лабораторії Гран Сассо (7 грудня 2004 р,

Італія), Інституті сцинтиляційних матеріалів НАНУ (11 березня 2005 р, Харків), кафедрі ядерної фізики Київського національного університету ім. Т.Шевченка (17 березня 2005 р), в Центрі дослідження темної матерії Сеульського Національного університету (10 жовтня 2005 р, Сеул, Корея), на об‘єднаному семінарі Інституту ядерних досліджень НАНУ (27 жовтня 2005 р.), Інституті теоретичної та експериментальної фізики (9 листопада 2005 р., Москва, Росія), Інституті неорганічної хімії ім. А.В. Ніколаєва СВ РАН (16 листопада 2005 р., Новосибірськ, Росія).

Основні результати дисертації опубліковані в роботах:

**РОЗДІЛ 1**

# 2β–РОЗПАД: СУЧАСНИЙ СТАН ТЕОРІЇ ТА ЕКСПЕРИМЕНТУ

## 1.1. Теорія 2β–розпаду

В 1935 році Марія Гіпперт–Майєр вперше вказала на можливість двохнейтринного подвійного бета–розпаду [ 6 ]. Робота була викликана дослідженнями геологічного віку Землі. Важливо було з'ясувати питання про можливу нестабільність ядер, які мали не найменшу серед ізобарів з певним атомним числом енергію, в той час як звичайний β–розпад цих ядер був заборонений законом збереження енергії. Приклад такого триплету ($^{116}$Cd – $^{116}$In – $^{116}$Sn) показаний на рис. 1.1.

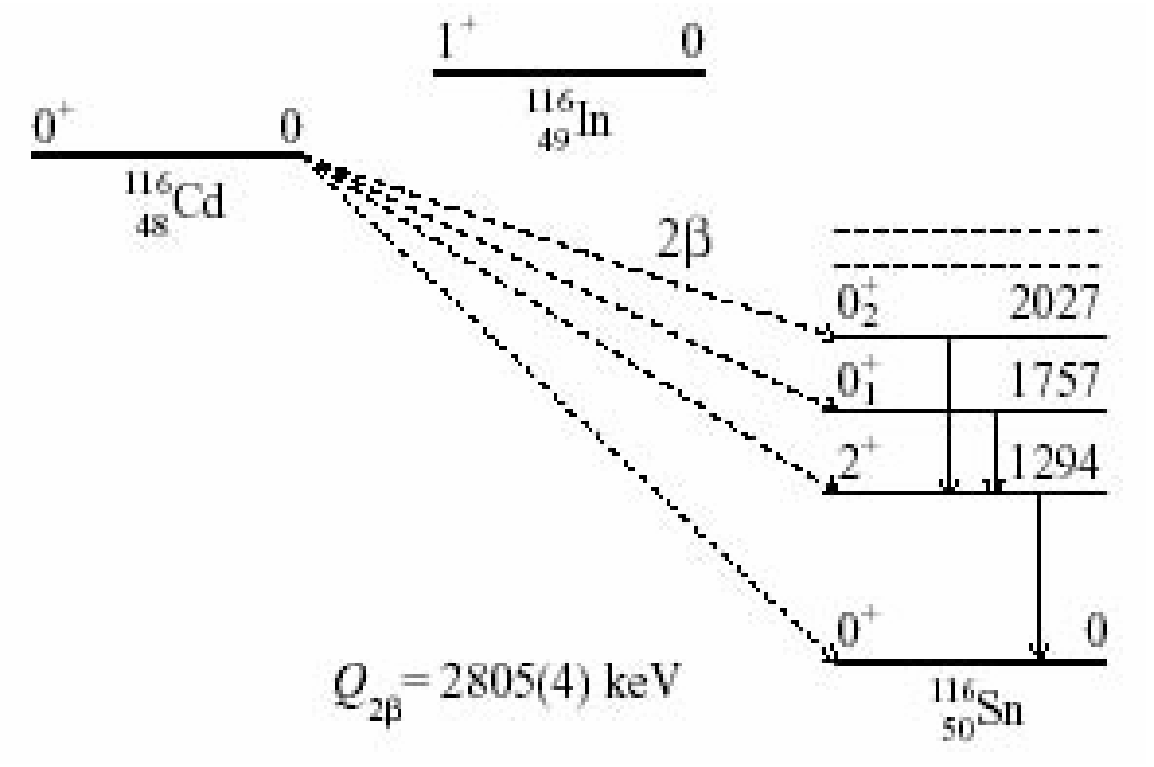

Рис. 1.1. Схема рівнів ізобаричного триплету $^{116}$Cd – $^{116}$In – $^{116}$Sn. Для ядра $^{116}$Cd енергетично дозволений перехід на основний та збуджені стани $^{116}$Sn з випромінюванням двох електронів (2β–розпад).

Гіпперт–Майєр провела розрахунки ймовірності 2β–розпаду, використовуючи теорію слабкої взаємодії Фермі, звідки слідувало, що період напіврозпаду для цього процесу перевищує $10^{17}$ років. В роботі Гіпперт–Майєр мова йшла про розпад ядра, в якому разом з двома електронами випромінюються антинейтрино (2ν2β–розпад). Цей процес може бути представлений як одночасне перетворення 2 нейтронів у ядрі на 2 протони:

$$(A,Z) \rightarrow (A,Z + 2) + e^{-} + e^{-} + \nu + \nu. \quad (1.1)$$

Двохнейтринний 2β–розпад розглядається як процес другого порядку малості у слабкій взаємодії. Так само, як і у випадку одиночного β–розпаду, щоб забезпечити збереження законів збереження енергії та кутового моменту, необхідно, щоб разом з двома електронами випромінювались два антинейтрино. В той же час, процес з випромінюванням двох позитронів (замість якого може відбуватись захват одного або двох електронів з атомної оболонки) супроводжується вильотом нейтрино. В схемі розпаду 1.1 вже закладено наявність ще одного закону збереження, а саме, закону збереження лептонного числа ($L$). Дійсно, можна приписати електрону і нейтрино позитивне лептонне число ($L = +1$), а позитрону і антинейтрино негативне ($L = –1$). Принаймні звичайний одиночний β–розпад, а також двохнейтринна мода 2β–розпаду відбувається без порушення закону збереження величини $L$.

Згідно теорії електрослабких взаємодій Глешоу, Вайнберга, Салама (так званої Стандартної Моделі частинок – СМ) нейтрино відрізняється від антинейтрино спіральністю. В СМ нейтрино завжди має лише ліву спіральність: спін паралельний імпульсу, в той час як антинейтрино – лише праву: спін антипаралельний імпульсу. Але таким нейтрино могло б бути лише за умови, якщо маса цієї частинки була б рівна нулю. У випадку ж хоча б і надзвичайно малої маси, завжди є така система координат, в якій нейтрино змінює свою спіральність. Нейтрино такого типу називають нейтрино майоранівської природи, оскільки саме Етторе Майорана в 1937 році [7] запропонував модель нейтрино, яке тотожне своїй античастинці, на відміну від моделі Дірака, в якій нейтрино і антинейтрино є різними частинками.

Існує цілий ряд теорій поза межами СМ, зокрема теорії об’єднання взаємодій, в яких нейтрино є саме майоранівською частинкою [8]. Першим на можливість встановити природу нейтрино (нейтрино Майорани чи Дірака) саме за допомогою пошуків 2β–розпаду без вильоту нейтрино вказав G. Racah. В роботі, опублікованій в 1937 році [9], було запропоновано схему безнейтринного 2β–розпаду (0ν2β), в якій антинейтрино, випромінене в розпаді одного нейтрона, може поглинутись іншим нейтроном, що приведе до перетворення нейтрона в протон і випромінення другого електрона. В результаті, ядро випромінює два електрони:

$$(A,Z) \rightarrow (A,Z + 2) + e^{-} + e^{-}. \quad (1.2)$$

Очевидно, що такий процес призводить до зміни лептонного числа $L$ на дві одиниці. Але, коли в 1957 році було спостережене явище незбереження парності, стало зрозуміло, що такий двохступінчатий процес неможливий, оскільки нейтрино з правою спіральністю, випромінене в першому розпаді, не може поглинутись нейтроном. Для цього нейтрино має бути ліво–поляризованою частинкою, тобто необхідно, щоб антинейтрино перетворилося в нейтрино. Ще в 1939 році W.H. Furry [10] вказав на можливість 0ν2β–розпаду за участю віртуального нейтрино. До того ж в цій роботі було показано, що оскільки віртуальне нейтрино має значно більшу енергію, безнейтринний процес міг би проходити із значно більшою ймовірністю, ніж у випадку двохступінчатого механізму, запропонованого G. Racah. Перетворення антинейтрино в нейтрино могло б відбуватися, зокрема, завдяки наявності у нейтрино маси майоранівської природи, або за рахунок домішок правих токів в слабкій взаємодії. Запропоновано багато інших механізмів безнейтринного процесу, але, як показали J. Schechter і J.W.F. Valle [11], з аналізу калібровочної теорії слабкої взаємодії слідує: щоб 0ν2β–розпад відбувався, майоранівська маса нейтрино має бути відмінною від нуля. І навпаки, у випадку ненульової маси нейтрино майоранівської природи, процес 0ν2β–розпаду має відбуватися.

Таким чином, є дві принципово відмінні моди подвійного бета–розпаду: з вильотом двох нейтрино (цей процес не порушує ніяких відомих законів і дозволений в рамках СМ) і без вильоту нейтрино. Подвійний бета–розпад без вильоту нейтрино порушує закон збереження лептонного числа і вимагає, щоб нейтрино було масивною частинкою майоранівської природи.

Теорія 2β–розпаду розроблена в роботах [12,13,14,15,16,17,18]. Досить повними і широкими оглядами теорії 2β–розпаду є роботи [19,1,20,21,22,2,3,23,4], де можна знайти посилки як на теоретичні, так і на експериментальні роботи, що стосуються подвійного бета–розпаду. Значна увага приділена процесам 2β–розпаду в книгах [24,25,26].

### 1.1.1. 2β–Розпад з вильотом нейтрино.

На відміну від звичайного β–розпаду, який є процесом першого порядку в класичній теорії слабкої взаємодії Фермі, 2β–розпад розглядається як процес другого порядку із фермієвською константою зв’язку $G_F=1.66\times10^{-5}$ ГеВ$^2$. В сучасній теорії електрослабких взаємодій точково–подібна струм–струмова взаємодія замінена на взаємодію шляхом обміну проміжними бозонами, і тепер β–розпад розглядається як процес другого порядку, а 2β–розпад – четвертого порядку з періодом напіврозпаду, пропорційним $G_F^{-4}$, а отже, дуже великим. 2β–розпад як процес вищого порядку можна спостерігати лише тоді, коли є або енергетична заборона на β–розпад, або імовірність β–процесу сильно придушена за рахунок великої різниці в кутових моментах материнського та дочірнього ядер ($^{48}$Ca, $^{96}$Zr). Аналіз стабільності ядер вказує на те, що $2\beta^-$–розпад можливий для 35 існуючих в природній суміші ізотопів парно–парних ядер. Подвійний β–розпад можливий також, як альтернативний канал розпаду, для великої кількості радіоактивних ядер. Але, як показано в роботі [27], важко реалізувати експеримент по пошуку 2β–розпаду радіоактивних ядер з чутливістю, порівняною з досягнутою в дослідженнях стабільних, або достатньо довгоживучих (як $^{232}$Th або $^{238}$U) ядер.

Для 2ν2β–процесу величина, обернена до періоду напіврозпаду, може бути виражена як

$$(T_{1/2}^{2\nu})^{-1} = G^{2\nu}(Q_{\beta\beta},Z)|M^{2\nu}|^2, \qquad (1.3)$$

де $G^{2\nu}(Q_{\beta\beta},Z)$ – функція, яка може бути точно обрахована і залежить від енергії 2β–переходу $Q_{\beta\beta}$ та заряду ядра $Z$. Функція $G^{2\nu}(Q_{\beta\beta},Z)$ одержується шляхом інтегрування по фазовому простору електронів і нейтрино, що випромінюються [14,17]. Величина $M^{2\nu}$ – ядерний матричний елемент 2ν2β–розпаду. В теорії збурень другого порядку матричний елемент подвійного бета–розпаду має таку форму [8]:

$$M(i \to f) = \Sigma\langle f\,|H_\beta|\,k\rangle\,(E_k - E_i)^{-1}\langle k\,|H_\beta|\,i\rangle, \qquad (1.4)$$

де $i$ означає початковий стан материнського ядра, $f$ – кінцевий стан дочірнього ядра плюс електронів і (у випадку 2ν–моди) двох нейтрино, $k$ – проміжний стан ядра плюс один електрон і одне нейтрино. Енергії цих станів позначені як $E_k$ і $E_i$. Сумування проводиться по проміжним станам $k$. $H_\beta$ – позначає функцію Гамільтона слабкої взаємодії.

Для переходів з основного в основний стан (які для всіх ядер, кандидатів на подвійний бета–розпад, мають нульовий спін і позитивну парність), тобто для переходів $0^+ \to 0^+$ основний внесок дає аксіально–векторна частина Гамільтоніана слабкої взаємодії. Тому матричний елемент для двохнейтринного 2β–розпаду має такий вигляд:

$$M^{2\nu} = \langle f\,|\,\Sigma\tau_j\tau_k\,\boldsymbol{\sigma}_j \cdot \boldsymbol{\sigma}_k|\,i\rangle, \qquad (1.5)$$

де сумування здійснюється по j та k, оператори τ перетворюють нейтрон в початковому стані в протон в кінцевому стані, $\boldsymbol{\sigma}_j$, $\boldsymbol{\sigma}_k$ – матриці Паулі.

Для обрахування ядерних матричних елементів (як для двохнейтринної, так і для безнейтринної мод 2β–розпаду) необхідно знайти необхідні хвильові функції, вирішивши задачу багатьох тіл і врахувавши якомога точніше потенціал взаємодії нуклонів. Перші розрахунки 2β–розпаду за оболонковою моделлю були здійснені Хакстоном та Стефенсоном [28]. На думку авторів [4] обрахунки матричних елементів за оболонковою моделлю можна провести достатньо коректно лише для таких легких ядер як $^{48}$Ca, $^{76}$Ge, а також для ядра $^{136}$Xe, яке має магічне число нейтронів (82).

Для інших ядер використовують модель нейтрон–протонного квазічастинкового наближення з довільною фазою (quasi–particle random phase approximation, QRPA), вперше застосовану до обрахунків матричних елементів 2β–розпаду Хуфманом [29] і розвинуту в роботах [30,31,32]. В цьому методі використовуються такі властивості парно–парних ядер (якими є всі потенційно 2β–активні ядра) як спарювання, що призводить до великої енергії зв’язку, та спін–ізоспінова кореляція, що призводить до виникнення резонансу Гамова–Теллера. В розрахунках використовуються параметри підгонки, пов’язані з енергіями спарювання та енергією гігантського резонансу, відомі з експерименту. Оскільки нуклон–нуклонна взаємодія може бути врахована лише наближено, отримувані в розрахунках значення виявляються надзвичайно чутливими до сили взаємодії між частинками. Результати розрахунків різних авторів відрізняються на два–три і більше порядків величини (див. огляди [20,22]). Це ініціювало роботу теоретиків, спрямовану на розвиток методу QRPA. Були розроблені різноманітні вдосконалення цього підходу [33,34,35,36,37,38,39,40,41,42,43,44,45,46,47]. Задача точних розрахунків ядерних матричних елементів подвійного бета–розпаду, особливо безнейтринної моди, залишається надзвичайно важливою для точного обрахування значення маси нейтрино (або обмежень на її величину) з експериментальних даних [48].

Так само як і у випадку звичайного β–розпаду, спектр енергії електронів, що випромінюються в 2β–розпаді, є неперервним (рис. 1.2), оскільки частину енергії забирають два нейтрино.

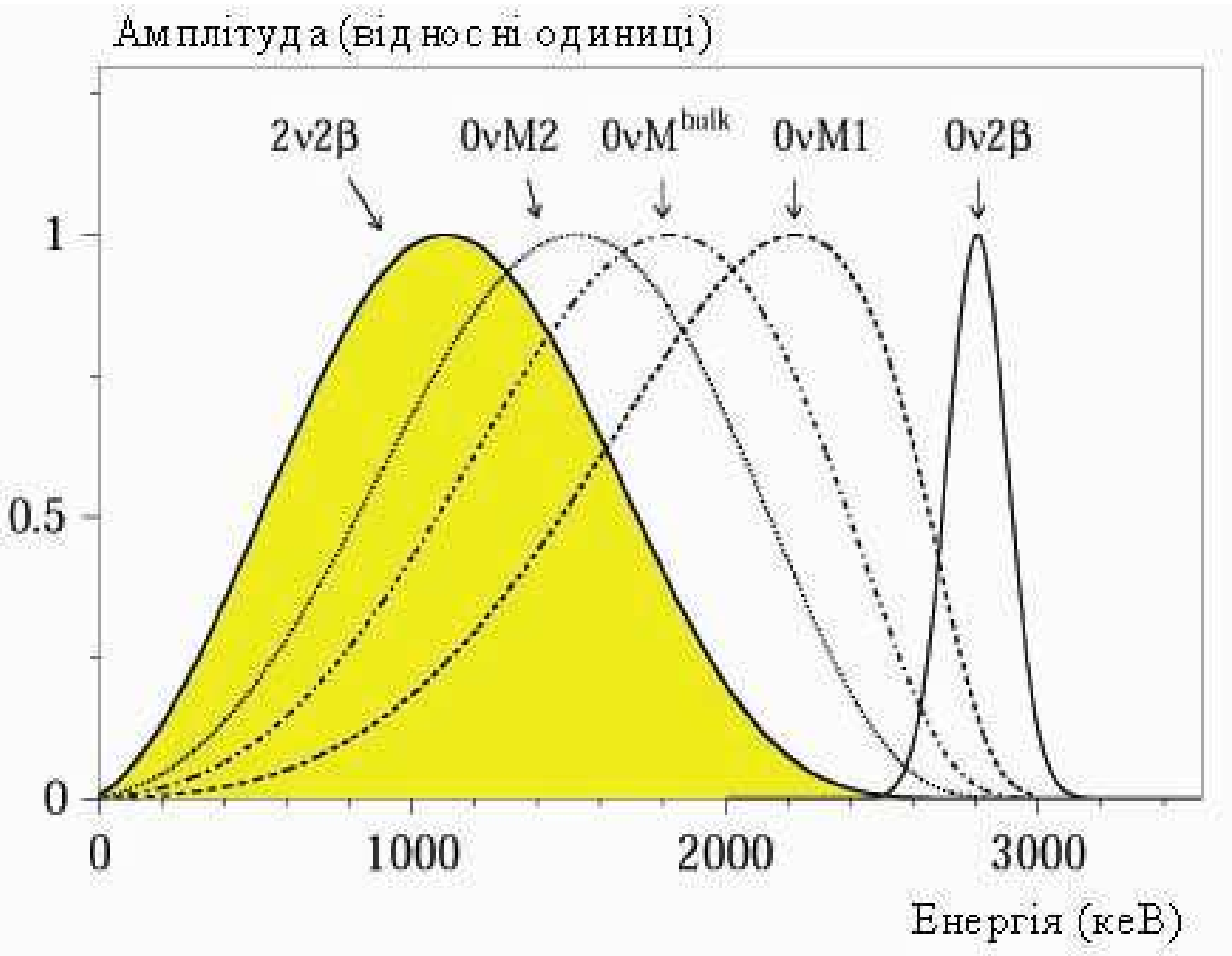

Рис. 1.2. Енергетичні спектри сумарної енергії електронів, що випромінюються в різних модах і каналах 2β–розпаду ядра $^{116}Cd$. Врахована енергетична роздільна здатність детектора.

1.1.2. Безнейтринний 2β–розпад.

У випадку 0ν2β–розпаду два електрони в кінцевому стані отримують всю енергію розпаду (не враховуючи незначну енергію віддачі ядра), а тому спектр суми їх енергій в детекторі з ідеальною роздільною здатністю має вигляд δ–функції на енергії 2β–переходу (рис. 1.2).

Період напіврозпаду 0ν2β–розпаду в припущенні про масовий механізм процесу із домішками правих струмів може бути виражений у наступній формі:

$$(T_{1/2}^{0\nu})^{-1} = C_{mn}^{0\nu}(\frac{\langle m_\nu \rangle}{m_e})^2 + C_{m\lambda}^{0\nu}\langle\lambda\rangle(\frac{\langle m_\nu \rangle}{m_e}) + C_{m\eta}^{0\nu}\langle\eta\rangle(\frac{\langle m_\nu \rangle}{m_e}) + C_{\lambda\lambda}^{0\nu}\langle\lambda\rangle^2 + C_{\eta\eta}^{0\nu}\langle\eta\rangle^2 + C_{\lambda\eta}^{0\nu}\langle\lambda\rangle\langle\eta\rangle, \qquad (1.6)$$

де $m_e$ – маса електрона, $\langle m_\nu \rangle$ – ефективна маса електронного нейтрино (див. нижче, формула 1.9), $\langle\lambda\rangle$ – ефективна константа зв'язку між правим лептонним струмом та правим кварковим струмом, а $\langle\eta\rangle$ – ефективна константа зв'язку між правим лептонним струмом та лівим кварковим струмом. Коефіцієнти $C_{ij}$ включають в себе матричні елементи та функції що одержуються шляхом інтегрування по фазовому простору електронів і нейтрино. Припускаючи, що всі значення *NME* можуть бути обчислені, отже, всі коефіцієнти $C_{ij}^{0\nu}$ відомі, формулу 1.6 можна інтерпретувати як рівняння еліпсоїда, який обмежує дозволені значення параметрів $\langle m_\nu\rangle$, $\langle\lambda\rangle$ і $\langle\eta\rangle$ для заданого значення чи межі періоду напіврозпаду. Нехтуючи внесками правих струмів ($\langle\lambda\rangle = 0$, $\langle\eta\rangle = 0$), отримаємо вираз, схожий на формулу 1.3:

$$(T_{1/2}^{0\nu})^{-1} = G^{0\nu}(Q_{\beta\beta},Z)\ |M^{0\nu}|^2\ (\langle m_\nu\rangle / m_e)^2, \tag{1.7}$$

де $G^{0\nu}$ – інтеграл по фазовому простору, а $M^{0\nu}$ – комбінація ядерних матричних елементів Гамова–Теллера і Фермі:

$$M^{0\nu} = |M_{0\nu}^{GT} - (g_V^2 / g_A^2)\ M_{0\nu}^{F}|, \tag{1.8}$$

де $g_V$ і $g_A$ позначають аксіально–векторну і векторну константи зв'язку слабких струмів. Проблема точності обрахунків матричних елементів $0\nu2\beta$–розпаду дуже важлива, оскільки від неї залежить точність обрахунків ефективної маси нейтрино $\langle m_\nu\rangle$.

Величина $\langle m_\nu\rangle$ – це ефективна маса електронного нейтрино, яка визначається як суперпозиція масових станів нейтрино:

$$\langle m_\nu\rangle = \left|\Sigma m_j\ U_{ej}^2\right| = |U_{e1}|^2\ m_1 + |U_{e2}|^2\ m_2\ e^{i(\alpha_2 - \alpha_1)} + |U_{e3}|^2\ m_3\ e^{i(-\alpha_1 - \delta)}, \tag{1.9}$$

де $m_j$ – масові стани нейтрино, $e^{i(\alpha_2 - \alpha_1)}$ і $e^{i(-\alpha_1 - \delta)}$ – фазові множники, пов'язані із порушенням *CP*–інваріантності (±1 у випадку збереження величини *CP*) через майоранівську природу нейтрино, $U_{ej}$ – елементи матриці змішування між масовими станами нейтрино і станами, відповідними різним ароматам нейтрино (електронному ($\nu_e$), мюонному ($\nu_\mu$), і $\tau$–лептонному ($\nu_\tau$)). Згідно роботі [49] матриця змішування Kobayashi–Maskawa [50] між трьома масовими ($\nu_1$, $\nu_2$, $\nu_3$) і трьома станами, які відповідають різним ароматам нейтрино, може мати наступний вигляд [5]:

$$\begin{pmatrix} |\nu_e\rangle \\ |\nu_\mu\rangle \\ |\nu_\tau\rangle \end{pmatrix} = \begin{pmatrix} c_3c_2 & s_3c_2 & s_2e^{-i\delta} \\ -s_3c_1 - c_3s_1s_2e^{i\delta} & c_3c_1 - s_3s_1s_2e^{i\delta} & s_1c_2 \\ s_3s_1 - c_3s_1s_2e^{i\delta} & -c_3s_1 - s_3c_1s_2e^{i\delta} & c_1c_2 \end{pmatrix} \begin{pmatrix} 1 & 0 & 0 \\ 0 & e^{i\alpha_2} & 0 \\ 0 & 0 & e^{i\alpha_1} \end{pmatrix} \begin{pmatrix} |\nu_1\rangle \\ |\nu_2\rangle \\ |\nu_3\rangle \end{pmatrix} \tag{1.10}$$

де $c_{ij} = \cos\theta_{ij}$, $s_i = \sin\theta_{ij}$; $\theta_{ij}$ – кути змішування; δ – діраківська фаза, $\alpha_1$, $\alpha_2$ – фази, що вводять частинки майоранівської природи. Як буде показано нижче, значна частина параметрів, що входять у вираз 1.10, можуть бути визначеними або обмеженими за допомогою аналізу даних експериментів, в яких досліджувались осциляції нейтрино. В той же час, лише з даних експериментів по пошуку безнейтринного подвійного бета–розпаду можна визначити шкалу маси найлегшого масового стану нейтрино, перевірити закон збереження лептонного числа та встановити природу нейтрино (частинка Майорани чи Дірака).

Ефективні константи зв’язку між правим лептонним струмом та лівим кварковим струмом $\langle\eta\rangle$, та між правим лептонним струмом та правим кварковим струмом $\langle\lambda\rangle$ в формулі 1.6, подібно до ефективної маси нейтрино $\langle m_\nu\rangle$, можуть бути записані наступним чином:

$$\langle\eta\rangle = \eta\,\Sigma\, U_{ej}\, V_{ei}, \quad (1.11)$$

$$\langle\lambda\rangle = \lambda\,\Sigma\, U_{ej}\, V_{ei}. \quad (1.12)$$

За допомогою цих параметрів вводяться відхилення від Стандартної Моделі електрослабких взаємодій. Тут сумування проводиться по індексу $i$ від 1 до $n$, де $n$ дорівнює кількості ароматів легких нейтрино (з масою $m_i$ < 100 МеВ), а матриці $V_{ei}$ мають розмірність $n \times 2n$. Нейтрино разом з електроном може утворювати правий заряджений лептонний струм, який може слабко взаємодіяти з адронними струмами. При цьому 0ν2β–розпад може відбуватися без зміни спіральності віртуального нейтрино.

Функція Гамільтона, яка описує слабкі взаємодії лептонних та адронних струмів при низьких (менших за масу W–бозона) енергіях, має такий вигляд [24]:

$$H_W = G[J_L^{\alpha}\,(M^{+}_{L\alpha} + k\,M^{+}_{R\alpha}) + J_R^{\alpha}\,(\eta\,M^{+}_{L\alpha} + \lambda\,M^{+}_{R\alpha})] + \text{e.c.} \quad (1.13)$$

де $J_{L(R)}^{\alpha}$ і $M^{+}_{L(R)\alpha}$ позначають лептонні та кваркові ліві ($L$) та праві ($R$) чотири–вектори, відповідно. За допомогою параметра $k$ вводиться взаємодія правих адронних струмів з лівими лептонними струмами. В той час як в СМ параметри $\eta$ і $\lambda$ дорівнюють нулю, а $k = 1$, в деяких ТВО, наприклад в ліво–правій симетричній моделі $SU(2)_L + SU(2)_R + U(1)$, вони мають такий вигляд: $k = \eta = -\tan\zeta$; $\lambda = (M_{WL} / M_{WR})^2$. Тут $\zeta$ – кут змішування проміжного векторного бозона W, $M_{WL}$ і $M_{WR}$ – маси ліво– і право–поляризованих W–бозонів. Параметр $k$ дає практично нульовий внесок в процес 0ν2β–розпаду, тому підходящими параметрами для оцінки вкладу правих струмів в слабку взаємодію є лише параметри $\lambda$ і $\eta$.

За умови існування важкого нейтрино майоранівської природи з масою на рівні кількох МеВ, 0ν2β–розпад за рахунок правих токів в слабкій взаємодії може бути домінуючим процесом. Така можливість реалізується в ліво–правих симетричних ТВО за рахунок обміну таким важким нейтрино.

Важливою особливістю 0ν2β–розпаду за рахунок правих струмів є те, що цей процес, на відміну від 0ν2β–розпаду за рахунок маси нейтрино, має йти також на $2^+$ збуджені стани дочірнього ядра. Як показано в роботах [13,14] для розрахунків періоду напіврозпаду ядра відносно 0ν2β–процесу на рівень $2^+$, що протікає внаслідок взаємодії правих лептонних струмів, необхідно обраховувати чотири матричних елементи. Наприклад, один з них має вигляд:

$$M^{0\nu}{}_2 = \langle 2f^+ \mid \Sigma\, H(r_{jk})\, \boldsymbol{\sigma}_j \cdot \boldsymbol{\sigma}_k\, [\boldsymbol{r}_{jk} \times \boldsymbol{r}_{jk}]^{(2)} / r_{jk}^2\, \tau_k^+ \tau_k^+ \mid 0^+_i \rangle, \qquad (1.14)$$

де сумування проводиться по індексам j та k, а оператори $\tau$ перетворюють нейтрон в початковому стані в протон в кінцевому стані. Спостереження 0ν2β–переходу на $2^+$ збуджені стани було б впевненою вказівкою на наявність правих лептонних токів в слабкій взаємодії.

Крім того, енергетичні спектри та кутові розподіли між електронами, що випромінюються, відрізняються від тих, що мають спостерігатися в 0ν2β–розпаді, ініційованому масивним нейтрино. Таким чином, експериментально можна визначити внесок правих струмів в швидкість 0ν2β–розпаду ядер і точніше обрахувати ефективну масу нейтрино. Враховуючи, що процес безнейтринного 2β–розпаду може йти за рахунок різних механізмів, на думку авторів [51] важливо виміряти періоди напіврозпаду для якнайбільшої кількості ядер.

1.1.3. 0ν2β–Розпад з вильотом майоронів.

Багато розширень СМ, зокрема ТВО, включають взаємодії із незбереженням лептонного заряду $L$ (більш точно, різниці між баріонним та лептонним зарядами $B-L$), оскільки у сучасних калібровочних теоріях збереження лептонного (баріонного) заряду розглядається як наближений закон. Це пов'язано з відсутністю будь–якої симетрії в основі цих квантових чисел, такої як, наприклад, калібровочна інваріантність, що забезпечує відсутність маси у фотона і абсолютне збереження електричного заряду. Тому цілком природним є припущення, що симетрія пов'язана із збереженням $B-L$ є наближеною і може бути порушеною. Порушення $B-L$ на 2 одиниці породжує масивні майоранівські нейтрино і тому веде до обмінного механізму нейтрино у 0ν2β–розпаді. Існує три можливості для цього: (1) точне порушення $B-L$, (2) спонтанне порушення локальної і (3) глобальної $(B-L)$–симетрій. В третьому випадку калібровочні моделі припускають існування фізичного бозона Намбу–Голдстоуна – так званого майорона (M) [52,53,54,55], що є гіпотетичною нейтральною псевдоскалярною частинкою з нульовою, або близькою до нуля масою, яка може випромінюватись у безнейтринному 2β–розпаді. Вимірювання ширини розпаду $Z^0$–бозона, проведені в експериментах на колайдерах, показали, що є всього три аромати легких нейтрино. Цей результат забороняє існування одиночного майорону типу Gelmini–Roncadelli [54], оскільки він мав би приводити до збільшення ширини, еквівалентного двом додатковим ароматам нейтрино [55]. Але як показано в роботі [56], цей результат не забороняє існування легкої скалярної частинки пов'язаної не з $Z^0$–бозоном, а з нейтрино майоранівської природи. При цьому, такі частинки в усьому будуть вести себе так, як майорони [54], в тому числі призведуть до процесів 0ν2β–розпаду.

Є кілька моделей майоронів. В 0ν2β–розпаді може випромінюватися одиночний майорон:

$$(A,Z) \rightarrow (A,Z + 2) + e^{-} + e^{-} + M, \qquad (1.15)$$

і два майорони:

$$(A,Z) \rightarrow (A,Z + 2) + e^{-} + e^{-} + M + M. \qquad (1.16)$$

Оскільки очікується, що майорони є дуже легкими або безмасовими, спектр суми енергій електронів є неперервним. Однак 2β–процеси з вильотом майоронів можуть бути відрізнені від 2ν2β–моди, оскільки їхні максимуми знаходяться при різних значеннях енергії. При цьому форма енергетичного спектру двох електронів відрізняються для мод 1.15 і 1.16 (див. рис. 1.2, де показано форми спектрів електронів в 0ν2β–розпадах з вильотом майоронів).

В роботі [57] запропоновано ще одну модель майорону (англійською мовою “bulk majoron”), який був запропонований в контексті теорії бран:

$$(A,Z) \rightarrow (A,Z + 2) + e^{-} + e^{-} + M^{bulk} \qquad (1.17)$$

Такий синглетний майорон виникає у випадку спонтанного порушення глобальної $B - L$ симетрії Стандартної Моделі частинок через взаємодію з калібровочним полем бозонів Хігса в багатовимірному просторі додаткових вимірів. Як видно з рис. 1.2, форма енергетичного спектру для $M^{bulk}$ відрізняється від 0ν2β–розпаду з випромінюванням одиночного і подвійного майоронів.

Формула для інтенсивності розпаду 0ν2β–моди з випромінюванням майоронів може бути отримана по аналогії з виразом 1.7 заміною $\langle m_\nu \rangle$ на $\langle g_M \rangle$, де $\langle g_M \rangle$ – ефективна константа зв’язку нейтрино з майороном, та заміною $G^{0\nu}$ на інтеграл по фазовому простору, що описує 2 електрони і безмасовий майорон у кінцевому стані:

$$(T_{1/2}^{0\nu M})^{-1} = G^{0\nu M}(Q_{\beta\beta},Z)\ |M^{0\nu}|^2\ (\langle g_M \rangle / m_e)^2, \qquad (1.18)$$

Ядерні матричні елементи для 0ν2β–розпаду з вильотом майоронів такі самі як і для 0ν2β–розпаду, що відбувається завдяки масі нейтрино. Інтеграли по фазовому простору для 2β–розпаду з майоронами були розраховані в роботі [16].

### 1.1.4. Інші можливі механізми 0ν2β–розпаду.

Крім механізму обміну лівими нейтрино, сучасні калібровочні теорії припускають багато інших можливостей для протікання 0ν2β–процесу. Наприклад, у ліво–право суперсиметричних моделях теорій великого об‘єднання безнейтринний 2β–розпад може бути спричинений важкими правими нейтрино [58,59]. Безнейтринний 2β–розпад може відбуватися через взаємодію за рахунок лептокварків (новий тип калібровочних бозонів, що передбачається деякими ТВО). Ці частинки можуть перетворювати кварки в лептони і спричиняти 0ν2β–розпад завдяки зв’язку з частинками Хігса [16,60,61].

Новий механізм 0ν2β–розпаду може дати гіпотетична субструктура кварків і лептонів. 0ν2β–Розпад може відбуватися через обмін композитними важкими майоранівськими нейтрино [62,63]. Більш того, існують можливі механізми 0ν2β–розпаду, основані на суперсиметричних (SUSY) взаємодіях: обмін скварків із незбереженням *R*–парності [64,65,66,67,68,69,70,71,72] та обмін суперсиметричним партнером нейтрино: снейтрино, коли *R*–парність зберігається [73,74]. *R*–парністю називається величина, яка визначається наступним чином: $R_p = (-1)^{3B+L+2S}$, де $B$ – баріонний заряд, $L$ – лептонний заряд, $S$ – спін. Важливо підкреслити, що як було показано в роботі [11], в „нормальних” (які включають групу SU(2) × SU(1)) калібровочних теоріях, 0ν2β–розпад, незалежно від того, завдяки якому механізму він відбувається, вимагає наявності майоранівської маси нейтрино. І навпаки, при наявності майоранівського масового компонента процес 0ν2β–розпаду має обов’язково протікати.

### 1.1.5. Процеси подвійного електронного захвату (2ε), електронного захвату і вильотом позитрона ($\varepsilon\beta^+$) та подвійного позитронного розпаду ($2\beta^+$).

Крім подвійного бета–розпаду з випромінюванням електронів, для багатьох ядер можливі процеси, в яких випромінюються два позитрони ($2\beta^+$), відбувається електронний захват з випроміненням позитрону ($\varepsilon\beta^+$), або подвійний електронний захват (2ε):

$$(A,Z) \rightarrow (A,Z-2) + e^+ + e^+ + \nu + \nu, \qquad (1.19)$$

$$e^- + (A,Z) \rightarrow (A,Z-2) + e^+ + \nu + \nu, \qquad (1.20)$$

$$e^- + e^- + (A,Z) \rightarrow (A,Z-2) + \nu + \nu. \qquad (1.21)$$

В процесах 1.19 – 1.21 випромінюються нейтрино і тому вони не порушують ніяких відомих законів збереження. Але якщо лептонний заряд не зберігається, можливі безнейтринні процеси, в яких заряд ядра зменшується на дві одиниці:

$$(A,Z) \rightarrow (A,Z-2) + e^+ + e^+, \qquad (1.22)$$

$$e^- + (A,Z) \rightarrow (A,Z-2) + e^+, \qquad (1.23)$$

$$e^- + e^- + (A,Z) \rightarrow (A,Z-2) + \gamma,\ X,\ e. \qquad (1.24)$$

Тут γ, X – гамма та рентгенівські промені, e – Оже–електрони і електрони внутрішньої конверсії. Оскільки в процесі 1.24 нейтрино не випромінюються, енергія розпаду, яка лише частково йде на захват електронів з внутрішніх електронних оболонок дочірнього ядра, може витрачатися на різні процеси: випромінювання γ–квантів (при цьому очікується існування γ–рівня з енергією $Q_{\beta\beta} - 2E_K$, де $E_K$ – гранична енергія зв'язку електрона на внутрішній оболонці дочірнього ядра), рентгенівських променів, Оже електронів, електронів внутрішньої конверсії.

Ймовірність процесів електронного захвату (2ε), електронного захвату і вильотом позитрона ($\varepsilon\beta^+$) та подвійного позитронного розпаду ($2\beta^+$) значно нижча. Це пов'язано із значно меншою енергією, яка передається частинкам, що випромінюються. Періоди напіврозпаду для таких розпадів для двохнейтринних мод очікуються більшими за $10^{20}$ років (див. огляди [20,22], де є посилки на теоретичні розрахунки ймовірності 2β–процесів) і дотепер 2β–розпади 1.19 – 1.21 не спостерігалися в експерименті. Так само ймовірності безнейтринних процесів 1.22 – 1.24 значно менша, ніж розпадів 1.2 з випромінюванням електронів.

В роботі [75] проведені розрахунки ймовірності безнейтринного подвійного електронного захвату різними ядрами. Автори оцінили, що у важких ядрах ($^{92}$Mo, $^{108}$Cd, $^{180}$W, $^{196}$Hg) очікувані періоди напіврозпаду знаходяться на рівні $10^{25}$ – $10^{31}$ років (для ефективної маси нейтрино $\langle m_\nu \rangle$ = 1 еВ). Найбільш сприятливим ядром є $^{180}$W, для якого автори отримали оцінку $T_{1/2}^{2\varepsilon 0\nu} = 2.5\times10^{25\pm1}$ років.

Ще один аргумент на користь проведення досліджень процесів 1.22 – 1.24 був запропонований авторами роботи [76]. Якщо безнейтринний 2β–розпад буде спостережено, постане питання, завдяки якому механізму, масі нейтрино чи домішкам правих токів в слабкій взаємодії, він відбувається. Дослідження різних каналів подвійних бета–процесів, що зменшують заряд ядра, дозволить прояснити це питання. Причому, навіть у випадку неспостереження процесів і лише встановлення обмежень на їх швидкість, такі експерименти дадуть цінну інформацію про механізми 0ν2β–розпаду.

### 1.1.6. 2β–Переходи на збуджені стани.

Для багатьох ядер подвійний бета–розпад може відбуватися на збуджені рівні дочірнього ядра (рис. 1.1). З огляду на деталі ядерної структури і розрахунків матричних елементів, переходи на рівні з нульовим спіном і позитивною парністю ($0^+$) не відрізняються від переходів на основні стани дочірніх ядер. Але їх ймовірність значно менша, оскільки меншою є енергія переходу. Так, в роботі [77] проведені розрахунки ядерних матричних елементів для 0ν2β–розпадів на збуджені $0^+$ рівні ядер $^{76}$Ge і $^{82}$Se. Показано, що очікувані періоди напіврозпаду становлять $10^{26}$ – $10^{27}$ років для маси нейтрино $\langle m_\nu \rangle \approx 0.7$ еВ. В роботі [78] були розраховані ядерні матричні елементи для 0ν2β–розпаду ядер $^{76}$Ge, $^{82}$Se, $^{100}$Mo, $^{136}$Xe на збуджені стани $0^+$ дочірніх ядер. Виявилося, що періоди напіврозпаду для таких переходів на найнижчі рівні в 10 – 100 разів більші, ніж для переходів на основні рівні. В роботі [79] здійснені систематичні дослідження ймовірності переходів на збуджені рівні $0^+$ ряду ядер. Причому розрахунки проведені не лише для $2\beta^-$–розпадів, але і для процесів подвійного електронного захвату та електронного захвату з вильотом позитрона та подвійного позитронного розпаду. Очікувані періоди напіврозпаду для ядер $^{124}$Sn, $^{130}$Te, $^{136}$Xe відносно $0\nu2\beta^-$–розпаду становлять $10^{26}$ – $10^{28}$ років для ефективної маси нейтрино $\langle m_\nu \rangle \approx 0.7$ еВ. Ймовірність $0\nu\beta^+\beta^+$, $0\nu\varepsilon\beta^+$ процесів в ядрах $^{92}$Mo, $^{96}$Ru, $^{106}$Cd, $^{124}$Xe, $^{130}$Ba, $^{136}$Ce для такого ж значення $\langle m_\nu \rangle$ також дуже низькі: періоди напіврозпаду знаходяться в межах $10^{26}$ – $10^{30}$ років. Незважаючи на досить малі теоретично обраховані ймовірності переходів на збуджені рівні дочірніх ядер, дослідження таких розпадів важливе для розробки методів розрахунків матричних елементів. Інтерес до цих розпадів зріс після повідомлень про спостереження 2ν2β–розпаду ядер $^{100}$Mo і $^{150}$Nd на $0^+$ збуджені рівні дочірніх ядер. Ці дані можуть бути використані для подальшої розробки методики розрахунків матричних елементів за допомогою QRPA, оскільки нуклон–нуклонна взаємодія у випадку розпадів на збуджені рівні відрізняється від переходів на основні стани [80,81]. В той же час, автори огляду [4] стверджують, що структура збуджених рівнів ядер зрозуміла не достатньо добре у порівнянні з основними станами ядер, і це утруднює теоретичний аналіз. Як уже зазначалося в п. 1.1.2, переходи на $2^+$ можливі лише за рахунок правих токів в слабкій взаємодії і можуть бути потужним методом вивчення механізмів 0ν2β–розпаду.

### 1.1.7. Спостереження осциляцій нейтрино і 0ν2β–розпад.

Безумовно, найбільш значущим результатом в фізиці частинок, одержаним протягом останнього десятиліття, є спостереження осциляцій нейтрино. Це перший

„удар” по Стандартній Моделі елементарних частинок, що була розроблена в результаті майже півстолітніх зусиль фізиків, головним чином на прискорювачах. Тим більше знаменним є те, що перший аргумент на користь фізики „за межами” СМ був одержаний не на колайдері, а в підземній лабораторії. А почалося все ще в кінці 60–х років, коли Р. Девіс із співробітниками в радіохімічному досліді, що проводився в шахті Homestake в Північній Дакоті (США) і мав на меті вимірювання потоку нейтрино від Сонця за допомогою реакції:

$$^{37}Cl + \nu_e \rightarrow {}^{37}Ar + e^- \qquad (1.25),$$

де $\nu_e$ – електронне нейтрино, спостерігали суттєво менший потік нейтрино, ніж очікуваний згідно теоретичних розрахунків [82]. Цей факт дістав назву „проблема сонячних нейтрино”. Подальші підтвердження дефіциту сонячних нейтрино були одержані в підземних радіохімічних експериментах GALLEX (Національна лабораторія Гран Сассо, Італія; з 1998 р – експеримент продовжувався під назвою GNO) [83,84,85] та SAGE (Баксанська нейтринна обсерваторія, Росія) [86]. В цих експериментах вимірювався потік низькоенергетичних, так званих *pp*–нейтрино, від Сонця в реакції захвату на ядрах галію:

$$^{71}Ga + \nu_e \rightarrow {}^{71}Ge + e^- \qquad (1.26),$$

і знову він виявився майже вдвічі меншим, ніж передбачуваний (причому, у випадку *pp*–нейтрино розрахунки їх потоку більш надійні). В якості можливого пояснення дефіциту сонячних нейтрино розглядалось явище осциляцій нейтрино, тобто взаємного перетворення нейтрино одного аромату в інші. Ефект осциляцій пояснює зменшення потоку електронних нейтрино за рахунок їх перетворення в нейтрино двох інших ароматів ($\nu_\mu$ і $\nu_\tau$), котрі не могли бути зареєстровані в реакціях 1.1 та 1.2.

Подальше підтвердження гіпотези нейтринних осциляцій було отримане в експериментах Kamiokande та Super–Kamiokande (Японія), у величезних підземних черенковських детекторах, в яких вимірювались події пружного розсіяння нейтрино від Сонця та з верхніх шарів атмосфери („атмосферні нейтрино”) на електронах:

$$\nu_x + e^- \rightarrow \nu_x + e^-. \quad (1.27)$$

Тут $\nu_x$ — нейтрино якого завгодно аромату, оскільки всі вони можуть розсіюватися на електронах, хоча $\nu_\mu$ і $\nu_\tau$ – з меншою ймовірністю. В експерименті Kamiokande також було зареєстровано менший, ніж теоретично розрахований потік сонячних нейтрино [87], що було підтверджено за допомогою детектора Super–Kamiokande [88,89], який, крім того, зафіксував асиметрію потоків атмосферних мюонних нейтрино „зверху” та „знизу”, що також може бути пояснене осциляціями нейтрино [90]. Дані Super–Kamiokande [91,92,93,94,95] та інших експериментів з атмосферними нейтрино (Soudan 2 [96], MACRO [97]) добре описуються в припущенні двохнейтринних осциляцій $\nu_\mu \rightarrow \nu_\tau$. При цьому в „осциляційних” експериментах вимірюється величина різниці квадратів масових станів нейтрино $\delta m_{ij}^2 = (m_j^2 - m_i^2)$ та кути змішування $\theta_{ij}$ в матриці 1.10. Із аналізу даних експериментів з атмосферними нейтрино знайдено, що значення цих параметрів, які відповідають найкращому фіту, рівні: $\delta m_{23}^2 = \delta m_{atm}^2 = 2.0^{+1.0}_{-0.7}\times10^{-3}$ еВ$^2$, $\theta_{23} = \theta_{atm} = \approx 45°$ [98].

В експериментах із сонячними та реакторними нейтрино (антинейтрино) вимірюються потоки електронної компоненти нейтрино, тому ці експерименти чутливі до різниці масових станів $\delta m_{12}^2 = \delta m_{sol}^2$. На першому етапі експерименту SNO (Sudbury Neutrino Observatory, Канада), реєструвались реакції захвату нейтрино від Сонця на ядрах дейтерію:

$$\nu_e + d \rightarrow e^- + p + p \quad (1.28),$$

$$\nu_x + d \rightarrow \nu_x + p + n \quad (1.29).$$

та розсіяння на електронах. З аналізу вимірювань та порівняння з результатами експерименту Super–Kamiokande, було зроблено висновок про можливу наявність в потоці нейтрино від Сонця неелектронної компоненти [99,100,101]. Пізніше цей результат було підтверджено завдяки більш надійній реєстрації процесу, що проходить за рахунок нейтральних струмів (тобто таких, які не призводять до зміни електричного заряду). Це було досягнуто завдяки розчиненню солі (*NaCl*) у важкій воді детектора [102], що дозволило реєструвати події захвату нейтронів на ядрах $^{35}Cl$.

Потоки антинейтрино від кількох десятків віддалених приблизно на 180 км (усереднена відстань) японських і південнокорейських реакторів, виміряні за допомогою детектора KamLAND (Японія) [ 103 , 104 ], та потік нейтрино від прискорювача, віддаленого на 200 км (експеримент K2K, Японія) [105], також виявилися меншими за розраховані. Осциляції мюонних нейтрино від прискорювача зареєстровані також в експерименті LSND [106]. Вся ця сукупність даних може бути найкраще описана у припущення про осциляції нейтрино, що в свою чергу можливе лише при наявності ненульової маси у нейтрино [107,108,109].

Дані експериментів, в яких вимірювались потоки сонячних та реакторних нейтрино, можна описати, якщо припустити присутність переходів $\nu_e \rightarrow \nu_{\mu,\tau}$. З глобального фіта методом $\chi^2$ для всіх даних по сонячним нейтрино та результатам реакторних осциляційних експериментів, отримані значення: $\delta m_{12}^2 = \delta m_{\mathrm{sol}}^2 = 7.1^{+1.2}_{-0.6} \times 10^{-5}$ еВ$^2$, $\theta_{12} = \theta_{sol} = \approx 32.5^{+2.4}_{-2.3}$ ° [4].

Таким чином, з сукупності експериментів, що вимірювали потоки нейтрино від різних джерел, слідує, що нейтрино має масу [110]. Проблема полягає в тому, що в осциляційних експериментах не можна виміряти абсолютну величину маси цієї частинки, оскільки вони чутливі до різниці квадратів мас масових станів нейтрино. Тому спостереження явища осциляцій нейтрино різко підвищило інтерес до пошуків 2β–розпаду, оскільки лише в цих експериментах можна виміряти масу нейтрино величиною ~0.1 – 0.01 еВ і встановити природу цієї частинки (нейтрино Майорани чи Дірака). Серед досить небагатьох експериментальних шляхів дослідження властивостей нейтрино, експерименти по пошуку безнейтринного подвійного бета–розпаду (0ν2β) дають найбільш жорсткі обмеження на ефективну майоранівську масу електронного нейтрино, ступінь незбереження лептонного заряду, присутність домішок правих струмів в слабкій взаємодії, константу зв'язку між майороном і нейтрино та інші параметри сучасних теорій елементарних частинок – розширень СМ. Пошуки безнейтринного подвійного бета–розпаду розглядаються зараз як одна з найбільш важливих задач нейтринної фізики [1,2,3,4,5,111]. Ці результати відіграють значну роль в астрофізиці та космології, оскільки припускається, що масивні нейтрино можуть давати внесок у релятивістську компоненту так званої темної матерії Всесвіту.

Параметри, одержані в осциляційних експериментах дозволяють встановити зв'язок між домінуючим станом електронного нейтрино $m_1$ і ефективною масою нейтрино $\langle m_\nu \rangle$, яка проявляється в 0ν2β–розпаді і визначається виразом 1.9. Цей зв'язок був проаналізований в багатьох роботах [112,113,114,115,116,117,118,119, 3,107,120,121,122,123,124,125,126,127]. Результати цього аналізу показані на рис. 1.3. Дві з трьох можливих схем масових станів нейтрино, нормальна (normal hierarchy) та інвертована (inverted hierarchy), зображені на рис. 1.4.

У випадку нормальної схеми, осциляції, що спостерігаються в потоці сонячних нейтрино, відбуваються між найлегшими масовими станами нейтрино, в той час як осциляції, що спостерігаються в потоках атмосферних нейтрино, відбуваються між найлегшим і найважчим масовими станами нейтрино. В інвертованій схемі осциляції сонячних нейтрино відбуваються між найважчими масовими станами нейтрино.

Можлива ще третя, так звана „вироджена" (degenerate hierarchy) схема, коли маса найлегшого масового стану нейтрино є значно більшою за різницю між масами окремих масових станів. Однак, ця схема виглядає не дуже привабливою з урахуванням космологічних [128,129,130] та отриманих в експериментах із β–спектром тритію обмежень на масу нейтрино [131,132].

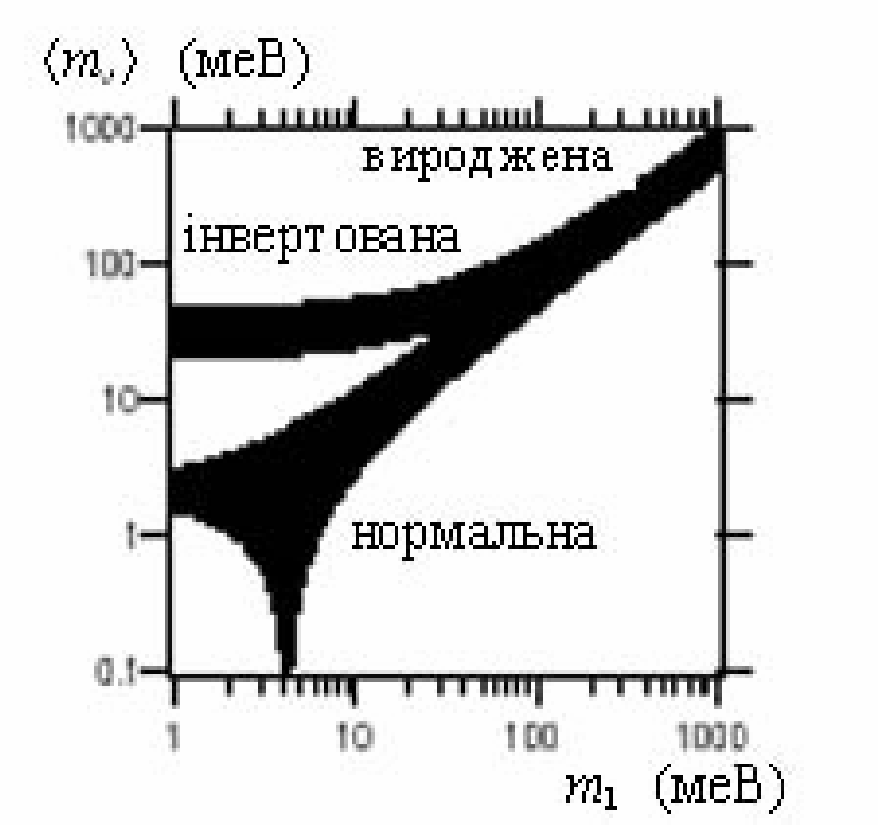

Рис. 1.3. Зв'язок між ефективною масою нейтрино $\langle m_\nu \rangle$, що проявляється в 0ν2β–розпаді і масою найлегшого масового стану нейтрино $m_1$ в різних схемах масових станів нейтрино: нормальній, інвертованій, і виродженій (рисунок взятий з роботи [4]).

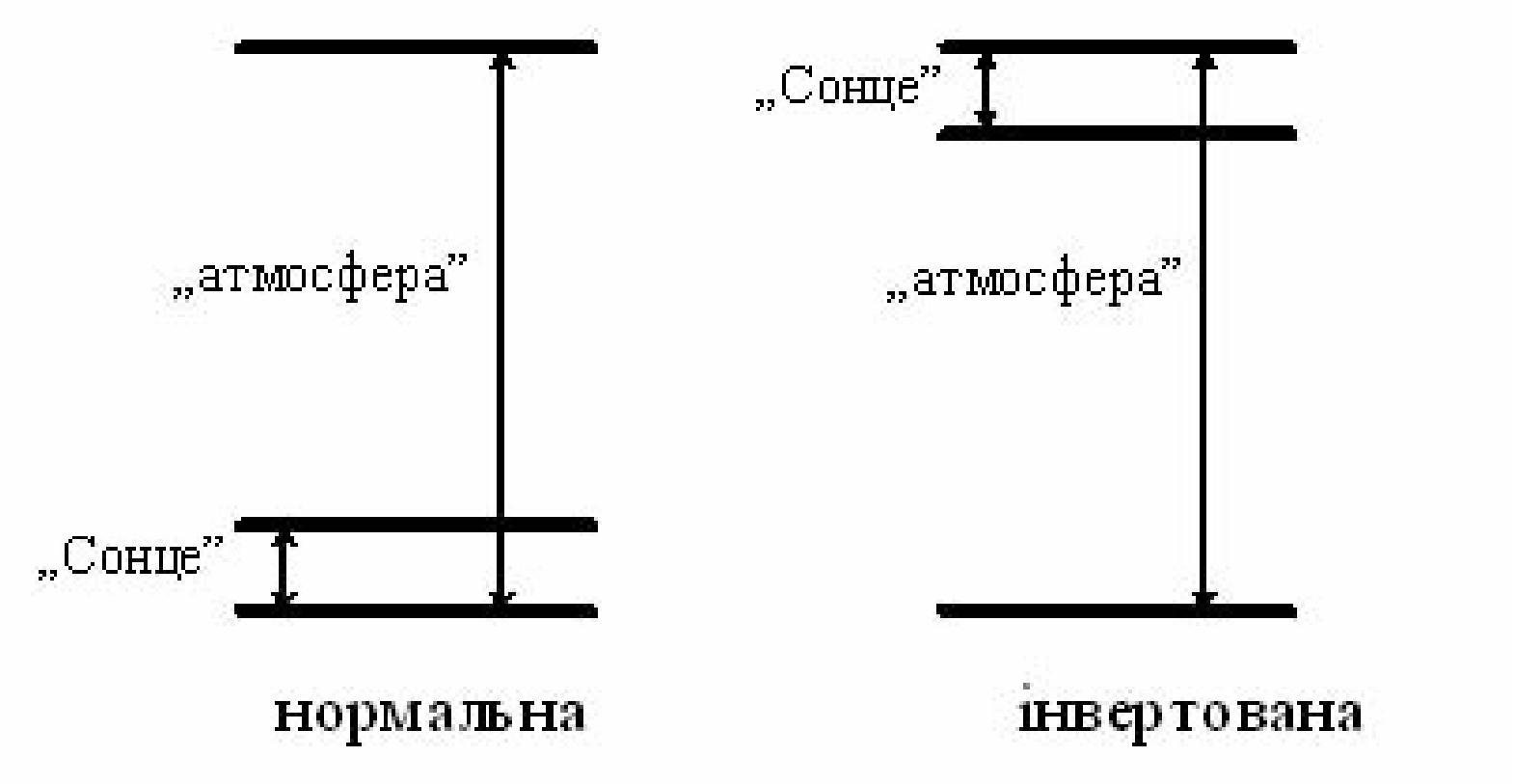

Рис. 1.4. Дві з трьох можливих схем масових станів нейтрино: нормальна та інвертована. У випадку нормальної схеми осциляції, що спостерігаються в потоці сонячних нейтрино, відбуваються між найлегшими масовими станами нейтрино. Осциляції, що спостерігаються в потоках атмосферних нейтрино, відбуваються між найлегшим і найважчим масовими станами нейтрино.

Спостереження безнейтринного 2β–розпаду дасть дуже важливу інформацію про фундаментальні властивості нейтрино: воно буде свідчити про те, що лептонний заряд не зберігається, а також що нейтрино є майоранівською частинкою. Але важливо також і те, що спостереження 0ν2β–розпаду дозволить виміряти ефективну масу нейтрино і встановити, яка із схем масових станів нейтрино має місце в природі. Як видно з рис. 1.3, щоб відрізнити „вироджену” схему від „інвертованої”, необхідно підняти чутливість експериментів до рівня $\langle m_\nu \rangle \sim 50$ меВ. Причому, навіть неспостереження процесу 0ν2β–розпаду, а встановлення обмеження на величину $\langle m_\nu \rangle$ на такому рівні, дасть можливість визначити, яка з цих схем реалізується в природі. Щоб відрізнити „інвертовану” схему від „нормальної”, чутливість експериментів необхідно буде підняти до рівня $\langle m_\nu \rangle \sim 10$ меВ. Як показує аналіз, виконаний в роботі [ 133 ], ефективна маса нейтрино буде близькою до $\langle m_\nu \rangle \approx 12$ меВ у випадку „інвертованої” схеми і $\langle m_\nu \rangle \approx 4$ меВ в „нормальній” схемі мас нейтрино. Як буде показано в розділі 7, досягнення такої чутливості в експериментах по пошуку 2β–розпаду є складною, але не неможливою задачею.

## 1.2. Сучасний стан експериментальних досліджень 2β–розпаду

Перший експеримент по пошуку 2β–розпаду був поставлений в 1948 році E. Fireman [ 134 ]. Дослід був спрямований на спостереження подвійного бета–розпаду в ядрах $^{94}$Zr і $^{124}$Sn за допомогою лічильників Гейгера включених в режимі збігів. Порівнюючи швидкість ліку детектора з ізотопами і із зразками стабільних до будь–якого виду активності матеріалів, автор встановив, що час життя $^{124}$Sn відносно 2β–розпаду перевищує $3 \times 10^{15}$ років.

З того часу було зроблено багато спроб зареєструвати це вкрай рідкісне явище. Об‘єктами сотень витончених експериментів стали близько півсотні ядер. В результаті цих зусиль були виміряні періоди напіврозпаду ряду ядер ($^{48}$Ca, $^{76}$Ge, $^{82}$Se, $^{96}$Zr, $^{100}$Mo, $^{116}$Cd, $^{128}$Te, $^{130}$Te, $^{150}$Nd, $^{238}$U) відносно 2ν2β–розпаду і встановлені нижні межі на періоди напіврозпаду відносно різних мод і каналів безнейтринного подвійного бета–розпаду (див. огляди Третяка і Здесенка [20,22], де наведені всі результати найбільш чутливих експериментів, опублікованих до квітня 2001 року). В найбільш чутливих прямих експериментах з ядрами $^{48}$Ca, $^{82}$Se, $^{100}$Mo, $^{116}$Cd, $^{130}$Te, $^{136}$Xe одержані жорсткі обмеження на період напіврозпаду, що перевищують $T_{1/2} \geq 10^{23}$ років, а в досліді з $^{76}$Ge досягнуто чутливості $T_{1/2} \geq 10^{25}$ років. Порівняння цих результатів з теоретичними розрахунками дозволяють обмежити ефективну масу нейтрино майоранівської природи на рівні $\langle m_\nu \rangle \leq 0.3$–2 еВ, в той час як з „осциляційних" експериментів слідує необхідність підвищити чутливість експериментів по пошуку 0ν2β–розпаду принаймні до 0.1 – 0.05 еВ.

Зараз обговорюються проекти експериментів з чутливістю навіть $T_{1/2} \geq 10^{28}$ років (що відповідає масі нейтрино ≈ 0.015 еВ), але зрозуміло, що реальною метою експериментаторів в наступному десятиріччі є досягнення „планки" $T_{1/2} \geq 10^{26}$ років, що дозволить наблизитись до рівня чутливості $\langle m_\nu \rangle \sim 0.1$–0.05 еВ . В той же час зрозуміло, що в разі, якщо масові стани нейтрино відповідають нормальній схемі ($m_1 < m_2 < m_3$), ефективна маса нейтрино, що проявляється в процесі 0ν2β–розпаду, може бути настільки малою (кілька меВ, період напіврозпаду $T_{1/2} \sim 10^{29} - 10^{30}$ років), що навряд чи той, хто її буде спостерігати, вже народився.

Поряд з тим, не треба забувати про наполегливі повідомлення групи з Гейдельбергу, в яких стверджується, що 0ν2β–розпад $^{76}$Ge спостерігається [135]. Цей результат було піддано жорсткій критиці в ряді робіт, але нещодавно група з Гейдельбергу опублікувала дані експерименту після повторної обробки та з більшою на 30 % статистикою і продовжує наполягати на спостереженні 0ν2β–розпаду $^{76}$Ge з $T_{1/2} \approx 1.2\times10^{25}$ років, звідки оцінена ефективна маса електронного нейтрино $\langle m_\nu \rangle \sim 0.4$ еВ [ 136 , 137 ]. Не дивлячись на скептичне ставлення до цього результату наукової спільноти, лише експеримент з більшою чутливістю може його спростувати або підтвердити.

Зараз ведеться інтенсивна робота сотень фізиків, спрямована на підвищення чутливості 2β–експериментів, на розробку нових експериментальних методів, розширення переліку досліджуваних ядер. Кожний наступний крок робиться, як правило, виходячи з досвіду вже проведених експериментів. Розглянемо сучасну експериментальну ситуацію в дослідженнях 2β–розпаду атомних ядер.

### 1.2.1. Класифікація експериментальних методів.

Експерименти, спрямовані на пошук і вимірювання характеристик подвійного бета–розпаду, можна поділити на два основних типи: методи прямого детектування ефекту і детектування продуктів розпаду. В свою чергу, в групі експериментів другого типу логічно окремо розглянути геохімічні та радіохімічні досліди. В той же час методи прямого детектування ефекту класифікують в залежності від того, в якому вигляді використовується досліджуваний ізотоп: у вигляді зразка досліджуваного матеріалу (пасивне джерело, “passive source”), або він входить в склад детектора (активне джерело, “active source”). Досліди другого типу ще називають калориметричними (calorimetric), оскільки вся енергія частинок, що випромінюються в 2β–розпаді, майже без втрат поглинається в детекторі.

#### 1.2.1.1 Геохімічні експерименти.

В основі геохімічного методу лежить процес накопичення в деяких мінералах за значні проміжки часу продуктів 2β–розпаду. Таким чином спостерігається надмірна, в порівнянні зі звичайною поширеністю, кількість дочірніх нуклідів, яка вимірюється за допомогою масс-спектрометричних методів. Завдяки дуже великому часу накопичення речовини (мільярди років) такі експерименти дуже чутливі. Імовірність розпаду $\lambda_{\beta\beta}$ обраховується з віку породи $T$ і виміряних поширеностей материнських $N(Z, A)$ і дочірніх $N(Z \pm 2, A)$ ядер з врахуванням експоненційного закону розпаду:

$$\lambda_{\beta\beta} \cong \frac{N(Z \pm 2, A)}{N(Z, A)} \cdot \frac{1}{T} \qquad (1.30)$$

Головний недолік цього методу полягає в тому, що детектуються лише продукти розпаду, при цьому канал і мода розпаду залишається невідомим. Двохнейтринну та безнейтринну моди розпаду неможливо відрізнити і $\lambda_{\beta\beta}$ визначає лише повну імовірність розпаду, тобто суму імовірностей двох мод:

$$\lambda_{\beta\beta}=\lambda_{2\nu}+\lambda_{0\nu} \quad (1.31)$$

Тому чутливість таких експериментів до 0ν2β–розпаду обмежується періодом напіврозпаду відносно 2ν2β–розпаду.

Крім того, лише ізотопи окремих елементів можуть бути досліджені за допомогою геохімічних експериментів. Це пов'язано з тим, що використовувані породи повинні задовольняти певним геологічним і хімічним вимогам. По-перше, мінерал повинен містити досліджуваний нуклід у великій концентрації. Крім того, необхідно бути впевненим в тому, що концентрація продукту розпаду змінювалась за рахунок лише 2β–переходів, а не завдяки іншим ефектам, таким як, наприклад, випаровування легких продуктів розпаду. До того ж вік породи необхідно надійно визначити по його геологічному оточенню.

Табл. 1.1. Періоди напіврозпаду відносно 2β–розпаду, отримані в геохімічних експериментах.

| Ізотоп | $T_{1/2}$, років | Посилка |
|---|---|---|
| $^{82}Se$ | $= (2.8\pm0.9)\times10^{20}$ | [138] |
| | $= (9.7^{+3.6}_{-4.5})\times10^{19}$ | [139] |
| | $< (1.7\pm0.2)\times10^{20}$ | [140] |
| | $= (1.0^{+0.3}_{-0.4})\times10^{20}$ | [141] |
| | $= (1.2\pm0.1)\times10^{20}$ | [142] |
| $^{128}Te$ | $\geq 8.0 \times 10^{24}$<br>$= (1.8 \pm 0.7) \times 10^{24}$<br>$= (7.7 \pm 0.4) \times 10^{24}$<br>$= (2.2 \pm 0.4) \times 10^{24}$ | [143]<br>[144]<br>[145]<br>[146] |
| $^{130}Te$ | $= (2.6 \pm 0.3) \times 10^{21}$<br>$= (1.0 \pm 0.3) \times 10^{21}$<br>$\leq (1.3 \pm 0.1) \times 10^{21}$<br>$= (7.5 \pm 0.3) \times 10^{20}$<br>$= (2.7 \pm 0.1) \times 10^{21}$<br>$= (7.9 \pm 1.0) \times 10^{20}$ | [142]<br>[147]<br>[140]<br>[142]<br>[145]<br>[146] |
| $^{96}Zr$ | $= (3.9 \pm 0.9) \times 10^{19}$<br>$= (9.4 \pm 3.2) \times 10^{18}$ | [148]<br>[149] |
| $^{130}Ba$ | $> 4.0 \times 10^{21}$<br>$= (2.2 \pm 0.5) \times 10^{21}$ | [150]<br>[151] |
| $^{132}Ba$ | $> 3.0 \times 10^{20}$<br>$= (1.3 \pm 0.9) \times 10^{21}$ | [150]<br>[151] |

В результаті таких обмежень використання геохімічного методу довгий час було практично можливе лише для селенових і телурових руд. В обох випадках дочірні речовини – хімічні інертні гази криптон і ксенон, початкова поширеність яких у мінералах дуже мала. Висока чутливість, якої можна досягнути при вивченні цих газів за допомогою мас–спектрометра, дозволяє зареєструвати мізерний надлишок дочірніх ядер, накопичених за тривалий геологічний період. Лише нещодавно з’явилися результати геохімічних експериментів, в яких досліджувався 2β–розпад ізотопів цирконію та барію. Результати найчутливіших геохімічних дослідів, здійснених за останні 25 років, подані в табл. 1.1 (більшість даних взята з огляду В.І. Третяка і Ю.Г. Здесенка [22]). Треба відмітити, що вперше 2β–розпад спостерігався саме в геохімічному досліді [ 152 ]. За допомогою мас–спектрометричного аналізу телурової руди віком 1.5±0.5 мільярдів років автори визначили період напіврозпаду ядра $^{130}$Te: $T_{1/2} = 1.4 \times 10^{21}$ років.

Хоча геохімічні експерименти і не спроможні дати пряму інформацію про канал та моду розпаду, із відношення імовірностей розпадів двох ізотопів телуру випливає, що спостережуване значення періоду напіврозпаду $^{130}$Te визначається 2ν–модою розпаду [153,24]. Цей результат оснований головним чином на різниці в енергетичній залежності інтегралів по фазовому простору ($G^{2\nu} \sim Q_{\beta\beta}^{11}$, $G^{0\nu} \sim Q_{\beta\beta}^{5}$). Для 0ν2β–розпаду з геохімічних експериментів можна отримати лише модельно залежну інформацію про верхню межу імовірності розпаду і практично неможливо отримати достатньо надійне свідоцтво існування 0ν2β–розпаду.

1.2.1.2. Радіохімічний метод.

Радіохімічні експерименти також основані на пошуку в досліджуваному матеріалі надлишку дочірніх ядер. Після ретельного очищення досліджуваної речовини, за декілька років накопичуються продукти розпаду, які потім хімічними методами екстрагуються із зразка. Важливо, щоб продукти розпаду самі були радіоактивними. Завдяки цьому може бути досягнута значно вища чутливість до реєстрації малої кількості атомів, ніж в мас–спектрометричних методах. Радіохімічні експерименти не потребують тривалого (геохімічного) накопичування речовини, історія зразка добре відома, а отже в цьому методі не має невизначеностей, пов’язаних з віком гірської породи, з початковою концентрацією дочірніх нуклідів та з можливими ефектами розсіяння або накопичення продуктів 2β–розпаду за тривалий геологічний період. Очевидними недоліками методу є обмежений перелік ізотопів, які можуть бути досліджені, а також неможливість визначити канал і моду 2β–розпаду. Підходящими кандидатами для радіохімічного пошуку 2β–розпаду є ядра $^{232}$Th ($Q_{\beta\beta} = 0.85$ МеВ) та $^{238}$U ($Q_{\beta\beta} = 1.15$ МеВ).

Вимірювання для ядра урану дали значення періоду напіврозпаду $T_{1/2}$ = (2.0±0.6)×$10^{21}$ років [154].

1.2.1.3. Методи прямого детектування ефекту.

Більшість експериментів результатів про 2β–розпад були одержані в експериментах, де безпосередньо детектувались події подвійного бета–розпаду. Важливо, що 2β–експерименти, в яких детектуються події 2β–розпаду, дають можливість відрізнити безнейтринну моду подвійного бета–розпаду від двохнейтринної. Такі експерименти, в яких реєструються події 2β–розпаду, можна в свою чергу, розділити на досліди з „пасивними” джерелами, в яких досліджуваний ізотоп розміщується у вигляді фольги всередині детектуючої системи, та на досліди з „активними” джерелами (їх ще називають калориметричними), коли ядра досліджуваного ізотопу безпосередньо входять до складу детектора.

Важливою перевагою детекторів з пасивним джерелом є можливість отримати інформацію про кінематику випромінюваних електронів: енергію частинок, точку, з якої випромінюються частинки, та треки, по яким вони рухаються, заряд (за кривизною треків в магнітному полі). Перевагою методу є можливість дослідження широкого кола ізотопів, оскільки єдиною вимогою є виготовлення досить тонких (десятки мг / см$^2$) фольг. Головним недоліком методу є низька ефективність реєстрації (≈ 10–30%), а також погіршення енергетичної роздільної здатності через втрати енергії в зразку. Крім того, необхідність розміщення в чутливому об’ємі детектора сотень кілограмів ізотопу вимагає створення дуже великих за розмірами установок.

В дослідах з активним джерелом вимірюється лише повне енерговиділення в детекторі. Вирішальною перевагою цієї методики є близька до 100% ефективність реєстрації ефекту, оскільки вся енергія β–частинок, що випромінюються в 2β–розпаді, без втрат поглинається в детекторі. Недоліком методу є неможливість довести, що спостережувана подія є дійсно двохелектронною.

Чутливість експерименту, метою якого є пошук 0ν2β–розпаду, згідно [2] визначається формулою:

$$T_{1/2} \sim \varepsilon \cdot \delta \cdot \sqrt{\frac{m \cdot t}{R \cdot Bg}}, \tag{1.32}$$

де $T_{1/2}$ – період напіврозпаду відносно 0ν2β–розпаду, $\varepsilon$ – ефективність реєстрації піка 0ν2β–розпаду, $\delta$ – збагачення детектора ядрами–кандидатами на 2β–розпад, $t$ – час вимірювань, $m$ та $R$ – загальна маса та енергетична роздільна здатність детектора, відповідно, $Bg$ – кількість відліків фону в області очікуваного піку.

З формули 1.32 слідують певні вимоги до ізотопів, з якими можна досягнути високої чутливості до періоду напіврозпаду відносно 0ν2β–розпаду. Важливими характеристиками детектуючої установки є ефективність реєстрації 2β–розпаду і збагачення ядрами–кандидатами на 2β–розпад, оскільки всі інші величини входять у вираз 1.32 під квадратним коренем. Умова $\varepsilon \approx 100\%$, як уже зазначалося, реалізується лише в калориметричних експериментах. Але ж для реалізації такого експерименту необхідно, щоб ядра певного елемента входили до складу детектора. В ідеальному випадку детектор має складатися лише з атомів досліджуваного ізотопу.

Рівень фону (який в ідеальному експерименті має бути рівним нулю) визначається багатьма факторами. Забрудненість детектора радіонуклідами (в першу чергу, ураном, торієм, калієм) відіграє важливу роль після того, як детектор захищений від випромінювання оточуючого середовища, матеріали захисту ретельно відібрані, а установка розміщена під землею. З огляду на рівень фону важливо, щоб енергія переходу була якнайбільшою, оскільки фон, як правило, зменшується з ростом енергії. Для ефективного придушення фону бажано, щоб енергія 2β–розпаду ($Q_{\beta\beta}$) перевищувала межу природної радіоактивності (інтенсивна γ–лінія $^{208}Tl$ з енергією 2615 кеВ). Ця умова виконується лише для $^{48}Ca$,

$^{82}$Se, $^{96}$Zr, $^{100}$Mo, $^{116}$Cd і $^{150}$Nd. Всього для чотирьох ізотопів з цього списку, а саме для $^{82}$Se, $^{100}$Mo, $^{116}$Cd і $^{150}$Nd вміст в природній суміші ізотопів перевищує 5%. Лише $^{130}$Te і $^{160}$Gd мають досить високу поширеність в природі ($\approx 34\%$ та $\approx 22\%$ відповідно), що є дуже сприятливим фактором, оскільки дозволяє здійснити експеримент з великою кількістю ядер без використання надзвичайно дорогих збагачених ізотопів.

Важливою характеристикою установки є також енергетична роздільна здатність. Значення роздільної здатності стає критичним, коли чутливість експерименту на порядки має перевищувати період напіврозпаду досліджуваного ізотопу відносно двохнейтринної моди 2β–розпаду. Як буде показано в розділі 7, при низькій роздільній здатності детектора, високоенергетичні електрони, що випромінюються в 2ν2β–розпаді, вносять фонові події в очікуваний пік 0ν2β–розпаду і їх неможливо ніяким чином відкинути.

### 1.2.2 Найбільш чутливі 2β–експерименти.

Вперше в прямому досліді 2β–розпад був спостережений М. Мое та його співробітниками в 1987 році за допомогою часово–проекційної камери [155], в якій було поміщено 14 г збагаченого до 97% ізотопу $^{82}$Se у вигляді фольги товщиною 7 мг/см$^2$. Камера була захищена від зовнішнього фону шаром свинцю товщиною 10–15 см. Свинцевий захист був оточений з усіх сторін лічильниками Гейгера, які виробляли сигнали антизбігів при проходженні космічних мюонів. Дослідники виміряли форму спектру і встановили, що спостерігають 2β–розпад $^{82}$Se з періодом напіврозпаду $T_{1/2} = (1.1\ ^{+0.8}_{-0.3}) \times 10^{20}$ років, який супроводжується вильотом нейтрино. Експеримент [155] був здійснений якраз за методикою пасивного джерела.

З того часу в лабораторних експериментах було зареєстровано двохнейтринну моду 2β–розпаду в ядрах $^{48}$Ca, $^{76}$Ge, $^{82}$Se, $^{96}$Zr, $^{100}$Mo, $^{116}$Cd, $^{150}$Nd (див. таблицю 1.2). Зокрема, період напіврозпаду ядра $^{116}$Cd був виміряний в експериментах ELEGANT [156], NEMO–II [157] і NEMO–III [158], а також у досліді, що проводився в Солотвинській підземній лабораторії ІЯД НАНУ з використанням сцинтиляційних кристалів вольфрамату кадмію, збагачених ізотопом $^{116}$Cd [159,160,161].

Табл. 1.2. Періоди напіврозпаду відносно 2ν2β–розпаду виміряні в прямих експериментах.

| Ізотоп | $T_{1/2}$, років | Посилка |
|---|---|---|
| $^{48}$Ca | $= (4.3^{+2.8}_{-1.8}) \times 10^{19}$ | [162] |
| | $= (4.2^{+3.3}_{-1.3}) \times 10^{19}$ | [163] |
| $^{76}$Ge | $=(9.0 \pm 1.0) \times 10^{20}$ | [164] |
| | $= (1.1^{+0.6}_{-0.3}) \times 10^{21}$ | [165] |
| | $=(8.4^{+1.0}_{-0.8}) \times 10^{20}$ | [166] |
| | $= (1.1 \pm 0.2) \times 10^{21}$ | [167] |
| | $= (1.8 \pm 0.1) \times 10^{21}$ | [168] |
| $^{82}$Se | $= (1.1^{+0.3}_{-0.1}) \times 10^{20}$ | [169] |
| | $= (8.3 \pm 1.2) \times 10^{19}$ | [170] |
| $^{96}$Zr | $= (2.1^{+0.8}_{-0.4}) \times 10^{19}$ | [171] |
| $^{100}$Mo | $= (1.2^{+0.5}_{-0.3}) \times 10^{19}$ | [172] |
| | $= (9.5 \pm 1.0) \times 10^{18}$ | [173] |
| | $= (7.6^{+2.2}_{-1.4}) \times 10^{18}$ | [174] |
| | $= (6.8^{+0.8}_{-0.9}) \times 10^{18}$ | [175] |
| | $=[7.2 \pm 0.9(\text{стат.}) \pm 1.8(\text{сист.})] \times 10^{18}$ | [176] |
| $^{116}$Cd | $= (2.6^{+0.9}_{-0.5}) \times 10^{19}$ | [156] |
| | $= (2.7^{+1.0}_{-0.7}) \times 10^{19}$ | [159] |
| | $= (3.8 \pm 0.4) \times 10^{19}$ | [158] |
| | $= (2.6^{+0.7}_{-0.4}) \times 10^{19}$<br>$= (2.9^{+0.4}_{-0.3}) \times 10^{19}$ | [160]<br>[161] |
| $^{150}$Nd | $= (1.9^{+0.7}_{-0.4}) \times 10^{19}$<br>$= (6.8 \pm 0.8) \times 10^{18}$ | [177]<br>[175] |

Не зважаючи на численні, починаючи з 1948 року, спроби зареєструвати безнейтринний 2β–розпад, він і досі залишається не спостережуваним. Найвищі границі на період напіврозпаду були встановлені у прямих експериментах з кількома нуклідами: $T_{1/2} > 10^{21}$ років для $^{96}$Zr, $^{160}$Gd, $^{150}$Nd, $^{186}$W; $T_{1/2} > 10^{22}$ років для $^{48}$Ca, $T_{1/2} > 10^{23}$ років для $^{82}$Se, $^{100}$Mo, $^{116}$Cd, $^{128}$Te, $^{130}$Te, $^{136}$Xe і $T_{1/2} > 10^{25}$ років для $^{76}$Ge. Ці експерименти дозволили встановити найбільш жорсткі обмеження на значення маси майоранівського нейтрино ($m_\nu \leq 0.3 \div 5$ еВ), домішки правих струмів у слабкій взаємодії ($\eta \approx 10^{-7}$, $\lambda \approx 10^{-5}$), константу зв'язку нейтрино з майороном ($g_\chi \approx 10^{-4}$) і параметр порушення R–парності мінімальної суперсиметричної СМ ($\xi \approx 10^{-4}$). Подвійний бета–розпад ядер $^{116}$Cd, $^{160}$Gd і $^{186}$W вивчався в Солотвинській підземній лабораторії ІЯД НАНУ і відповідні експерименти будуть описані в розділах 3 і 4. Нижче будуть описані найбільш чутливі експерименти по пошуку 0ν2β–розпаду ядер $^{48}$Ca, $^{76}$Ge, $^{82}$Se і $^{96}$Zr, $^{100}$Mo, $^{130}$Te, $^{136}$Xe, $^{150}$Nd. Стан експериментальних досліджень процесів 2β–розпаду описаний в багатьох оглядах, зокрема в роботах [178,179,180,20,22,181,3,182,4,183,5].

$^{48}$Ca.

Активований європієм сцинтиляційний детектор фториду кальцію $CaF_2$(Eu) розмірами ∅0.8 × 0.7 дюймів, виготовлений із збагаченого до 96.59% ізотопу $^{48}$Ca, був застосований в одному з найперших 2β–експериментів [184]. Всього в кристалі було 10.6 г ізотопу $^{48}$Ca. В якості активного захисту був використаний пластиковий сцинтилятор, пасивний захист був виготовлений із свинцю. В досліді було встановлене обмеження на 0ν2β–розпад $^{48}Ca$ на рівні $T_{1/2} > 2 \times 10^{20}$ років.

Незбагачені кристали $CaF_2$ загальною масою 28.7 кг були застосовані в експерименті [185]. В якості детекторів активного захисту були застосовані пластикові сцинтилятори товщиною 2.5 – 5 см. Пасивний захист складався із сталі (2 см) і свинцю (8 – 10 см). Експеримент проводився впродовж 7588 годин у вугільній шахті на глибині 1300 м.в.е. Автори отримали аномально великий коефіцієнт придушення фону активним захистом і дали верхню межу періоду напіврозпаду $^{48}Ca$ відносно 0ν2β–розпаду $T_{1/2} > 9.5 \times 10^{21}$ років з довірчою ймовірністю 76%, звідки було оцінено нижню межу на масу нейтрино $m_{\nu} < 8.3$ еВ. Цей результат вже був підданий критиці в роботах [186,22], де такий низький рівень фону пояснюється некоректним врахуванням мертвого часу, пов'язаного сигналами антизбігів основного детектора з сигналами активного захисту.

В роботі [163] пошук 2β–розпаду $^{48}Ca$ здійснювався за допомогою методики пасивного джерела. В експерименті використовувався всього 1 грам збагаченого ізотопу $^{48}Ca$, який у вигляді зразка $^{48}CaCO_3$ товщиною 42 мг/см$^3$ був розміщений між планарними германієвими детекторами установки TGV [187]. Експеримент був проведений в підземній лабораторії Modane у Франції на глибині 4800 м.в.е. В результаті близько 8700 годин вимірювань було встановлене обмеження на період напіврозпаду $^{48}Ca$ на рівні $T_{1/2} > 1.5 \times 10^{21}$ років з 90% CL.

В недавній роботі [188,189] були застосовані сцинтиляційні кристали $CaF_2$, встановлені в установку ELEGANT VI. Експеримент проводився в підземній лабораторії Oto Cosmo Observatory. З аналізу даних експозиції 4.23 кг × років автори отримали обмеження на 0ν2β–розпад $^{48}Ca$ на рівні $T_{1/2} > 1.4 \times 10^{22}$ років з 90% CL. В досліді досягнутий дуже низький рівень фону в околі очікуваного піку 0ν2β–розпаду. В таблиці 1.3 приведені рівні фону, виміряні із сцинтиляційними детекторами $CaF_2$ та $CaWO_4$ в різних експериментах. Не зважаючи на порівняно високий рівень радіоактивної забрудненості цих кристалів, рівень фону, досягнутий в околі піка 0ν2β–розпаду $^{48}Ca$, знаходиться на рівні фону германієвих детекторів. Це є ілюстрацією того, як велика енергія розпаду дозволяє досягнути низького рівня фону.

Таблиця 1.3. Фон, виміряний із сцинтиляційними детекторами $CaF_2$ та $CaWO_4$ в різних експериментах.

| Експеримент | Радіоактивний вміст кристалів, мБк/кг | | Фон (відліків / рік / кеВ / кг) в енергетичному інтервалі | |
|---|---|---|---|---|
| | $^{226}$Ra | $^{228}$Th | 2.8 – 3.0 МеВ | 4 – 5 МеВ |
| $CaF_2$(Eu) 4.23 кг × рік [189] | 1.11(1) | 0.098(2) | 0.5 | 0.00047 |
| $CaF_2$(Eu) 0.046 кг × років [190] | 2.4(10) | 1.2(5) | 5–8 | |
| $CaF_2$ 28.7 кг × років [185] | | | 0.8 | 0.02 |
| $CaWO_4$ 0.037 кг × років [191] | 5.6(5) | 0.6(2) | 6.4 | 0.11 |

$^{76}$Ge.

Напівпровідникові германієві детектори, завдяки високій роздільній здатності і надзвичайно низькому вмісту радіонуклідів в кристалах германію, успішно використовуються для пошуку 2β–розпаду ядра $^{76}$Ge. Подвійний бета–розпад ізотопу $^{76}$Ge на рівні чутливості до періоду напіврозпаду $T_{1/2} > 10^{25}$ років досліджувався двома колабораціями: Heidelberg–Moscow та IGEX.

В експерименті IGEX було використано 3 детектори з надчистого германію (HPGe по 2 кг кожен, збагачення $^{76}$Ge до ≈ 88%), встановлених у підземній лабораторії Канфранк (Іспанія). Захист складався з 2.5 тон археологічного свинцю і 10 тон 70–річного свинцю з низькою активністю, а також із пластикових сцинтиляторів для захисту від космічних мюонів. Для аналізу експериментальних даних застосовувались методики дискримінації за формою імпульсу. Рівень фону складав ≈ 0.06 відліків/(рік × кг × кеВ) в енергетичному інтервалі 2.0 – 2.5 МеВ. Енергетична роздільна здатність для очікуваного піку від 0ν2β–розпаду ($Q_{\beta\beta}$ = 2038.5 кеВ) була оцінена в ході енергетичних калібровок і становила 4 кеВ [192]. Аналіз даних із загальною експозицією 116.75 моль × рік (чи 8.87 кг × рік для $^{76}$Ge) дозволив встановити нижню межу $T_{1/2}^{0\nu} \geq 1.57 \times 10^{25}$ років при 90% CL [193].

Експеримент по пошуку 2β–розпаду $^{76}$Ge, що був здійснений колабораціє Heidelberg–Moscow, проводився в підземній лабораторії Гран Сассо (Італія). Використовувалося 5 HPGe детекторів (збагачення $^{76}$Ge до 86%) із загальною активною масою 10.96 кг. Пасивний та активний захист, а також аналіз даних за формою імпульсу забезпечили зменшення фону до ≈ 0.06 відліків/(рік × кг × кеВ). Енергетична роздільна здатність на енергії 2038.5 кеВ становила 3.9 кеВ. Проміжні результати експерименту опубліковані, зокрема, в роботах [194,195,196]. З аналізу даних експозиції 35.5 кг × рік було встановлено нижню межу періоду напіврозпаду ядра $^{76}$Ge відносно безнейтринного подвійного бета–розпаду $T_{1/2}^{0\nu} \geq 1.9\ (3.1) \times 10^{25}$ років при 90% (68%) CL [197]. Використовуючи розрахунки матричних елементів [198], автори встановили обмеження на масу нейтрино $\langle m_\nu \rangle \leq 0.35\ (0.27)$ еВ. Крім того, в цьому експерименті були встановлені жорсткі обмеження на 0ν2β–розпад з вильотом синглетного майорона [199] $T_{1/2}^{0\nu2\beta M} \geq 1.66\ (1.99) \times 10^{22}$ років. Ці обмеження були покращені з використанням більшої статистики в роботі [200].

Згодом група з Гейдельбергу повідомила про реєстрацію 0ν2β–розпаду $^{76}$Ge

[135]. Це повідомлення піддано сумнівам в ряді робіт [107, 201 , 202 ], але H.V. Klapdor–Kleingrothaus і кілька його співавторів продовжують наполягати на тому, що $0\nu2\beta$–розпад $^{76}Ge$ спостерігається в експерименті [203]. Нещодавно група авторів під керівництвом H.V. Klapdor-Kleingrothaus опублікувала дані експерименту з більшою на 30% статистикою і продовжує наполягати на спостереженні $0\nu2\beta$–розпаду $^{76}Ge$ з $T_{1/2}^{0\nu} \approx 1.2\times10^{25}$ років, звідки оцінена ефективна маса електронного нейтрино $\langle m_\nu \rangle \approx 0.4$ еВ [136,137].

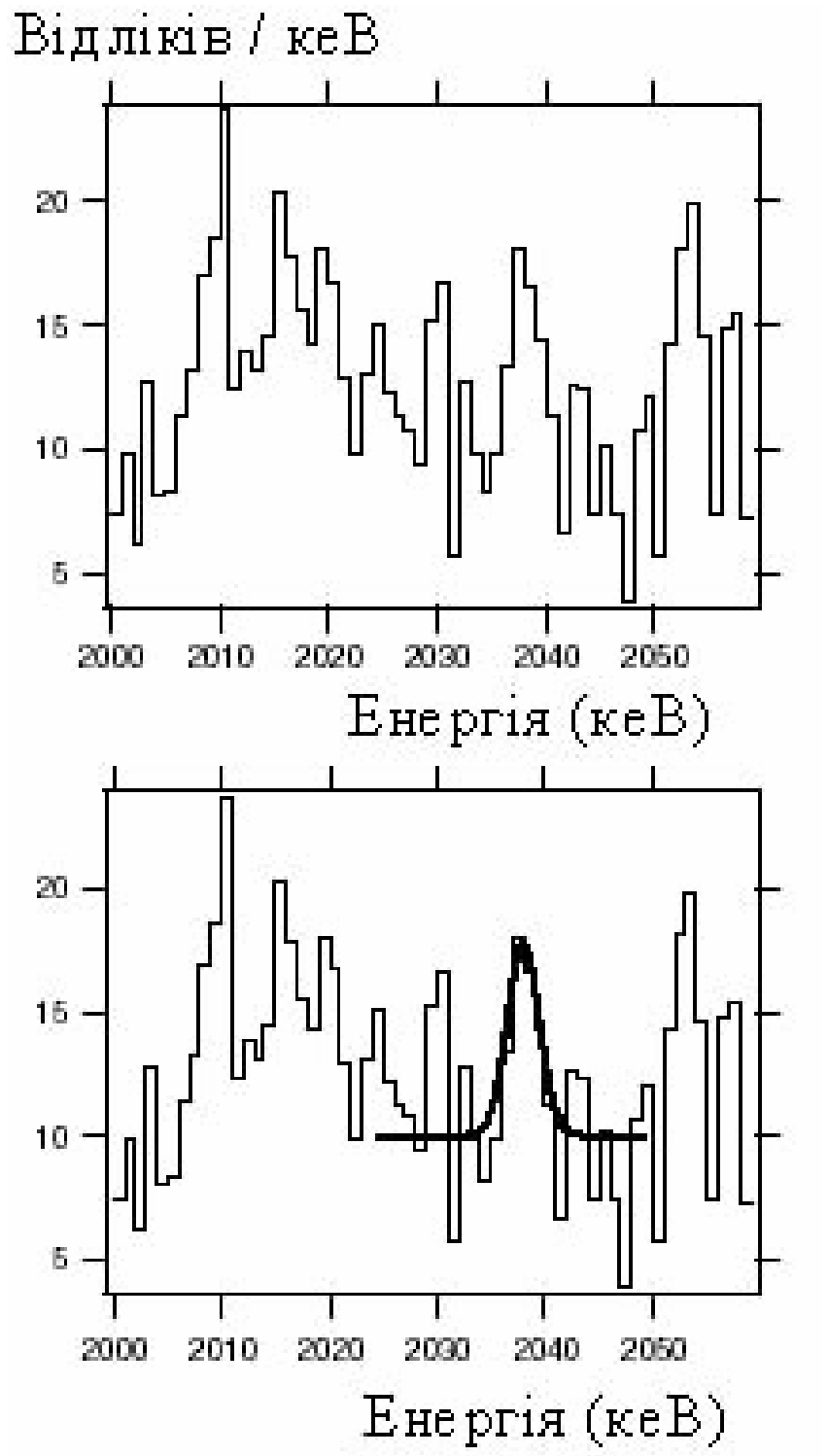

Рис. 1.5. Енергетичний спектр, накопичений в експерименті Heidelberg–Moscow з експозицією 71.7 кг × рік. Група з Гейдельбергу наполягає на тому, що в спектрі є пік від $0\nu2\beta$–розпаду $^{76}Ge$ з площею $28.8 \pm 6.9$ відліків [135,136] (Рисунок взятий з огляду [4]).

Частина учасників колаборації з Інституту атомної енергії ім. Курчатова з Москви опублікувала результати свого власного аналізу даних експерименту

Heidelberg–Moscow [204], в якому стверджує, що події в піку з енергією близько 2039 кеВ (котрі група з Гейдельбергу інтерпретує як безнейтринний 2β–розпад $^{76}$Ge) асоціюються з подіями нижче певного енергетичного порогу, що може свідчити про не зовсім коректну роботу детектора. Група з Гейдельбергу відповіла, що дані такого типу не були включені в аналіз [136]. Ефективна маса нейтрино $\langle m_\nu \rangle \approx 0.4$ еВ, яка слідує з заявленого в роботі [135,136,137] періоду напіврозпаду $^{76}$Ge, може відповідати виродженій (degenerate) схемі масових станів нейтрино, в якій найлегше нейтрино значно важче за різницю мас між масовими станами. Як вже було зазначено, лише експеримент з більшою чутливістю може або спростувати, або підтвердити повідомлення [135,136,137].

$^{82}$Se, $^{96}$Zr, $^{100}$Mo.

Подвійний бета–розпад ряду ядер, і зокрема, $^{82}$Se, $^{96}$Zr, $^{100}$Mo, досліджувався за допомогою детекторів NEMO 2 [ 205 ] і NEMO 3 [ 206 ]. Експерименти проводились в підземній лабораторії (Laboratoire de Souteran de Modane, LSM) у Франції. Глибина лабораторії становить 4800 м.в.е. Установка NEMO 3, яка була збудована замість детектора NEMO 2, зараз продовжує набір статистики.

Дослідження $^{96}$Zr [170] проводилось за допомогою детектора NEMO 2. Детектор являє собою ≈ 1 м$^3$ камеру з гейгерівськими лічильниками, наповнену гелієм, в якій розміщено зразок досліджуваного ізотопу у вигляді фольги. Гейгерівські лічильники (всього було 10 площин, кожна з яких складалась із 32 перпендикулярно розміщених лічильників) дозволяють реконструювати треки електронів, в той час як дві стінки пластикових сцинтиляторів, по 25 детекторів з кожної сторони, реєстрували енергію та час прильоту електронів. Точність визначення часу була достатньою для відокремлення за часом прольоту подій від двох електронів, що вилетіли з фольги, від електрона, що пролітав через увесь детектор. Пасивний захист детектора складався із міді (5 см) і заліза (20 см).

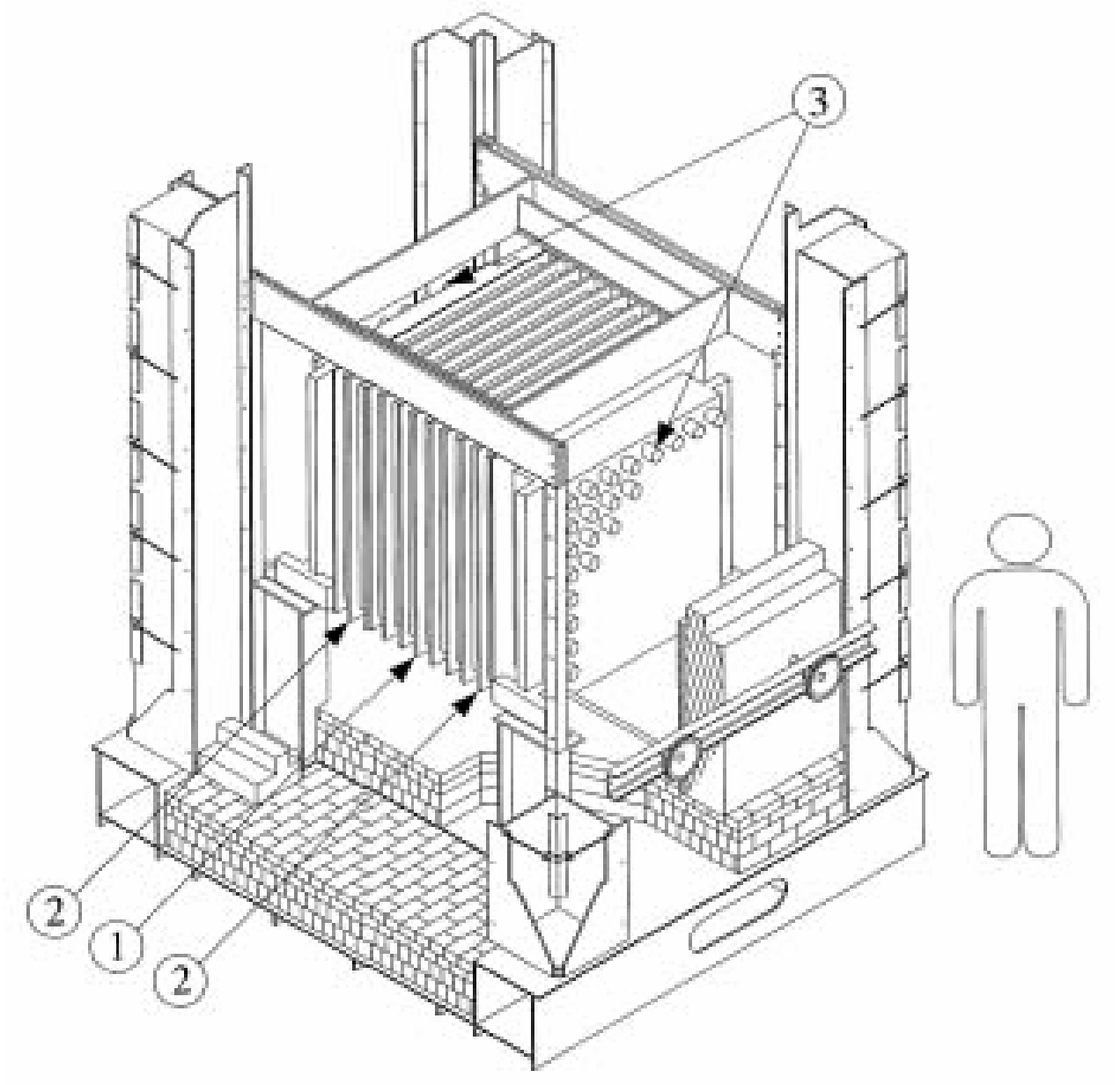

Рис. 1.6. Схематичний вигляд детектора NEMO 2. 1 – площина, в якій було розміщено зразок досліджуваного ізотопу; 2 – трекові об'єми з лічильниками Гейгера; 3 – дві площини пластикових сцинтиляторі по 25 детекторів в кожній.

Просторова роздільна здатність в поперечних напрямках становить 0.5 мм і 4.7 мм в повздовжньому напрямку. Енергетична роздільна здатність системи сцинтиляційних детекторів становила (FWHM) 18% на енергії 1 МеВ. Збагачення джерела цирконію ізотопом $^{96}$Zr становило 57%. Загальна маса $^{96}$Zr в збагаченому зразку становила 6.8 г. Крім збагаченого зразка, в детекторі була також встановлена фольга з цирконію з природним вмістом ізотопів. Товщина зразків складала 50 мг/см$^2$ ізотопно збагаченого і 45 мг/см$^2$ природного цирконію. Радіоактивна забрудненість збагаченого джерела становила 16 мБк/кг ($^{228}$Th), 11 мБк/кг ($^{226}$Ra), 1130 мБк/кг ($^{40}$K). Після 10 357 годин вимірювань в експерименті був виміряний період напіврозпаду $^{96}$Zr відносно 2ν2β–розпаду на основний стан $^{96}$Mo: $T_{1/2}^{2\nu} = (2.1^{+0.8}_{-0.4})\times10^{19}$ років та встановлене обмеження на 0ν2β–розпад на основний стан дочірнього ядра: $T_{1/2}^{0\nu} \geq 1.0 \times 10^{21}$ років з 90% C.L. Були також встановлені обмеження на переходи на збуджений рівень $2^+$ $^{96}$Mo ($T_{1/2}^{0\nu} \geq 3.9 \times 10^{20}$ років з 90% C.L) та розпад з випроміненням майорону ($T_{1/2}^{0\nu M} \geq 3.5 \times 10^{20}$ років з 90% C.L).

В 2003 р колаборація розпочала вимірювання на детекторі NEMO–3, який являє собою розвиток попередньої установки NEMO–2. В установці NEMO–3 треки електронів вимірюються 6180 дрейфовими лічильниками, а енергія частинок – за допомогою 1940 блоків пластикових сцинтиляторів. В детекторі у вигляді фольг встановлені ізотопи $^{100}$Mo (6 914 г), $^{82}$Se ( 932 г), $^{130}$Te (454 г), $^{116}$Cd (405 г), $^{150}$Nd (36.6 г), $^{96}$Zr (9.4 г), $^{48}$Ca (7 г). Товщина зразків ≈ 50 мг/см$^3$. Фольги з міді та телуру використовуються для оцінок фону. Роздільна здатність становить 8.8% на енергії 3 МеВ (15% на енергії 1 МеВ). Після 5797 годин вимірювань встановлені нові обмеження на період напіврозпаду ядер $^{82}$Se і $^{100}$Mo відносно 0ν2β–розпаду: $T_{1/2}^{0\nu} \geq 1.4{\times}10^{23}$ років і $T_{1/2}^{0\nu} \geq 3.1{\times}10^{23}$ років, відповідно [ 207 ], що дозволило встановити обмеження на масу нейтрино $\langle m_\nu \rangle \leq$ (0.8–1.2) еВ та $\langle m_\nu \rangle \leq$ (1.5–3.1) еВ, відповідно.

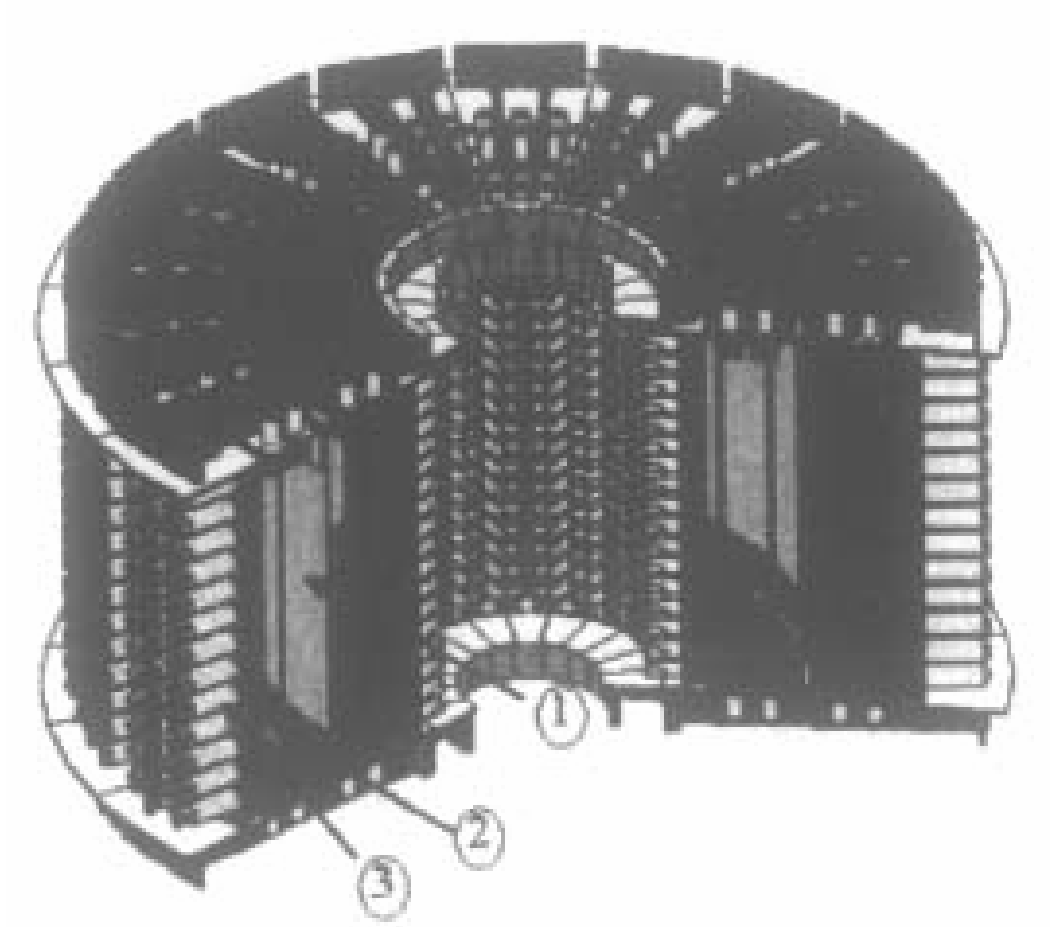

Рис. 1.7. Схематичний вигляд детектора NEMO 3. 1 – джерело у вигляді фольги; 2 – трекові об’єми з лічильниками Гейгера; 3 – пластикові сцинтилятори.

Дослідження 2β–розпаду $^{100}$Mo було також здійснене за допомогою спектрометра ELEGANT V (Університет міста Осака, Японія). Цей детектор складався з трьох дрейфових камер для вимірювання траєкторій обох електронів, що випромінюються при 2β–розпаді, пластикових сцинтиляторів для визначення моменту появи і енергії β–випромінювання та системи із сцинтиляційних кристалів NaI(Tl) для активного захисту та детектування γ–променів при переходах на збуджені стани дочірніх ядер. Два пасивних джерела $^{100}$Mo (збагачених до ≈ 95%) із товщинами 20 мг/см$^2$ та загальною масою 171 г були встановлені у центральній трековій камері. Радіоактивна забрудненість збагаченого джерела $^{100}$Mo становила 0.2 мБк/кг ($^{228}$Th), 8 мБк/кг ($^{226}$Ra). Після 7582 години вимірювань в обсерваторії Ото у поєднанні з попередніми результатами (7333 години у підземній лабораторії Каміока [ 208 ]) було досягнуто межі $T_{1/2}^{0\nu} \geq 5.5\,(10) \times 10^{22}$ років з довірчою імовірністю 90% (68%) CL [209]. З цього обмеження було отримано нижні межі на масу нейтрино та домішки правих струмів в слабкій взаємодії $\langle m_\nu \rangle \leq 2.1$ (1.5) еВ, $\langle \lambda \rangle \leq 3.2\ (2.5) \times 10^{-6}$, та $\langle \eta \rangle \leq 2.4\ (1.9) \times 10^{-8}$ з 90% (68%) CL.

Слід відмітити також експеримент по пошуку 0ν2β–розпаду $^{100}$Mo за допомогою 145 кремнієвих напівпровідникових детекторів [210,211]. Експеримент проводився в підземній лабораторії, влаштованій в срібній шахті Consolidated Solver mine Osburn в штаті Айдахо на глибині 3300 м.в.е. В експерименті було застосовано збагачений до 97% ізотоп $^{100}$Mo у вигляді фольги товщиною 34.4 мг/см$^2$. За допомогою цієї установки після набору експозиції 0.2664 моль × рік було встановлене обмеження на безнейтринний 2β–розпад $^{100}$Mo $T_{1/2}^{0\nu} \geq 2.2 \times 10^{22}$ років з 68% CL, а також виміряний період напіврозпаду ядра $^{100}$Mo відносно 2ν2β–розпаду: $T_{1/2}^{2\nu} = 0.76 \times 10^{19}$ років.

$^{130}$Te.

Важливою перевагою ядра $^{130}$Te є порівняно високий ізотопний вміст в природній суміші ізотопів (≈ 34%), що дозволяє обійтися без дуже дорогих збагачених ізотопів. Для пошуку 2β–розпаду $^{130}$Te використовуються кристали двооксиду телуру в якості кріогенних детекторів (болометрів). Принцип дії низькотемпературних болометрів оснований на дуже низькій теплоємності деяких кристалів при наднизьких температурах (~$10^{-2}$ К). Важливою умовою є діамагнітні властивості кристалів, оскільки при наявності парамагнітних властивостей кристал не вдається охолодити до необхідних температур. Крім того, бажаним є якнайбільше значення температури Дебая. Згідно закону Дебая, при низьких температурах теплоємкість *C* описується виразом: $C(T) \approx \left(\frac{T}{\theta_D}\right)^3$, де $\theta_D$ – температура Дебая. При Т $\to$ 0 теплоємкість така мала, що навіть енергія в кілька десятків еВ може бути виміряна. Енергія, *E*, яка виділяється в кристалі, викликає вимірюване зростання температури: $\Delta T \sim E / C(T)$. Теоретичні оцінки показують, що з кріогенними болометрами можна досягнути дуже високої роздільної здатності на рівні FWHM ~ 3 еВ для енергії рентгенівських променів 6 кеВ, в той час як теоретична межа для напівпровідникових детекторів становить ~ 110 еВ [212].

Дослідження 2β–розпаду $^{130}$Te проводиться групою під керівництвом проф. Е. Фіоріні вже багато років, починаючи з початку 80–х років [213,214,215]. Зараз експеримент проводиться з установкою “CUORICINO” [216,217], яка складається з 62 низькотемпературних болометрів з кристалами $TeO_2$ (44 кубічних з основою 5 см та 18 з об’ємом 3×3×6 см$^3$ кожен) загальною масою близько 40 кг. Кристали охолоджуються до температури ≈ 10 мК за допомогою холодильної установки в підземній лабораторії Гран Сассо (Італія). Ззовні і всередині кріостату розміщений низькорадіоактивний свинцевий захист із археологічного свинцю. Установка розміщена в герметичному контейнері з плексигласу, заповненому азотом для зменшення вмісту радону. Шар поліетилену встановлений для зменшення нейтронного фону. В області очікуваного піку від 0ν2β–розпаду (2528 кеВ) рівень фону становить $0.26 \pm 0.06$ відліків / (кг × кеВ × рік), енергетична роздільна здатність по γ–лінії $^{232}$Th з енергією 2615 кеВ дорівнює 9 кеВ. В цьому експерименті встановлена межа на період напіврозпаду $^{130}Te$ на рівні $T_{1/2} \geq 5.5 \times 10^{23}$ років з 90% CL. Звідси, з урахуванням розрахунків матричних елементів в кількох різних роботах, отримане обмеження на ефективну масу майоранівського нейтрино $\langle m_\nu \rangle \leq (0.37 - 1.9)$ еВ [216]. Очевидні переваги експерименту полягають у високій енергетичній роздільній здатності детектора і майже 100% ефективності реєстрації. Недоліками болометричних детекторів є низька надійність і малий живий час вимірювань (але треба віддати належне дослідникам з групи проф. Е.Фіоріні, їм вдалося досягнути дуже високого живого часу установки, який зараз в спектрометрі "Cuoricino" становить не менше 70–80% [218]). Крім того, зниження рівня фону обмежене складністю установок. Основним джерелом фону, який не вдається придушити, є альфа–активність урану, торію з їх дочірніми на поверхні кристалів оксиду телуру та в безпосередньо біля цих кристалів розташованих деталях установки [219]. Значна частина фонових відліків пов'язана з радіонуклідами космогенного походження. Зокрема в фоні спостережені $\gamma$–піки від радіонуклідів $^{121}Te$, $^{121m}Te$, $^{123m}Te$, $^{125m}Te$, $^{127m}Te$, $^{57}Co$, $^{58}Co$, $^{60}Co$, $^{54}Mn$, які утворилися в результаті космогенної активації кристалів оксиду телуру та деталей мідного кріостату.

На основі експериментів з болометричними детекторами з кристалами оксиду телуру запропонований проект "CUORE", де планується збільшити масу кристалів $TeO_2$ до 750 кг [220]. Зараз цей проект вже отримав фінансування і знаходиться на стадії підготовки. Автори оцінюють чутливість проекту до періоду напіврозпаду відносно $0\nu2\beta$–розпаду $^{130}Te$ на рівні $T_{1/2} \approx 2 \times 10^{26}$ років. Детектор буде також застосований для пошуків сонячних аксіонів, частинок темної матерії Всесвіту, рідкісних ядерних $\alpha$– та $\beta$–розпадів.

$^{136}Xe$.

Енергія $2\beta$–переходу становить 2468(7) кеВ, ізотопна розповсюдженість $^{136}Xe$ – 8.87(0.16)%. Ізотоп $^{136}Xe$ є одним з найбільш перспективних для пошуку $0\nu2\beta$–розпаду. Це пов'язано, в першу чергу, з порівняно низькою ціною збагачення ізотопів ксенону. Крім того, ксенон, подібно до інших інертних газів, може бути ефективно очищений від радіоактивних домішок. Ксенон може бути застосований в пропорційних лічильниках, часово–проекційних і дрейфових камерах, сцинтиляційних детекторах. Хоча роздільна здатність газових і сцинтиляційних детекторів приблизно на порядок гірше роздільної здатності напівпровідникових детекторів, перевагою $^{136}$Xe є більше, у порівнянні з $^{76}$Ge, значення $Q_{\beta\beta}$.

Колаборацією Caltech–Neuchatel–PSI було побудовано часово–проекційну камеру з активним об’ємом 180 л [221], що містила 24.2 моль (3.3 кг) ксенону (збагаченого $^{136}$Xe до 62.5%) під тиском в 5 атм (рис. 1.8). Енергетична роздільна здатність детектора (FWHM) становила 6.6% на енергії 2β–переходу. Реконструкція треків у камері забезпечувала ефективне придушення фону до 0.02 відліків/(рік × кг × кеВ) в околі енергії 2.47 МеВ. Після 12 843 годин (що відповідає експозиції 4.9 кг × рік) вимірювань у підземній лабораторії Готхард (Швейцарія) було встановлено межу $T_{1/2}^{0\nu} \geq 4.4 \times 10^{23}$ років для масового механізму 0ν2β–розпаду і $T_{1/2}^{0\nu} \geq 2.6 \times 10^{23}$ років для 0ν2β–розпаду за рахунок правих струмів в слабкій взаємодії (з 90% CL).

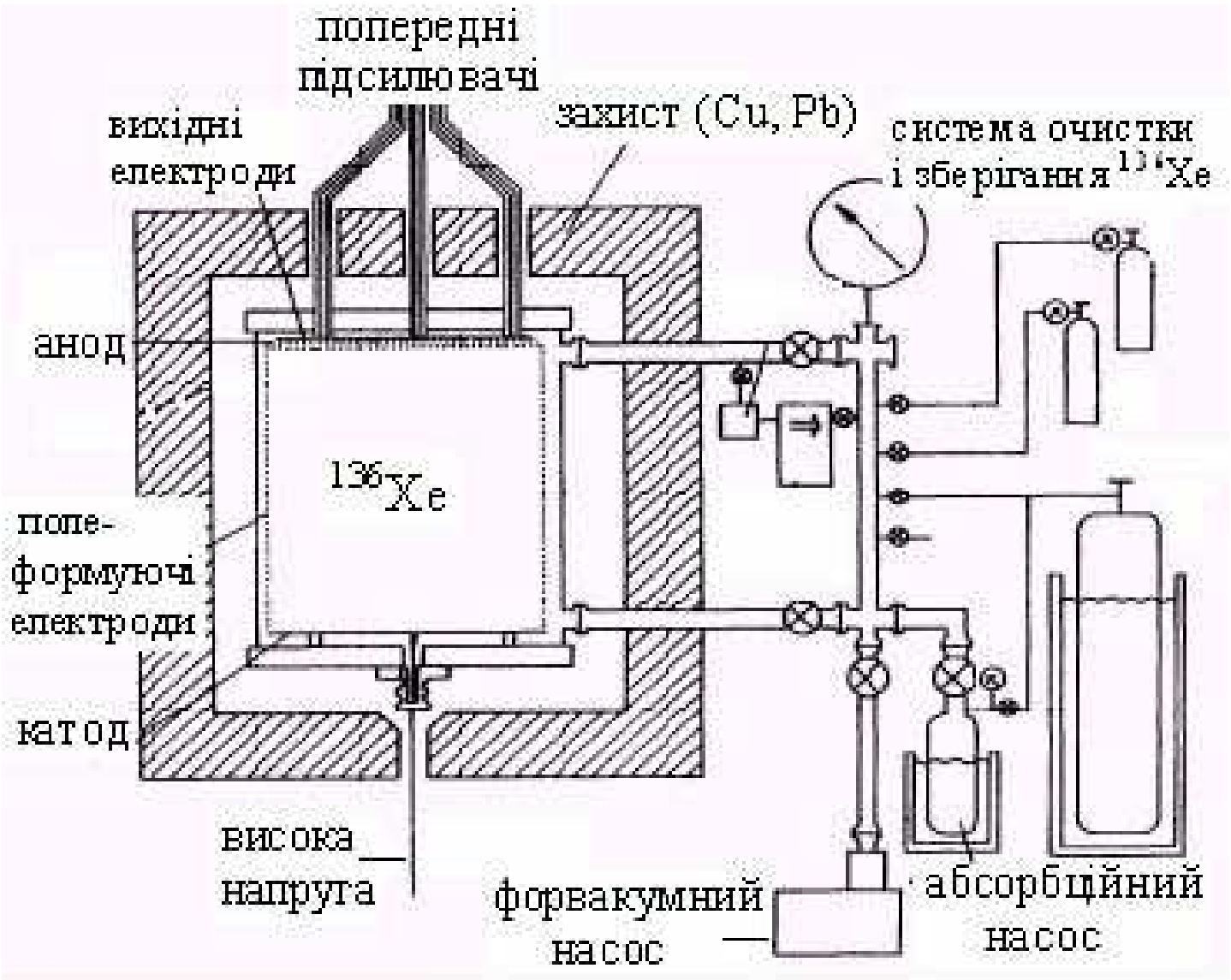

Рис. 1.8. Схематичний вигляд детектора для дослідження 2β–розпаду $^{136}$Xe, побудованого колаборацією Caltech–Neuchatel–PSI.

Логічним розвитком експерименту є проект EXO, в якому запропоновано збільшити масу збагаченого $^{136}Xe$ до 1 – 2 тон [222]. Енергія і треки електронів будуть вимірюватися за допомогою часово–проекційної камери високого (5 – 10 атм) тиску, можливо, що також будуть реєструватися сцинтиляційні сигнали. Хоча, оскільки спектр емісії ксенону знаходиться в ультрафіолетовій області, необхідно буде застосовувати спеціальні ФЕП з вікнами із фториду магнію, а також забезпечити постійну очистку ксенону, оскільки сцинтиляційні властивості і прозорість такого детектора надзвичайно чутливі до чистоти ксенону. В експерименті запропоновано застосувати метод лазерної резонансної спектроскопії для спостереження окремих іонів $^{136}Ba^{2+}$, що дозволить значно придушити фон. Автори оцінюють чутливість установки до маси нейтрино на рівні $\langle m_\nu \rangle \approx 0.02$ еВ. Зараз колаборація працює над спорудженням прототипу установки з 200 кг збагаченого $^{136}Xe$.

$^{150}Nd$.

Ядро $^{150}Nd$ виглядає дуже привабливим для пошуків $0\nu2\beta$–розпаду в зв’язку з великою енергією $2\beta$–переходу 3367.5 (2.2) кеВ (розповсюдженість цього ізотопу в природній суміші ізотопів неодиму становить 5.6 (0.2)%). Значення функції $G^{0\nu}(Q_{\beta\beta},Z)$, що входить в формулу для періоду напіврозпаду, є найбільшим серед усіх $2\beta$–ізотопів. Велике значення енергії розпаду є дуже сприятливим фактором для експерименту, оскільки фон, в загальному випадку, зменшується з ростом енергії. Але оскільки ядро $^{150}Nd$ сильно деформоване, деякі розрахунки ядерних матричних елементів дають менш оптимістичні прогнози ймовірності процесу (див. дискусію в розділі 6). Крім того, лантаноїди, як правило, суттєво забруднені радіонуклідами уранових і торієвого рядів.

Найбільш чутливий дослід по пошуку $0\nu2\beta$–розпаду $^{150}Nd$ був проведений за допомогою часово–проекційної камери [223]. В експерименті були використані 15.5 г ізотопно збагаченого до 91% оксиду неодиму–150: $^{150}Nd_2O_3$. Установка була розміщена на глибині 72 м в греблі Hoover Dam в США. Камера була захищена від зовнішнього фону 15 сантиметрами свинцю. В результаті 6287 годин вимірювань була встановлена межа на період напіврозпаду ядра $^{150}$Nd відносно 0ν2β–розпаду $T_{1/2}^{0\nu} \geq 1.2 \times 10^{21}$ років з 90% CL. Використовуючи найбільш оптимістичні розрахунки матричних елементів з роботи [198], автори встановили обмеження на масу нейтрино $\langle m_\nu \rangle \leq 3$ еВ. Двохнейтринний бета–розпад $^{150}$Nd був спостережений в досліді [223] з періодом напіврозпаду $T_{1/2}^{2\nu} = 6.8 \times 10^{18}$ років.

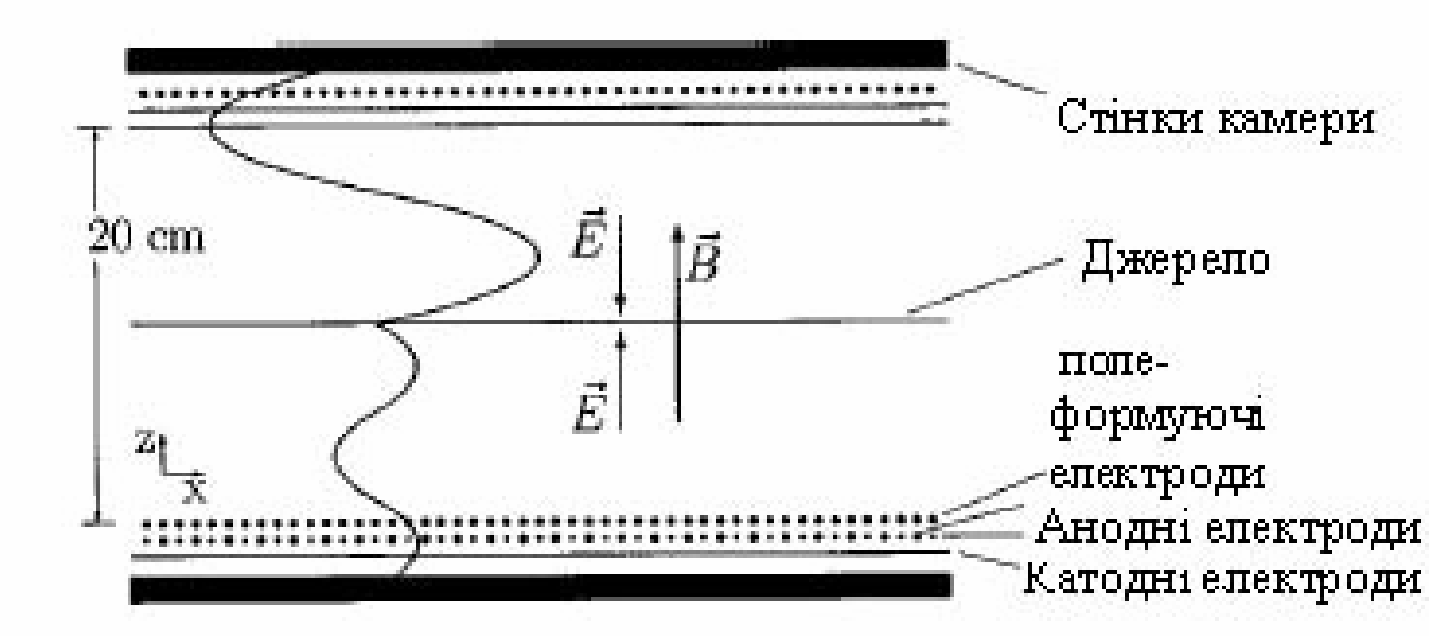

Рис. 1.9. Схема часово–проекційної камери, що була застосована для пошуку 2β–розпаду $^{150}$Nd. Кривими показані треки електронів в постійному електромагнітному полі.

**Висновки розділу.** Описані основні положення теорії 2β–розпаду, зокрема, різних мод: двохнейтринної та безнейтринної та механізмів цього процесу. Показано зв'язок між результатами експериментів, в яких спостерігаються осциляції нейтрино і які однозначно свідчать про наявність у нейтрино маси, та експериментів по пошуку 0ν2β–розпаду атомних ядер. Лише в прямих експериментах, спрямованих на пошуки безнейтринної моди 2β–розпаду, можна встановити природу нейтрино (частинка Майорани чи Дірака), визначити абсолютне значення шкали мас і схему масових станів нейтрино на рівні десятих, сотих і, можливо тисячних еВ, перевірити закон збереження лептонного заряду.

Подана класифікація експериментальних методів, за допомогою яких вже більше як півстоліття ведуться пошуки і дослідження цього вкрай рідкісного процесу. За цей час, з використанням різних експериментальних підходів: геохімічного, радіохімічного, прямої реєстрації подій, $2\nu2\beta$–розпад спостережений в ядрах $^{48}$Ca, $^{76}$Ge, $^{82}$Se, $^{96}$Zr, $^{100}$Mo, $^{116}$Cd, $^{128}$Te, $^{130}$Te, $^{150}$Nd, $^{238}$U, в тому числі, виміряні періоди напіврозпаду $2\nu2\beta$–переходів $^{100}$Mo і $^{150}$Nd на збуджені стани дочірніх ядер. Найвищі границі на період напіврозпаду були встановлені у прямих експериментах з кількома нуклідами: $T_{1/2} > 10^{21}$ років для $^{96}$Zr, $^{160}$Gd, $^{150}$Nd, $^{186}$W; $T_{1/2} > 10^{22}$ років для $^{48}$Ca, $T_{1/2} > 10^{23}$ років для $^{82}$Se, $^{100}$Mo, $^{116}$Cd, $^{128}$Te, $^{130}$Te, $^{136}$Xe і $T_{1/2} > 10^{25}$ років для $^{76}$Ge. Ці експерименти дозволили встановити найбільш жорсткі обмеження на значення маси майоранівського нейтрино ($m_\nu \leq 0.3 \div 5$ еВ), домішки правих струмів у слабкій взаємодії ($\eta \approx 10^{-7}$, $\lambda \approx 10^{-5}$), константу зв’язку нейтрино з майороном ($g_\chi \approx 10^{-4}$) і параметр порушення R–парності мінімальної суперсиметричної СМ ($\xi \approx 10^{-4}$).

# РОЗДІЛ 2

# РОЗРОБКА СЦИНТИЛЯЦІЙНОГО МЕТОДУ ДЛЯ ДОСЛІДЖЕНЬ 2β–РОЗПАДУ

## 2.1. Фактори, що визначають чутливість експерименту по пошуку 2β–розпаду

Чутливість експерименту до реєстрації 0ν2β–розпаду може бути оцінена за формулою 1.32. Концентрація ядер досліджуваного ізотопу, ефективність реєстрації подій 0ν2β–розпаду, рівень фону та енергетична роздільна здатність – це ті параметри, значення яких можуть бути поліпшені шляхом вибору і вдосконалення детектора, конструкції установки, застосуванням методів відбору фонових подій в ході аналізу даних вимірювань. Час вимірювань залежить від надійності і стабільності роботи установки. Практично важливою є і дешевизна детектора, установки, електроніки.

Перевагами сцинтиляційних детекторів, які містять в своєму складі досліджувані 2β–активні ізотопи, є, в першу чергу, висока ефективність реєстрації, простота і надійність, що дозволяє проводити експеримент протягом років і навіть десятків років, здатність до дискримінації частинок за формою сцинтиляційних сигналів, притаманна більшості сцинтиляторів, швидкий відгук. Сцинтиляційні детектори здатні забезпечити енергетичну роздільну здатність на рівні кількох відсотків, що достатньо для планування експериментів на рівні чутливості $T_{1/2}^{0\nu} \sim 10^{26} – 10^{27}$ років. Енергетична роздільна здатність і, особливо, рівень фону сцинтиляційних детекторів, залежить від багатьох факторів. Як показує досвід кількох десятиріч історії пошуків і вимірювань 2β–розпаду, фон сцинтиляційного детектора може бути зменшений до дуже низького рівня. Оптимізація конструкції детектора дозволяє отримати високу роздільну здатність навіть в умовах вимог, які протирічать одна одній: з одного боку до максимального збору світла, а з іншого – необхідності екранувати сцинтилятор від випромінювання з ФЕП, порівняно сильно забрудненого радіонуклідами природного походження. Розглянемо загальні властивості сцинтиляційних детекторів, які визначають їх енергетичну роздільну здатність та рівень фону.

### 2.1.1. Енергетична роздільна здатність сцинтиляційних детекторів.

Основними факторами, що визначають енергетичну роздільну здатність сцинтиляційних детекторів є світловихід, відповідність спектрів емісії сцинтилятора спектральній чутливості фотоелектронних помножувачів, коефіцієнт світлозбору, однорідність світлозбору, нелінійна залежність світловиходу від енергії. В конструкції сцинтиляційного низькофонового детектора обов'язково треба передбачувати світловоди, оскільки ФЕП містять багато радіоактивних домішок (уран, торій, калій).

З деяким наближенням можна вважати, що флуктуації величин, які визначають роздільну здатність сцинтиляційного детектора ($R$), складаються як вектори:

$$R^2 = R_i^2 + R_p^2 + R_{PMT}^2 \qquad (2.1)$$

Тут $R_i$ – внутрішня роздільна здатність сцинтилятору; $R_p$ та $R_{PMT}$ – роздільні здатності, пов'язані з розповсюдженням світла та з процесами в ФЕП відповідно [224,225,226].

Внутрішня роздільна здатність кристала $R_i$ пов'язана з багатьма ефектами, але найголовнішими є два фактори. Суттєво впливає на внутрішню роздільну здатність непропорційний відгук сцинтилятора в залежності від енергії γ–квантів. Як правило, для енергій менших ~ 0.1 МеВ, світловихід нелінійно зменшується із зменшенням енергії. Як результат, сумарна кількість сцинтиляційних фотонів залежить від флуктуацій розподілу енергії вторинних, так званих δ–електронів, що утворюються в результаті взаємодії первинних електронів, які виникають в результаті передачі енергії γ–кванту через фотоефект, комптонівське розсіяння і народження пар. Цей ефект дає основний внесок у внутрішню роздільну здатність при енергіях ≤ 1 МеВ. Другий внесок пов'язаний з неоднорідністю світловиходу в різних точках сцинтилятора. В детекторах з активаторами, як наприклад NaI(Tl), CsI(Tl), GSO(Ce) ($Gd_2SiO_5$:Ce), неоднорідність світловиходу пов'язана з градієнтами концентрації центрів люмінесценції в сцинтиляційному кристалі. При цьому, кількість фотонів залежить від точки в кристалі, де відбувається сцинтиляційний спалах.

Величина компоненти $R_p$ залежить від оптичних властивостей сцинтилятора, його геометричної форми та розмірів, якості оптичного контакту між кристалом і ФЕП, наявності світловодів, неоднорідності квантової ефективності фотокатоду і її залежності від довжини хвилі фотона, кута падіння фотона на фотокатод. Наприклад, в сцинтиляторах вольфрамату кадмію ($CdWO_4$) ця компонента може бути досить суттєвою, оскільки вольфрамат кадмію має високий коефіцієнт заломлення (2.2 – 2.3), невисоку прозорість (10–30 см$^{-1}$), значну анізотропію, помітну кількість макро– і мікро–неоднорідностей.

Величина $R_{PMT}$ характеризує статистичні флуктуації сигналу на виході ФЕП, а отже залежить від ефективності збору електронів на першому диноді, флуктуацій підсилення електронів в системі динодів, рівня шуму та стабільності апаратури.

Для „ідеального" сцинтиляційного детектора, в якому роздільна здатність визначається лише флуктуаціями кількості фотоелектронів, роздільну здатність можна вважати залежною лише від кількості фотоелектронів, що потрапляють на перший динод „ідеального" фотопомножувача:

$$R = 2.35 / N_{phe}^{1/2} \qquad (2.2)$$

де $N_{phe}$ – середня кількість фотоелектронів, що збираються на першому диноді. Формула 2.2 дає оцінку нижньої межі роздільної здатності сцинтиляційного детектора. Оцінки зроблені по формулі 2.2 тим точніші, чим більшою є енергія γ–квантів (електронів), яка поглинається в сцинтиляторі (в цьому випадку внесок, пов'язаний з ФЕП стає несуттєвим), а також у випадку, коли забезпечена висока однорідність збору світла на фотокатод ФЕП.

2.1.2. Джерела фону детекторів по пошуку 2β–розпаду.

Пошуки і дослідження такого надзвичайно рідкісного процесу, як 2β–розпад, можливі лише за умови значного придушення фону установки, призначеної для його реєстрації. Ідеальний детектор має бути взагалі безфоновим. Основними джерелами фону детектора, призначеного для реєстрації подій 2β–розпаду, є:

1) гамма–кванти від розпадів радіонуклідів природного походження;
2) космічні промені;
3) забрудненість радіонуклідами природного та антропогенного (тобто таких, які виникли внаслідок людської діяльності) походження конструкційних матеріалів установки (у випадку застосування напівпровідникових, сцинтиляційних, газових, рідинних детекторів, низькотемпературних болометрів – доцільно розглядати окремо забрудненість робочого матеріалу детектора);
4) розпади радону, присутнього в повітрі, а також в робочих газах або рідинах детектора;
5) нейтрони;
6) радіонукліди, що утворилися внаслідок опромінення матеріалів установки космічними променями (космогенна активація);
7) процеси двохнейтринного 2β–розпаду ядра (коли мова йде про пошук безнейтринної моди 2β–розпаду);
8) радіонаводки, збої в системі реєстрації.

Розглянемо ці компоненти фону детальніше.

2.1.2.1. Гамма–кванти від розпадів радіонуклідів в оточуючому середовищі.

Всі речовини, що є в природі або створені людиною, містять, в тій чи іншій кількості, радіонукліди. Найнебезпечнішими при проведенні пошуків 2β–розпаду є ті, що випромінюють гамма–кванти великої енергії. Це, в першу чергу, дочірні продукти розпаду $^{232}$Th ($^{228}$Ac з максимальною енергією гамма–квантів ≈2 МеВ та $^{208}$Tl, що випромінює інтенсивну гамма–лінію 2615 кеВ), $^{238}$U ($^{214}$Bi з максимальною енергією гамма–квантів ≈ 3.2 МеВ), а також $^{40}$K ($E_\gamma$ = 1462 кеВ). Для ефективного захисту від гамма–променів з такими енергіями необхідні досить значні шари речовини (як правило, сотні грамів на см$^2$). Матеріали для спорудження пасивного захисту (захисні екрани з різних матеріалів ми назвемо “пасивним” захистом, на відміну від “активного” захисту, що являє собою детектуючу систему, здатну реєструвати фонове випромінювання і, таким чином, відрізнити його від корисного сигналу) мають бути достатньо чистими відносно вмісту радіонуклідів (менше ніж $10^{-11}$ – $10^{-12}$ г/г торію і урану), недорогими, мати якнайбільшу густину. Крім того, бажано, щоб вони підлягали механічній обробці і могли бути елементами конструкції низькофонової установки. Цим вимогам найкраще відповідають свинець і мідь. Щоправда, свинець містить бета–радіоактивний ізотоп $^{210}$Pb ($Q_\beta$ = 63.5 кеВ), який з періодом напіврозпаду $T_{1/2}$ = 22.3 р. розпадається в $^{210}$Bi ($Q_\beta$ = 1162.7 кеВ). Гальмівне випромінювання $^{210}$Bi спричинює фон до енергії $Q_\beta$ $^{210}$Bi. Активність ізотопу $^{210}$Pb у свинці, виплавленому недавно у порівнянні з періодом напіврозпаду, як правило, досить значна (сотні і навіть тисячі Бк/кг). Єдиним шляхом є використання археологічного свинцю, який був виплавлений сотні і тисячі років тому. Але такий свинець практично малодоступний, принаймні, на території України. Тому, для захисту від гальмівного випромінювання $^{210}$Bi використовують електролітичну мідь високої електропровідності, що є одним з найбільш чистих металів відносно домішок урану, торію і калію. Як вже було відмічено, активний захист також дозволяє зменшити фон від зовнішнього випромінювання, хоча основне його призначення – придушення фону від космічних променів та радіоактивності матеріалів, з яких виготовлено детектор.

2.1.2.2. Космічні промені.

Космічні промені на поверхні Землі являють собою потоки вторинних частинок, що утворились в верхніх шарах атмосфери в результаті взаємодії високоенергетичних первинних частинок (в основному, протонів) з ядрами атомів газів. На висотах ≤5 км над рівнем моря вторинні космічні промені складаються, головним чином, з мюонів ($\mu^{\pm}$), електронів ($e^{\pm}$), а також протонів і $\pi^{\pm}$–мезонів. Електрони породжують каскади $\gamma$–квантів і електрон–позитронних пар, формуючи так звану м‘яку компоненту космічних променів. Протони і $\pi^{\pm}$–мезони, в результаті глибоко непружних реакцій на ядрах, приводять до появи нейтронів (адронна компонента). Якщо від м‘якої та адронної компоненти детектор можна захистити шаром речовини (скажімо, будівельних матеріалів, ґрунту) товщиною порядку метрів, то мюони, які біля поверхні Землі мають середню енергію близько 4 ГеВ, здатні проникати значно глибше. Єдиним способом захисту від мюонної компоненти космічних променів є розміщення детектора глибоко під землею. Тому практично всі сучасні експерименти по пошуку подвійного бета–розпаду проводяться в підземних лабораторіях на глибинах кількох сотень і навіть тисяч метрів. Як правило, оскільки породи мають різну густину, глибину лабораторії приводять в метрах водяного еквіваленту. Солотвинська підземна лабораторія Інституту ядерних досліджень НАНУ розташована в діючій шахті №9 по видобутку солі (NaCl) на глибині близько 450 м від поверхні землі [227]. Враховуючи товщину і склад порід, що розміщені над лабораторією, це відповідає приблизно 1000 м водного еквіваленту. Така товща речовини послаблює потік космічних мюонів приблизно в $10^4$ разів. Крім того, для придушення фону від залишкових мюонів використовуються детектори активного захисту, що дозволяють реєструвати час проходження мюонів через установку і таким чином відкидати спричинені ними фонові події. Оскільки мюони, внаслідок взаємодії з ядрами, можуть породжувати нейтрони, необхідно враховувати час, за який нейтрони термалізуються. Наприклад, в наднизькофоновому спектрометрі для пошуку 2β–розпаду $^{116}$Cd, після кожного спрацювання лічильника мюонів, реєстрація подій заборонялась протягом $\approx 0.4$ с. Важливою перевагою Солотвинської лабораторії є також чистота солі відносно домішок торію, урану і калію. В результаті, гамма–фон в лабораторії приблизно в 30 – 50 разів менший, ніж в звичайному лабораторному приміщенні на поверхні землі (рис. 2.1).

Фон від нейтрино (потік нейтрино, в основному, має космічне походження, і тому цей тип фону логічно виділити окремо, оскільки „сховатися” від нейтрино в підземній лабораторії неможливо) на нинішньому рівні чутливості не впливає на результати експериментальних досліджень 2β–розпаду. Але ця компонента фону може стати суттєвою, коли маса досліджуваного ізотопу наблизиться до 10 – 100 тон і всі інші компоненти фону будуть придушені.

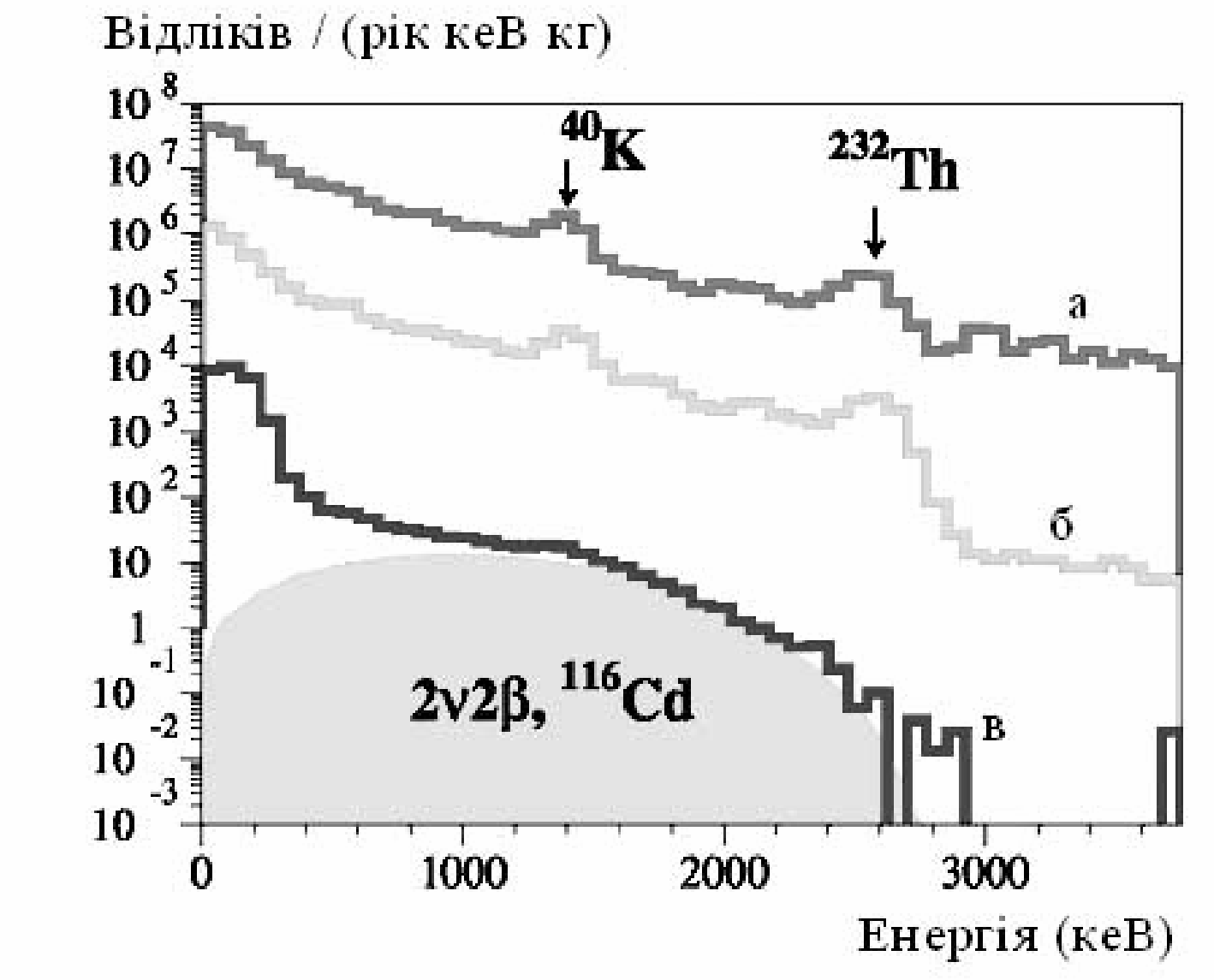

Рис. 2.1. Спектри фону, виміряні сцинтиляційним детектором вольфрамату кадмію в (а) лабораторії в Києві; (б) Солотвинській підземній лабораторії ІЯД; (в) в низькофоновій установці, розташованій в Солотвинській підземній лабораторії.

2.1.2.3. Забрудненість радіонуклідами конструкційних матеріалів установки.

Забрудненість радіонуклідами конструкційних матеріалів установки є наступним важливим фактором зменшення фону установки. Очевидно, що чим ближче до детектора знаходиться матеріал і чим менший шар речовини знаходиться між матеріалом і детектором, тим більшою є ефективність реєстрації фонового гамма–випромінювання. А отже, тим більш жорсткі вимоги ставляться до чистоти цих матеріалів. Для матеріалів, що безпосередньо оточують чутливий об‘єм (у випадку сцинтиляційних спектрометрів це можуть бути, наприклад, світловідбиваючі матеріали, або речовини, що забезпечують оптичний контакт сцинтилятора із світловодами), важливо враховувати також бета– і поверхневу альфа–активність. Всі конструкційні матеріали мають бути попередньо відібрані за рівнем радіоактивних домішок. Для цього використовують спеціальні низькофонові радіометри, а також враховують результати попередніх експериментів. Крім того, особливості технології виробництва тих чи інших матеріалів часто дозволяють прогнозувати рівень їхньої чистоти відносно радіонуклідів. Наприклад, органічні сполуки, як правило, мають досить низькі рівні забрудненості радіонуклідами рядів торію та урану. В той час різні види керамік – містять ці радіонукліди в дуже значних концентраціях. Перелік матеріалів, що можуть бути застосовані в наднизькофонових детекторах для пошуків і реєстрації процесів подвійного бета–розпаду, досить обмежений. Крім того, важливу увагу слід приділяти ретельній очистці деталей спектрометра від можливих поверхневих забруднень.

2.1.2.4. Радіоактивна забрудненість детектора.

Забрудненість радіонуклідами матеріалів, з яких виготовлено чутливий елемент детектора (напівпровідник або сцинтилятор, газ або рідина, кристал в низькотемпературних болометрах), є, в решті решт, найважливішим фактором досягнення мінімального рівня фону. Це обумовлено високою, часто практично 100%, ефективністю реєстрації подій радіоактивних розпадів всередині детектора.

Більше того, оскільки при цьому реєструються також бета– і альфа–частинки, енерговиділення в детекторі є значно більшим, ніж коли лише гамма–кванти можуть досягти чутливого об‘єму останнього. Так, наприклад, розпади $^{228}$Th (дочірній продукт $^{232}$Th) в оточуючих матеріалах призводять до появи фонових подій з енергіями до 2.6 МеВ (інтенсивна гамма–лінія $^{208}$Tl). Енерговиділення при внутрішній забрудненості цим радіонуклідом досягає 5 МеВ (енергія бета–розпаду $^{208}$Tl). В той же час, саме ця компонента фону може бути ефективно придушена завдяки використанню різних методик аналізу подій, таких як часово–амплітудний або аналіз за формою сигналів (див. нижче п. 2.3.1 і 2.3.2).

Радіоактивні забруднення в сцинтиляторах можна умовно поділити на три типи. Найбільшою мірою на проведення низькофонових вимірювань впливають радіонукліди природного походження з великими, у порівнянні з часом виробництва детектора, періодами напіврозпаду. Ці ізотопи, в тій чи іншій кількості, завжди присутні у всіх матеріалах, а отже, і в сцинтиляторах. Найбільший внесок в фон детектора дають, як правило, дочірні продукти розпаду урану ($^{235}$U та $^{238}$U) та торію ($^{232}$Th), а також $^{40}$K.

Крім того, в природі є ряд радіоактивних ізотопів, які, в певних випадках, можуть впливати на результати низькофонових експериментів. Їх вплив суттєвий тоді, коли вони мають хімічні властивості близькі до тих елементів з яких складається детектор, або ж входять до складу ізотопів основного елемента детектора. До них можна віднести довгоживучі бета–активні ізотопи рубідію, кадмію та індію:

$^{87}$Rb, ($T_{1/2} = 4.75\times10^{10}$ р., $Q_\beta = 283$ кеВ),

$^{113}$Cd, ($T_{1/2} = 7.7\times10^{15}$ р., $Q_\beta = 316$ кеВ),

$^{115}$In, ($T_{1/2} = 4.1\times10^{14}$ р., $Q_\beta = 495$ кеВ),

Рідкоземельні елементи лантан, самарій, гадоліній, лютецій також мають радіоактивні ізотопи:

$^{138}$La ($T_{1/2} = 1.05\times10^{11}$ р., $Q_\beta = 1044$ кеВ),

$^{147}$Sm ($T_{1/2}$ = 1.06×10$^{11}$ р., $Q_\alpha$ = 2310 кеВ),

$^{152}$Gd ($T_{1/2}$ = 1.08×10$^{14}$ р., $Q_\alpha$ = 2205 кеВ),

$^{176}$Lu ($T_{1/2}$ = 3.78×10$^{10}$ р., $Q_\beta$ = 1192 кеВ).

Вплив цих радіонуклідів особливо помітний в детекторах, як наприклад $Gd_2SiO_5$(Ce), $CeF_3$, Nd:YAG, до складу яких входять рідкоземельні елементи.

Радіоактивний ізотоп вуглецю $^{14}$C породжує фон в пластикових та рідких сцинтиляторах, основою яких є органічні сполуки [228,229]. Радіоактивний аргон $^{39}$Ar з періодом напіврозпаду $T_{1/2}$ = 269 років призводить до значного фону в детекторах, робочим тілом яких є аргон (пропорційні камери на рідкому аргоні), в той час як радіоактивний кремній $^{32}$Si періодом напіврозпаду $T_{1/2}$ = 132 років є джерелом фону в кремнієвих напівпровідникових детекторах. Щоправда, $^{14}$C, $^{39}$Ar, $^{32}$Si є радіонуклідами, головним чином, космогенного походження, тобто утвореними в результаті взаємодії космічних променів з речовиною.

Зменшити фон від цих нуклідів можна, в першу чергу, зменшивши їх концентрацію в детекторі шляхом очистки сировини та вдосконалення технології росту кристалів (очевидно, це не стосується $^{14}$C, $^{113}$Cd, $^{115}$In, $^{152}$Gd, $^{210}$Pb, для видалення яких необхідно здійснювати надзвичайно дорогий процес розділення ізотопів).

Радіонукліди природного походження, але такі, які виникають головним чином в результаті взаємодії космічних променів з ядрами стабільних ізотопів, слід віднести до наступного класу радіонуклідів внутрішнього забруднення, яке називається космогенним (cosmogenic activation).

Третім типом радіонуклідів, які часто присутні в детекторах, є радіонукліди техногенного (тобто утворені внаслідок діяльності людини) походження. Особливо актуальною проблема уникнення впливу цього типу забрудненостей стала після катастрофи на Чорнобильській АС. Зокрема, в деяких експериментах, що проводились в Солотвинській підземній лабораторії, було зареєстровано вплив ізотопів цезію: $^{134}$Cs ($T_{1/2}$ = 2.062 р., $Q_\beta$ = 2059 кеВ) та $^{137}$Cs ($T_{1/2}$ = 30.07 р., $Q_\beta$ = 1176 кеВ). Присутність обох цих ізотопів спостерігалась в сцинтиляторах CsI(Tl), що використовувались в якості детекторів активного захисту в одному з варіантів установки для пошуку 2β–розпаду $^{116}Cd$ [186]. Не зважаючи на те, що $^{90}Sr$ ($T_{1/2}$ = 28.78 р., $Q_β$ = 546 кеВ) не був зареєстрований в ході солотвинських експериментів (це пов‘язано з тим, що в розпадах цього ізотопу практично не випромінюються гамма–кванти, а отже його ідентифікація є досить непростим завданням), не можна виключити впливу цього ізотопу на визначення періоду напіврозпаду $^{116}Cd$ відносно розпаду з вильотом двох нейтрино. Ізотоп $^{90}Sr$ розпадається в $^{90}Y$ з досить великою енергією бета–частинок $Q_β$ = 2282 кеВ ($T_{1/2}$ = 64.1 годин), порівняною з енергією 2β–розпаду $^{116}Cd$. Особливо суттєвою була систематична похибка визначення періоду напіврозпаду $^{116}Cd$ відносно 2ν2β–розпаду на ранніх етапах експерименту, коли експозиція була незначною і важко було відрізнити ці два процеси за формою накопиченого енергетичного спектру [159]. Крім впливу цезію та можливого впливу стронцію, в ході низькофонових експериментів, що проводились в Солотвинській та Національній підземній лабораторії Гран Сассо (Італія) ми не спостерігали впливу яких–небудь інших радіонуклідів – продуктів катастрофи на ЧАЕС. Це пояснюється тим, що був здійснений певний комплекс мір, метою яких було зменшення такого впливу. До них, зокрема можна віднести строге виконання вимог чистоти в Солотвинській лабораторії, заходи по упаковці під час транспортування та ретельній очистці поверхонь всіх матеріалів, що були задіяні не лише в експериментальних установках, а й у складі допоміжного обладнання.

Очевидно, що крім розповсюджених повсюди, принаймні на територіях, забруднених в результаті Чорнобильської катастрофи, радіонуклідів, техногенне забруднення низькофонового детектора може відбутись при необережному поводженні з якими завгодно радіоактивними матеріалами. Тому в процесі підготовки і монтажу низькофонових установок завжди здійснювались виключні міри для попередження такого забруднення.

В ході низькофонових експериментів, метою яких був пошук рідкісних процесів розпаду атомних ядер, було оцінено ступінь чистоти сцинтиляційних кристалів $CdWO_4$ (як збагачених ізотопом $^{116}Cd$, так і вирощених з кадмію з природним ізотопним складом), $CaWO_4$, $ZnWO_4$, $Gd_2SiO_5$(Ce), $CeF_3$, YAG:Nd відносно домішок радіоактивних ізотопів. Для оцінки активності радіонуклідів було застосовано, головним чином, такі три підходи, як аналіз форми енергетичних спектрів із застосуванням моделей, розрахованими методом Монте–Карло; часово–амплітудний аналіз даних; аналіз форми сцинтиляційних сигналів. Останні два методи не лише дозволяють виміряти активність деяких радіонуклідів уранових та торієвих рядів, але і ефективно придушувати спричинений ними фон. Детальніше результати вимірювань радіочистоти сцинтиляційних детекторів будуть описані в розділі 7.

Ефективним способом придушення фону від забрудненості бета–активними ізотопами (у випадках, коли переходи йдуть на збуджені рівні дочірніх ядер, а отже супроводжуються гамма–квантами) матеріалів установки і самого чутливого елементу детектора є оточення останнього допоміжними детекторами активного захисту. Такі детектори повинні мати максимально високу ефективність і мінімальний енергетичний поріг реєстрації гамма–квантів, якомога нижчий рівень забрудненості радіонуклідами, не потребувати герметизації і механічної обробки, тобто виступати в якості конструкційних матеріалів.

#### 2.1.2.5. Вплив радону.

Радіоактивний газ радон, який є продуктом розпаду торію і урану, завжди присутній в повітрі. Радон складається з трьох альфа–активних ізотопів. Ізотоп $^{220}$Rn з енергією альфа–розпаду $Q_\alpha$ = 6405 кеВ і періодом напіврозпаду $T_{1/2}$ = 55.6 с є продуктом розпаду $^{232}$Th. Ізотоп $^{219}$Rn ($Q_\alpha$ = 6946 кеВ, $T_{1/2}$ = 3.96 с) є дочірнім $^{235}$U. І, нарешті, $^{222}$Rn ($Q_\alpha$ = 5590 кеВ, $T_{1/2}$ = 3.83 доби) потрапляє в повітря внаслідок розпадів $^{238}$U. Останній ізотоп радону є найнебезпечнішим для низькофонових досліджень внаслідок порівняно великого періоду напіврозпаду. При цьому основним джерелом фону в області енергій ≈ 1 – 3 МеВ є бета–активний ізотоп $^{214}$Bi ($Q_{\beta}$ = 3272 кеВ, $T_{1/2}$ = 19.9 хв.). Фон з енергією до 1 МеВ продукується внаслідок розпадів $^{210}$Pb ($Q_{\beta}$ = 63.5 кеВ, $T_{1/2}$ = 22.3 р.) $^{210}$Bi ($Q_{\beta}$ = 1163 кеВ, $T_{1/2}$ = 5.013 діб). Активність радону в повітрі на поверхні землі становить в середньому кілька десятків Бк/м$^3$. В шахтах активність радону, внаслідок обмеженого об‘єму повітря і, як правило, значного вмісту урану в оточуючих породах, значно вища і може досягати тисяч Бк/м$^3$. В багатьох підземних лабораторіях активність радону сильно залежить від ефективності роботи вентиляційних систем. Тому перевагою Солотвинської лабораторії є досить низька активність радону, що є наслідком низької концентрації урану в солі Солотвинського родовища. Радон, внаслідок своєї інертності, має дуже високу здатність проникати через різні матеріали. Тому для захисту від його впливу установка має бути ретельно ізольована і/або в ній не має бути пустот. Але навіть недостатньо ретельна ізоляція від доступу повітря деталей низькофонових детекторів під час їх зберігання призводить до забрудненості поверхонь радіоактивним ізотопом $^{210}$Pb, що є продуктом розпаду радону ($^{222}$Rn).

2.1.2.6.Фон, спричинений нейтронами.

Нейтрони в підземних лабораторіях виникають в результаті спонтанного поділу $^{238}$U, а також в результаті непружних реакцій мюонів на ядрах в оточуючих породах та матеріалах детектора. Потоки нейтронів в підземних лабораторіях нижчі, ніж на поверхні землі. В Солотвинській лабораторії для потоку теплових нейтронів була встановлена межа на рівні $3\times10^{-6}$ см$^{-2}\times$сек$^{-1}$ [230]. Для захисту від нейтронів використовуються матеріали, що містять ядра водню (парафін, поліетилен), а також екрани з матеріалів з великим перерізом захвату теплових нейтронів (кадмій, бор).

2.1.2.7.Космогенна активація.

Опромінення матеріалів детектора потоками високоенергетичних протонів, нейтронів, піонів призводить, внаслідок реакцій зриву, до виникнення широкого спектру радіоактивних ядер. Переважна більшість цих ядер розпадається під час переміщення детектора в підземну лабораторію і протягом перших кількох тижнів перебування під землею. Але ядра, що мають періоди напіврозпаду близько року і більше, призводять до появи фону, який ми назвемо космогенним. Під час перебування детектора під землею космогенна активація практично припиняється, оскільки адронна компонента космічних променів поглинається в кількох метрах – десятках метрів від поверхні землі. Питома активність космогенної компоненти дуже мала у порівнянні з типовою забрудненістю навіть найбільш чистих матеріалів. І все–таки, космогенна активація вже була спостережена в експериментах по пошуку 2β–розпаду $^{76}$Ge [136] і $^{130}$Te [5]. Є підстави стверджувати, що в експерименті, що проводився в Солотвинській лабораторії, також спостерігалась космогенна активація детектора із сцинтиляційними кристалами $^{116}CdWO_4$. Ця компонента фону буде однією з головних проблем в наступному поколінні надчутливих детекторів для пошуку процесів 2β–розпаду. Очевидною мірою зменшення впливу космогенної радіації є витримка детектора в підземних умовах впродовж достатньо довгого часу. Але вже зараз в деяких проектах розглядається можливість виготовлення детектора безпосередньо в підземних лабораторіях, або з використанням захисту від адронної компоненти космічних променів під час всього процесу виготовлення і транспортування детектора в підземну лабораторію [231,232].

2.1.2.8.Процеси двохнейтринного 2β–розпаду.

Процеси двохнейтринного 2β–розпаду, не дивлячись на великі періоди напіврозпаду ($^{150}$Nd і $^{100}$Mo — найбільш “короткоживучі” з виміряних на сьогодні 2β–активних ізотопів мають періоди напіврозпаду відносно 2ν2β–розпаду на рівні $T_{1/2} \approx 10^{19}$ р [22]), можуть, при низькій енергетичній роздільній здатності детектора, продукувати фонові події, які неможливо відрізнити від подій безнейтринної моди 2β–розпаду [233]. Єдиним способом уникнути цієї компоненти фону є покращення енергетичної роздільної здатності детектора. Звичайно, вибір ядер, у яких двохнейтринна мода значно придушена, а безнейтринна – ні, також, певною мірою, дозволить вирішити цю проблему. Більш детально ця компонента фону буде розглянута у 7–му розділі.

#### 2.1.2.9.Радіонаводки, збої в системі реєстрації

Радіонаводки та збої в системі реєстрації також можуть бути джерелом фону. Тому завжди необхідними є ретельне заземлення та екранування детектора. Крім того, події такого типу, на відміну від шуканих подій 2β–розпаду, як правило, мають не пуасонівський часовий розподіл. Радіонаводки призводять до значних збільшень темпів набору, що дозволяє виявити їх в ході вимірювань, контролюючи частоти слідування сигналів. Ефективним способом є заборона реєстрації подій що співпадають у часі з сигналами від спеціальних електронних систем, чутливих до радіонаводок. Крім того, фонові події такого типу можна знаходити в масивах даних за допомогою відповідного аналізу і, таким чином, відкидати "зіпсовані" частини експозиції. Але найбільш ефективним способом боротьби з цим типом фону є використання методик аналізу сигналів за формою.

## 2.2. Низькофоновий сцинтиляційний спектрометр з кристалами $^{116}CdWO_4$ в Солотвинській підземній лабораторії

### 2.2.1. Відбір конструкційних матеріалів низькофонових установок за рівнем радіоактивної забрудненості.

Як уже зазначалося в підпункті 2.1.2.4, матеріали, з яких виготовляються деталі низькофонової установки, мають бути ретельно відібрані за рівнем вмісту радіоактивних домішок. Причому, чим ближче до чутливого об‘єму детектора знаходиться та чи інша деталь, тим більш жорсткими є вимоги до радіочистоти. При виборі матеріалів ми, по–перше, використовуємо опубліковані дані, по–друге, відбір матеріалів здійснюється за допомогою спеціальних вимірювань в допоміжних низькофонових установках і, нарешті, аналіз даних довгострокових експериментів дозволяє оцінити, в яких деталях спектрометра містяться домішки радіонуклідів.

Аналіз наявних в літературі даних [234,235,236,237,238,239] дозволяє зробити деякі загальні висновки про рівень забрудненості матеріалів. Виявляється, що різні види кераміки та скла мають досить високий рівень вмісту радіонуклідів урано–торієвих рядів та калію. Рівень забрудненості цих матеріалів може досягати десятків і сотень Бк/кг. Цей факт має бути врахований при використанні елементів, виготовлених з цих матеріалів: ізоляторів, радіодеталей, фотопомножувачів. Забрудненість пластмас, металів і сплавів значно нижча, але має досить широкий діапазон значень ($\sim 10^{-3}$ і менше Бк/кг). Врешті решт, перелік радіочистих матеріалів (активність торію, урану, калію не перевищує рівня $\sim 10^{-6}$ Бк/кг) досить вузький: це безкиснева електролітична мідь, фторопласт, деякі пластмаси, плавлений кварц, органічні рідини. Дуже чистим матеріалом є надчистий германій, який використовується для виготовлення напівпровідникових германієвих детекторів. Рівень забрудненостей цього германію менший за $\sim 10^{-10}$ Бк/кг. Рідини, зокрема рідкі сцинтилятори, а також вода, рідкий азот, можуть бути очищені до надзвичайного високого рівня чистоти ($\sim 10^{-10}$ Бк/кг) [229].

Для відбору матеріалів за рівнем активності в Солотвинській підземній лабораторії була побудована спеціальна низькофонова сцинтиляційна установка. В якості детектора використовується кристал вольфрамату кадмію великого об’єму ($\approx 125$ см$^3$, масою $\approx 1$ кг). Кристал проглядається спеціальним низькофоновим ФЕП (D724KFL, THORN EMI) через світловод $\varnothing 10 \times 33$ см, виготовлений з надчистого кварцу марки КУ–1. Світловод виконаний у формі логарифмічної спіралі, що дозволяє ефективно збирати сцинтиляційне світло. Детектор оточений шарами пасивного захисту: фторопластом (3 – 5 см), поліметилметакрилату (6 – 13 см), міді марки М0 (3 – 6 см), свинцю (15 см), поліетилену (8 см). Над установкою встановлені два сцинтиляційні детектори активного захисту з пластиковими сцинтиляторами для реєстрації космічних мюонів. Енергетична роздільна здатність спектрометра складає ≈13% по γ–лінії $^{137}$Cs з енергією 662 кеВ. Слід підкреслити, що конструкція установки дозволяє використовувати різні сцинтиляційні детектори.

Завдяки високій густині сцинтилятора $CdWO_4$ і можливості його використання без герметизації досягається досить висока ефективність реєстрації γ–квантів. Необхідним є лише шар світловідбиваючих матеріалів, який складається з дуже тонкої тефлонової плівки та майлару. Другою перевагою цього детектора є його дуже високий рівень радіочистоти. Ефективність реєстрації детектора була розрахована за допомогою пакета приграм GEANT3.14. Наприклад, чутливість детектора при реєстрації $^{40}$K і $^{232}$Th в зразку масою ≈ 1 кг з точністю ±30% за 24 години вимірювань становить 0.04 і 0.007 Бк/кг, відповідно.

За допомогою цієї установки було проведено вимірювання зразків матеріалів. Результати вимірювань деяких зразків представлені в таблиці 2.1.

Таблиця 2.1. Активність радіонуклідів (в мБк/кг, окрім ФЕП, для яких приведені активності в Бк/ФЕП та майлару – мБк/дм$^2$), присутніх як домішки в матеріалах, що були застосовані в низькофонових установках.

| Матеріал | $^{40}$K | $^{137}$Cs | $^{226}$Ra | $^{228}$Th |
|---|---|---|---|---|
| Пластиковий сцинтилятор | ≤10 | | ≤2.3 | ≤0.09 |
| Мідь М0 | ≤28 | | ≤3.7 | ≤0.25 |
| Свинець | ≤4 | ≤1.4 | ≤2 | ≤1 |
| Борований поліетилен С1 | ≤10 | 70(10) | 12(2) | 0.8(0.4) |
| Тефлонова плівка | ≤4000 | ≤500 | ≤1500 | ≤120 |
| Оптична змазка (Франція) | ≤500 | ≤80 | ≤130 | ≤50 |
| Оптична змазка (СРСР) | ≤1000 | ≤90 | ≤600 | ≤6 |
| Майлар | | ≤0.05 | ≤0.02 | ≤0.005 |
| ФЕП ФЭУ–110 | 3.0(3) | | 0.8(2) | 0.17(7) |
| ФЕП ФЭУ–125нф | 3.0(2) | | 2.2(9) | 0.20(3) |
| ФЕП EMI 9390KFLB53 | 4.(1) | | 0.9(4) | 0.21(5) |

2.2.2. Детектор, активний та пасивний захист.

Дослід, метою якого було дослідження подвійного бета–розпаду ядра $^{116}Cd$ проводився в Солотвинській підземній лабораторії ІЯД НАНУ, спорудженій на глибині 430 м в шахті №9 Солотвинського солерудника в с. Солотвино Тячівського району Закарпатської області. Враховуючи склад і густину геологічних порід над лабораторією, глибина лабораторії, в перерахунку на еквівалентну товщину води, становить близько 1000 м. Потік космічних мюонів послаблюється приблизно в $10^4$ разів і становить близько $1.5\times10^3$ $м^{-2}$ за добу, потік теплових нейтронів не перевищує $3\times10^{-6}$ $см^{-2}\times сек^{-1}$, рівень радону становить $\approx$ 30 Бк $\times$ $м^{-3}$. Завдяки чистоті солі відносно радіоактивних елементів торію, урану, калію, гамма–фон в лабораторії в 30 – 50 разів нижчий, ніж в звичайній лабораторії на поверхні Землі. Температура в лабораторії, незалежно від пори року, становить 24 ± 1 ° С.

В експерименті були використані сцинтилятори $^{116}CdWO_4$, збагачені $^{116}Cd$ до 83%. Було використано чотири сцинтилятори $^{116}CdWO_4$, виготовлені з однієї булі, загальною масою 330 г. Сцинтилятори, обгорнуті трьома шарами тефлонової плівки (BICRON, BC – 642) та шаром алюмінізованого майлару були встановлені на світловод склеєний з пластикового сцинтилятора (BICRON, BC–412, довжиною 30 см) та надчистого кварцу (КУ–1, 25 см). Діаметр світловода становив 10 см. Світловод був виготовлений у формі логарифмічної спіралі [240], що забезпечило передачу близько 65% світла на фотопомножувач з рубідій–цезієвим фотокатодом, виготовлений з низькорадіоактивного скла (EMI D724KFL). Квантова чутливість цього ФЕП, виготовленого компанією THORN EMI спеціально для Солотвинського експерименту, сягала 30% в максимумі чутливості (≈ 400 нм). Оптичні контакти між сцинтиляційними кристалами світловодами та ФЕП забезпечувались силіконовою змазкою (Dow Corning, Q2–3067).

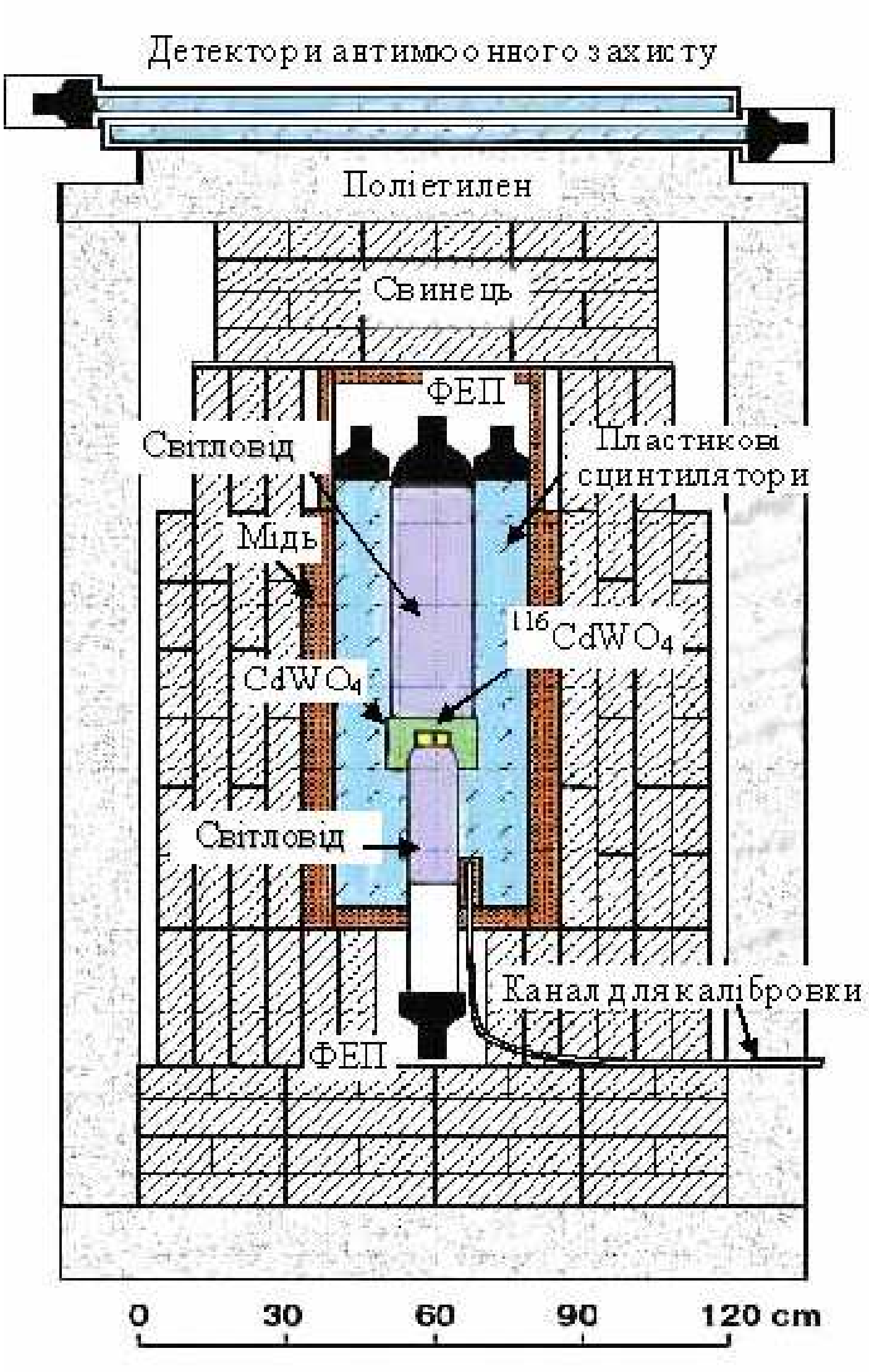

Рис. 2.2. Схема низькофонового сцинтиляційного спектрометра з кристалами $^{116}CdWO_4$.

Кристали $^{116}CdWO_4$ були оточені 15 сцинтиляторами $CdWO_4$ загальною масою 20.6 кг. Ці сцинтилятори проглядалися як одне ціле низькофоновим ФЕП (ФЭУ–125нф) діаметром 18 см через активний пластиковий світлопровід діаметром 17 см та довжиною 49 см (BICRON, BC–412). Обидва детектори, з кристалів $^{116}CdWO_4$ та $CdWO_4$, були розташовані всередині додаткового активного захисту, зробленого з чотирьох блоків пластикових сцинтиляторів на основі полістиролу. Розміри пластикового захисту становили 40×40×95 см. Світло від цих пластикових сцинтиляторів реєструвалося двома низькорадіоактивними ФЕП (ФЭУ–164). Таким чином, разом з обома світловодами детекторів $^{116}CdWO_4$ та $CdWO_4$, було досягнуто 4π–геометрії активного захисту основного детектора ($^{116}CdWO_4$).

Пасивний захист складався з електролітичної міді марки М0 (шар завтовшки 3–6 см), свинцю (22.5–30 см) та поліетилену (8 см чистого та 8 см насиченого бором). Для захисту від радону мідний пасивний захист був герметизований силіконовим клеєм від проникнення повітря.

Два сцинтиляційних детектори були встановлені над установкою й використовувались як лічильники космічних мюонів. В детекторах, корпус яких був виготовлений з титану, пластикові сцинтилятори розмірами $120 \times 130 \times 3$ см проглядалися чотирма ФЕП (ФЭУ–137).

Схема низькофонового сцинтиляційного спектрометра з кристалами $^{116}CdWO_4$ зображена на рис. 2.2.

Всі матеріали, використані в спектрометрі, були ретельно відібрані за рівнем домішок урану, торію, калію. Вимірювання були здійснені на спеціальному низькофоновому детекторі (п. 2.2.1). Крім того, був зроблений аналіз всіх попередніх експериментів по пошуку рідкісних процесів, що були проведені у відділі фізики лептонів ІЯД. Використовувалися також дані, опубліковані в наукових джерелах.

### 2.2.3. Електроніка.

Імпульси з аноду ФЕП EMI D724KFL через повторювач, виконаний на диференціальному підсилювачі (HARRIS, HA–5033), надходили на попередній підсилювач з трьома виходами. З першого виходу імпульс надходив на спеціально розроблений блок (блок «АТ»), де формувались «часові» (точність часової прив'язки для енергій > 50 кеВ становила 0.3 мкс) та спектрометричні сигнали. Для вироблення спектрометричних сигналів вхідні імпульси (нагадаємо, що постійна висвічування сцинтиляторів вольфрамату кадмію становить близько 13 мкс) інтегрувались протягом ≈ 35 мкс, потім з проінтегрованого імпульсу вироблявся короткий імпульс (2 мкс), що надходив на блок аналогово–цифрового перетворювача (АЦП) БПА2–97. Крім того, в блоці «АТ» здійснювався попередній відбір шумів ФЕП та імпульсів, утворених накладанням сцинтиляцій від пластикового світловоду на сцинтиляційні сигнали від $^{116}CdWO_4$ (такі події могли бути викликані, наприклад, комптонівським розсіянням гамма–кванта в пластиковому світловоді, а потім його поглинанням в кристалі $^{116}CdWO_4$). У випадку наявності таких накладань блок виробляв логічний сигнал «пластик + $CdWO_4$», що забороняв реєстрацію форми сигналу (див. нижче). З другого виходу імпульси надходили на вхід 12–тибітного оцифровщика форми, що працював з частотою 20 МГц (записувались 2048 каналів по 50 нс). Сигнал з третього виходу попереднього підсилювача з постійною формування близько 200 мкс надходив через спектрометричний підсилювач БУС2–97 на диференційний дискримінатор БСА2–95, що дозволяв вибирати діапазон енергій імпульсів від сцинтиляторів $^{116}CdWO_4$ для оцифровки. Тригер для запуску оцифровщика форми вироблявся у випадку збігів «часових» та «енергетичних» (з виходу блока БСА2–95) сигналів, а також за умови відсутності сигналів «пластик+$CdWO_4$». Таким чином, форми сцинтиляційних імпульсів від детектора з кристалами $^{116}CdWO_4$ записувались в діапазоні енергій приблизно 0.1–6 МеВ, за умови відсутності домішок сигналів від пластикового світловода.

Сигнали від фотопомножувача ФЭУ–125нф, що реєстрував сцинтиляційні сигнали від захисного $CdWO_4$ детектора, через повторювач, аналогічний тому, що використовувався в каналі основного детектора, надходили на попередній підсилювач з 2–ма виходами. З першого виходу сигнали надходили на блок «АТ», в якому вироблявся спектрометричний сигнал, з іншого – на вхід підсилювача (БУИ2–92) з постійними диференціювання та інтегрування 5 і 4 мкс, відповідно. Формування сигналу часової прив‘язки захисного $CdWO_4$ детектора здійснювалось за допомогою формувача з відслідковуючим порогом (constant fraction discriminator) з точністю часової прив‘язки ±0.2 мкс і ефективним енергетичним порогом $\approx 100$ кеВ. Сигнал часової прив‘язки захисного $CdWO_4$ детектора надходив на блок збігів (БСВ2–90). У випадку збігів з часовими сигналами від детектора $^{116}CdWO_4$, логічний сигнал (“$^{116}CdWO_4$ & $CdWO_4$”) був тригером для АЦП (БПА–97), куди надходив спектрометричний сигнал від детектора $CdWO_4$. Крім того, сигнал збігів “$^{116}CdWO_4$ & $CdWO_4$” надходив на вхід блоку ознак, що функціонував в крейті САМАС. Це дозволяло системі реєстрації (див. нижче) записувати для всіх подій інформацію про збіги сигналів в основному детекторі ($^{116}CdWO_4$) з сигналами від детекторів активного захисту і в подальшому використовувати цю інформацію в ході “off line” аналізу.

Сигнали від ФЕП ФЭУ–167, що проглядали пластикові сцинтилятори, після підсилення в попередніх підсилювачах надходили на порогові дискримінатори. Ефективний енергетичний поріг пластикових сцинтиляторів становив $\approx 200$ кеВ. Логічні сигнали з виходів дискримінаторів сумувались і надходили на схему збігів з сигналами основного детектора. Сигнали збігів подавались на блок ознак системи реєстрації.

Сигнали від детекторів космічних мюонів після підсилення в попередніх підсилювачах надходили на порогові дискримінатори. Енергетичні пороги дискримінаторів були вибрані на рівні близько 1.5 МеВ. При цьому мюони, енергетичні втрати яких становлять близько 2 МеВ на грам речовини (отже в троьхсантиметровому пластиковому сцинтиляторі виділялась енергія близько 6 МеВ) надійно реєструвались при проходженні через пластикові сцинтилятори. Для надійного виділення мюонних подій детектори працювали в режимі збігів. З цих сигналів збігів (частота таких подій була дуже малою, близько 0.01 Гц) формувався логічний сигнал довжиною 470 мкс, який надходив на схему збігів з часовими сигналами основного детектора. Таким чином, всі події, які надходили протягом цього часу від основного детектора, записувались з ознакою „збігів з мюоном” і відкидались в ході подальшого аналізу як фонові. Такий значний час був вибраний для того, щоб надійно відкинути події від захоплення в матеріалах пасивного захисту установки теплових нейтронів, які можуть виникати під дією мюонів.

Контроль живого часу спектрометра здійснювався за допомогою світловипромінюючого діода. Світовий сигнал від діода з частотою 1 сигнал за приблизно 5 хвилин по кварцовому світловоду надходив на ФЕП основного детектора. В ході аналізу даних можна було точно визначити кількість сигналів, що надійшли на світлодіод і оцінити мертвий час системи реєстрації. В залежності від порогу реєстрації, він становив близько 4%. Мертвий час детектора визначався, головним чином, часом передачі оцифрованих імпульсів з блока оцифровщика в комп'ютер (≈10 мс).

Оскільки сіль має досить високий опір, заземлення електронної апаратури в лабораторії є неможливим. Експериментальним шляхом було встановлено, що найнижчий рівень радіонаводок забезпечується при умові ізоляції установки від соляного оточення лабораторії. Корпус установки (свинцевий захист) і стійка з електронікою були електрично з'єднані. Кабелі, що з'єднували установку зі стійкою з електронними блоками, були обгорнуті мідною фольгою, приєднаною до стійки. Комп'ютери системи реєстрації були розміщені в герметичних, електрично ізольованих від електронної апаратури боксах. Це значно знизило рівень наводок від терміналів комп'ютерів.

В ході вимірювань, частоти надходження сигналів від основного та детекторів активного захисту один раз на добу контролювались оператором і, в разі відхилення від контрольних значень, проводилось регулювання порогів дискримінаторів та високої напруги на ФЕП.

В зв'язку з тим, що на шахті №9 йде промислова добича солі, аварійні відключення електропостачання є звичним явищем. Щоб уникнути втрат інформації і виходів з ладу електроніки, живлення наднизькофонового спектрометра здійснювалось від системи безперебійного живлення. Система складалась із джерела постійної напруги, батареї акумуляторів на 400 ампер-годин та перетворювача постійного струму в змінний (220 В, 50 Гц). У випадку відключення шахтного електропостачання спектрометр міг працювати 10–15 годин за рахунок акумуляторів, що дозволяло проводити довгострокові вимірювання безперервно.

### 2.2.4. Система реєстрації даних.

Система реєстрації спектрометра складалась з двох IBM–сумісних персональних комп'ютерів та крейта CAMAC з електронним блоками. Для кожної події в сцинтиляторах $^{116}CdWO_4$, та кожної події в сцинтиляторах $CdWO_4$, яка збігалась у часі з подією в детекторі $^{116}CdWO_4$, на жорсткий диск першого комп'ютера записувалась наступна інформація: амплітуда та час появи сигналу, допоміжні ознаки (збіг між основним та захисними детекторами, сигнал запуску світловипромінюючого діода та тригер зчитування сигналу від оцифровувача форми імпульсів). Спектрометрична інформація, що надходила з аналого–цифрових перетворювачів БПА2–97, а також вміст регістрів блоку ознак та блоку таймера зчитувались через контролер крейта CAMAC (типу U2) і плату зв'язку (ЕС1842.0102) в персональний комп'ютер (486 IBM, 66 МГц).

Другий ПК записував форму імпульсів (2048 каналів, шириною 50 нс кожний) від детектора $^{116}CdWO_4$ в інтервалі енергій 0.3–5.4 МеВ (в додаткових вимірюваннях нижній поріг було встановлено на рівні $\approx$ 80 кеВ). Оцифровувач форми імпульсів був побудований на основі 12–бітного аналого–цифрового перетворювача (Analog Devices AD9022). Оцифрована форма імпульсу передавалась через плату зв'язку PC–DIO–24. Обидва комп'ютери були з'єднані між собою також за допомогою таких плат. Другий ПК працював під керуванням першого і зчитував оцифровану форму сигналів лише у випадку, коли параметри події (номер каналу АЦП, значення регістру ознак, тощо) відповідали заданим. При цьому перший ПК передавав, по–перше, всю інформацію про подію, яка записувалась на жорсткий диск другого ПК разом з оцифрованою формою сигналу, і, по–друге, порядковий номер події. Таким чином, за порядковим номером можна було встановити відповідність між інформацією про подію, записаною на різних ПК.

Використання двох ПК дало змогу записувати на диск першого ПК енергію, час та інформацію з блока ознак для тих подій, котрі слідували після події, для якої записувалась вся інформація (тобто, і форма сигналів), навіть коли час після першої події був меншим ніж 10 мс (час передачі оцифрованої форми імпульсу). Мінімальний час між такими подіями складав 95 мкс. Інформацію про події, час між виникненням яких був меншим 95 мкс, можна було одержати, аналізуючи записану форму сигналів. Це дало змогу аналізувати події від розпадів короткоживучих радіонуклідів торієвого та уранових рядів, що присутні в сцинтиляторах $^{116}CdWO_4$.

2.2.5. Проведення калібровочних вимірювань.

2.2.5.1.Енергетична роздільна здатність.

Енергетична калібровка та енергетична роздільна здатність спектрометра із сцинтиляційними кристалами $^{116}CdWO_4$ визначались за допомогою джерел γ–квантів ($^{22}Na$, $^{40}K$, $^{60}Co$, $^{137}Cs$, $^{207}Bi$, $^{226}Ra$, $^{232}Th$ та $^{241}Am$) у відкритій установці. Залежність роздільної здатності від енергії в діапазоні 60–2615 кеВ можна записати наступним чином: $FWHM_\gamma = -44 + (2800 + 23.4 \times E_\gamma)^{-1/2}$, де величини $E_\gamma$ та $FWHM_\gamma$ виражені в кеВ. Наприклад, енергетична роздільна здатність для γ–ліній з енергіями 60, 662, 1064 та 2615 кеВ становило, 33.7%, 13.5%, 11.5% та 8.0%, відповідно.

В ході експерименту калібровки проводились раз на тиждень (а також після кожного відключення високої напруги на ФЕП) за допомогою джерел γ–квантів $^{207}Bi$ та $^{232}Th$. Джерела вводились в установку через спеціальний канал в мідний коліматор. Опромінення детектора $^{116}CdWO_4$ здійснювалось через пластиковий світловод, тому частота слідування сигналів була досить високою (≈ 5 кГц). Оскільки час інтегрування сигналу для отримання спектрометричної інформації складав близько 40 мкс, це призводило до деякого зміщення енергетичної шкали спектрометра. Тому виникла необхідність урахування цього ефекту і коригування енергетичної шкали детектора. Були проведені дві серії вимірювань. Спочатку енергетичну калібровку було проведено у відкритій установці, коли сцинтиляційні кристали опромінювались зверху і з такої відстані, коли частота сигналів була достатньо малою, щоб не викликати помітного зміщення піків. Потім вимірювання були проведені з джерелами $\gamma$–квантів, введеними в канал для калібровки, тобто в режимі, в якому в ході вимірювань потім здійснювалась калібровка. Таким чином були визначені поправки, які дали змогу отримати вірну енергетичну калібровку детектора в повному захисті (коли частота слідування подій була дуже малою $\approx 1 - 2$ Гц) по даним калібровки „через пластиковий світловод”.

Проведення систематичних енергетичних калібровок дозволило визначити і стабілізувати енергетичну шкалу спектрометра. Протягом довгострокових вимірювань спостерігався деякий дрейф підсилення в спектрометричному каналі. Крім того, зміни підсилення спостерігались після аварійних відключень та, очевидно, після корекції високої напруги на ФЕП оператором. Але ці нестабільності були усунуті в ході аналізу даних “off line” шляхом зведенням спектрів до однієї стандартної енергетичної калі бровки за допомогою процедури пересипки спектрів [241]. Про стабільність спектрометра свідчить той факт, що енергетичне розділення піку $^{137}$Cs у фоновому спектрі (накопиченому протягом 13 316 годин) дорівнює 14.0%, що практично співпадає із значенням (13.5%), отриманим в ході калібровочних вимірювань протягом 2 годин.

2.2.5.2. Відгук до альфа–частинок.

В ході калібровочних вимірювань був також виміряний відносний світловихід при опроміненні сцинтиляторів $^{116}CdWO_4$ $\alpha$–частинками, так зване $\alpha/\beta$–співвідношення. $\alpha/\beta$–Співвідношення можна визначити як відношення енергії $\alpha$–частинок в $\gamma$–шкалі детектора ($E_\alpha^\gamma$) до їх дійсної енергії ($E_\alpha$).

$\alpha/\beta$–Співвідношення було виміряне в діапазоні енергій $\alpha$–частинок 0.5–8.8 МеВ за допомогою колімованого джерела $\alpha$–частинок $^{241}Am$ та поглиначів, що складались з набору майларових плівок товщиною ≈ 0.65 мг/см$^2$. Енергія $\alpha$–частинок після проходження шару повітря в коліматорі (1.0 мм) вираховувалась за допомогою

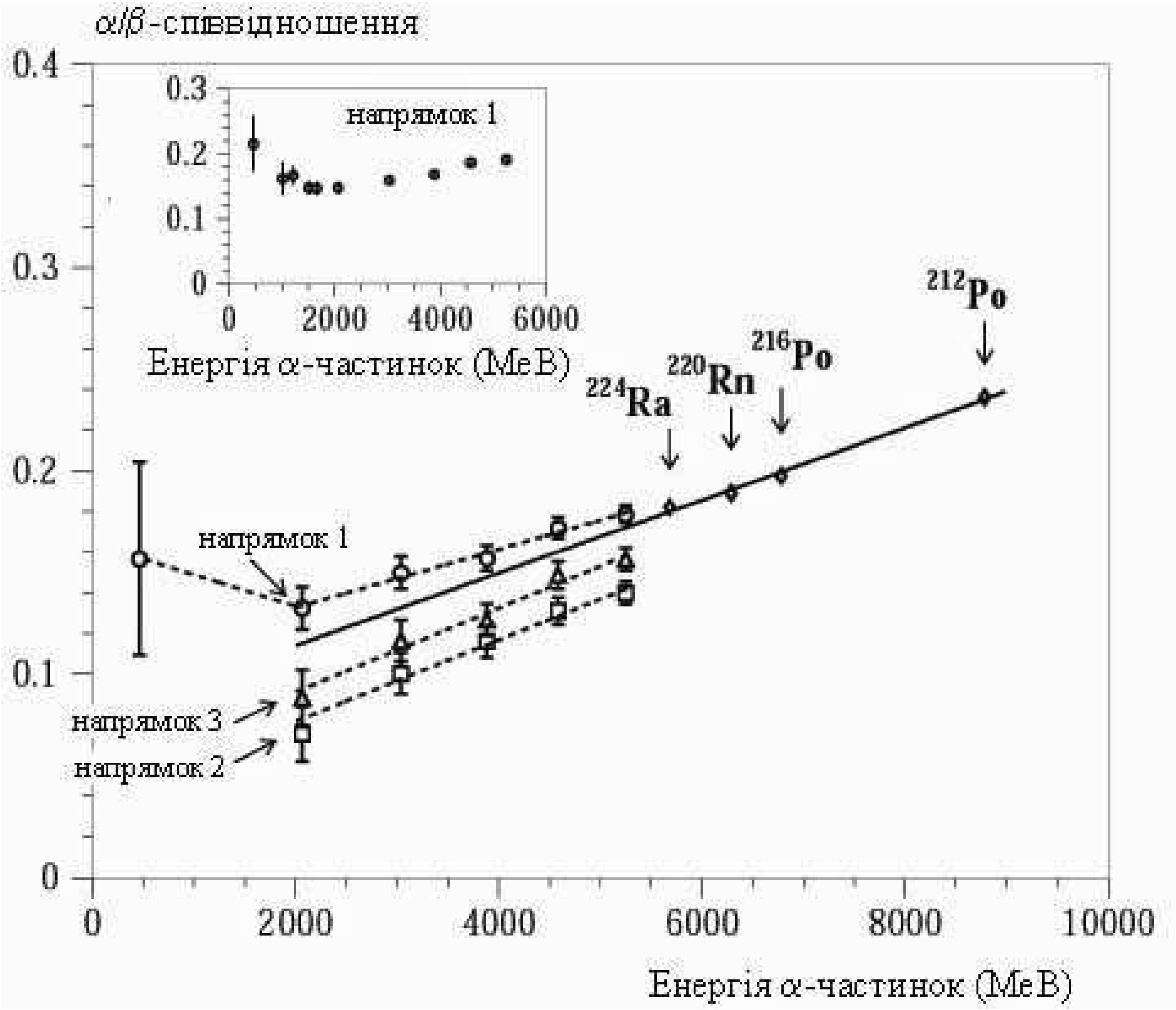

Рис. 2.3. Залежність αβ–співвідношення від енергії та напрямку α–частинок відносно основних кристалографічних осей кристала $^{116}CdWO_4$. Кристал опромінювався перпендикулярно до (010), (001) і (100) кристалічних площин (напрямки 1, 2 і 3, відповідно). Альфа–піки $^{224}Ra$, $^{220}Rn$, $^{216}Po$, $^{212}Po$ були виділені з даних довгострокових вимірювань за допомогою часово–амплітудного аналізу та аналізу за формою сцинтиляційного сигналу. На вставці показана залежність α/β–співвідношення від енергії α–частинок, виміряна з кристалом $CdWO_4$ розмірами ∅ 25 × 0.9 мм в напрямку 1.

пакета GEANT, а після проходження поглиначів – вимірювалась поверхнево–бар’єрним детектором. Крім того, $\alpha/\beta$–співвідношення було визначено для $\alpha$–частинок, що випромінювались всередині кристалів $^{116}CdWO_4$ радіонуклідами $^{224}Ra$, $^{220}Rn$, $^{216}Po$ (енергії $\alpha$–частинок 5685, 6288 і 6778 кеВ, відповідно) від слідового забруднення кристалів торієм. Події $\alpha$–розпадів цих нуклідів були відібрані за допомогою часового–амплітудного аналізу та аналізу форми імпульсу. Була також виявлена залежність $\alpha/\beta$–співвідношення від напрямку $\alpha$–частинок відносно головних кристалічних осей кристалів $^{116}CdWO_4$. Виміряна залежність $\alpha / \beta$–співвідношення від енергії та напрямку $\alpha$–частинок відносно основних кристалографічних осей кристалу $CdWO_4$ показана на рис. 2.3. Для вольфраматів кадмію не було літературних даних про такі залежності. На основі проведених вимірювань було визначено таку залежність $\alpha/\beta$–співвідношення та енергетичної роздільної здатності від енергії: $\alpha/\beta = 0.083(9) + 1.68(13) \times 10^{-5} E_\alpha$, та $FWHM_\alpha$(кеВ) $= 33 + 0.247 \times E_\alpha$, де $E_\alpha$ – енергія $\alpha$–частинок в кеВ. Слід зазначити, що явище залежності світловиходу для $\alpha$–частинок від енергії та напрямку опромінення давно відоме для інших сцинтиляційних кристалів (наприклад, стильбену [242]).

2.2.5.3. Вимірювання форми сцинтиляційних сигналів.

Калібровочні вимірювання включали в себе також визначення форми сцинтиляційних сигналів та дослідження ефективності ідентифікації за формою для детектора з кристалами $^{116}CdWO_4$ при опромінені $\gamma$–квантами та $\alpha$–частинками в широкому діапазоні енергій. Було також вивчено залежність форми сцинтиляційних сигналів та якості розділення сигналів від $\gamma$–квантів та $\alpha$–частинок в діапазоні температур –5 – +30 °C.

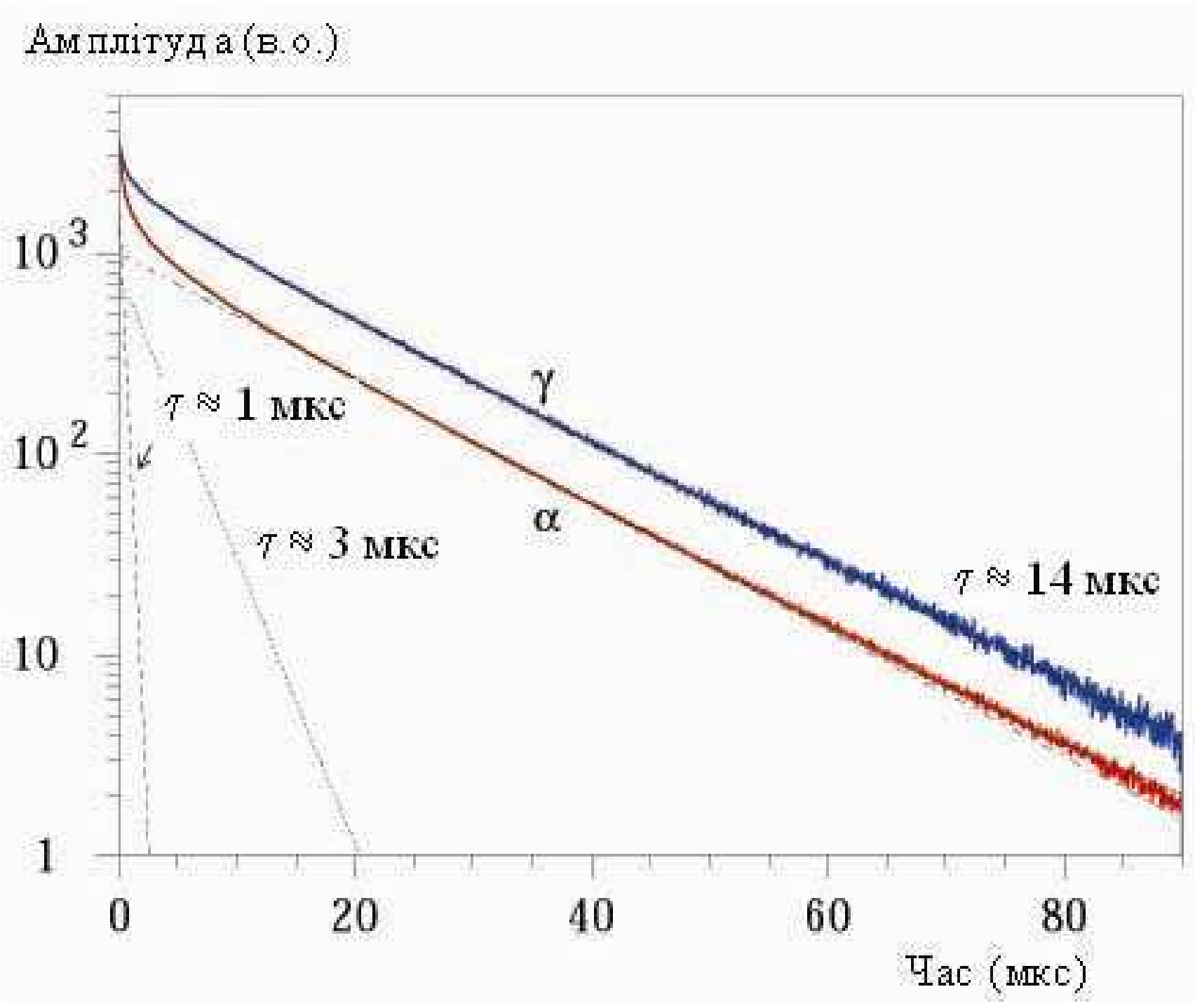

Рис. 2.4. Форма сцинтиляційних сигналів в кристалі $CdWO_4$ при опроміненні α–частинками та γ–квантами. Можна виділити принаймні три компоненти сцинтиляційного спалаху з часом спаду ≈ 1 мкс, ≈ 3 мкс і ≈14 мкс.

Форма сцинтиляційних сигналів кристалів $^{116}CdWO_4$ може бути описана сумою експоненційних функцій:

$$f(t) = \Sigma A_i /( \tau_i - \tau_0)(e^{-\tau/\tau i} - e^{-\tau/\tau 0}), \qquad (2.3)$$

де $A_i$ – інтенсивності, $\tau_i$ – постійні затухання різних складових сцинтиляційного спалаху, а $\tau_0$ – час інтегрування електронного тракту (≈0.2 мкс). Для аналітичного опису α та γ сигналів, форму імпульсу було підігнано сумою трьох (для α–частинок) і двох (для γ–квантів) експоненціальних функцій, і, таким чином, отримано аналітичні вирази $f_\alpha(t)$ та $f_\gamma(t)$ (рис. 2.4). Для сцинтиляторів $^{116}CdWO_4$, було отримані такі значення параметрів, що описують форми імпульсів для альфа–частинок та гамма–квантів: $A_1^\alpha$=80.9(1.9)%, $\tau_1^\alpha$ = 12.7(0.6) мкс, та $A_2^\alpha$ = 13.4(1.3)%, $\tau_2^\alpha$=3.3(1.1) мкс, $A_3^\alpha$=5.7(1.0)%, $\tau_3^\alpha$ = 0.96(0.08) мкс для

α–частинок з енергією ≈5 МеВ, та $A_1^{\gamma}$=94.3(0.3)%, $\tau_1^{\gamma}$=13.6(0.2) мкс та $A_2^{\gamma}$=5.7(0.2)%, $\gamma_2^{\gamma}$=2.1(0.1) мкс для γ–квантів ≈1 МеВ.

Відомо, що час люмінесценції і, зокрема, час висвічування сцинтиляторів скорочується з підвищенням температури. Були проведені вимірювання залежності від температури усередненого часу висвічування (τ) сцинтиляторів $CdWO_4$ при опроміненні α–частинками та γ–квантами (рис. 2.5). В діапазоні температур від 0 до 30 °C величина τ лінійно спадає, в межах похибки однаково для обох типів випромінювань, з коефіцієнтом –0.05 мкс/(градус С).

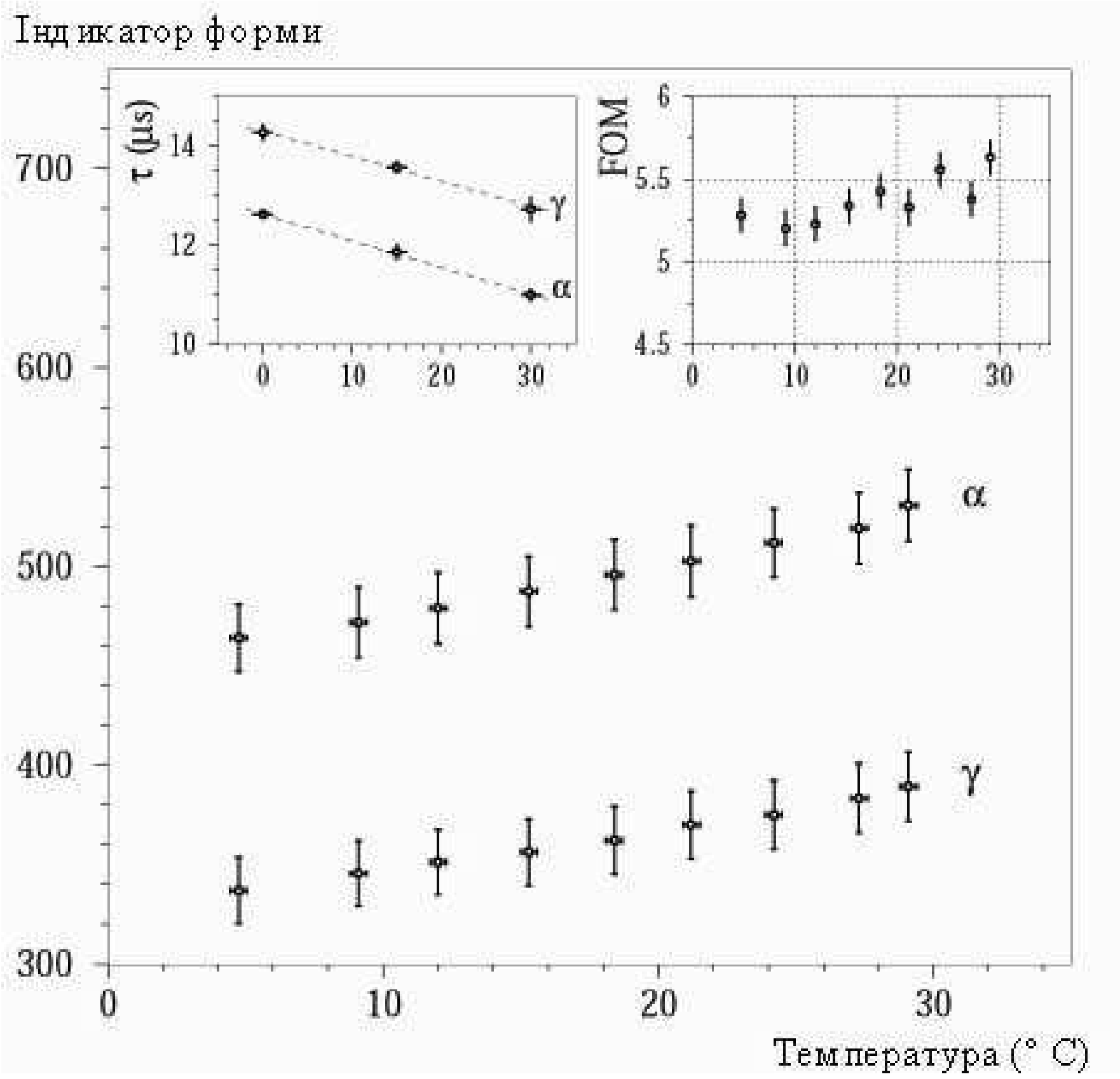

Рис. 2.5. Залежність індикатора форми, усередненого часу світіння (τ), та показника якості дискримінації (FOM) від температури кристала $CdWO_4$.

2.2.6. Дискримінація подій за формою сцинтиляційних сигналів.

Альфа–розпади радіонуклідів торієвого та уранових рядів всередині чутливого об‘єму детектора або на його поверхні можуть бути суттєвою компонентою фону детектора. Скажімо, автори експерименту по пошуку 2β–розпаду $^{130}$Te за допомогою низькотемпературних болометрів з кристалами оксиду телуру вважають, що саме поверхневі забруднення кристалів альфа–активними ізотопами є головним джерелом фону в цьому експерименті [ 243 ]. На жаль, техніка низькотемпературних болометричних вимірювань, застосована у вищезгаданому експерименті, не дозволяє відрізнити події α–розпадів від шуканих подій 2β–розпаду. В той же час, ця задача, як правило, може бути розв‘язана у випадку використання сцинтиляційних детекторів. Треба зазначити, що для успішного здійснення високочутливого експерименту по пошуку 2β–розпаду необхідно, крім ідентифікації подій від α–частинок та γ–квантів, вирішити значно ширше коло задач розпізнавання сигналів за формою. Це, в першу чергу, вдосконалення методів виявлення подій ланцюжків розпадів короткоживучих радіонуклідів торієвого та уранових рядів, таких як $^{212}$Bi та $^{214}$Bi:

$^{212}$Bi ($Q_\beta$ =2.25 MeB) → $^{212}$Po ($E_\alpha$ = 8.79 MeB, $T_{1/2}$ = 0.3 мкс)→ $^{208}$Pb,

$^{214}$Bi ($Q_\beta$ =3.27 MeB) → $^{214}$Po ($E_\alpha$ = 7.83 MeB, $T_{1/2}$ = 164 мкс)→ $^{210}$Pb.

Річ у тім, що розпади цих радіонуклідів (наприклад в сцинтиляторах $CdWO_4$) можуть давати фонові події з енергіями до 4 – 5 МеВ.

Крім того, є ще багато інших типів фонових подій, які можуть бути відкинуті за допомогою аналізу форми сигналів. Це випадкові збіги різної природи подій радіоактивних розпадів, накладення імпульсів від пластикового сцинтилятора (якщо цей матеріал використовується в якості світловода), радіонаводки та збої в електронних системах реєстрації, шуми фотопомножувача та інші події, форма яких відрізняється від форми подій досліджуваного ефекту. Компонента фону, пов‘язана з шумами фото помножувача, особливо відчутна при пошуку і вимірюваннях низькоенергетичних процесів, таких як двохнейтринний подвійний електронний захват, бета–розпад з великими періодами напіврозпаду, рідкісні альфа–розпади. Шуми фотопомножувачів є головним джерелом фону в сцинтиляційних експериментах направлених на пошуки “темної матерії” Всесвіту. В той же час, прості методи дискримінації за формою (такі як метод найменших квадратів, “середнього часу”, метод максимальної правдоподібності і т.д.) не здатні ефективно виявляти складні за часовими та амплітудними властивостями події (як, наприклад, розпади $^{212}$Bi та $^{214}$Bi), а також визначати їх точні характеристики (вимірювати бета– і альфа–спектри, визначати часові властивості цих розпадів). Це важливо не лише для їх відкидання, але і для коректної інтерпретації виміряного спектра, що завжди суттєво в наднизькофонових експериментах. Тому, на наш погляд, необхідною є розробка більш досконалих методів розпізнавання образів, таких, зокрема, як методи штучних нейронних мереж [244].

Різниця у формі сцинтиляційних сигналів при реєстрації $\alpha$–частинок та $\gamma$–квантів була використана для ідентифікації подій в сцинтиляторах вольфрамату кадмію [245]. Для цього був застосований метод оптимального цифрового фільтра, запропонований в [246]. У процесі обробки даних цифровий фільтр застосовувався до кожного експериментального сигналу $f(t_k)$ з метою отримання числової характеристики його форми — так званого індикатора форми ($SI$):

$$SI = \Sigma f(t_k) \times P(t_k) / \Sigma f(t_k), \qquad (2.4)$$

де сумування проводиться по часовим каналам k, починаючи з початку сигналу і до 50 мс, $f(t_k)$ – оцифрована амплітуда даного сигналу. Вагова функція $P(t)$ описується так:

$$P(t) = [f_\alpha(t) - f_\gamma(t)]/[f_\alpha(t) + f_\gamma(t)], \qquad (2.5)$$

де $f_\alpha(t)$ і $f_\gamma(t)$ – форми імпульсів для $\alpha$–частинок і $\gamma$–квантів відповідно. На рис. 2.6 показані розподіли індикаторів форми для сцинтиляційних сигналів від $\gamma$–квантів та $\alpha$–частинок в кристалі $CdWO_4$. Одержані розподіли добре описуються функціями

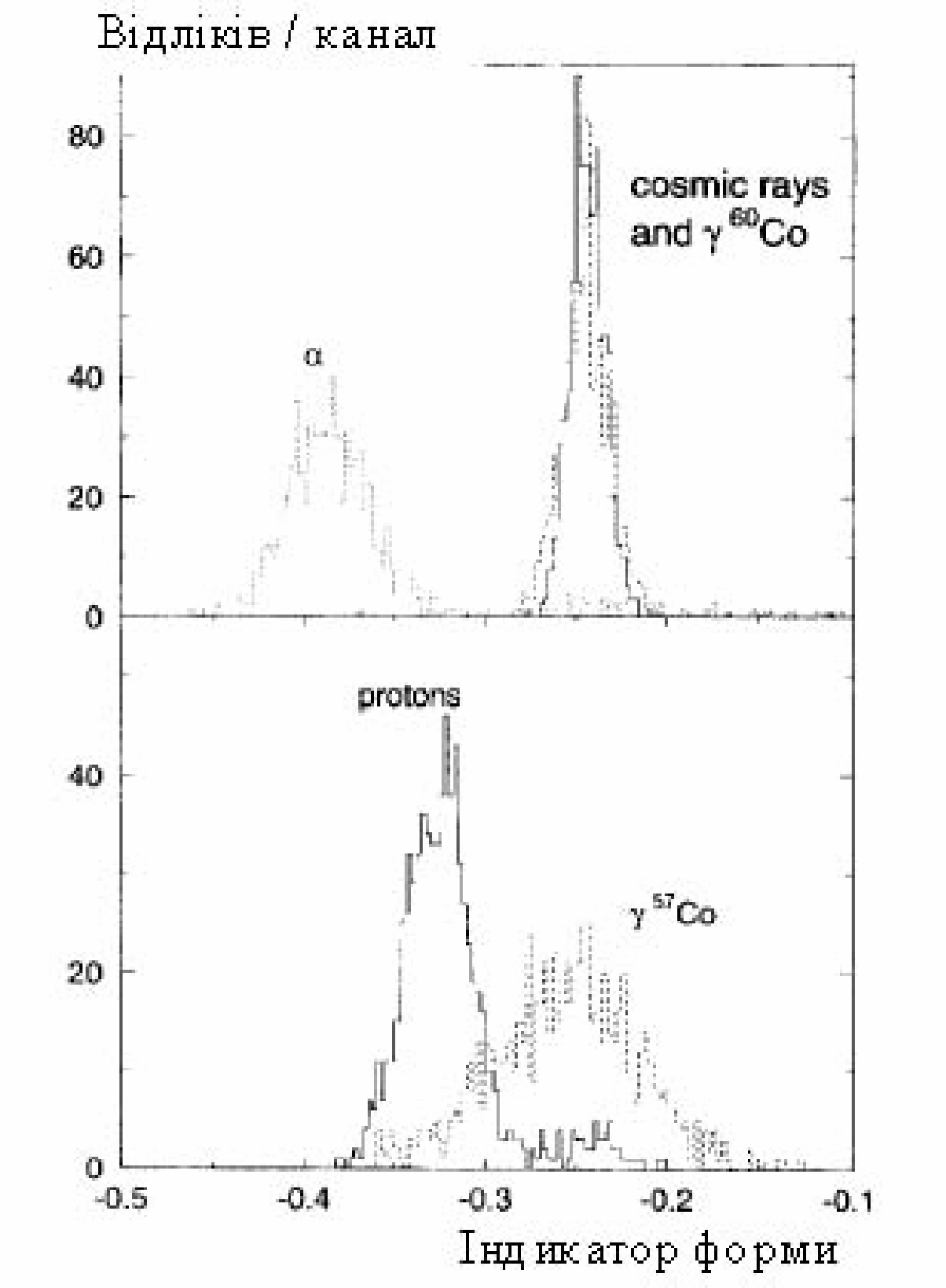

Рис. 2.6. Розподіли індикаторів форми, виміряних із сцинтиляційним кристалом вольфрамату кадмію для α–частинок, γ–квантів, космічних мюонів та протонів.

Гауса, тому для оцінки ефективності дискримінації можна використовувати таку величину (FOM – factor of merit):

$$FOM = |SI_\alpha - SI_\gamma|/(\sigma_\alpha^2 + \sigma_\gamma^2)^{1/2} \tag{2.6}$$

яку ми назвемо якістю дискримінації. Тут $SI_\alpha$, $\sigma_\alpha$, $SI_\gamma$, $\sigma_\gamma$ — значення центрів ваги та дисперсій розподілів індикаторів форм для α–частинок та γ–квантів відповідно. З формули 2.4 видно, що флуктуації індикатора форми (а отже значення величин $\sigma_\alpha$ та $\sigma_\gamma$ у формулі 2.6) визначаються флуктуаціями оцифрованого сигналу. Таким чином, величина FOM, при певній різниці у формах сигналів, має характер оберненої кореневої залежності від кількості фотоелектронів, що досягають першого динода фотопомножувача. А отже, якість дискримінації для даного сцинтилятора буде кращою за умови найбільш повного збору сцинтиляційних фотонів на фотокатод фотопомножувача, високої квантової ефективності фотокатода, максимального збору фотоелектронів на першому диноді.

Для забезпечення найбільшого значення величини FOM важливо, враховуючи залежність часу висвічування сцинтиляторів від температури, стабілізувати температуру кристалів протягом експерименту. На рис. 2.5 показано залежності індикаторів форми для $\alpha$–частинок та $\gamma$–квантів від температури. Очевидно, що варіація температури сцинтилятора призведе до збільшення ширини розподілів індикаторів форми, а отже, до погіршення якості дискримінації. В той же час, при сталій температурі величина FOM практично не залежить від температури в діапазоні від 0 до 30$^{o}$ C.

Крім того, на якість дискримінації $\alpha$–частинок та $\gamma$–квантів в сцинтиляторах $CdWO_4$ впливає той факт, що форма сцинтиляційних сигналів для $\alpha$–частинок залежить від енергії та напрямку опромінення відносно кристалічних осей кристалу. Форма сцинтиляційних сигналів була виміряна в діапазоні енергій $\alpha$–частинок 0.5–8.8 МеВ за допомогою колімованого джерела $\alpha$–частинок $^{241}$Am та поглиначів з майлару. Так само як і у випадку $\alpha$ / $\beta$–співвідношення, була виявлена залежність форми сцинтиляційних сигналів для $\alpha$–частинок від напрямку опромінення відносно головних кристалічних осей сцинтиляторів $^{116}CdWO_4$ (рис. 2.7). На цьому рисунку показані також точки, що відповідають $\alpha$–частинкам від розпадів дочірніх нуклідів торієвого ряду: $^{224}$Ra, $^{220}$Rn і $^{216}$Po, одержані за допомогою часово–амплітудного аналізу даних довгострокових фонових вимірювань. Енергетична залежність форми сцинтиляційних сигналів для $\alpha$–частинок може бути пояснена енергетичною залежністю щільності іонізації для $\alpha$–частинок і притаманна всім сцинтиляторам, в той час як відмінність форми сигналів при опроміненні в різних напрямках спостерігається лише в окремих кристалах [242]. Була спостережена залежність форми сцинтиляційних сигналів від напрямку опромінення α–частинками в сцинтиляційних монокристалах вольфраматів кадмію та цинку ($ZnWO_4$), в той час як в кристалах вольфрамату кальцію ($CaWO_4$) та фториду церію ($CeF_3$) такої залежності не було виявлено.

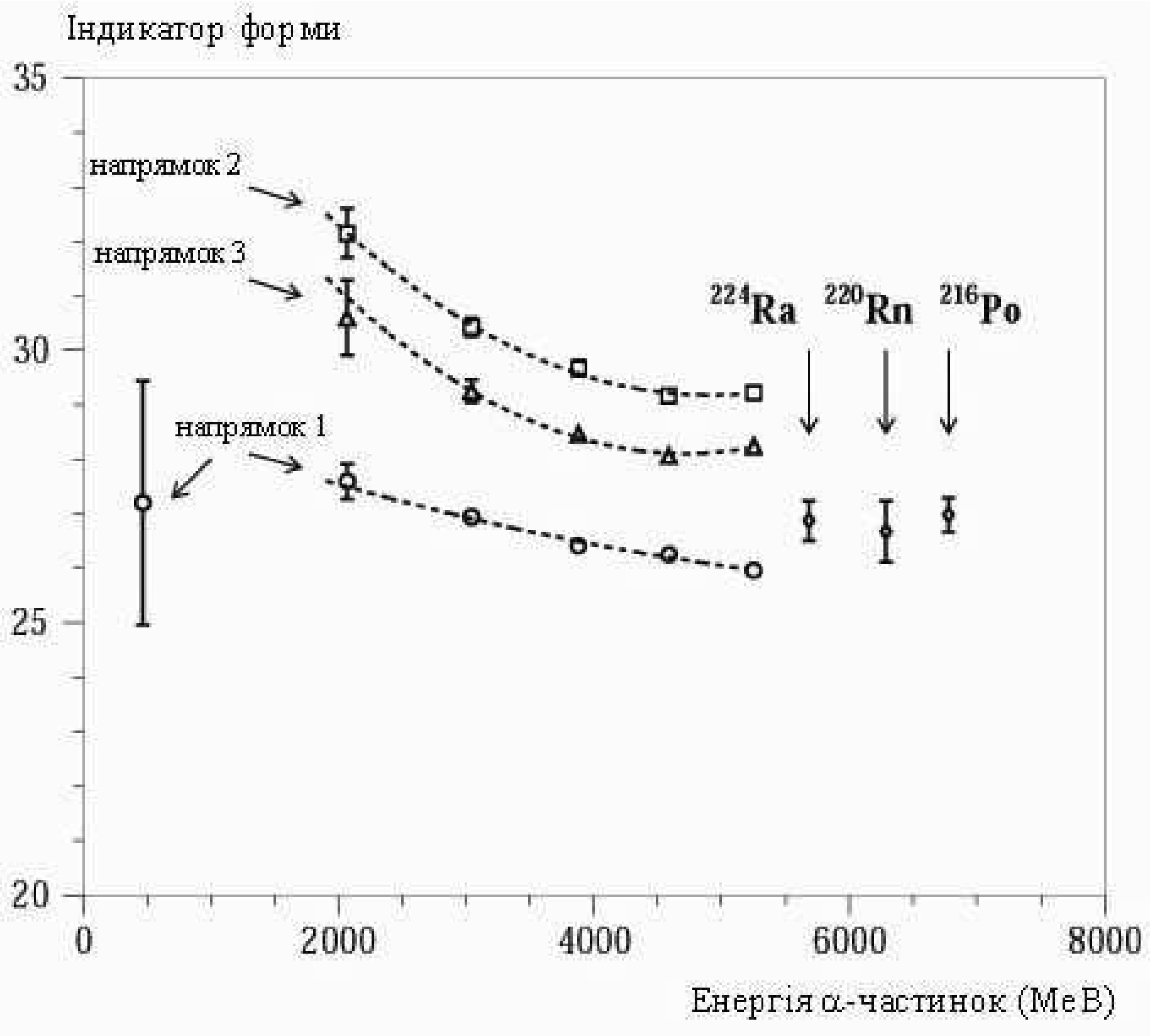

Рис. 2.7. Залежність індикатора форми від енергії та напрямку опромінення відносно головних кристалографічних осей кристала $^{116}CdWO_4$.

Як можна бачити на рис. 2.8, події від гамма–квантів (електронів) та альфа–частинок добре відрізняються при енергіях вище $E_\gamma \geq 0.8$ МеВ (що відповідає енергіям альфа–частинок $E_\alpha \geq 4$ МеВ). Хоча дискримінація по формі імпульсу і погіршується при низьких енергіях, але навіть α–частинки з енергією 2 МеВ ($E_\alpha^\gamma \geq 0.3$ МеВ) все ще можуть бути відокремлені від гамма–фону.

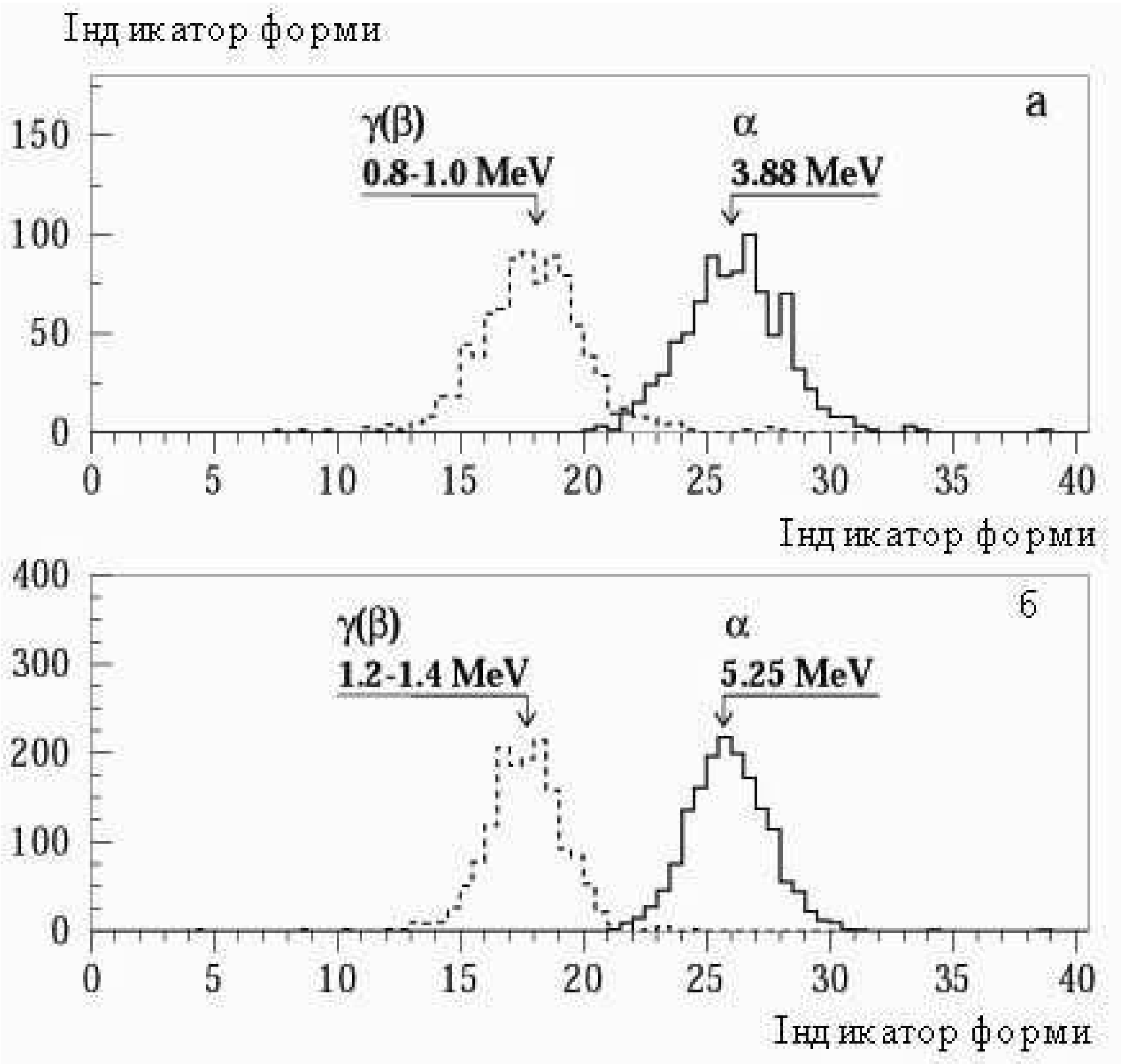

Рис. 2.8. Розподіли індикаторів форми для альфа–частинок та гамма–променів, виміряні з кристалом $^{116}CdWO_4$ (∅32×19 мм): а) $E_\gamma$=0.8–1.0 МеВ, $E_\alpha$=3.88 МеВ; б) $E_\gamma$ = 1.2 – 1.4 МеВ, $E_\alpha$ = 5.25 МеВ. Кристали опромінювались $\alpha$–частинками в напрямку, перпендикулярному до кристалічної осі (010) сцинтиляційного кристала.

Ідентифікація частинок за допомогою аналізу форми сцинтиляційних імпульсів є, поряд з часово–амплітудним аналізом, потужним засобом зменшення фону та визначення активності радіонуклідів в сцинтиляторах.

### 2.2.7. Калібровочні вимірювання для розробки та перевірки моделювання подій в спектрометрі.

Достовірність розрахунків методом Монте–Карло була перевірена в серії спеціальних вимірювань з джерелами $\gamma$–квантів $^{22}$Na, $^{54}$Mn, $^{60}$Co, $^{137}$Cs, $^{207}$Bi, $^{232}$Th, та β–частинок – $^{90}Sr$, $^{210}Bi$ та сцинтиляторами вольфрамату кадмію різних об’ємів та розмірів. Розрахунки були проведені для різних конфігурацій детекторів, починаючи з найпростішої "детектор плюс джерело" і закінчуючи повністю спорудженою низькофоновою установкою, описаною вище. Було отримано добру узгодженість модельних та виміряних спектрів, що дозволило використовувати метод Монте–Карло для моделювання та реконструкції даних експерименту та побудови моделей різних шуканих 2β–процесів. Всі розрахунки методом Монте–Карло за допомогою програм GEANT3.21 [ 247 ] та генератора подій DECAY4 [248] були зроблені Олегом Анатолійовичем Понкратенком.

## 2.3. Аналіз даних вимірювань

### 2.3.1. Часово–амплітудний аналіз подій.

Розпади атомних ядер є випадковим процесом і відбуваються за експоненційним законом радіоактивного розпаду. Тому потік в часі фонових подій, що спричинені розпадами радіонуклідів зовні або всередині детектора, описується розподілом Пуассона. Інша ситуація виникає, коли детектор здатний зареєструвати послідовні розпади ядер (що можливо, як правило, лише тоді, коли такі ядра знаходяться в чутливому об‘ємі детектора). У випадку, коли час між надходженням фонових подій більший за період напіврозпаду одного з членів радіоактивного ряду, такі “ланцюжки” розпадів можна реєструвати, а отже визначити їх природу і активність за допомогою аналізу часу, амплітуди (тому ми назвемо такий аналіз “часово–амплітудним), а інколи і природи розпадів (можна, наприклад, відрізнити альфа– і бета–розпади, використовуючи методи аналізу форми сигналів). Типовими ланцюжками, що підлягають часово–амплітудному аналізу в низькофонових експериментах, є розпади короткоживучих радіонуклідів торієвого та уранових рядів:

$^{224}Ra$ ($Q_\alpha$ = 5789 кеВ) $\rightarrow$ $^{220}Rn$ ($Q_\alpha$ = 6405 кеВ, $T_{1/2}$ = 55.6 с) $\rightarrow$ $^{216}Po$ ($Q_\alpha$ = 6907 кеВ, $T_{1/2}$ = 0.145 с) $\rightarrow$ $^{212}Pb$ (ряд $^{232}Th$) (2.7);

$^{212}Bi$ ($Q_\beta$ =2254 кеВ) $\rightarrow$ $^{212}Po$ ($Q_\alpha$ = 8954 кеВ, $T_{1/2}$ = 0.3 мкс)$\rightarrow$$^{208}Pb$ ($^{232}Th$) (2.8);

$^{223}Ra$ ($Q_\alpha$ = 5979 кеВ) $\rightarrow$ $^{219}Rn$ ($Q_\alpha$ = 6946 кеВ, $T_{1/2}$ = 3.96 с) $\rightarrow$ $^{215}Po$ ($Q_\alpha$ = 7526 кеВ, $T_{1/2}$ = 1.781 мс) $\rightarrow$ $^{211}Pb$ ($^{235}U$) (2.9);

$^{214}Bi$ ($Q_\beta$ =3272 кеВ) $\rightarrow$ $^{214}Po$ ($Q_\alpha$ = 7834 кеВ, $T_{1/2}$ = 164 мкс)$\rightarrow$ $^{210}Pb$ ($^{238}U$) (2.10).

Часово–амплітудний аналіз був застосований в декількох низькофонових експериментах з використанням сцинтиляційних детекторів $CaWO_4$, $CdWO_4$, $ZnWO_4$, GSO, $CeF_3$ [159,230,249, 359]. Розглянемо можливості цього методу на прикладі експерименту по пошуку 2β–розпаду $^{116}Cd$ за допомогою кристалів $^{116}CdWO_4$.

Враховуючи значення α/β–співвідношення для сцинтиляторів вольфрамату кадмію, α–частинки $^{220}Rn$ з енергіями ≈ 6.3 МеВ призводять до сцинтиляційних сигналів з енергіями ≈ 1.2 МеВ (в шкалі γ–квантів), в той час як α–розпади $^{216}Po$ з енергіями ≈ 6.8 МеВ дають сигнали з енергією близько 1.35 МеВ. Тому всі події в енергетичному інтервалі 0.6 – 2.0 МеВ були використані в якості тригера для пошуку (в такому ж діапазоні енергій) наступних за ними подій в часовому інтервалі 10 – 1000 мс. Враховуючи період напіврозпаду $^{216}Po$, в це часове вікно попадає 95% подій цих розпадів. Таким чином в масиві даних, що були одержані протягом 14 745 годин вимірювань, було знайдені події розпадів $^{216}Po$. Наступним кроком був здійснений пошук попередніх розпадів $^{224}Ra$. Для цього були відібрані всі події з енергіями 0.5 –1.5 МеВ в часовому вікні 1 – 111 с (74% розпадів $^{220}Rn$). Енергії одержаних в результаті такого аналізу піків, розподіли часових інтервалів

(див. рис. 2.9), та природа подій (за допомогою аналізу форми сцинтиляційних сигналів було встановлено, що це α–частинки) відповідають очікуваним значенням для ланцюжка розпадів $^{224}Ra \to {}^{220}Rn \to {}^{216}Po \to {}^{212}Pb$. В ході цього аналізу була відібрана також певна кількість фонових подій, які утворюють неперервний розподіл поза межами знайдених α–піків.

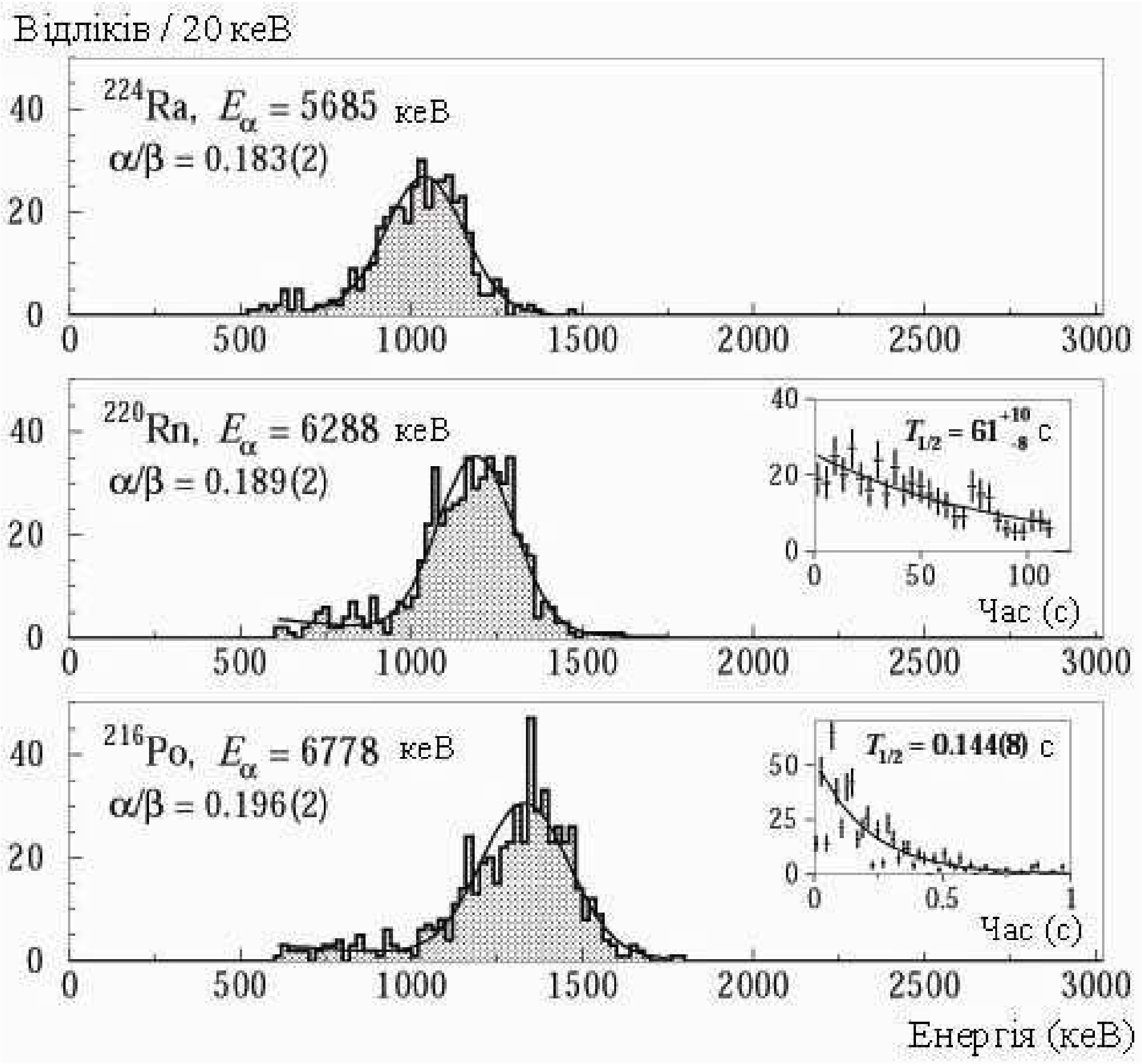


Рис. 2.9. Альфа–піки $^{224}Ra$, $^{220}Rn$, $^{216}Po$, виділені за допомогою часово–амплітудного аналізу з даних вимірювань протягом 14 745 годин з детектором $^{116}CdWO_4$ в низькофоновій установці в Солотвинській підземній лабораторії. На вставках показані часові спектри, які дають можливість визначити періоди напіврозпадів ядер $^{220}Rn$ і $^{216}Po$. Отримані значення періодів напіврозпадів узгоджуються з табличними значеннями 55.6(1) с і 0.145(2) с, відповідно.

Таким чином була визначена активність $^{228}Th$ в кристалах $^{116}CdWO_4$: 39(2) мкБк/кг. Оскільки рівновага радіоактивних рядів, як правило, порушується в процесі виробництва монокристалів, активність материнського ізотопу $^{232}Th$ може бути відмінною від знайденої активності $^{228}Th$. Щоб визначити активність $^{232}Th$, були проаналізовані результати часово–амплітудного аналізу даних вимірювань впродовж приблизно 9 років, починаючи з 5 років після росту кристалів. Як видно з рис. 2.10, активність $^{228}Th$ зростала протягом цього часу. Підгонка даних була здійснена сумою трьох функцій. Одна з них відповідає розпаду $^{228}Th$, що спочатку був у кристалі, ще одна описує поведінку в часі активності $^{228}Th$, який утворювався в результаті розпаду $^{228}Ra$, і остання відображає процес накопичення $^{228}Th$ в результаті розпадів $^{232}Th$. Таким чином було визначено активність $^{232}Th$ в кристалах: 53(9) мкБк/кг. Можна передбачити, що в майбутньому активність $^{228}Th$ досягне цієї величини. З цього ж аналізу слідує обмеження на активність $^{228}Ra$ на рівні $\leq 4$ мкБк/кг.

Вищеописана методика часово–амплітудного аналізу була застосована також до ланцюжка розпадів $^{219}Rn \rightarrow {}^{215}Po \rightarrow {}^{211}Pb$ з ряду $^{235}U$. Для пошуку послідовних розпадів $^{219}Rn$ і $^{215}Po$ були проаналізовані події з енергіями 0.6 – 2.2 МеВ в часовому інтервалі 1 – 10 мс. Цей часовий інтервал містить 66% розпадів $^{215}Po$. Одержані $\alpha$–піки, а також розподіл часових інтервалів відповідають дуже малій активності $^{227}Ac$ в кристалах: 1.4 (9) мкБк/кг.

Пошук розпадів $^{214}Bi \rightarrow {}^{214}Po \rightarrow {}^{210}Pb$ з ряду $^{238}U$, здійснений за допомогою вищеописаної методики, дав лише обмеження ($\leq 4$ мкБк/кг) на активність ізотопу $^{226}Ra$ в кристалах. Враховуючи, що $^{214}Bi$ має досить малий період напіврозпаду ($T_{1/2} = 164$ мкс), порівняний з часом висвічування сцинтиляторів вольфрамату кадмію (середня постійна спаду $\approx 13$ мкс), значна частина подій послідовних розпадів $^{214}Bi$ і $^{214}Po$ попадала в часовий інтервал $\approx 100$ мкс. Форма сцинтиляційних сигналів таких події була записана за допомогою оцифровщика форми. Це дало

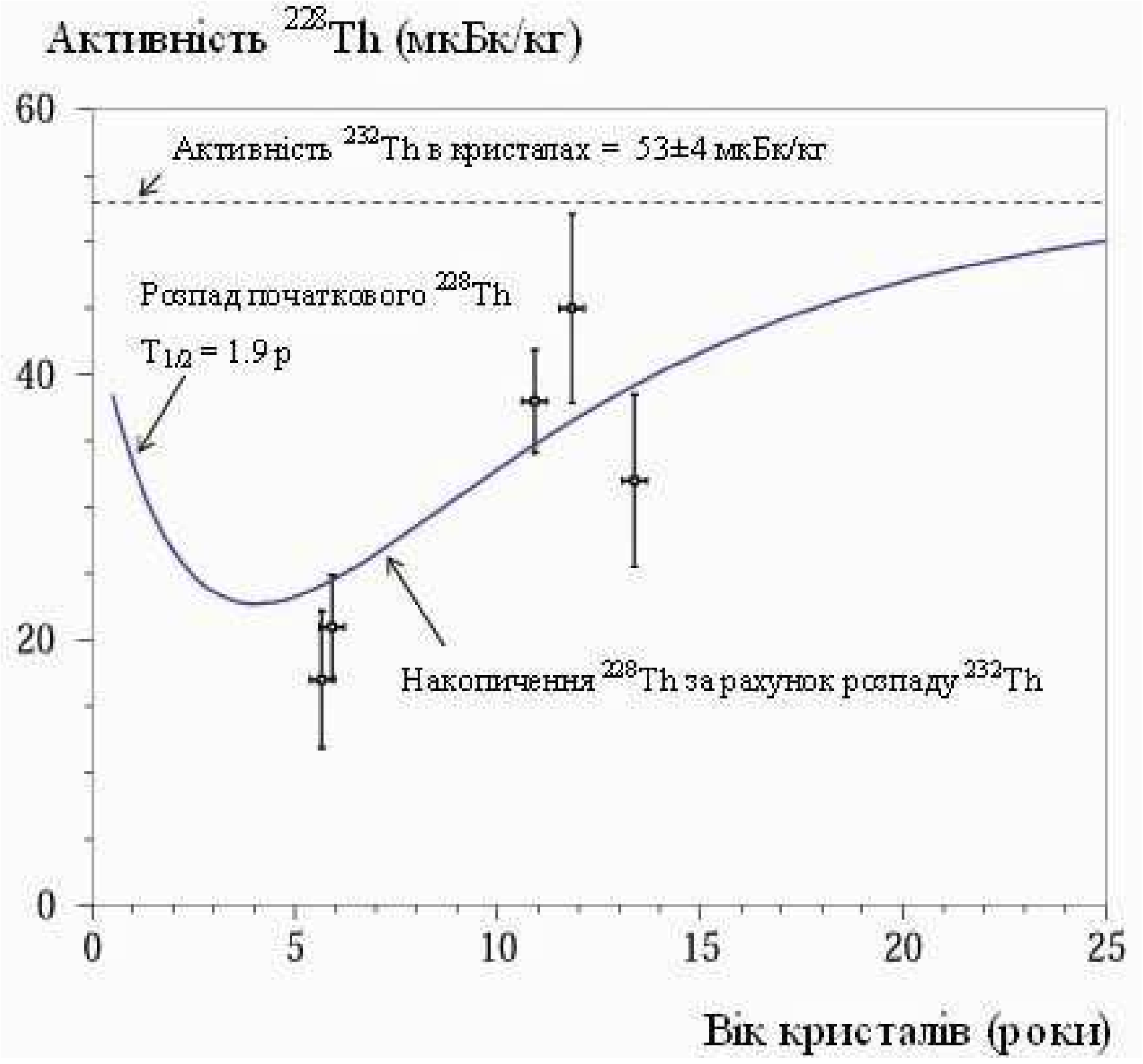


Рис. 2.10. Залежність активності $^{228}$Th в кристалах $^{116}$CdWO$_4$ від їх віку.

змогу провести комплексний часово–амплітудний аналіз та аналіз форми сцинтиляційних сигналів для цього ланцюжка розпадів. В той же час, ланцюжок розпадів $^{212}$Bi $\rightarrow$ $^{212}$Po ($T_{1/2}$ = 0.3 мкс)$\rightarrow$ $^{208}$Pb був проаналізований лише завдяки аналізу форми сигналів (див. п. 2.3.2).

Часово–амплітудний аналіз дозволяє по–перше, відкинути знайдені події як фонові і, по–друге, досить точно визначити активності радіонуклідів рядів торію та урану всередині сцинтиляторів, а отже, інтерпретувати спектр фону детектора, накопиченого в ході експерименту.

Програми для часово–амплітудного аналізу даних були розроблені В.В. Кобичевим.

2.3.2. Аналіз форми сцинтиляційних сигналів.

Ефективність застосування аналізу за формою імпульсу до фонових подій детектора $^{116}CdWO_4$ проілюстрована на рис. 2.11, на якому показаний тривимірний розподіл фонових подій в залежності від енергії та форми імпульсу. На цьому рисунку ясно виділяються дві групи подій: α–події, що спричинені розпадами радіонуклідів сімейств U/Th, та γ (β) події.

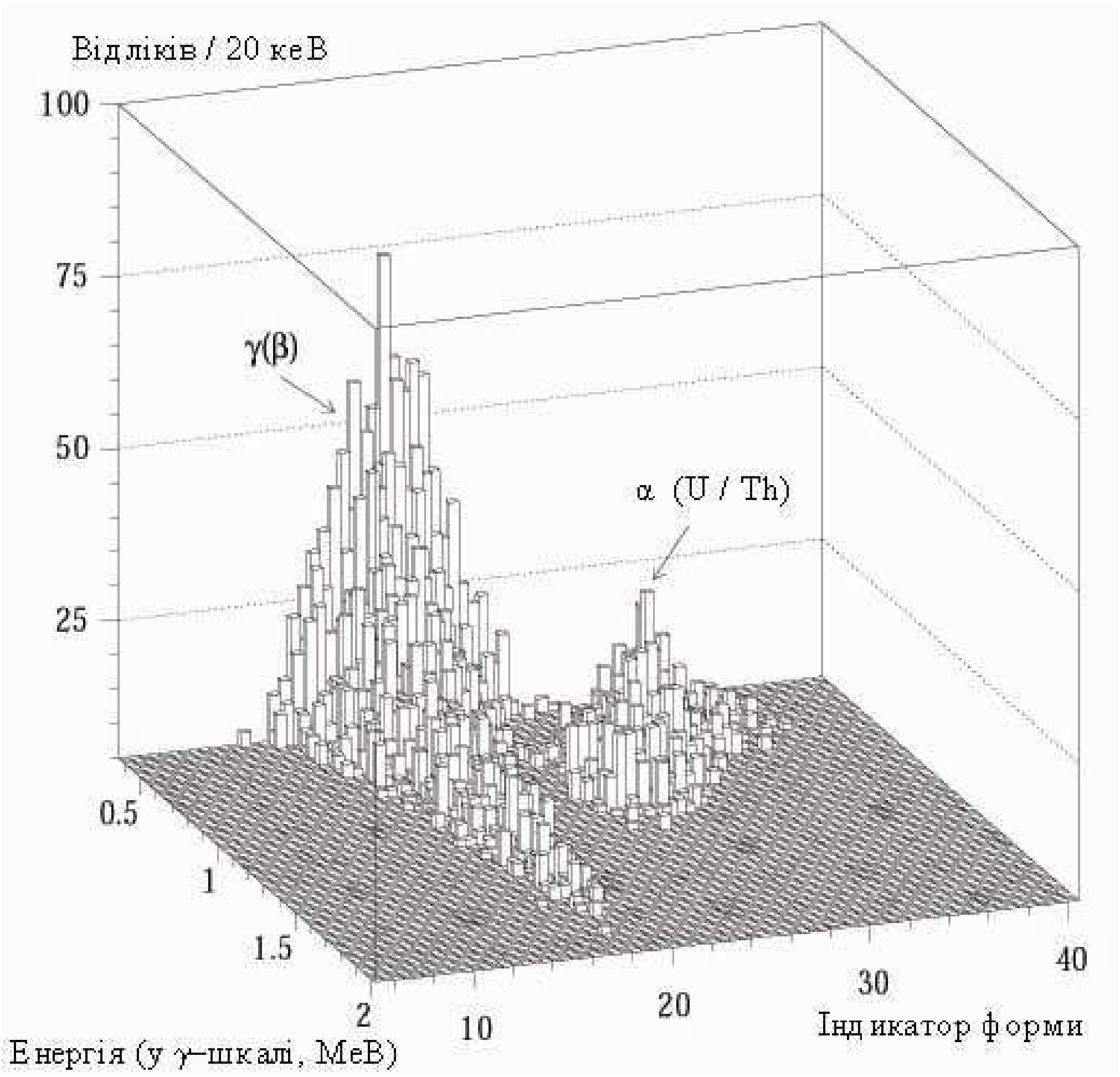


Рис. 2.11. Тримірний розподіл фонових подій (експозиція 2975 години з кристалами $^{116}CdWO_4$) в залежності від їх енергії та форми імпульсу. Події від α–частинок U/Th чітко відокремлюються від γ(β)–подій.

Треба відмітити, що аналіз форми сцинтиляційних сигналів дозволяє не лише відрізняти події від альфа–частинок, але також відкидати фонові відліки, викликані подвійними імпульсами, перекриттям з сигналом від пластикового сцинтилятора, радіонаводками, тощо. Але найголовнішим виграшем від використання аналізу форми сигналів стала можливість знайти і відкинути події, спричинені розпадами в ланцюжку $^{212}$Bi ($Q_{\beta}$ =2254 кеВ) $\rightarrow$ $^{212}$Po ($Q_{\alpha}$ = 8954 кеВ, $T_{1/2}$ = 0.3 мкс)$\rightarrow$$^{208}$Pb, дочірніх радіонуклідів $^{232}$Th, що в слідовій кількості міститься в кристалах $^{116}CdWO_4$. Енергетичні та часові розподіли $\beta$–розпаду $^{212}$Bi та $\alpha$–розпаду $^{212}$Po, знайдені в фонових даних, зображено на рис. 2.12 (а–в), а типовий приклад такої події показано на рис. 2.12 (г). Подібним чином також можна відрізняти події з послідовності розпадів $^{214}$Bi $\rightarrow$ $^{214}$Po $\rightarrow$ $^{210}$Pb. Завдяки використанню такого аналізу, події, викликані двома швидкими розпадами в обох ланцюжках ($^{212}$Bi $\rightarrow$ $^{212}$Po $\rightarrow$ $^{208}$Pb та $^{214}$Bi $\rightarrow$ $^{214}$Po $\rightarrow$ $^{210}$Pb), які можуть давати внесок в фон аж до енергій 4 – 4.5 МеВ, відкинуті з дослідних даних. Це дало змогу знизити фон в області $0\nu2\beta$–розпаду $^{116}$Cd в 15 – 20 разів.

### 2.3.3. Інтерпретація фону детектора.

#### 2.3.3.1. Фон, спричинений $\alpha$–частинками.

Фонові спектри ($\beta+\gamma$) та $\alpha$–подій, виміряні чотирма кристалами $^{116}CdWO_4$ (330 г, експозиція 2975 г) зображені на рис. 2.13. $\beta(\gamma)$–Спектр, показаний на рис. 2.13, був побудований при умові відбору наступного інтервалу значень індикатора форми *SI*: $SI_{\gamma} - 2.4\sigma_{\gamma} < SI < SI_{\gamma} + 2.4\sigma_{\gamma}$, який містить 98% $\gamma(\beta)$–подій. Низькоенергетична область фону детектора $^{116}CdWO_4$ детектора обумовлена, головним чином, чотирикратно забороненим $\beta$–розпадом $^{113}$Cd ($T_{1/2}$=7.7×10$^{15}$ р., $Q_{\beta}$=316 кеВ), який присутній в збагачених кристалах з розповсюдженістю $\sigma \approx 2\%$. Розподіл вище ≈350 кеВ описується наявністю слідових забруднень збагачених та захисних кристалів

$^{40}$K, $^{137}$Cs, та $^{113m}$Cd, двохнейтринним подвійним бета–розпадом $^{116}$Cd з $T_{1/2}$=2.9$\times 10^{19}$ років, та зовнішніми $\gamma$–променями.

Енергетичний спектр $\alpha$–частинок (рис. 2.13) було отримано за рахунок відбору подій з наступними значеннями індикатора форми: $SI_{\gamma} + 4\ \sigma_{\gamma} < SI < SI_{\alpha} + 2.4\ \sigma_{\alpha}$. При таких обмеженнях ефективність аналізу по формі імпульсів ($\varepsilon_{PSA}$) залежить від

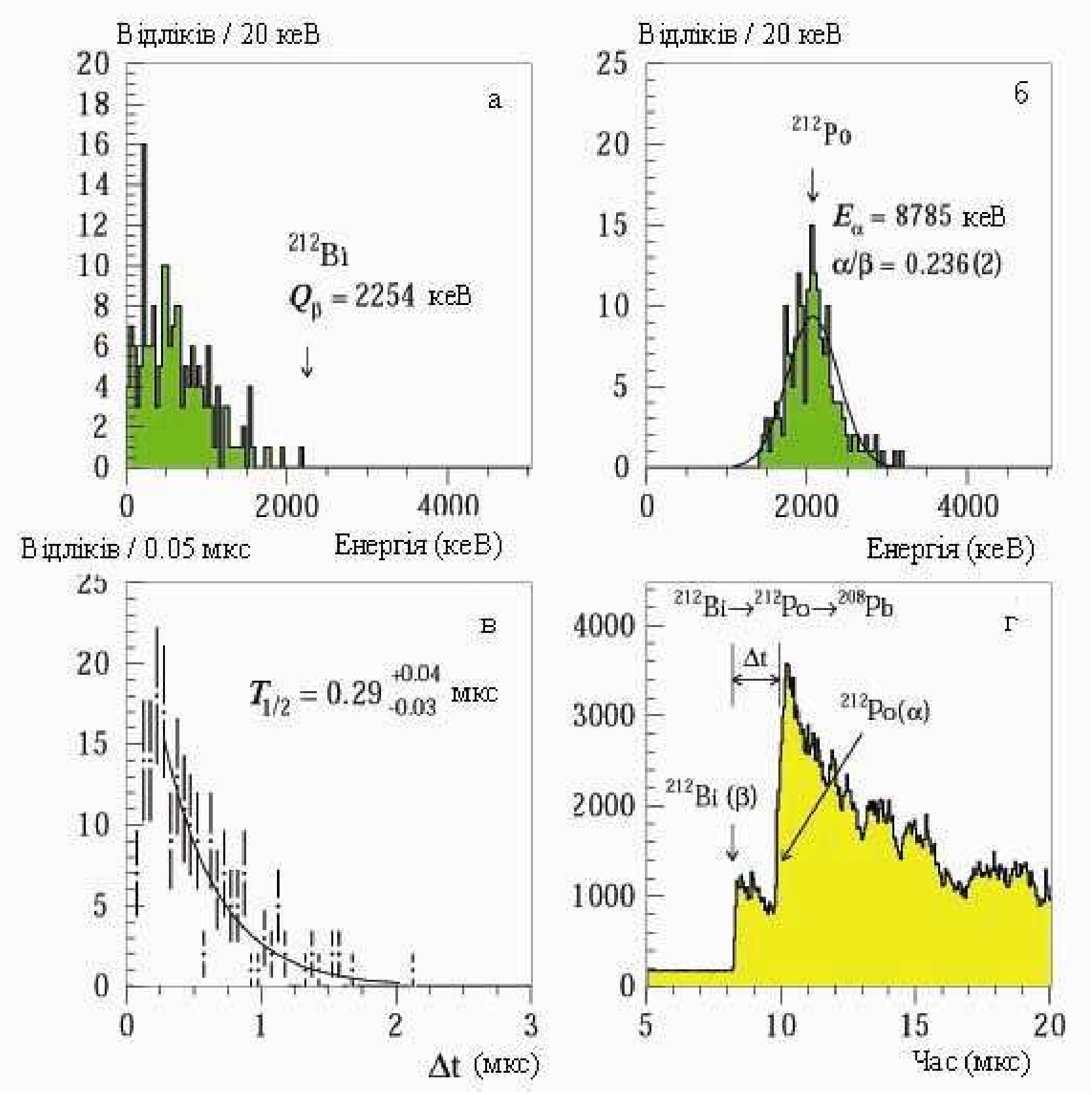

Рис. 2.12. Енергетичний (а, б) та часовий (в) розподіли швидкої послідовності $\beta$– ($^{212}$Bi, $Q_{\alpha}$=2254 кеВ) та $\alpha$– ($^{212}$Po, $E_{\beta}$=8785 кеВ, $T_{1/2}$=0.299(2) мкс) розпадів, відокремлені від фону за допомогою аналізу форми імпульсів. (г) Приклад форми сцинтиляційного сигналу для подібної події в сцинтиляторі $^{116}$CdWO$_4$.

енергії α частинок. Наприклад, для α–піку $^{180}W$ ця ефективність складає 49.5%, в той час як застосування допоміжного фільтру для сигналів пластикового сцинтилятора зменшує це значення до 47%.

Приймаючи до уваги той факт, що вікова рівновага в кристалах звичайно порушена, розподіл α–подій добре відтворюється моделлю, яка включає α–розпади ядер із родин $^{232}Th$ та $^{238}U$. Загальна α–активність, асоційована з α–піком, який в γ–шкалі має енергію 400–1500 кеВ, становить1.40(10) мБк/кг.

Однак із–за меншої роздільної здатності для α–частинок та невизначеності α/β–співвідношення можна дати лише обмеження на присутність нуклідів з родин урану та торію в сцинтиляторах $^{116}CdWO_4$, отримані з підгонки енергетичного спектру в діапазоні 400 – 1500 кеВ: $^{232}Th$ < 0.15 мБк/ кг, $^{238}U$ ($^{234}U$) < 0.6 мБк/кг, $^{230}Th$ < 0.5 мБк/кг, та $^{210}Pb$ < 0.4 мБк/кг.

Низькоенергетичну частину α–спектру (менше 200 кеВ) можна пояснити шумами ФЕП, остаточним γ(β)–фоном, α–розпадами на поверхні сцинтиляторів $^{116}CdWO_4$.

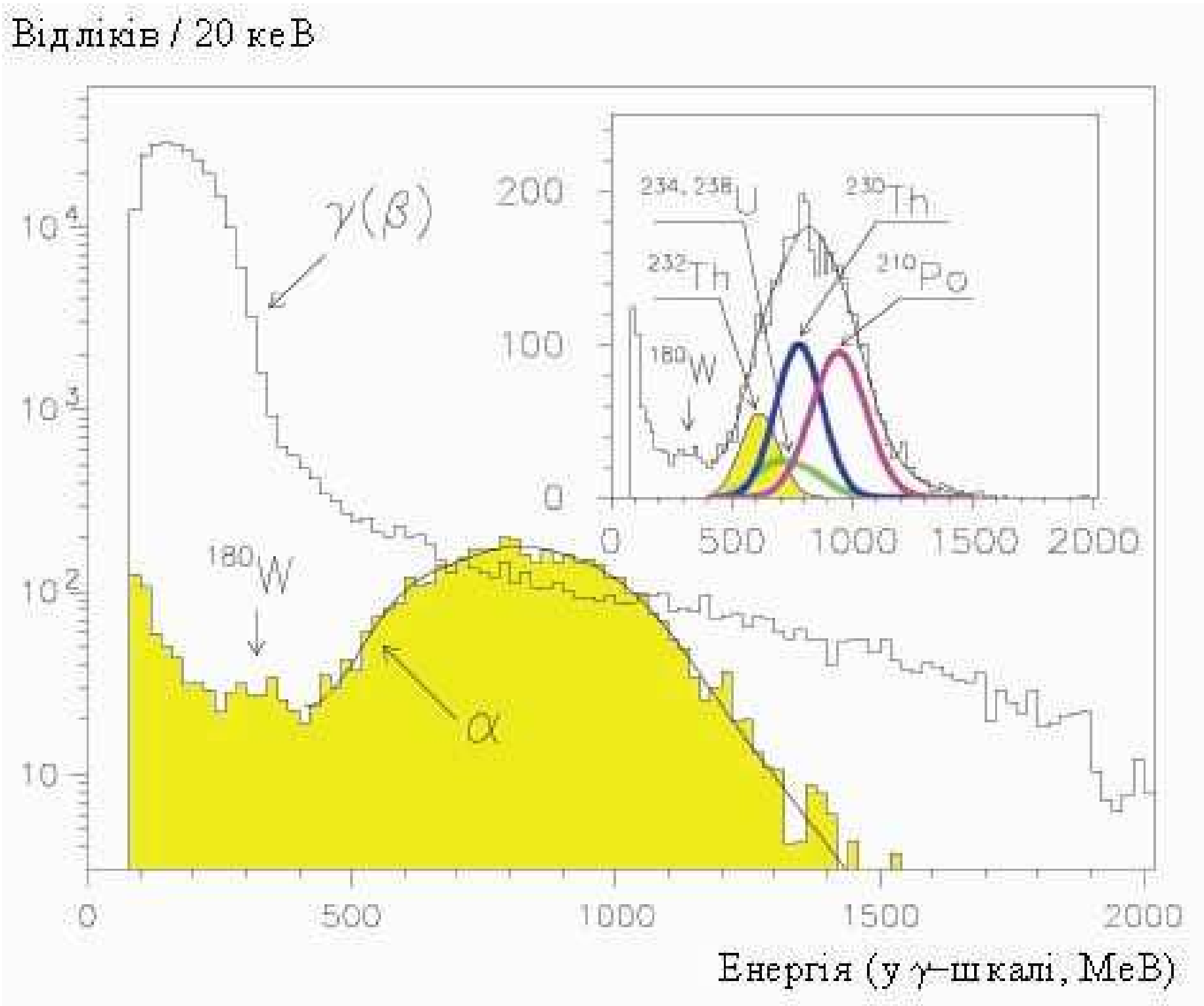

Рис. 2.13. Енергетичний розподіл β(γ) та α–подій, що буди відібрані за допомогою аналізу форми імпульсу з даних, накопичених кристалами $^{116}CdWO_4$ (330 г) за 2975 г. На вкладці приведено α–спектр разом з моделлю, яка включає α–розпади нуклідів з родин $^{232}Th$ та $^{238}U$. Їх загальна активність в кристалах $^{116}CdWO_4$ становить 1.40(10) мБк/кг.

В спектрі на рис. 2.13 є слабкий α–пік при енергії приблизно 300 кеВ, який може бути пояснений α–розпадом $^{180}W$. Очікуване положення α–піку $^{180}W$ з енергією α-частинок 2.46 МеВ становить 307±24 кеВ (з FWHM = 110 кеВ). Нами була проаналізована можливість імітації α–розпаду вольфраму α–активністю платини ($^{190}Pt$), оскільки для росту кристалів були використані платинові тиглі. Ізотопна розповсюдженість $^{190}Pt$ становить 0.014% [ 250 ], $T_{1/2} = 6.5 \times 10^{11}$ р, $E_\alpha$ = 3164(15) кеВ [251], або 435 ± 30 кеВ в шкалі γ–квантів. Розрахунки показують, що зареєстрований пік може бути пояснений забрудненням кристалів платиною (однорідно розподіленою в об’ємі кристалів $^{116}CdWO_4$) на рівні ≈ 3 ppm.

Для оцінки реального вмісту платини в кристалах $^{116}CdWO_4$, були використані результати попередніх низькофонових вимірювань, які проводились колаборацією Мілан–Київ з кристалом $CdWO_4$ вагою 58 г [ 252 ], який був виготовлений в тих же умовах, що і збагачені кристали. Дослід [252] виконувався в лабораторії Gran Sasso. Кристал $CdWO_4$ був охолоджений до температури ≈ 25 мК. Енергетична роздільна здатність (FWHM) детектора становила 5 кеВ при енергії 2.6 МеВ. Не було зареєстровано ніяких подій в області 3100 – 3300 кеВ протягом 340 годин вимірювань, що дозволило встановити лише обмеження на активність $^{190}Pt$. У відповідності з роботами [253,254] для оцінки граничної кількості подій, гіпотеза про наявність яких може бути відкинута з рівнем довірчої імовірності 68% C.L., слід взяти 1.3 відліки. Таким чином можна встановити обмеження на концентрацію платини в кристалі на рівні ≤1.2 ppm. Крім того, два зразки кристалів $CdWO_4$ розмірами $1.5 \times 1.5 \times 0.1$ см сканувалися (за допомогою електронного мікроскопу) на предмет наявності включень, елементний склад яких відрізняється від складу $CdWO_4$. Для цього сканувалася поверхня (≈ 2 см$^2$) кожного кристалу, і, якщо подібні включення знаходились, електронний пучок фокусувався на них, а рентгенівське випромінювання аналізувалося кристалічним спектрометром (з енергетичною роздільною здатністю кращою 0.1 еВ), налаштованим на характеристичне рентгенівське випромінювання платини. Для всіх без винятку спостережених включень (діаметром 2 – 30 мкм) не було спостережено рентгенівських променів платини. Це дозволило встановити обмеження на концентрацію платини в кристалах $CdWO_4$ на рівні 0.1 ppm (або 0.03 ppm) для включень платини з діаметром, меншим за 5 мкм (≤ 3 мкм). Подібний аналіз, який було повторено із зразками збагачених кристалів $^{116}CdWO_4$, дав ще більш жорсткі обмеження: ≤ 0.07 ppm (≤ 0.02 ppm), завдяки більшій сканованій площі та кращому енергетичному розділенню електронного мікроскопу, який використовувався в останньому випадку.

Таким чином, можна зробити висновок, що всі обмеження на гомогенні включення платини в кристалах $CdWO_4$ знаходяться нижче рівня 3 ppm, на якому міг бути імітований α–пік $^{180}W$. До того ж відомо, що в кристалах, які вирощувались у платинових тиглях, домішки платини розподіляються не гомогенно, а у вигляді включень розмірами близько ≈ 20 мкм [255], отже має спостерігатися розмитий енергетичний розподіл замість піку. Останнє було підтверджено за допомогою моделювання по методу Монте–Карло α–розпадів $^{190}Pt$ в кристалах $CdWO_4$ за допомогою програм GEANT [256] та генератору подій DECAY4 [248]. Розрахунки підтвердили, що ефект α–розпаду $^{180}W$ міг бути імітований α–розпадами $^{190}Pt$ лише у випадку наявності скупчень розмірами 2 – 3 мкм, із середньою концентрацією платини в кристалах на рівні 4 – 6 ppm, що набагато вище отриманих експериментальних обмежень.

На основі даних вимірювань з детектором $^{116}CdWO_4$ не можна виключити інші пояснення піку ≈ 0.3 МеВ в α–спектрі, тому цей результат був інтерпретований як перша вказівка на α–розпад $^{180}W$ [ 257 ]. Підтвердження спостереження альфа–активності природного вольфраму було отримане у вимірюваннях з кристалами вольфрамату кальцію ($CaWO_4$) [191]. В роботі [191] використовувався сцинтиляційний детектор з $CaWO_4$ (див. 6 розділ). Остаточне підтвердження α–розпаду $^{180}W$ було отримане за допомогою кристала $CaWO_4$, що працював в якості кріогенного детектора з високою енергетичною роздільною здатністю [258].

#### 2.3.3.2. Фон від γ–квантів та β–частинок. Радіочистота сцинтиляторів $^{116}CdWO_4$ та $CdWO_4$.

Енергетичний спектр γ(β)–подій, накопичений за 13 316 годин вимірювань в низькофоновій установці з кристалами $^{116}CdWO_4$, показано на рис. 2.14. Події були відібрані за допомогою аналізу сигналів за формою. В області низьких енергій фон обумовлений, головним чином, β–розпадом $^{113}$Cd четвертого ступеню заборони ($T_{1/2}$=7.7·10$^{15}$ р. [259], $Q_β$ = 316 кеВ) та β–розпадом $^{113m}$Cd ($T_{1/2}$=14.1 р., $Q_β$=580 кеВ). Концентрація $^{113}$Cd в збагачених кристалах була виміряна за допомогою мас–спектрометра і становить ε = 2.15(20)% [260]. Можлива присутність $^{113m}$Cd в кристалах $CdWO_4$ була підтверджена у вимірюваннях з двома кристалами вирощеними із незбагаченого кадмію, в яких спостерігався β–спектр $^{113m}$Cd [261]. Для визначення активності $^{113}$Cd фон детекторів $^{116}CdWO_4$ був виміряний протягом 692 годин з енергетичним порогом ≈ 80 кеВ, що дозволило застосувати в цьому енергетичному діапазоні техніку аналізу форми імпульсів. Підгонка даних сумою модельного β–спектру $^{113}$Cd та експоненціальної функції (для опису залишкового фону) дала змогу оцінити активність $^{113}$Cd у збагачених кристалах 91 (5) мБк/кг. Така активність відповідає ізотопній розповсюдженості цього ізотопу в кристалах $^{116}CdWO_4$ ε = 1.9 (2)%, що співпадає з даними вимірювань на мас–спектрометрі.

Спектр при енергіях вище 0.5 МеВ описується 2ν2β–розпадом $^{116}$Cd з $T_{1/2}^{2β}$ = 2.9·10$^{19}$ р. (див. нижче), дуже незначним забрудненням збагачених та захисних кристалів $^{137}$Cs та $^{40}$K і зовнішніми γ–променями. Енергетичні розподіли зазначених компонентів фону були промодельовані методом Монте–Карло за допомогою програми GEANT3.21 [247] та генератора подій DECAY4 [248]. Для опису фону від зовнішнього γ–випромінювання було взято експоненційну функцію. Крім того, внесок від $^{40}$K та нуклідів з сімейств $^{232}$Th та $^{238}$U, які містяться в ФЕП, був також промодельований методом Монте–Карло. Активності цих радіонуклідів в ФЕП були визначені за допомогою вимірювань в установці з кристалом $CdWO_4$, що була розміщена в Солотвинській підземній лабораторії (див. п. 2.2.1).

Підгонка за допомогою методу найменших квадратів ($χ^2$/ч.с.с. = 119 / 108 = 1.1) експериментального спектру в інтервалі енергій 0.34 – 2.7 МеВ сумою вказаних складових дала такі значення для активності радіонуклідів в кристалах $^{116}CdWO_4$ (в мБк/кг): 1.1(1) для $^{113m}$Cd, 0.43(6) для $^{137}$Cs та 0.3(1) для $^{40}$K.

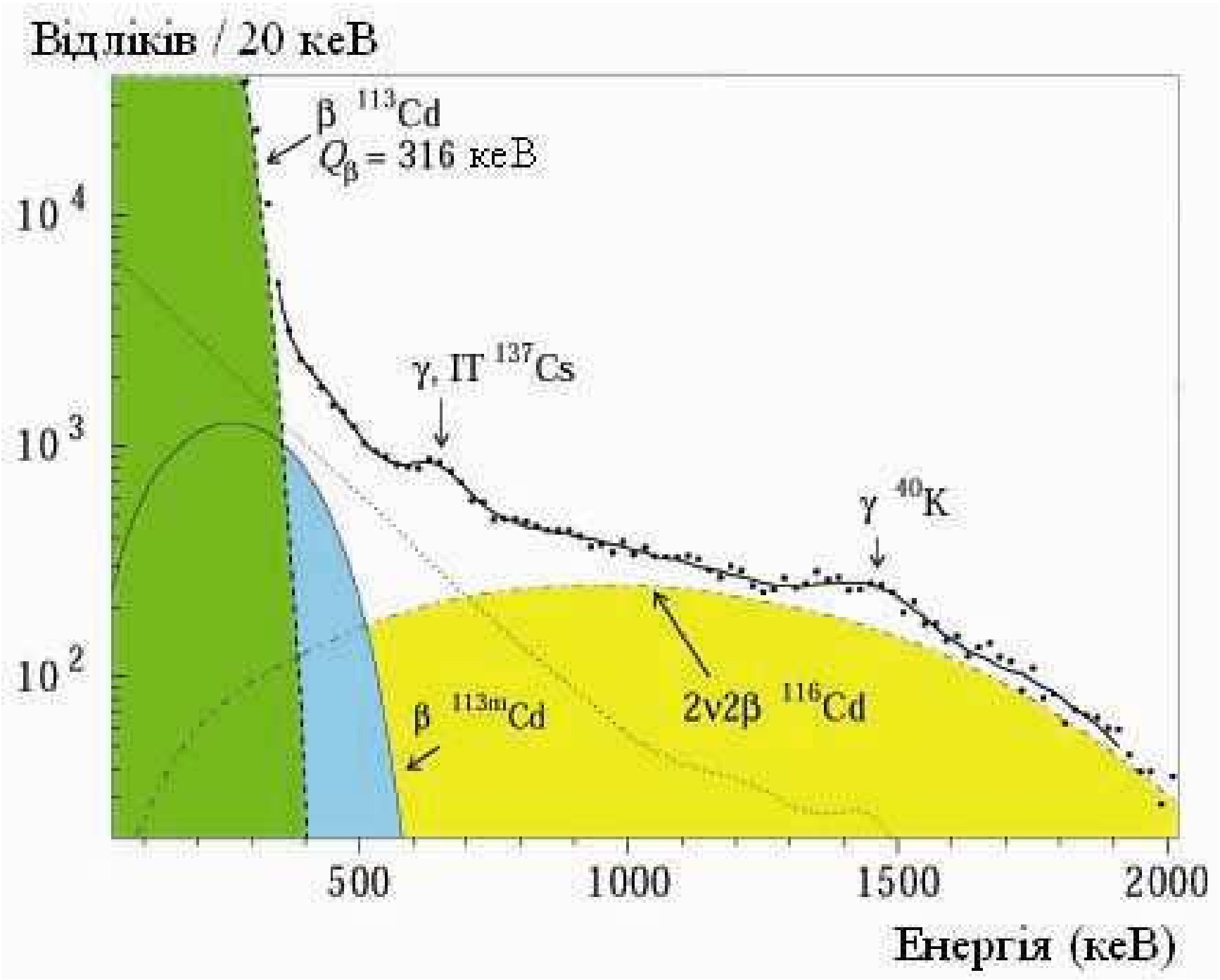

Рис. 2.14. Енергетичний спектр фону γ(β)–подій, відібраних з даних низькофонових вимірювань протягом 13 316 годин в наднизькофоновому спектрометрі з кристалами $^{116}CdWO_4$ в Солотвинській підземній лабораторії ІЯД НАНУ. Показані основні компоненти фону: β–спектри $^{113}$Cd та $^{113m}$Cd, 2ν2β–спектр $^{116}$Cd, фон γ–квантів від ФЕП (точкова лінія).

Забрудненість $^{40}$K захисних кристалів $CdWO_4$ становить 1.5(3) мБк/кг. Компоненти фону детектора $^{116}CdWO_4$ зображені на рис 2.14. Оцінки активностей радіонуклідів, що присутні в слідових кількостях в кристалах $^{116}CdWO_4$ і $CdWO_4$, наведені в таблиці 2.2.

Завдяки активному і пасивному захисту, часово–амплітудному аналізу та аналізу форми імпульсів, інтенсивність фону детекторів $^{116}CdWO_4$ в області енергій 2.5–3.2 МеВ (енергія 0ν2β–розпаду $^{116}$Cd становить 2.8 МеВ) була зменшена до 0.04 відліків/(рік × кг × кеВ). Це один з найнижчих рівнів фону, що будь–коли був досягнутий у вимірюваннях з кристалічними сцинтиляторами.

Таблиця 2.2. Активність різних нуклідів, присутніх як домішки в сцинтиляційних кристалах $^{116}CdWO_4$.

| Сімейство | Нуклід | Активність, мБк/кг |
|---|---|---|
| $^{232}Th$ | $^{232}Th$ | 0.053(9) |
| | $^{228}Ra$ | <0.004 |
| | $^{228}Th$ | 0.039(2) |
| $^{238}U$ | $^{238}U$+$^{234}U$ | <0.6 |
| | $^{234m}Pa$ | <0.2 |
| | $^{230}Th$ | <0.5 |
| | $^{226}Ra$ | <0.004 |
| | $^{210}Pb$ | <0.4 |
| $^{235}U$ | $^{227}Ac$ | 0.0014(9) |
| $^{40}K$ | | 0.3(1) |
| $^{90}Sr$ | | <0.2 |
| $^{113}Cd$ | | 91(5) |
| $^{113m}Cd$ | | 1.1(1) |
| $^{137}Cs$ | | 0.43(6) |

**Висновки розділу.** Проаналізовані фактори, що визначають чутливість експерименту по пошуку 2β–розпаду за допомогою сцинтиляційних детекторів: концентрація ядер досліджуваного ізотопу, ефективність реєстрації подій 0ν2β–розпаду, рівень фону та енергетична роздільна здатність. Розглянуті джерела фону детекторів по пошуку 2β–розпаду, такі як γ–кванти від розпадів радіонуклідів в оточуючому середовищі, космічні промені, забрудненість радіонуклідами конструкційних матеріалів установки, радіоактивна забрудненість матеріалу детектора, вплив радону, фон, спричинений нейтронами, космогенна активація, процес двохнейтринного 2β–розпаду ядра, радіонаводки та збої в системі реєстрації. В Солотвинській підземній лабораторії ІЯД НАНУ споруджена низькофонова сцинтиляційна установка для вимірювання наднизьких активностей радіонуклідів в зразках об'ємом до 30 дм$^3$ з чутливістю при реєстрації $^{40}$K і $^{232}$Th в зразку масою ≈ 1 кг з точністю ±30% за 24 години вимірювань 0.04 і 0.007 Бк/кг, відповідно, що порівняно з чутливістю низькофонових напівпровідникових детекторів з надчистого германію. На установці проведено відбір конструкційних матеріалів та сцинтиляційних кристалів для наднизькофонових експериментів, спрямованих на пошук 2β–розпаду та інших рідкісних розпадів атомних ядер. В Солотвинській підземній лабораторії споруджений наднизькофоновий сцинтиляційний спектрометр з кристалами вольфрамату кадмію, збагаченими ізотопом $^{116}$Cd. Спектрометр включає комплекс пасивного та активного захисту з відібраних за рівнем радіочистоти матеріалів. Рівень фону детектора, 0.04 відліків / (рік кеВ кг) на енергії 2β–розпаду $^{116}$Cd, є одним з кращих серед досягнутих в експериментах по пошуку 2β–розпаду. Розроблено багатоканальну систему реєстрації даних для наднизькофонових установок, яка дозволяє записувати енергію та час події, інформацію про збіги подій в основному детекторі та детекторах активного захисту, форму сцинтиляційних сигналів. Аналіз даних вимірювань за допомогою часово–амплітудний аналізу подій та аналізу форми сцинтиляційних сигналів дозволили інтерпретувати та суттєво знизити фон детектора, оцінити рівень радіочистоти сцинтиляторів $^{116}CdWO_4$ та $CdWO_4$.

Основні результати, викладені в 2–му розділі, опубліковані в роботах:

РОЗДІЛ 3

# ПОДВІЙНИЙ 2β–РОЗПАД КАДМІЮ ТА ВОЛЬФРАМУ

## 3.1. Подвійний β–розпад ядра $^{116}Cd$

Ядро $^{116}Cd$, завдяки великій енергії переходу (2805±4 кеВ [262]), порівняно високому ізотопному розповсюдженню (7.9%), а також сприятливим теоретичним оцінкам добутку $T_{1/2}^{0\nu} \times \langle m_\nu \rangle^2 = 4.9 \times 10^{23}$ [198], є одним з найбільш зручних для пошуків 2β–розпаду. Не менш важливим є те, що кадмій входить до складу сцинтиляційних кристалів вольфрамату кадмію, що дає змогу поставити калориметричний експеримент з високою ефективністю реєстрації ефекту.

Схема рівнів триплету $^{116}Cd - {}^{116}In - {}^{116}Sn$ показана на рис.1.1. Функції відгуку детектора $^{116}CdWO_4$ для різних каналів 2ν2β та 0ν2β розпаду $^{116}Cd$ (змодельовані при допомозі програм GEANT3.21 та DECAY4) зображені на рис. 1.2.

### 3.1.1. Двохнейтринний 2β–розпад $^{116}Cd$.

#### 3.1.1.1. 2ν2β–розпад ядра $^{116}Cd$ на основний стан $^{116}Sn$.

При 2ν2β–переході на основний стан ядра $^{116}Sn$ спектр повної енергії двох електронів є неперервним з максимумом на енергії близько 1 МеВ (див. рис. 1.2).

Вперше надлишок подій в експериментальному спектрі сцинтиляційного кристала $^{116}CdWO_4$, в порівнянні із спектром, виміряним детектором з природним вмістом ізотопів кадмію, було зареєстровано ще в 1995 р. [159].

На першому етапі експерименту, що проводився в колаборації з вченими Флорентійського університету та Флорентійського відділення Національного інституту ядерної фізики (Італія), після накопичення даних досліду протягом 4629 г., було одержане значення періоду напіврозпаду відносно двохнейтринного подвійного бета–розпаду $^{116}Cd$: $T_{1/2}{}^{2\nu} = 2.6 \pm 0.1$ (стат.)$_{-0.4}{}^{+0.7}$(сист.)$\cdot 10^{19}$ р. [160]. З метою уточнення значення $T_{1/2}{}^{2\nu}$, було використано більшу статистику (12 649 г.) даних, набраних після останньої модернізації установки у 1999 році, коли була покращена енергетична роздільна здатність спектрометра та якість дискримінації за формою імпульсів.

Кристали $^{116}CdWO_4$ містять $4.54 \times 10^{23}$ ядер $^{116}Cd$, отже загальна експозиція експерименту склала $6.56 \times 10^{23}$ ядер × рік. Ефективність реєстрації процесу $2\nu 2\beta$–розпаду детектором з кристалами $^{116}CdWO_4$ становила $\eta = 0.93$ (моделювання ефективності по методу Монте–Карло дало значення $\eta_{mc}$=0.96, а ефективність відбору по формі імпульсів подій подвійного бета–розпаду була визначена за допомогою $\gamma$–квантів $\eta_{psa}$=0.97).

Частина експериментального спектру, що була використана для аналізу, зображена на рис. 3.1 Дані в інтервалі енергій 800–2800 кеВ були промодельовані методом Монте–Карло за допомогою програми GEANT3.21 та генератора подій DECAY4. На додаток до спектру двохнейтринного подвійного бета–розпаду $^{116}Cd$ розглядалися три компоненти фону. Це забруднення $^{40}K$ збагачених і незбагачених сцинтиляторів вольфрамату кадмію та зовнішній $\gamma$–фон, викликаний домішками $^{40}K$, $^{232}Th$ та $^{238}U$ у ФЕП. Радіоактивна забрудненість ФЕП була визначена попередньо як (2–4) Бк/ФЕП для $^{40}K$, (0.4–2.2) та (0.1–0.2) Бк/ФЕП для активності $^{226}Ra$ та $^{228}Th$, відповідно [263].

Підгонка експериментальних даних в енергетичному інтервалі 860 – 2700 кеВ ($\chi^2$/ч.с.с. = 64 / 86 = 0.7) дала наступні результати: активність $^{40}K$ всередині збагачених та натуральних кристалів $CdWO_4$ дорівнює 0.4(2) та 1.6(4) мБк/кг, відповідно; значення періоду напіврозпаду $^{116}Cd$ відносно $2\nu 2\beta$–розпаду становить $2.93 \pm 0.06$(стат.)$\times 10^{19}$ років (відповідна активність $^{116}Cd$ в збагачених кристалах становить близько 1 мБк/кг). Інтервал підгонки вибирався за таким вимогами:

(i) якомога більша накопичена статистика;

(ii) найбільше значення відношення ефект/фон;

(iii) максимальна частина спектру $2\nu2\beta$–розпаду;

(iv) якість підгонки (критерій $\chi^2$/ч.с.с.);

(v) якомога менші помилки параметрів, що включалися в підгонку.

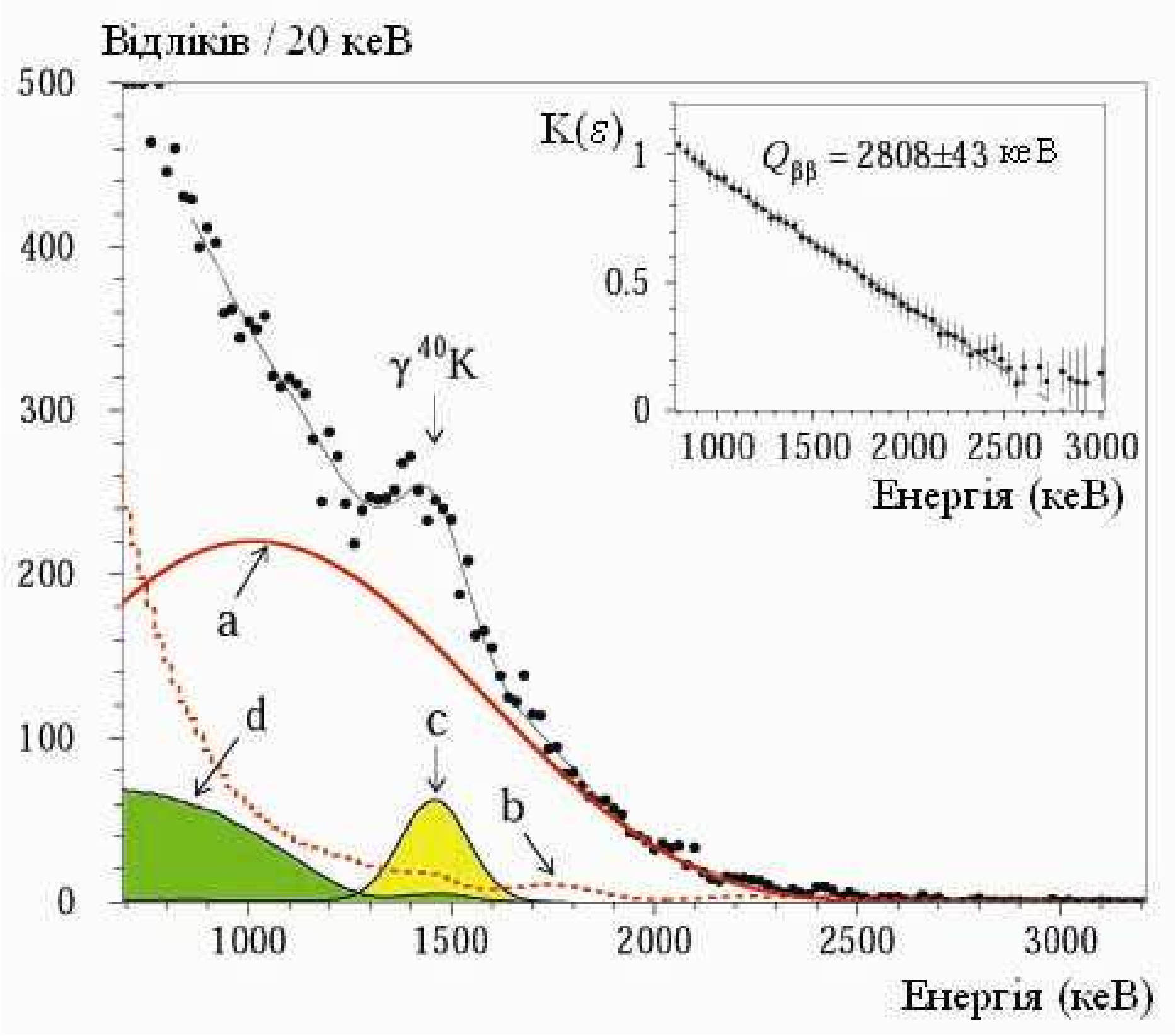

Рис. 3.1. Частина спектру $\gamma$ і $\beta$ подій, виміряного детекторами $^{116}CdWO_4$ протягом 12 649 годин, яка використовувалась для визначення періоду напіврозпаду $^{116}Cd$ відносно $2\nu2\beta$–розпаду. Показані найважливіші компоненти фону: (а) $2\nu2\beta$–спектр $^{116}Cd$; (b) зовнішній $\gamma$–фон, викликаний забрудненням ФЕП радіонуклідами $^{40}K$, $^{232}Th$ та $^{238}U$; внески від домішок $^{40}K$ в натуральних (с) та збагачених (d) сцинтиляторах $^{116}CdWO_4$. Суцільною лінією показано модель енергетичного спектра, отриману в результаті підгонки даних в енергетичному інтервалі 860–2700 кеВ. На вкладці: графік Кюрі для $2\nu2\beta$–розпаду та його підгонка прямою в області енергій 900–2500 кеВ.

Необхідно зазначити, що результати підгонки були досить стабільними для різних інтервалів в межах діапазону (800–2800) кеВ; а відповідні значення періоду напіврозпаду змінювались в межах $(2.6 - 3.1) \times 10^{19}$ р. Ці зміни були обумовлені, головним чином, невизначеністю моделі фону.

Про якість отриманих результатів можна судити по побудованому графіку Кюрі для 2ν2β–розпаду: $K(\varepsilon) = [S(\varepsilon)/\{(\varepsilon^4+10\varepsilon^3+40\varepsilon^2+60\varepsilon+30)\cdot\varepsilon\}]^{1/5}$, де $S$ – число подій з енергією ε (в одиницях маси електрона) в експериментальному спектрі після віднімання фону. Для 2ν2β–розпаду графік Кюрі має бути прямою лінією $K(\varepsilon) \sim (Q_{2\beta} - \varepsilon)$. Отриманий з експериментальних даних графік показано на рис. 3.1. З нього видно, що в області енергій 0.9–2.5 МеВ він насправді добре співпадає з прямою з граничною енергією $Q_{2\beta}$=2808(43) кеВ (табличне значення $Q_{2\beta}$=2805(4) кеВ). Приймаючи до уваги енергетичну роздільну здатність детектора, процедуру підгонки було повторено з використанням згортки теоретичного розподілу 2ν2β $\rho(\varepsilon) = A\varepsilon(\varepsilon^4+10\varepsilon^3+40\varepsilon^2+60\varepsilon+30)\cdot(Q_{2\beta} - \varepsilon)^5$ з функцією, яка описує відгук детектора (функція Гауса). При цьому значення величин $A$, (яка обернено пропорційна величині періоду напіврозпаду $T_{1/2}^{2\nu2\beta}$), та енергію переходу $Q_{2\beta}$ були вільними параметрами. Підгонка даних в інтервалі енергій 1.2 – 2.8 МеВ дала такі значення енергії розпаду та періоду напіврозпаду: $Q_{2\beta}$ = 2748 (42) кеВ; $T_{1/2}$=2.9 (1) $\times 10^{19}$ р.

Необхідно наголосити, що статистика, накопичена в нашому експерименті (9846 подій 2ν2β–розпаду $^{116}$Cd в енергетичному діапазоні 800–2800 кеВ) та відношення сигнал/фон (3:1 в енергетичному інтервалі 1.2–2.8 МеВ та 8:1 для області енергій 1.9–2.2 МеВ) є одними з найвищих серед досягнутих в дослідах по вивченню 2β–розпаду.

Значення періоду напіврозпаду ядра $^{116}$Cd відносно подвійного бета–розпаду на основний стан (о.с.) ядра $^{116}$Sn становить:

$$T_{1/2}^{2\nu2\beta} \text{ (о.с. – о.с.)} = 2.9 \pm 0.06\text{(стат.)}_{-0.3}^{+0.4}\text{(сист.)} \times 10^{19} \text{ років.}$$

Це значення узгоджується з попереднім результатом $T_{1/2}^{2\nu2\beta}=(2.6_{-0.4}^{+0.7}) \times 10^{19}$ р [160], і з отриманим раніше в інших дослідах: $T_{1/2}^{2\nu2\beta} = (2.6_{-0.5}^{+0.9}) \times 10^{19}$ р [156], $T_{1/2}^{2\nu2\beta} = [2.7_{-0.4}^{+0.5}(\text{стат.})_{-0.6}^{+0.9}(\text{сист.})] \times 10^{19}$ р [159], та $T_{1/2}^{2\nu2\beta} = [3.75 \pm 0.35(\text{стат.}) \pm 0.21(\text{сист.})] \times 10^{19}$ р [157].

Систематична похибка виміряного періоду напіврозпаду має декілька джерел (див. табл. 3.1). Головною причиною невизначеності періоду напіврозпаду при підгонці в різних енергетичних інтервалах могли бути β–розпади ядер $^{234m}$Pa та $^{90}$Y (дочірнє ядро $^{90}$Sr) в кристалах $^{116}CdWO_4$. З верхнього обмеження на вміст $^{226}$Ra, отриманого за допомогою часово–амплітудного аналізу даних, було визначено, що активність $^{238}$U (отже і $^{234m}$Pa) в збагачених кристалах менша ніж 0.7 мБк/кг. Ця оцінка була отримана з обмеження на активність $^{226}$Ra (<4 мБк/кг) у припущенні, що усі ядра $^{226}$Ra є результатом розпадів $^{238}$U. Аналіз α–спектру [264] дає значення <0.6 мБк/кг для сукупної активності $^{238}$U та $^{234}$U в кристалах $^{116}CdWO_4$. Для оцінки систематичної похибки обидва β–активні радіонукліди ($^{234m}$Pa та $^{90}$Sr–$^{90}$Y) були включені в процедуру підгонки, що призвело до більш жорсткого обмеження на їх сукупну активність: < 0.3 мБк/кг.

Таблиця 3.1. Різні причини систематичної похибки та їх внесок в систематичну значення періоду напіврозпаду ядра $^{116}$Cd відносно подвійного бета–розпаду на основний стан ядра $^{116}$Sn.

| Причина систематичної похибки | Точність визначення | Внесок до значення $T_{1/2}$ ($\times 10^{19}$ років) |
|---|---|---|
| Живий час вимірювань | (96±2)% | ±0.06 |
| Ефективність відбору подій 2ν2β–розпаду за формою імпульсів | $97_{-3}^{+1}$ % | +0.03, –0.09 |
| Ефективність реєстрації | (96±3)% | ±0.09 |
| 2ν2β–розпаду | | |
| Підгонка в різних енергетичних інтервалах | | +0.2, –0.3 |
| Можлива забруднення кристалів $^{116}CdWO_4$ радіонуклідами $^{90}Sr$–$^{90}Y$ та $^{234m}Pa$ | <0.3 мБк/кг | +0.35 |

3.1.1.2.Обмеження на процеси 2ν2β–розпаду $^{116}Cd$ на збуджені рівні $^{116}Sn$.

У випадку 2ν2β–розпаду $^{116}Cd$ на збуджені рівні ядра $^{116}Sn$ (були здійснені пошуки переходів на рівні $2^+$ з енергією $E_{lev}$=1294 кеВ, $0^+_1$ з $E_{lev}$=51757 кеВ та $0_2^+$ з $E_{lev}$=2027 кеВ) очікувані енергетичні спектри (рис. 3.3) матимуть зовсім інший характер, ніж у випадку переходу на основний стан дочірнього ядра. Це пов‘язано з випроміненням γ–квантів, що можуть втратити лише частину своєї енергії в детекторі, а також з ненульовим спіном рівня $2^+$ з енергією $E_{lev}$=1294 кеВ.

Імовірність таких 2β–переходів значно придушена із–за меншого (у порівнянні з переходами на основний стан) енерговиділення. Теоретичні розрахунки періоду напіврозпаду $^{116}Cd$ на збуджені рівні ядра $^{116}Sn$ знаходяться в межах $10^{22}$–$10^{24}$ р [22]). Оскільки в експериментальних даних відсутні будь–які особливості, які могли б бути викликані шуканими розпадами, можна лише встановити обмеження на їх імовірність за допомогою формули:

$$\lim T_{1/2} = \ln 2 \times N \times t \times \eta \,/\, \lim S, \qquad (3.1)$$

де $N$ – кількість ядер $^{116}Cd$, $t$ – час вимірювань, η – загальна ефективність реєстрації, lim$S$ – кількість подій шуканого ефекту, яка може бути виключена з даним довірчим рівнем імовірності. Значення ефективності реєстрації були розраховані за допомогою програми GEANT3.21 та генератора подій DECAY4: $\eta_{mc}(2^+) = 0.18$, $\eta_{mc}(0_1^+) = 0.09$ та $\eta_{mc}(0_2^+) = 0.06$. Приймаючи до уваги ефективність аналізу форми імпульсів $\eta_{psa} = 0.97$, отримаємо загальну ефективність $\eta(2^+) = 0.18$, $\eta(0_1^+) = 0.09$ та $\eta(0_2^+)$ =0.06. Для оцінки значення lim$S$ було проведено підгонку спектра в енергетичному інтервалі 1.7–2.7 МеВ. Модель фону включала функції відгуку для шуканого ефекту та фону. Підгонка дала $S = -33 \pm 108$ ($2^+$), $S = -53 \pm 59$ ($0_1^+$) та $S = -3 \pm 47$ відліків ($0_2^+$). Оцінки величин ефекту, гіпотеза про наявність якого може бути відкинута з заданою довірчою імовірністю, були зроблені з використанням процедури, рекомендованої Particle Data Group [253]. Були отримані такі обмеження на періоди напіврозпаду відносно 2ν2β–розпаду $^{116}$Cd на збуджені рівні $^{116}$Sn (з довірчою імовірністю 90% (68%)):

$$T_{1/2}^{2\nu}(\text{о.с.} \rightarrow 2^+) > 0.6\ (1.1) \times 10^{21}\ \text{р.},$$

$$T_{1/2}^{2\nu}\ (\text{о.с.} \rightarrow 0_1^+) > 0.8\ (2.2) \times 10^{21}\ \text{р.},$$

$$T_{1/2}^{2\nu}\ (\text{о.с.} \rightarrow 0_2^+) > 0.4\ (0.6)\ \times 10^{21}\ \text{р.}$$

### 3.1.2. Результати пошуків безнейтринного 2β–розпаду $^{116}$Cd.

#### 3.1.2.1. Обмеження на 0ν2β–розпад $^{116}$Cd на основний стан $^{116}$Sn.

Частина спектру фону, виміряного детектором з кристалами $^{116}CdWO_4$ протягом 14 183 годин у режимі антизбігів із захисними детекторами та після відбору γ(β)–подій за формою сцинтиляційних сигналів, показана на рис. 3.2. Експозиція в цьому експерименті становить $7.4 \times 10^{23}$ (ядер $^{116}$Cd)×р. Цей спектр включає також дані, отримані в першій частині досліду [160]. Енергетична роздільна здатність та ефективність відбору γ(β)–подій за формою імпульсів були обчислені для всієї експозиції, приймаючи до уваги результати калібровочних вимірювань. Так, енергетична роздільна здатність при енергії 2.8 МеВ дорівнює FWHM = 8.9%.

Інтенсивність фону в енергетичному інтервалі 2.5–3.2 МеВ становить 0.037(10) відліків/(рік × кг × кеВ). Оскільки в спектрі відсутній пік від 0ν2β–розпаду $^{116}$Cd, можна лише отримати обмеження на період напіврозпаду для цього процесу.

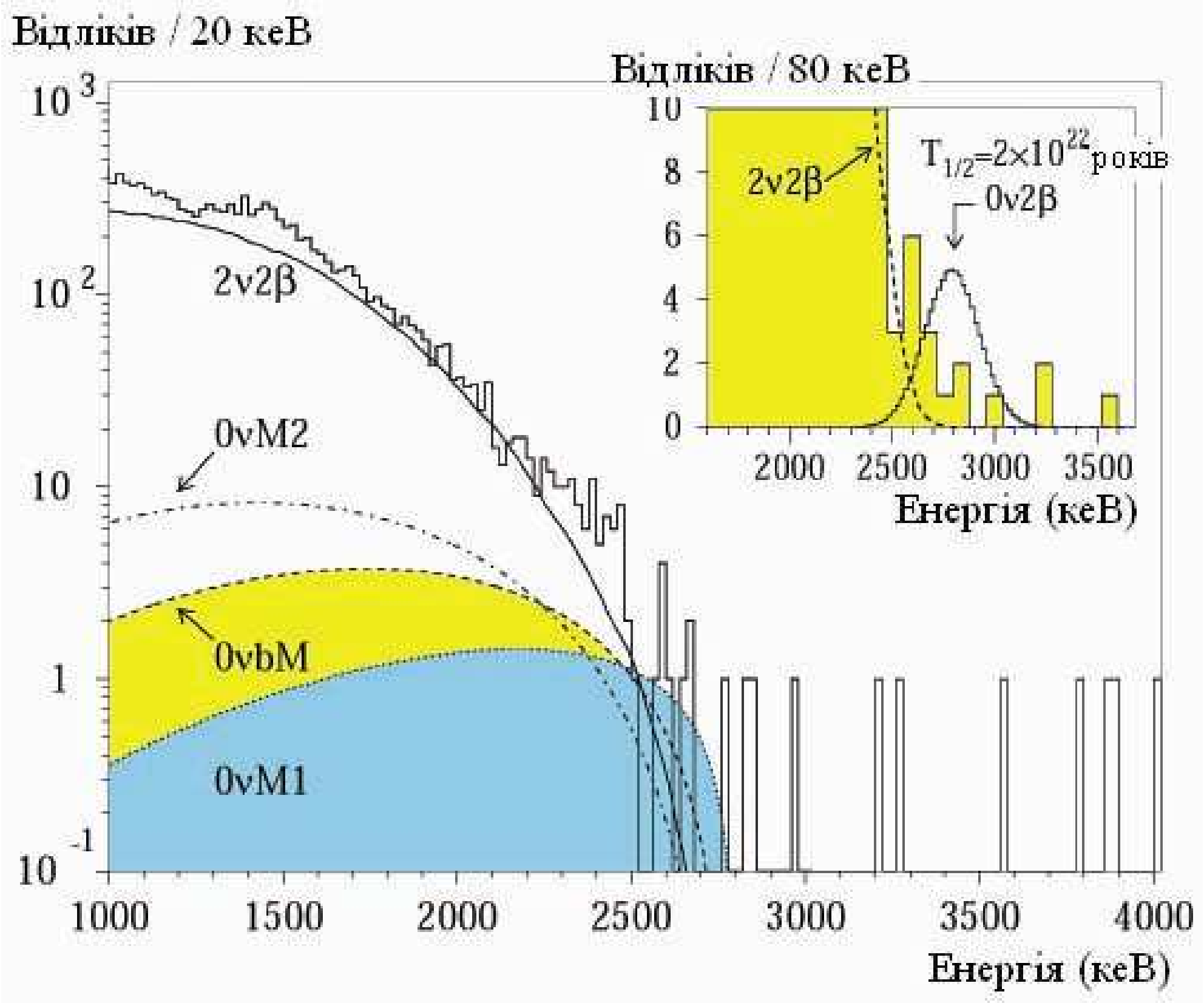

Рис. 3.2. Частина експериментального спектру детектора $^{116}CdWO_4$, виміряного за 14 183 годин (гістограма) разом з розподілом від 2ν2β–розпаду ($T_{1/2}^{2\nu2\beta}$ = 2.9 × $10^{19}$ років). Гладкі криві відповідають ефектам від розпадів з вильотом майоронів: 0ν2β*M*1, 0ν2β*M*2 та 0ν2β$M^{bulk}$, виключені з імовірністю 90% C.L.: $T_{1/2}^{0\nu2\beta M1}$ = 8.0×$10^{21}$ років, $T_{1/2}^{0\nu2\beta M2}$ = 8.0×$10^{20}$ років, та $T_{1/2}^{0\nu2\beta Mbulk}$ = 1.7×$10^{21}$ років. На вкладці показано очікуваний пік від 0ν2β–розпаду з $T_{1/2}^{0\nu2\beta}$ = 2×$10^{22}$ років.

Ефективність реєстрації 0ν2β–розпаду в піку повного поглинання була обчислена за методом Монте–Карло (за допомогою програм GEANT3.21 та DECAY4) і склала $\eta_{mc}$=0.83. Приймаючи до уваги ефективність дискримінації по формі імпульсу $\eta_{psa}$=0.96, отримуємо загальну ефективність η = 0.80. Для оцінки кількості подій ефекту, гіпотезу про наявність яких можна відкинути (lim*S*), частина спектру в енергетичному інтервалі 2.0 – 3.6 МеВ підганялася сумою таких компонент: промодельованої функції відгуку детектора до 0ν2β–розпаду та трьох фонових функцій: $2\nu2\beta$–розпаду (внесок у вказаному енергетичному інтервалі в експериментальний спектр становить ≈83%), зовнішніх $\gamma$–променів від ФЕП (≈14%) та фону від розпадів $^{228}$Th, що присутній в кристалах (≈3%).

В ході підгонки площі цих компонент фону були обмежені з урахуванням похибок. Метод найбільшої правдоподібності ($\chi^2$/ч.с.с. = 37/33 = 1.1) дав площу $0\nu2\beta$–піку $S = 0.3 \pm 1.3$ відліки, що дає lim$S$ = 2.4 (1.6) відліків з довірчою імовірністю 90% (68%) C.L. З цієї оцінки слідує обмеження на період напіврозпаду $^{116}$Cd відносно $0\nu2\beta$–розпаду на основний стан ядра $^{116}$Sn:

$$T_{1/2}^{0\nu2\beta}\text{ (о.с. – о.с.)} \geq 1.7\ (2.6) \times 10^{23}\text{ років.}$$

Як уже було описано раніше, енергетична роздільна здатність спектрометра ретельно перевірялась протягом досліду. Для оцінки можливого впливу роздільної здатності на оцінки lim$S$, підгонку було повторено для різних значень FWHM (8%, 9%, 10% та 11% при енергії 2.8 МеВ). Ці результати були враховані у всіх моделях фону. Було визначено, що варіація lim$S$ не перевищує ≈ 15%. Крім того, підгонка проводилась в різних енергетичних інтервалах в області 1900–3800 кеВ. Результуючі значення lim$S$ знаходились в межах ± 25%.

3.1.2.2.$0\nu2\beta$–розпад $^{116}$Cd на збуджені стани $^{116}$Sn.

При $0\nu2\beta$–розпаді $^{116}$Cd можуть бути також заселені збуджені рівні $^{116}$Sn з $E_{lev} < Q_{2\beta}$ (відповідні функції відгуку показані на рис. 3.3 б). Повне поглинання всіх випромінених частинок приведе до появи піку з енергією $Q_{2\beta}$. За допомогою програм GEANT3.21 та DECAY4 були розраховані ефективності реєстрації подій $0\nu2\beta$–розпаду на перший, другий та третій збуджені рівні $^{116}$Sn: $\eta_{mc}(2^+) = 0.14$, $\eta_{mc}(0_1^+) = 0.07$ та $\eta_{mc}(0_2^+) = 0.03$ в піку з енергією $2\beta$–розпаду $^{116}$Cd. Ці значення відповідають наступним обмеженням на процеси $0\nu2\beta$–розпаду $^{116}$Cd на збуджені рівні $^{116}$Sn при 90% (68%) C.L.:

$$T_{1/2}^{0\nu2\beta}\text{(о.с.} - 2^+) \geq 2.9\ (4.3) \times 10^{22}\text{ р,}$$

$$T_{1/2}^{0\nu2\beta}(\text{о.с.} - 0_1^+) \geq 1.4\ (2.2)\ \times 10^{22}\ \text{р.},$$

$$T_{1/2}^{0\nu2\beta}(\text{о.с.} - 0_2^+) \geq 0.6\ (0.9) \times 10^{22}\ \text{р.}$$

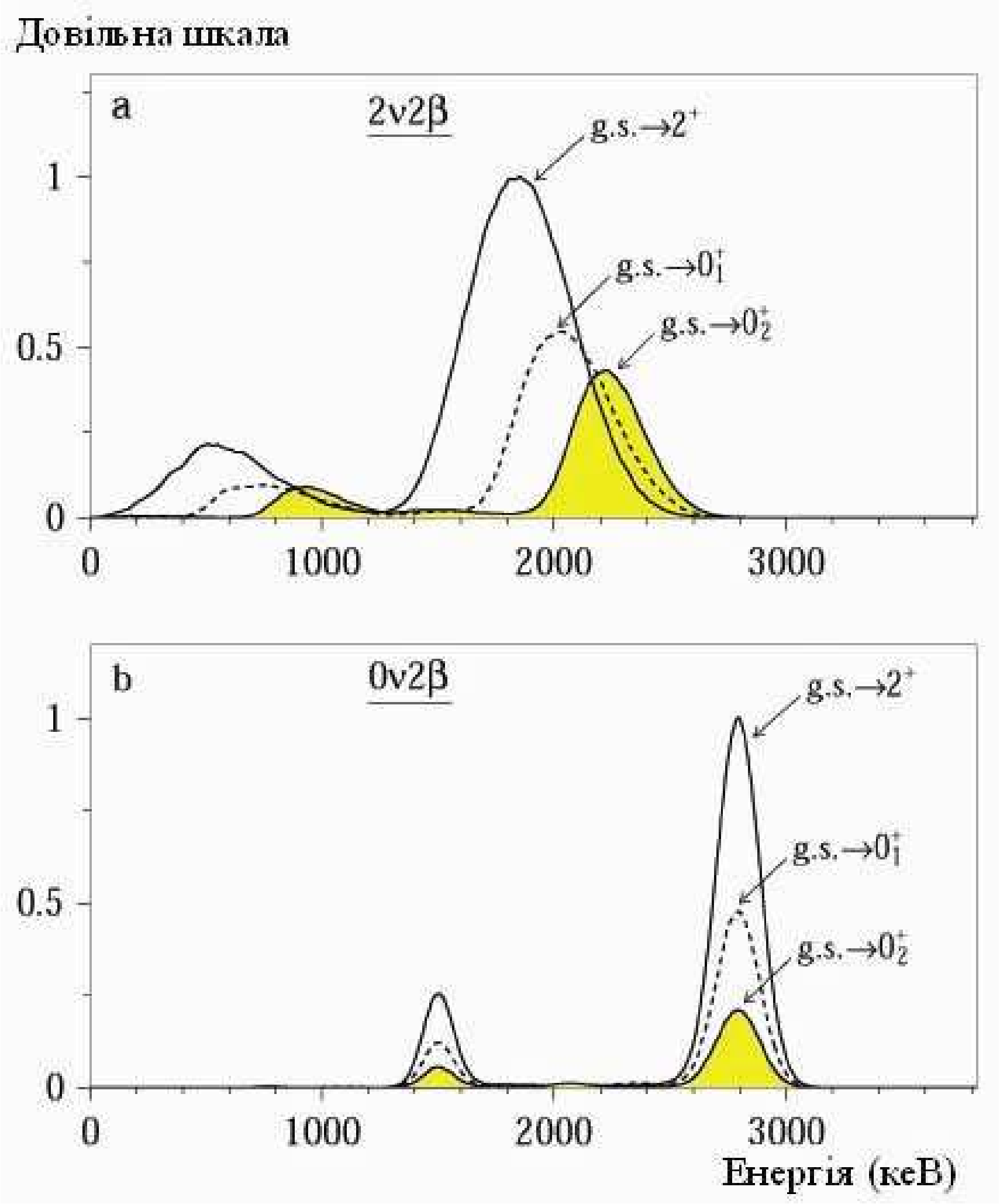

Рис. 3.3. Промодельовані функції відгуку сцинтиляційного детектора з кристалами $^{116}CdWO_4$ до процесів 2β–розпаду ядра $^{116}$Cd на збуджені стани $^{116}$Sn.

3.1.2.3. Пошук процесів з вильотом майоронів.

Для отримання обмежень на 0ν2β–розпад ядра $^{116}$Cd з випроміненням майоронів дослідний спектр в енергетичному інтервалі 1.6 – 2.8 МеВ підганявся з використанням тої ж самої моделі фону, що і для пошуку 0ν2β–розпаду. В результаті підгонки було отримане таке число подій під теоретичною кривою можливого $0\nu2\beta$–розпаду з вильотом одного майорона (M1): –37 ± 56. Відповідно до рекомендацій Particle Data Group [253], це значення дало верхнє обмеження на число подій lim$S$ = 59 (25) з 90% (68%) C.L., що з урахуванням значенням ефективності $\eta_{mc}$=0.905 дає обмеження:

$$T_{1/2}^{0\nu2\beta M1} \geq 0.8\ (1.8)\times10^{22} \text{ років} \qquad \text{з 90\% (68\%)C.L.}$$

Подібна процедура для $0\nu2\beta$–розпаду з випроміненням двох майоронів, а також так званих "bulk" майоронів [57] дала такі обмеження (з 90% (68%)C.L.):

$$T_{1/2}^{0\nu2\beta M2} \geq 0.8\ (1.4)\times10^{21} \text{ років,}$$

$$T_{1/2}^{0\nu2\beta Mbulk} \geq 1.7\ (2.3)\times10^{21} \text{ років.}$$

Виключені з 90% C.L. розподіли, що очікуються в процесах $0\nu2\beta M1$, $0\nu2\beta M2$ та $0\nu2\beta$–розпаду з випроміненням "bulk" майоронів показані на рис. 3.2.

## 3.2. Пошук $2\beta$–процесів в ядрах $^{106}$Cd, $^{108}$Cd, $^{114}$Cd, $^{180}$W, та $^{186}$W

### 3.2.1. Експериментальні обмеження на процеси $2\beta$–розпаду ядер $^{114}$Cd та $^{186}$W.

Експериментальний спектр фону, виміряний детектором з кристалами $^{116}CdWO_4$, було використано для пошуку $2\beta$–розпаду ядер $^{114}$Cd та $^{186}$W, що входять до складу цього детектора. Енергія $2\beta$–переходу $^{114}$Cd становить 536 кеВ, а у ядра $^{186}$W – 487.9 кеВ. В обох цих ядер досить висока розповсюдженість у природній суміші ізотопів: 28.73% ($^{114}$Cd) і 28.6% ($^{186}$W). В той час як $^{114}$Cd може розпадатися з випроміненням двох електронів лише на основний стан $^{116}$Sn, $2\beta$–розпад $^{186}$W може відбуватися як на основний, так і на перший збуджений стан ($2^+$) $^{186}$Os (рис. 3.5).

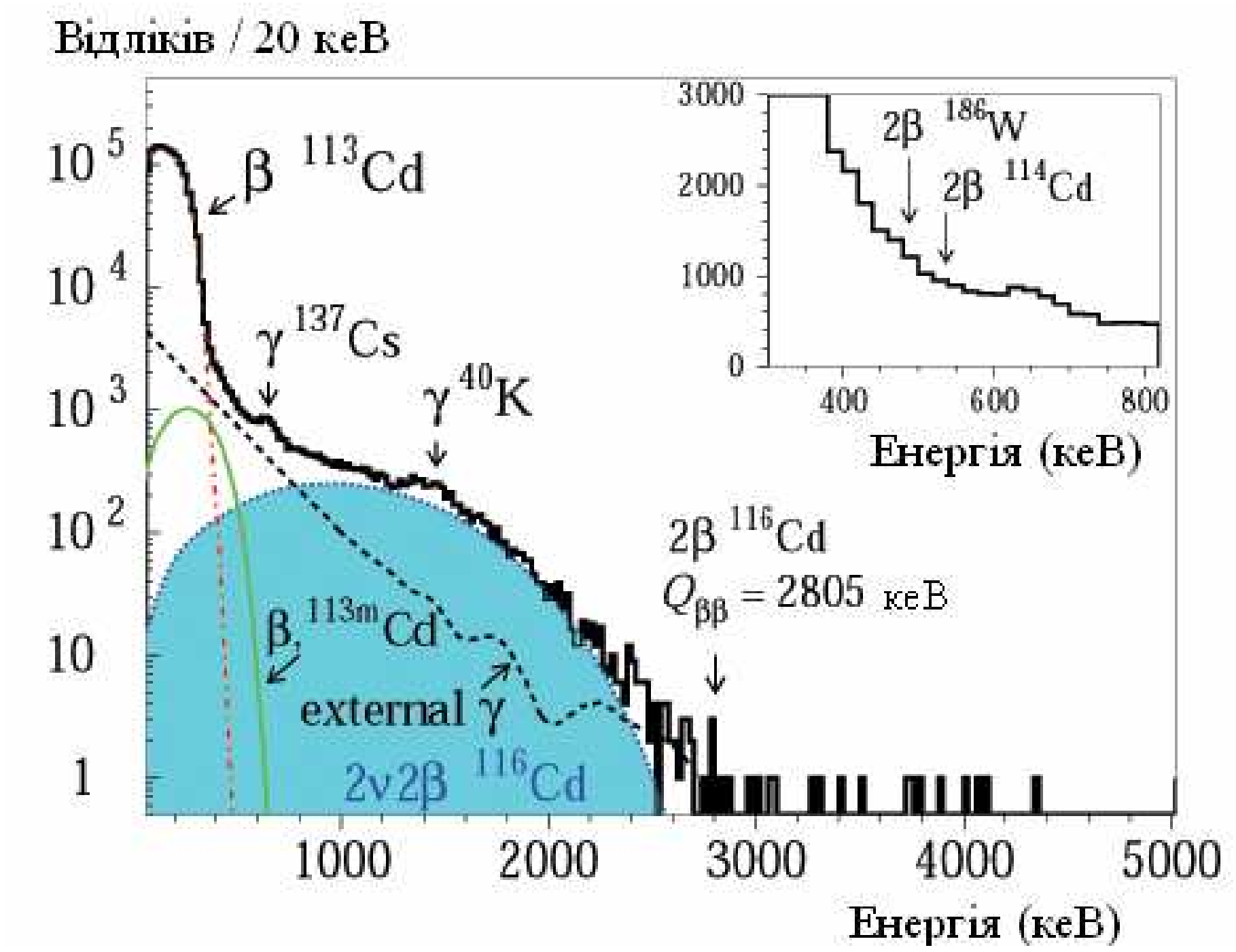

Рис. 3.4. Спектр γ(β)–подій, виміряний з детектором $^{116}CdWO_4$ протягом 13 316 годин, після відбору за допомогою дискримінації за формою сцинтиляційних сигналів. Показані основні компоненти фону. На вставці зображена ділянка спектру в області, де очікуються піки від 0ν2β–розпаду ядер $^{186}W$ та $^{114}Cd$.

Інтенсивність фону в області 440 – 540 кеВ дорівнює 0.333(4) відліки/(добу × кеВ × кг). В дослідних даних нема вказівок на можливий сигнал від 0ν2β–розпаду $^{186}W$, а отже можна встановити лише обмеження на імовірність цього процесу за допомогою формули 3.1. Число ядер $^{186}W$ в кристалі становить $1.56 \times 10^{23}$, час вимірювань – 13 316 годин. Загальна ефективність реєстрації включає ефективність реєстрації подій 0ν2β–розпаду детекторами $^{116}CdWO_4$ ($\eta_R$), та ефективність відбору γ(β) подій за допомогою дискримінації частинок за формою сцинтиляційних сигналів ($\eta_{PSA}$). Перше значення визначалось за допомогою моделювання по методу Монте–Карло і склало $\eta_R = 99.4\%$. Таким чином, приймаючи до уваги ефективність $\eta_{PSA} = 95\%$, загальна ефективність становить $\eta = 94\%$.

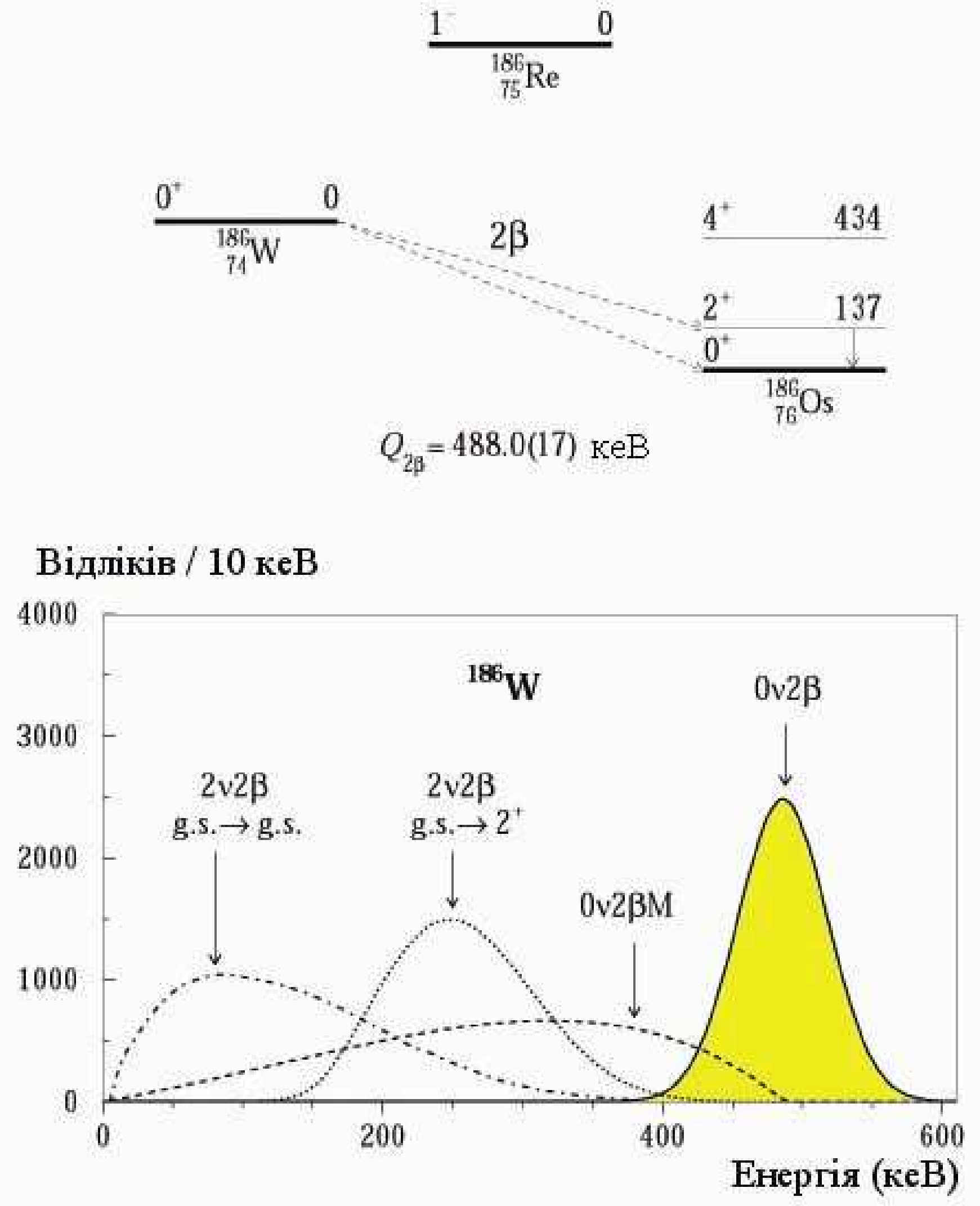

Рис. 3.5. Схема рівнів $^{186}W$ – $^{186}Re$ – $^{186}Os$ та енергетичні спектри різних мод 2β–розпаду $^{186}W$.

Значення lim$S$ було визначене двома шляхами. По–перше, воно було оцінене просто як квадратний корінь з числа фонових подій в даному енергетичному вікні. Незважаючи на простоту цього методу, він дає вірну оцінку чутливості експерименту. В інтервалі 440–540 кеВ, який містить 89% очікуваного 0ν2β–піку (повна ефективність реєстрації ефекту $\eta$=84%), налічується 6091 подій, що дає lim$S$ = 128 відліків з 90% C.L. Використовуючи це значення, було отримано обмеження:

$$T_{1/2} \geq 1.1 \times 10^{21} \text{ років} \qquad \text{з 90\% C.L.}$$

Значення lim$S$ було також визначене із застосуванням підгонки по методу найменших квадратів. Дослідний спектр в інтервалі 380–1200 кеВ підганявся сумою розподілів, які описували фон ($^{113m}$Cd, $^{137}$Cs, експонента від зовнішніх $\gamma$–квантів та $2\nu2\beta$–розпад $^{116}$Cd) і шуканого $0\nu2\beta$–піку. Підгонка ($\chi^2$/ч.с.с. = 1.06) дала площу піку $-21 \pm 95$ відліків, тобто не виявила ефекту. Консервативно приймаючи від‘ємне значення за нульове, ми можемо отримати таку оцінку граничної площі піку: lim$S$ = 156 (95) відліків з 90% (68%) C.L. Використовуючи цю величину та значення ефективності ($\eta$ = 94%), було встановлене обмеження на період напіврозпаду $^{186}$W відносно $0\nu2\beta$–розпаду на основний рівень $^{186}$Os:

$$T^{0\nu}_{1/2}(^{186}\text{W}) \geq 1.0\ (1.6)\times10^{21} \text{ років} \qquad \text{з 90\% (68\%) C.L.}$$

У випадку $0\nu2\beta$–розпаду $^{186}$W на перший збуджений стан ($2^+$) $^{186}$Os повинен випромінюватися $\gamma$–квант з енергією 137 кеВ (зняття збудження рівня $2^+$ $^{186}$Os). Завдяки практично повному його поглинанню детекторами $^{116}CdWO_4$, очікувана функція відгуку майже не відрізняється від випадку $2\beta$–переходу на основний стан (рис. 3.5). З урахуванням ефективності реєстрації ефекту $\eta$=92%, одержимо обмеження на період напіврозпаду ядра $^{186}$W відносно $0\nu2\beta$–розпаду на перший збуджений рівень ядра $^{186}$Os:

$$T^{0\nu}_{1/2}(^{186}\text{W, о.с.}\rightarrow2^+) \geq 1.0\ (1.6)\times10^{21} \text{ років} \quad \text{з 90\% (68\%) C.L.}$$

Подібним чином, для $0\nu2\beta$–розпаду $^{186}$W з емісією майорону підгонка спектру дала наступні обмеження:

$$\text{T}^{0\nu\text{M}}_{1/2}\ (^{186}\text{W}) \geq 1.2\ (1.4)\times10^{20} \text{ років} \quad \text{з 90\% (68\%) C.L.}$$

Для оцінки нижньої границі на період напіврозпаду відносно двохнейтринної моди $2\beta$–розпаду $^{186}$W аналізувався спектр, показаний на рис. 3.6. Оскільки практично 100% $\beta$–частинок від очікуваного $2\nu2\beta$–розпаду поглинається в кристалах, загальна ефективність складає $\eta = 98\%$. Всього 2% $\gamma(\beta)$ подій з цих даних були відкинуті в результаті аналізу за формою сигналів. Для підгонки експериментальних даних було використано просту модель, яка включає $2\nu2\beta$ спектр $^{186}W$, $\beta$–спектр $^{113}Cd$ та експоненційну функцію. Остання компонента моделі описує зовнішній $\gamma$–фон. Найкращий фіт ($\chi^2$/ч.с.с.=1.9), досягнутий в інтервалі 100–500 кеВ, дав 60±1962 події для шуканого ефекту. Це дозволяє відкинути гіпотезу про наявність 3277 (2022) подій шуканого ефекту з довірчою імовірністю 90% (68%) і отримати наступні обмеження на $2\nu2\beta$–розпад $^{186}W$ на основний стан дочірнього ядра:

$$T^{2\nu}_{1/2}(^{186}W) \geq 2.6\ (4.1) \times 10^{18} \text{ років} \qquad \text{з } 90\%\ (68\%)\ \text{C.L.}$$

Розподіл $2\nu2\beta$–розпаду $^{186}W$, виключений з 90% C.L., зображено на рис 3.6. Така ж методика дала обмеження на $2\nu2\beta$–переходи на перший збуджений стан ($2^+$) $^{186}Os$:

$$T^{2\nu}_{1/2}(^{186}W, \text{о.с.}\to 2^+) \geq 1.0\ (1.3)\times 10^{19} \text{ років} \quad \text{з } 90\%\ (68\%)\ \text{C.L.},$$

та на процеси подвійного бета–розпаду ядра $^{114}Cd$. Всі отримані обмеження приведені в таблиці 3.2.

Отримані для ядра $^{186}W$ результати все ще далекі від теоретичних передбачень, які, наприклад, для $0\nu2\beta$–розпаду на основний стан $^{186}Os$ знаходяться в межах $T_{1/2} \times \langle m_\nu \rangle^2 = (6 \times 10^{24}$ [198] – $5 \times 10^{25}$ років×еВ$^2$ [265]). В той же час, згідно теоретичних оцінок двохнейтринна мода $2\beta$–розпаду ядра $^{186}W$ дуже сильно придушена. Це значно полегшує пошуки безнейтринної моди, оскільки саме двохнейтринна мода $2\beta$–розпаду стає одним з найнебезпечніших джерел фону при спробі підвищити чутливість експериментів до рівнів $T_{1/2} \geq 10^{26}$ років. Це особливо важливо для пошуків $0\nu2\beta$–розпаду з вильотом майоронів, оскільки цей процес має дати неперервний розподіл електронів, а отже, навіть у випадку використання детектора з дуже високою роздільною здатністю, проблема фону від $2\nu2\beta$–моди може бути вирішена лише завдяки вибору ядер з суттєво придушеною ймовірністю $2\nu2\beta$–розпаду. Одним з таких ядер і є ядро $^{186}$W, досліджене в даній роботі.

3.2.2. Пошук $2\varepsilon$, $\varepsilon\beta^+$, $2\beta^+$ процесів в ядрах $^{106}$Cd, $^{108}$Cd та $^{180}$W.

Для пошуку процесів подвійного електронного захвату в ядрі $^{180}$W був використаний спектр, накопичений з детекторами $^{116}CdWO_4$ за 692 годин з низьким енергетичним порогом ≈80 кеВ. Цей спектр зображено на рис. 3.6.

Підгонка спектру методом найменших квадратів в інтервалі 90–240 кеВ дала 91±177 відліків для шуканого піку ($\chi^2$/ч.с.с. = 1.3), що може бути інтерпретовано як відсутність ефекту на даному рівні чутливості. Ці дані дозволяють оцінити величину $\lim S = 381$ (268) відлік з 90% (68%) C.L. Приймаючи до уваги 100% ефективність реєстрації цього процесу (отже, сукупна ефективність визначається ефективністю відбору подій від $\gamma$–квантів, рентгенівських квантів та $\beta$–частинок за методом аналізу форми сцинтиляційних сигналів: $\eta = \eta_{PSA} = 98\%$) та число ядер $^{180}$W ($6.57 \times 10^{20}$), можна вирахувати обмеження на процес безнейтринного подвійного електронного захвату в ядрі $^{180}$W:

$$T^{0\nu2\varepsilon}_{1/2}(^{180}\text{W}) \geq 0.9\ (1.3)\times10^{17}\ \text{років} \qquad \text{з 90\% (68\%) C.L.}$$

Така ж методика дала обмеження на двохнейтринний подвійний електронний захват з К–оболонки в ядрі $^{180}$W:

$$T^{2\nu2K}_{1/2}\ (^{180}\text{W}) \geq 0.7\ (0.8)\cdot10^{17}\ \text{років} \qquad \text{з 90\% (68\%) C.L.}$$

Піки $0\nu2\varepsilon$– та $2\nu2K$–розпадів $^{180}$W, виключені з 90% C.L., зображені на вкладці рис. 3.6.

Кристали $^{116}CdWO_4$, окрім ядер $^{116}$Cd, містять інші ізотопи кадмію, які є потенціальними кандидатами на $2\beta$–розпад: $^{106}$Cd з ізотопною розповсюдженістю $\varepsilon = 0.16\%$ і енергією розпаду $Q_{\beta\beta} = 2771$ кеВ та $^{108}$Cd($\varepsilon = 0.11\%$, $Q_{\beta\beta} = 5269$ кеВ). Це дозволило встановити обмеження на періоди напіврозпаду цих нуклідів відносно процесів $2\beta$–розпаду.

У відповідності до схеми рівнів триплету $^{108}Cd$–$^{108}Ag$–$^{108}Pd$ дозволений подвійний електронний захват $^{108}Cd$ ($Q_{2\beta}$ = 269 кеВ). У випадку двохнейтринного подвійного електронного захвату з *K*–оболонки (2ν2*K*) загальна вивільнена у

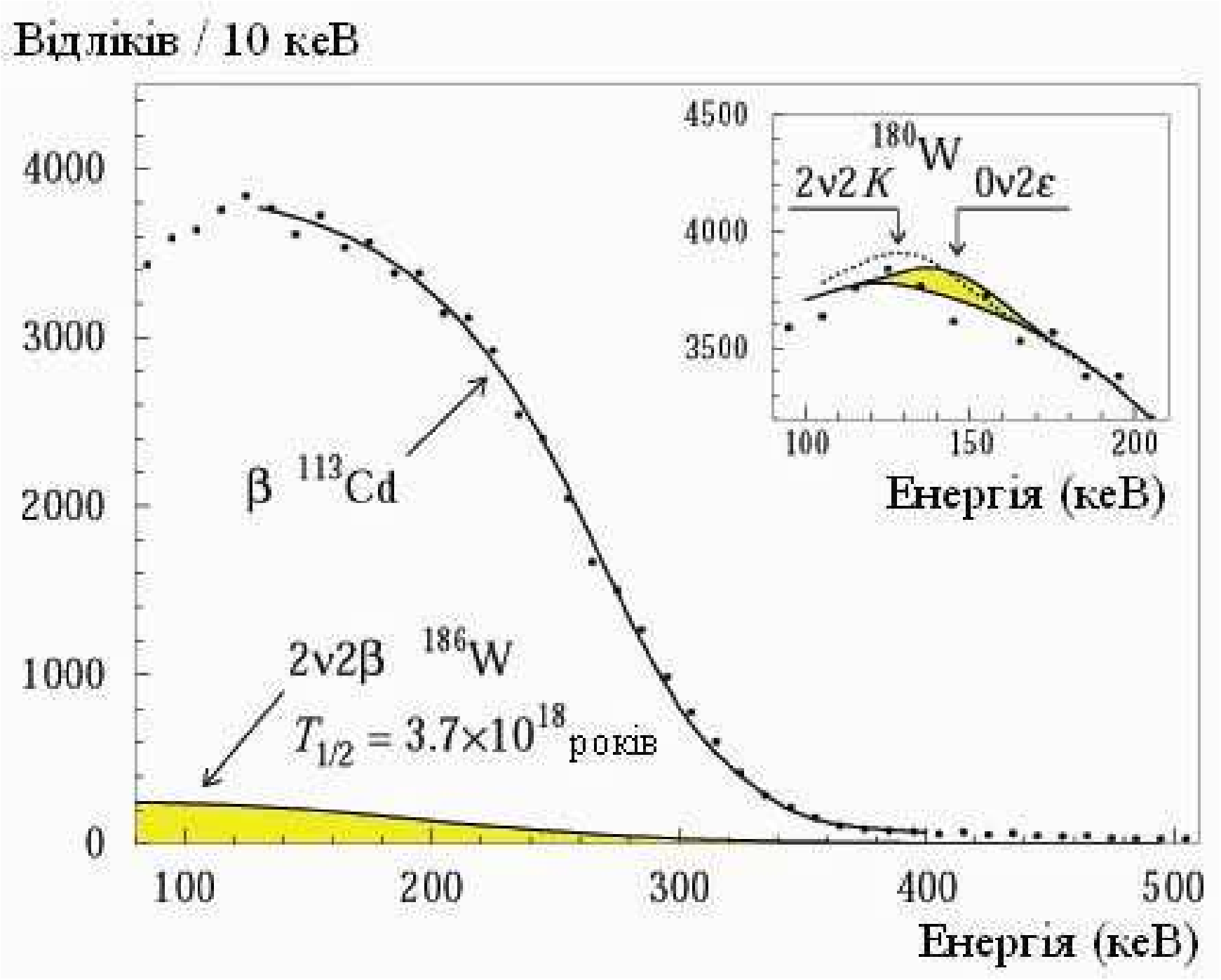

Рис. 3.6. Низькоенергетична частина спектру, виміряного детекторами $^{116}CdWO_4$ за 692 годин з енергетичним порогом 80 кеВ (γ(β)–події були відібрані за допомогою аналізу форми сигналів з ефективністю 98%). Суцільною лінією показана модель спектру, отримана в результаті підгонки. Заштрихована гістограма – спектр 2ν2β–розпаду $^{186}W$ з $T_{1/2}^{2\nu}$ = 3.7×10$^{18}$ р, виключений на рівні 90% C.L. На вкладці показана частина спектру разом з піком 2ν2K–захвату $^{180}W$ з $T_{1/2}^{2\nu2K}$=0.7×10$^{17}$ р та від процесу 0ν2ε–захвату $^{180}W$ з $T_{1/2}^{0\nu2\varepsilon}$ = 0.9×10$^{17}$ р (заштриховано). Обидва піки виключені на рівні 90% C.L.

детекторі енергія становить 2$E_K$ (де $E_K$ = 24.4 кеВ – енергія зв'язку електронів на *K*–оболонці атома паладію), в той час як інша частина енергії ($Q_{2\beta}$–2$E_K$ ≈ 220 кеВ) забирається нейтрино. При безнейтринному подвійному електронному захваті (0ν2ε) уся доступна енергія (яка передається *X*–променям, Оже–електронам, $\gamma$–квантам чи конверсійним електронам) повинна привести до реєстрації в детекторі піку з енергією $Q_{2\beta}$ = 269 кеВ.

Для встановлення обмежень на $0\nu2\varepsilon$–процес в ядрі $^{108}Cd$ була здійснена підгонка фонового спектра, виміряного детекторами $^{116}CdWO_4$ протягом 692 годин (рис. 3.6). Підгонка дозволила отримати обмеження:

$T_{1/2}^{0n2\varepsilon} \geq 1.5\ (2.5) \times 10^{17}$ років з 90% (68%)C.L.

Таким же чином, з аналізу спектра фону, виміряного з незбагаченим (а отже із значно більшим ізотопним вмістом $^{106}Cd$ та $^{108}Cd$ – 1.25% та 0.89%, відповідно) детектором $CdWO_4$ масою 433 г [266], були одержані обмеження на процеси подвійного електронного захвату в ядрах $^{106}Cd$ та $^{108}Cd$ з вильотом нейтрино. Оскільки в цьому випадку енерговиділення в детекторі досить мале (49 кеВ), важливо було використовувати дані вимірювань з достатньо низьким порогом, який якраз і був досягнутий ($\approx$ 40 кеВ) в досліді [266]. Енергетичне розділення детектора при енергії 60 кеВ ($\gamma$–промені $^{241}Am$) становило FWHM = 25 кеВ. Цей кристал містив $9.5 \times 10^{21}$ та $6.8 \times 10^{21}$ ядер $^{106}Cd$ та $^{108}Cd$, відповідно. Виміряний спектр підганявся сумою очікуваного піку при енергії 49 кеВ та моделі фону ($\beta$–спектр $^{113}Cd$). Це дало наступні обмеження на $2\nu2K$–захват в ядрі $^{106}Cd$:

$T_{1/2}^{2\nu2K} \geq 5.8\ (9.5) \times 10^{17}$ років з 90% (68%) C.L.

Аналогічно були встановлені границі імовірності $2\nu2K$–процесу в $^{108}Cd$:

$T_{1/2}^{2\nu2K} \geq 4.6(6.7) \times 10^{17}$ років з 90%(68%) C.L.

Як видно із схеми рівнів триплету $^{106}Cd$ – $^{106}Ag$ – $^{106}Pd$ (рис. 3.7 (а)), окрім подвійного електронного захвату, для $^{106}Cd$ дозволені подвійний позитронний розпад ($\beta^+\beta^+$) та процес захвату електрону з одночасним випроміненням позитрону ($\varepsilon\beta^+$) як на основний, так і на збуджений стани $^{106}Pd$. Розрахована за методом Монте–Карло функція відгуку детектора $^{116}CdWO_4$ по відношенню до безнейтринних мод цих процесів показана на рис. 3.7 (б). Видно, що кожний з вищезгаданих процесів призведе до появи піку повного поглинання з енергією 2771

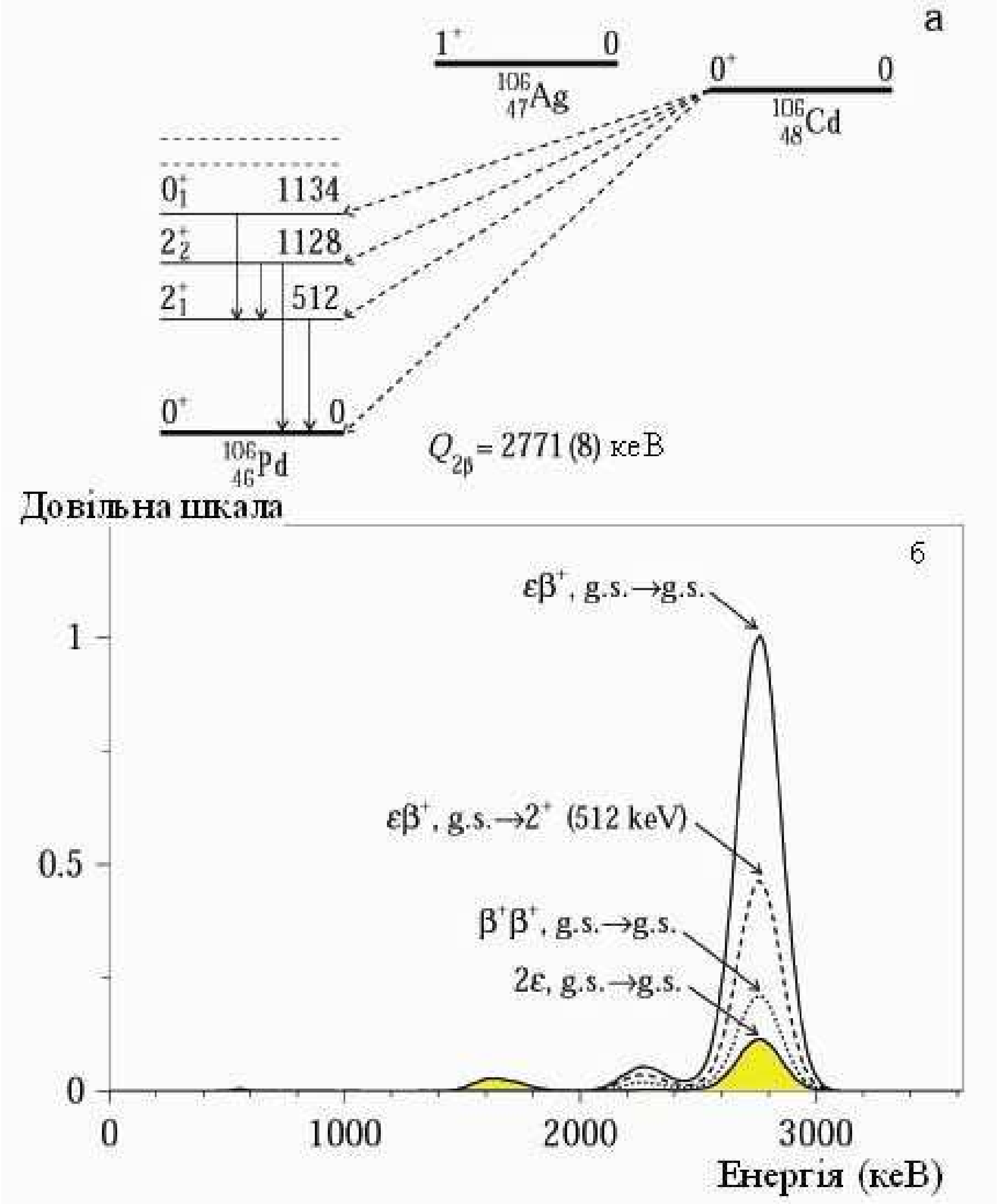

Рис. 3.7. Схема рівнів триплету $^{106}$Cd – $^{106}$Ag – $^{106}$Pd (а); функція відгуку детектора $^{116}CdWO_4$ по відношенню до різних мод безнейтринних процесів в $^{106}$Cd (б).

кеВ з шириною на половині висоти піку FWHM = 216 кеВ. Обчислені значення ефективності реєстрації подій в цьому піку становлять: $\eta_{mc}(0\nu 2\varepsilon$, о.с. → о.с.) = 1.5%; $\eta_{mc}(0\nu\varepsilon\beta^+$, о.с. → о.с.) = 12.8%; $\eta_{mc}(0\nu\varepsilon\beta^+$, о.с. → $2^+$, 512 кеВ) = 5.9%; $\eta_{mc}(0\beta^+\beta^+$,о.с. → о.с.) = 2.7%; $\eta_{mc}(0\nu\beta^+\beta^+$, о.с. →$2^+$, 512 кеВ) = 1.2%; і так далі.

Підгонка енергетичного спектру показаного на рис. 3.2 в інтервалі 2.2 – 3.4 МеВ дала можливу площу шуканого піку $S = –1.0±1.6$ відліків. З останнього випливає, наприклад, період напіврозпаду $^{106}$Cd по каналу 0ν2ε–захвату:

$$T_{1/2}^{0n2\varepsilon}(\text{o.c.} \rightarrow \text{o.c.}) \geq 0.8\ (1.7)\times10^{19} \text{ років} \qquad \text{з } 90\%\ (68\%)\ \text{C.L.}$$

Таким же чином було розраховано обмеження на інші 0ν та 2ν подвійні β–процеси в $^{106}$Cd, які приведені в таблиці 3.2.

## 3.3. Результати експерименту із сцинтиляційними кристалами $^{116}CdWO_4$

Всі обмеження на імовірність 2β–процесів, отримані в досліді з изькофоновими кристалами $^{116}CdWO_4$, підсумовані в таблиці 3.2. Слід наголосити, що більшість обмежень для $^{106}$Cd, $^{108}$Cd, $^{114}$Cd, $^{116}$Cd, $^{180}$W та $^{186}$W вищі за попередні результати або отримані вперше. Наприклад $T_{1/2}^{0\nu2\beta} \geq 10^{21}$ р для 0ν2β–розпаду $^{186}$W майже на порядок більше за попередній результат [260], а обмеження на 0ν–розпад з випроміненням Майорона та 2β–розпад на перший ($2^+$) збуджений стан $^{186}$Os були встановлені вперше. Підкреслимо, що на сьогодні рівень чутливості $T_{1/2}^{0\nu2\beta} > 10^{21}$ р досягнутий лише для 10 нуклідів [22].

Головними результатами досліду є виміряне значення періоду напіврозпаду $^{116}$Cd відносно двохнейтринної моди подвійного бета–розпаду $T_{1/2}^{2\nu} = (2.9_{-0.3}^{+0.4}) \times 10^{19}$ р. (графік Кюрі для 2ν2β–розпаду добре описується прямою, а вирахуване значення $Q_{\beta\beta}(2808 \pm 43$ кеВ) співпадає з табличною величиною 2805(4) кеВ) та обмеження на різні канали безнейтринного подвійного бета–розпаду $^{116}$Cd. Слід підкреслити, що всі отримані для $^{116}$Cd обмеження найбільш жорсткі для цього ядра, наприклад, граничне значення для 0ν2β–розпаду $^{116}$Cd на основний стан $^{116}$Sn: $T_{1/2}^{0\nu} > 1.7\ (2.6) \times 10^{23}$ років з 90% (68%)C.L. є одним з найкращих обмежень на процеси 0ν2β–розпаду.

Таблиця 3.2. Загальний перелік результатів, отриманих для процесів 2β–розпаду ядер кадмію та вольфраму.

| Нуклід | Процес | Мода розпаду | Перехід (енергія рівня) | Обмеження або значення $T_{1/2}$, років, 90% (68)% C.L. |
|---|---|---|---|---|
| $^{106}$Cd | 2ε | 0ν | o.c. – o.c. | $\geq 0.8(1.7)\times10^{19}$ |
| | $\varepsilon\beta^+$ | 0ν | o.c. – o.c. | $\geq 0.7(1.6)\times10^{20}$ |
| | | 0ν | o.c. – $2_1^+$(511.9 кеВ) | $\geq 3.1(7.2)\times10^{19}$ |
| | | 0ν | o.c. – $2_2^+$(1128.0 кеВ) | $\geq 1.4(3.3)\times10^{19}$ |
| | | 0ν | o.c. – $0_1^+$(1133.8 кеВ) | $\geq 1.4(3.2)\times10^{19}$ |
| | $2\beta^+$ | 0ν | o.c. – o.c. | $\geq 1.4(3.3)\times10^{19}$ |
| | | 0ν | o.c. – $2_1^+$(511.9 кеВ) | $\geq 0.6(1.5)\times10^{19}$ |
| | 2*K* | 2ν | o.c. – o.c. | $\geq 5.8(9.5)\times10^{17}$ |
| | $\varepsilon\beta^+$ | 2ν | o.c. – o.c. | $\geq 1.2(2.0)\times10^{18}$ |
| | $2\beta^+$ | 2ν | o.c. – o.c. | $\geq 5.0(8.2)\times10^{18}$ |
| $^{108}$Cd | 2ε | 0ν | o.c. – o.c. | $\geq 1.5(2.5)\times10^{17}$ |
| | 2*K* | 2ν | o.c. – o.c. | $\geq 4.1(6.7)\times10^{17}$ |
| $^{114}$Cd | $2\beta^-$ | 0ν | o.c. – o.c. | $\geq 2.5(4.1)\times10^{20}$ |
| | $2\beta^-$ | 2ν | o.c. – o.c. | $\geq 6.0(9.3)\times10^{17}$ |
| $^{116}$Cd | $2\beta^-$ | 0ν | o.c. – o.c. | $\geq 1.7(2.6)\times10^{23}$ |
| | | 0ν | o.c. – $2^+$ (1293.5 кеВ) | $\geq 2.9(4.3)\times10^{22}$ |
| | | 0ν | o.c. – $0_1^+$(1756.8 кеВ) | $\geq 1.4(2.2)\times10^{22}$ |
| | | 0ν | o.c. – $0_2^+$(2027.3 кеВ) | $\geq 0.6(0.9)\times10^{22}$ |
| | | 0ν*M*1 | o.c. – o.c. | $\geq 0.8(1.8)\times10^{22}$ |
| | | 0ν*M*2 | o.c. – o.c. | $\geq 0.8(1.4)\times10^{21}$ |
| | | 0ν$M^{bulk}$ | o.c. – o.c. | $\geq 1.7(2.3)\times10^{21}$ |
| | $2\beta^-$ | $2\nu$ | о.с. – о.с. | $= 2.9_{-0.3}^{+0.4}\times10^{19}$ |
| | | $2\nu$ | о.с. – $2^+$ (1293.5 кеВ) | $\geq 0.6(1.1)\times10^{21}$ |
| | | $2\nu$ | о.с. – $0_1^+$ (1756.8 кеВ) | $\geq 0.8(2.2)\times10^{21}$ |
| | | $2\nu$ | о.с. – $0_2^+$(2027.3 кеВ) | $\geq 0.4(0.6)\times10^{21}$ |
| $^{180}$W | $2\varepsilon$ | $0\nu$ | о.с. – о.с. | $\geq 0.9(1.3)\times10^{17}$ |
| | $2K$ | $2\nu$ | о.с. – о.с. | $\geq 0.7(0.8)\times10^{17}$ |
| $^{186}$W | $2\beta^-$ | $0\nu$ | о.с. – о.с. | $\geq 1.1(2.1)\times10^{21}$ |
| | | $0\nu$ | о.с. – $2^+$ (137.2 кеВ) | $\geq 1.1(2.0)\times10^{21}$ |
| | | $0\nu M1$ | о.с. – о.с. | $\geq 1.2\ (1.4)\times10^{20}$ |
| | $2\beta^-$ | $2\nu$ | о.с. – о.с. | $\geq 3.7(5.3)\times10^{18}$ |
| | | $2\nu$ | о.с. – $2^+$ (137.2 кеВ) | $\geq 1.0(1.3)\times10^{19}$ |

### 3.3.1. Маса нейтрино майоранівської природи та параметри домішок правих токів в слабкій взаємодії.

Використовуючи обмеження на безнейтринний подвійний бета–розпад $^{116}$Cd на основний стан $^{116}$Sn та обчислення [198], можна отримати обмеження на майоранівську масу нейтрино та домішки правих струмів у слабкій взаємодії:

$$\langle m_\nu\rangle \leq 1.9 \text{ eВ},$$

$$\eta \leq 2.5\cdot10^{-8},$$

$$\lambda \leq 2.2\cdot10^{-6}$$

з 90% C.L. Якщо не приймати до уваги внеску правих струмів, отримаємо:

$$\langle m_\nu\rangle \leq 1.7\ (1.4)\text{ eВ} \qquad \text{з 90\% (68\%) C.L.,}$$

а на основі розрахунків [157] обмеження становить $\langle m_\nu\rangle \leq 1.5$ (1.2) eВ. Ці результати, разом з кращими обмеженнями на $T_{1/2}^{0\nu}$, отриманими в прямих дослідах (та відповідні границі значення майоранівської маси нейтрино), приведені в таблиці 3.3.

Обмеження на $\langle m_\nu \rangle$ були обраховані на основі розрахунків [198], які були вибрані по тій причині, що містять найбільш широкий перелік ядер 2β–кандидатів, що дозволяє порівняти чутливість різних дослідів до маси нейтрино. З таблиці 3.3 видно, що отриманий нами результат для $^{116}$Cd є одним з найкращих. При порівнянні обмежень на ефективну масу нейтрино слід приймати до уваги, що в зв'язку з досить низькою точністю розрахунків матричних ядерних елементів (принаймні ± один порядок величини), обмеження на $\langle m_\nu \rangle$ слід розглядати як такі, які мають точність в межах множника ×3.

Таблиця 3.3. Найкращі опубліковані значення $T_{1/2}^{0\nu2\beta}$ та обмеження на $\langle m_\nu \rangle$, отримані в прямих дослідах з 2β–розпаду.

| Нуклід | Експериментальне обмеження $T_{1/2}^{0\nu}$ (р) | | Робота | Обмеження на $\langle m_\nu \rangle$ (еВ) на основі роботи [198] | |
|---|---|---|---|---|---|
| | 68% C.L. | 90% C.L. | | 68% C.L. | 90% C.L. |
| $^{76}$Ge | $3.1\times10^{25}$ | $1.9\times10^{25}$ | [116] | 0.27 | 0.35 |
| | $1.6\times10^{25}$ | | [192,193] | 0.38 | |
| | $4.2\times10^{25}$ | $2.5\times10^{25}$ | [267] | 0.24 | 0.31 |
| $^{82}$Se | | $1.4\times10^{23}$ | [207] | | 2.1 |
| $^{100}$Mo | | $3.1\times10^{23}$ | [207] | | 2.1 |
| $^{116}$Cd | $2.6\times10^{23}$ | $1.7\times10^{23}$ | [161] | 1.4 | 1.7 |
| $^{130}$Te | $5.5\times10^{23}$ | | [219] | 0.94 | |
| $^{136}$Xe | $4.4\times10^{23}$ | | [221] | 2.2 | |

Таблиця 3.4. Найкращі опубліковані обмеження (90% C.L.) на параметри домішок правих токів (з огляду [268]).

| Нуклід | Обмеження на параметр | | Робота |
|---|---|---|---|
| | $\langle\lambda\rangle\times10^{-6}$ | $\langle\eta\rangle\times10^{-8}$ | |
| $^{76}$Ge | ≤ 1.1 | ≤ 0.64 | [269] |
| $^{116}$Cd | ≤ 2.2 | ≤ 2.5 | [161] |
| $^{130}$Te | ≤ (1.6 – 2.4) | ≤ (0.9 – 5.3) | [270] |
| $^{136}$Xe | ≤ 4.4 | ≤ 2.3 | [271] |
| $^{128}$Te | | ≤ 5.3 | [145][2] |

### 3.3.2. Обмеження на константу зв'язку нейтрино з майороном та параметр мінімальної суперсиметричної теорії з порушенням R–парності.

Використовуючи обмеження на 0ν2β–розпад з вильотом Майорона: $T_{1/2}^{0\nu} > 0.8\ (1.8) \times 10^{22}$ років з 90% (68%) C.L. та обчислення [60] можна встановити обмеження на ефективну константу зв'язку майорон–нейтрино $g_M < 8.1\ (5.4)\times10^{-5}$, а на основі [157] $g_M < 4.6\ (3.1)\times10^{-5}$. Ці результати знаходяться серед кращих, отриманих до цього часу в прямих експериментах по пошуку 2β–розпаду $^{76}$Ge, $^{82}$Se, $^{100}$Mo, $^{130}$Te та $^{136}$Xe (див. табл. 3.5).

Таблиця 3.5. Найбільш жорсткі обмеження (90% C.L.) на періоди напіврозпаду ядер відносно 0ν2β–розпаду з вильотом майоронів (з огляду [268]).

| Нуклід | Експериментальне обмеження $T_{1/2}^{0\nu2\beta M1}$ ($\times10^{21}$) років | Обмеження на параметр $\langle g_M\rangle$ ($\times10^{-5}$) | Робота |
|---|---|---|---|
| $^{76}$Ge | ≥ 17 | | [199] |

[2] Результат геохімічного експерименту.

| | | | |
|---|---|---|---|
| $^{82}$Se | ≥ 2.4 | | [170] |
| $^{96}$Zr | ≥ 0.31 | | [171] |
| $^{100}$Mo | ≥ 5.8 | | [272] |
| $^{116}$Cd | ≥ 8 | ≤ (4.6 – 8.1) | [161] |
| $^{130}$Te | ≥ 2.2 | ≤ (17 – 33) | [270] |
| $^{136}$Xe | ≥ 7.2 | | [271] |

У відповідності з роботою [66] параметр порушення R–парності в мінімальній суперсиметричній стандартній моделі обмежується отриманим значенням $\lim T_{1/2}^{0\nu2\beta}$ до $\varepsilon \leq 7.0\ (6.3)\times10^{-4}$ з 90% (68%) C.L. Розрахунки згідно з роботою [ 273 ] дають ще більш жорсткі обмеження на цей параметр: $\varepsilon < 2.7\ (2.4)\times10^{-4}$.

**Висновки розділу.** В розділі представлені результати досліджень подвійного β–розпаду ядер кадмію та вольфраму. Виміряний період напіврозпаду ядра $^{116}$Cd відносно 2ν2β–розпаду $T_{1/2} = 2.9^{+0.4}_{-0.3} \times 10^{19}$ років. Встановлене нове обмеження на період напіврозпаду відносно 0ν2β розпаду $^{116}$Cd на основний стан ядра $^{116}$Sn (з довірчою імовірністю 90%): $T_{1/2}\ (0^+ \rightarrow 0^+) \geq 1.7 \times 10^{23}$ років. Встановлені нові обмеження на періоди напіврозпаду $^{116}$Cd відносно 0ν2β–розпаду на збуджені стани ядра $^{116}$Sn (з 90% CL): $T_{1/2}(0^+ \rightarrow 2^+) \geq 2.9 \times 10^{22}$ років, $T_{1/2}(0^+ \rightarrow 0_1^+) \geq 1.4 \times 10^{22}$ років, $T_{1/2}(0^+ \rightarrow 0_2^+) \geq 0.6 \times 10^{22}$ років. Встановлені нові обмеження на періоди 0ν2β–розпаду ядра $^{116}$Cd з випроміненням майоронів – одного, двох і т.зв. bulk майорону (з 90% CL): $T_{1/2}(\mathrm{M1}) \geq 0.8 \times 10^{22}$ років, $T_{1/2}(\mathrm{M2}) \geq 0.8 \times 10^{21}$ років, $T_{1/2}(\mathrm{M^{bulk}}) \geq 1.7 \times 10^{21}$ років. З експериментального обмеження на 0ν2β–розпад $^{116}$Cd отримані одні з кращих у світі обмежень на ефективну масу нейтрино майоранівської природи: $\langle m_\nu\rangle \leq 1.7$ еВ, параметри домішок правих токів в слабкій взаємодії: $\langle\eta\rangle \leq 2.5 \times 10^{-8}$, $\langle\lambda\rangle \leq 2.2 \times 10^{-6}$, та параметр порушення $R$–парності в мінімальній суперсиметричній СМ з порушенням *R*–парності: $\varepsilon \leq 7 \times 10^{-4}$. З експериментального обмеження на $0\nu2\beta$–розпад $^{116}Cd$ з вильотом майорона отримане одне з найбільш жорстких обмежень на константу зв'язку нейтрино з майороном: $\langle g_M \rangle \leq 4.6 \times 10^{-5}$. За допомогою сцинтиляторів вольфрамату кадмію було також здійснено пошук $2\beta$–процесів в ядрах $^{106}Cd$, $^{108}Cd$, $^{114}Cd$, $^{180}W$, $^{186}W$ та $2\varepsilon$, $\varepsilon\beta^+$, $2\beta^+$ процесів в ядрах $^{106}Cd$, $^{108}Cd$ та $^{180}W$.

Результати, викладені в цьому розділі, опубліковані в роботах:

1) F.A.Danevich, A.Sh.Georgadze, J.Hellmig, M.Hirsch, H.V.Klapdor–Kleingrothaus, V.V.Kobychev, B.N.Kropivyansky, V.N.Kuts, A.Muller, A.S.Nikolaiko, F.Petry, O.A.Ponkratenko, H.Strecker, V.I.Tretyak, M.Vollinger, Yu.Zdesenko.
   Investigation of $\beta^+\beta^+$ and $\beta^+$/EC decay of $^{106}Cd$.
   Z. Physik A 355(1996)433–437.
2) F.A.Danevich, A.Sh.Georgadze, V.V.Kobychev, B.N.Kropivyansky, A.S.Nikolaiko, O.A.Ponkratenko, V.I.Tretyak, Yu.Zdesenko.
   Limits on Majoron modes of $^{116}Cd$ neutrinoless $2\beta$ decay.
   Nucl. Phys. A 643(1998)317–328.
3) F.A.Danevich, A.Sh.Georgadze, V.V.Kobychev, B.N.Kropivyansky, A.S.Nikolaiko, O.A.Ponkratenko, V.I.Tretyak, S.Yu.Zdesenko, Yu.G.Zdesenko, P.G.Bizzeti, T.F.Fazzini, P.R.Maurenzig.
   New results of $^{116}Cd$ double $\beta$ decay study with $^{116}CdWO_4$ scintillators.
   Phys. Rev. C 62(2000)045501, 9 p.
4) F.A.Danevich, A.Sh.Georgadze, V.V.Kobychev, B.N.Kropivyansky, A.S.Nikolaiko, O.A.Ponkratenko, V.I.Tretyak, S.Yu.Zdesenko, Yu.G.Zdesenko, P.G.Bizzeti, T.F.Fazzini, P.R.Maurenzig.
   Search for $2\beta$ decay of cadmium and tungsten isotopes: Final results of the

РОЗДІЛ 4

# ПОШУКИ 2β–РОЗПАДУ ЯДЕР ГАДОЛІНІЮ ТА ЦЕРІЮ ЗА ДОПОМОГОЮ СЦИНТИЛЯТОРІВ $Gd_2SiO_5$(Ce) та $CeF_3$

## 4.1. Експеримент по пошуку 2β–процесів із сцинтиляторами $Gd_2SiO_5$(Ce)

Найбільш чутливими являються експерименти по вивченню 0*ν*2β–розпаду, здійснені за допомогою детекторів, що містять в своєму складі досліджувані ядра. Перш за все, ця методика („активного джерела") дозволяє реєструвати ефект з високою ефективністю. Тому важливо розробляти детектори, що містять в своєму складі потенційно 2β–активні ядра.

Кристалічні сцинтилятори силікату гадолінію, леговані церієм, $Gd_2SiO_5$:Ce (GSO), були розроблені в кінці 80–х, на початку 90–х років минулого століття [274,275,276,277]. Вони негігроскопічні і мають високу питому вагу (6.71 г/см$^3$), малий час висвітлювання (близько 30–60 нс), достатньо високий світловихід (≈ 20% від NaI(Tl)), та максимум спектру емісії при 440 нм. Вперше кристали GSO були застосовані для пошуку 2β–розпаду $^{160}$Gd у відділі фізики лептонів Інституту ядерних досліджень НАНУ в 1993 році [278]. Ядро $^{160}$Gd є одним з найбільш перспективних ядер для пошуку 0ν2β–розпаду. По–перше, незважаючи на порівняно невисоку енергію 2β–розпаду ($Q_{2\beta}$ = 1729.7(13) кеВ), теоретичне значення $T^{0\nu}_{1/2}\times\langle m_\nu\rangle^2 = 8.6\times10^{23}$ років × еВ$^2$ майже втричі менше, ніж для $^{76}$Ge чи $^{136}$Xe [198], отже таке саме обмеження $T_{1/2}^{0\nu}$ в досліді з $^{160}$Gd дасть більш високі обмеження на масу нейтрино та інші теоретичні параметри. По–друге, теоретичні обчислення [279] показують, що двохнейтринний 2β–розпад $^{160}$Gd являється забороненим або дуже придушеним у зв'язку з великою деформацією цього ядра. Тим часом, послаблення моди 0*ν*2β не таке сильне із–за наявності високоенергетичних станів проміжного ядра, які у випадку безнейтринного процесу, через значно більшу енергію віртуального нейтрино, впливають на збільшення ймовірності 0ν2β–розпаду. Отже енергетичний інтервал 0ν2β–розпаду $^{160}$Gd має бути вільним від фону від 2ν2β–розпаду, який становить серйозну проблему для 2β–детекторів з низькою роздільною здатністю. І по–третє, природна розповсюдженість $^{160}$Gd досить висока (21.86%), що дозволяє конструювати детектор із застосуванням звичайних, незбагачених кристалів GSO.

### 4.1.1. Низькофоновий детектор із сцинтилятором $Gd_2SiO_5$(Ce).

Для досліду був використаний кристал GSO (довжиною 5.4 см та діаметром 4.7 см), вирощений по методу Чохральського. Маса його становила 635 г, а число ядер $^{160}$Gd – $3.951\times10^{23}$. Перші 630 годин вимірювань були проведені з кристалом вагою 698 г, а потім його бокові поверхні були зішліфовані на глибину 1 – 1.5 мм. Це було зроблено для зменшення активності урану і торію, але, на відміну від кристалів $^{116}CdWO_4$, ця міра не привела до зменшення питомої активності радіонуклідів в сцинтиляторі GSO. Це свідчить про те, що в цих кристалах домішки урану і торію не концентруються в поверхневому шарі.

Дослід проводився у Солотвинській лабораторії Інституту ядерних досліджень НАН України. Світло з кристала GSO збиралося на ФЕП–110 через пластиковий світловід діаметром 8.6 см та довжиною 18.2 см. Енергетична роздільна здатність детектора вимірювалась в області 60 – 2615 кеВ за допомогою γ–променів $^{22}$Na, $^{137}$Cs, $^{207}$Bi, $^{226}$Ra, $^{232}$Th та $^{241}$Am. Роздільна здатність (FWHM) становила: 16.8%, 13.5%, 11.2% та 10.7% при енергіях 662, 1064, 1770 та 2615 кеВ, відповідно. Калібровка детектора в ході вимірювань проводилась за допомогою γ–джерела $^{207}$Bi, що вводилось всередину захисту через фторопластову трубку. Пасивний захист було споруджено з високочистої електролітичної міді марки М0 (товщина 5 см), високочистої ртуті марки Р0У, налитої в титанові контейнери (7 см), та свинцю (15 см). Записувалися енергія та час подій в діапазоні енергій 0.05 – 8 МеВ.

4.1.2. Вимірювання, аналіз даних.

Загальна протяжність досліду склала 13 949 годин (експозиція 1.015 р × кг). У фоновому спектрі кристалу GSO зображеному на рис. 4.1, можна бачити наступні особливості: чіткий пік при енергії ≈ 420 кеВ, відносно широкий пік при енергії ≈ 1050 кеВ, та два широких розподіли, що тягнуться до енергій ≈ 2.4 і ≈ 5.5 МеВ. Фон при більших енергіях може бути пояснений проходженням залишкових космічних мюонів. Спектрометр на мав антимюонного захисту, оскільки із–за порівняно високої внутрішньої забрудненості використання такого захисту було недоцільним.

Враховуючи відносний світловихід при реєстрації в сцинтиляторі GSO α–частинок по відношенню до електронів та γ–квантів (α/β–співвідношення), перший пік пояснюється α–активністю $^{152}$Gd ($T_{1/2}$ = 1.08 × $10^{14}$ р; $E_\alpha$ = 2140 кеВ; ізотопна розповсюдженість δ = 0.20%) та $^{147}$Sm ($T_{1/2}$ = 1.06×$10^{11}$ р; $E_\alpha$ = 2233 кеВ; δ = 15%). Самарій присутній в кристалі GSO в якості забруднення на рівні ≈ 8 ppm.

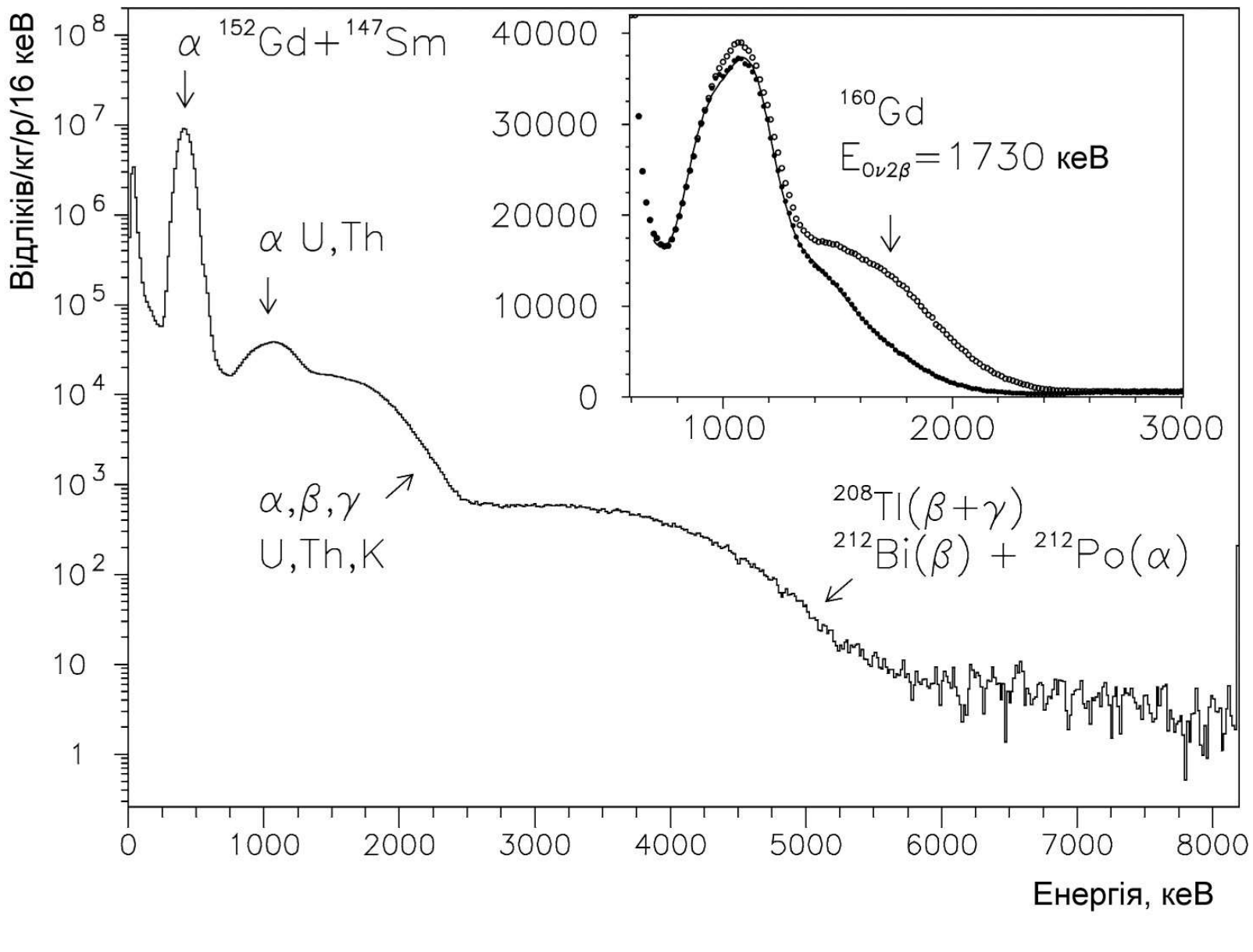

Рис. 4.1. Фон кристалу GSO (95 см$^3$), накопичений протягом 13 949 г. На вкладці: частина спектру в інтервалі 600 – 3000 кеВ разом із спектром, отриманим після відкидання розпадів $^{226}$Ra, $^{227}$Ac та $^{228}$Th від внутрішнього забруднення кристалу.

Пік з енергією ≈1050 кеВ, так само як широке розподілення до 2.4 МеВ, спричинені радіоактивною забрудненістю кристалу нуклідами $^{232}$Th, $^{235}$U та $^{238}$U. Розподіл до енергії ≈ 5.5 МеВ викликаний, головним чином, розпадами дочірніх радіонуклідів сімейства $^{232}$Th: (а) β–розпадами $^{208}$Tl ($Q_\beta$ = 5.00 МеВ); (б) β–розпадами $^{212}$Bi ($Q_\beta$ = 2.25 МеВ) та слідуючими за ними α–розпадами ядра $^{212}$Po ($T_{1/2}$ = 0.3 мкс; $E_\alpha$ = 8.78 МеВ). α–Частинки $^{212}$Po, враховуючи α/β–співвідношення, дають енергію 2.7 МеВ в γ–шкалі.

Для зменшення фону від внутрішнього забруднення кристалів було застосовано часово–амплітудний аналіз подій. За допомогою цього методу шукались швидкі ланцюжки розпадів, наприклад, послідовність двох α–розпадів із родини $^{232}$Th: $^{220}$Rn ($Q_\alpha$ = 6.41 МеВ, $T_{1/2}$ = 55.6 с) → $^{216}$Po ($Q_\alpha$ = 6.91 МеВ, $T_{1/2}$ = 0.145 с) → $^{212}$Pb. Оскільки енергія α–частинок $^{220}$Rn становить ≈1.7 МеВ в шкалі γ–квантів, події з енергіями 1.2 – 2.2 МеВ використовувались в якості тригера. Потім відбирались події в інтервалі часу 10–1000 мс (вони містять $\eta_t$ = 0.945 загальної кількості розпадів $^{216}$Po). α–Спектри $^{220}$Rn та $^{216}$Po, а також розподіл часових проміжків між першою та другою подіями, отримані за допомогою часово–амплітудного аналізу подій, зображені на рис. 4.2. Приймаючи до уваги ефективність часово–амплітудного аналізу, кількість випадкових збігів та внесок від ланцюжка $^{219}$Rn → $^{215}$Po з сімейства $^{235}$U (ефективність 2.04%), була визначена активність $^{228}$Th в кристалі GSO: 2.287(13) мБк/кг. На наступному етапі аналізу знайдені пари розпадів $^{220}$Rn та $^{216}$Po були використані в якості тригера для пошуку попередніх α–розпадів $^{224}$Ra ($Q_\alpha$ = 5.79 МеВ, $T_{1/2}$ = 3.66 д). Для пошуку подій α–розпадів $^{224}$Ra → $^{220}$Rn було обрано часовий проміжок 1–30 с (у цей часовий інтервал потрапляє 30% подій $^{220}$Rn). Отриманий розподіл, який включає випадкові збіги (обчислене їх відношення до фонових подій становить 0.825), також знаходиться у добрій відповідності з очікуваним для α–піку $^{224}$Ra.

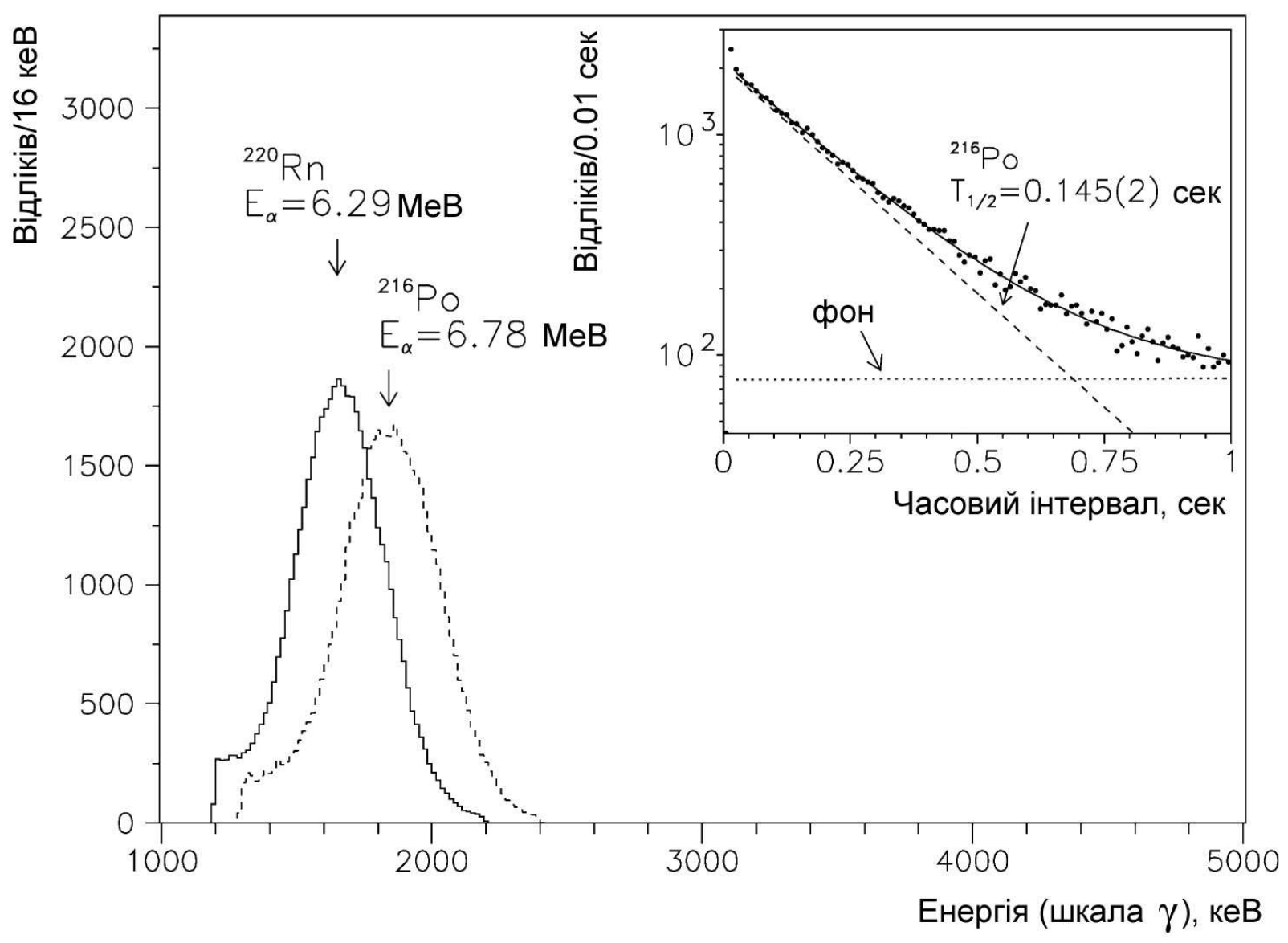

Рис 4.2. Енергетичні спектри першої та другої α–частинок з ланцюжка $^{220}$Rn → $^{216}$Po → $^{212}$Pb, знайдені за допомогою часово–амплітудного аналізу даних, накопичених за 8609 годин. На вкладці показаний розподіл часових інтервалів між першою та другою подіями разом з підгонкою (суцільна лінія) сумою експоненти (пунктир) з $T_{1/2}$ = 0.145 с (табличне значення $T_{1/2}$ = 0.145(2) с) та фону (точки).

Техніка часово–амплітудного аналізу була також застосована для послідовності α–розпадів ядер із сімейства $^{235}$U: $^{223}$Ra ($Q_\alpha$ = 5.98 МеВ, $T_{1/2}$ = 11.44 д) → $^{219}$Rn ($Q_\alpha$ = 6.95 МеВ, $T_{1/2}$ = 3.96 с) → $^{215}$Po ($Q_\alpha$ = 7.53 МеВ, $T_{1/2}$ = 1.78 мс) → $^{211}$Pb. Для відбору пар подій α–розпадів $^{219}$Rn та $^{215}$Po, в якості тригера використовувались події в області 1.3 – 2.4 МеВ і було вибрано часовий інтервал 0.5 – 10 мс (який містить 80.3% розпадів $^{215}$Po з енергіями від 1.5 до 2.6 МеВ). В ході такого аналізу відбирались також розпади $^{220}$Rn → $^{216}$Po та $^{214}$Bi → $^{214}$Po з ефективностями, відповідно, 4.64% та 12.08%. Цей внесок враховувався при визначенні активності $^{235}$U. Отримані α–піки відповідають забрудненості кристалу GSO ізотопом $^{227}$Ac (дочірній радіонуклід з сімейства актинію $^{235}$U) на рівні 0.948(9) мБк/кг.

Для оцінки активності $^{226}$Ra (сімейство $^{238}$U), використовувалась наступна послідовність β– та α–розпадів: $^{214}$Bi ($Q_\beta$ = 3.27 МеВ, $T_{1/2}$ = 19.9 хв.) → $^{214}$Po ($Q_\alpha$ = 7.83 МеВ, $T_{1/2}$ = 164.3 мкс) → $^{210}$Pb. Для першої події було встановлено поріг 0.5 МеВ, а для другої вибране енергетичне вікно 1.3 – 3.0 МеВ. Використовувався часовий інтервал 2 – 500 мкс (87.2% подій $^{214}$Po). Ефективність відбору була трохи зменшена завдяки високому порогу відбору подій β–розпадів $^{214}$Bi. Моделювання спектру $^{214}$Bi методом Монте–Карло дало значення 79.4% для частини подій з енергією вище порогу 0.5 МеВ. Отримані спектри $^{214}$Bi та $^{214}$Po показані на рис. 4.3. Активність $^{226}$Ra в кристалах становить 0.271(4) мБк/кг. Пік в β–спектрі $^{214}$Bi з енергією ≈ 1.8 МеВ можна пояснити α–активністю $^{219}$Rn (сімейство $^{235}$U). Оскільки активність $^{227}$Ac була визначена окремо, внесок від розпадів $^{219}$Rn→$^{215}$Po (як і подій від ланцюжка $^{220}$Rn→$^{216}$Po) було враховано при визначенні активності $^{226}$Ra.

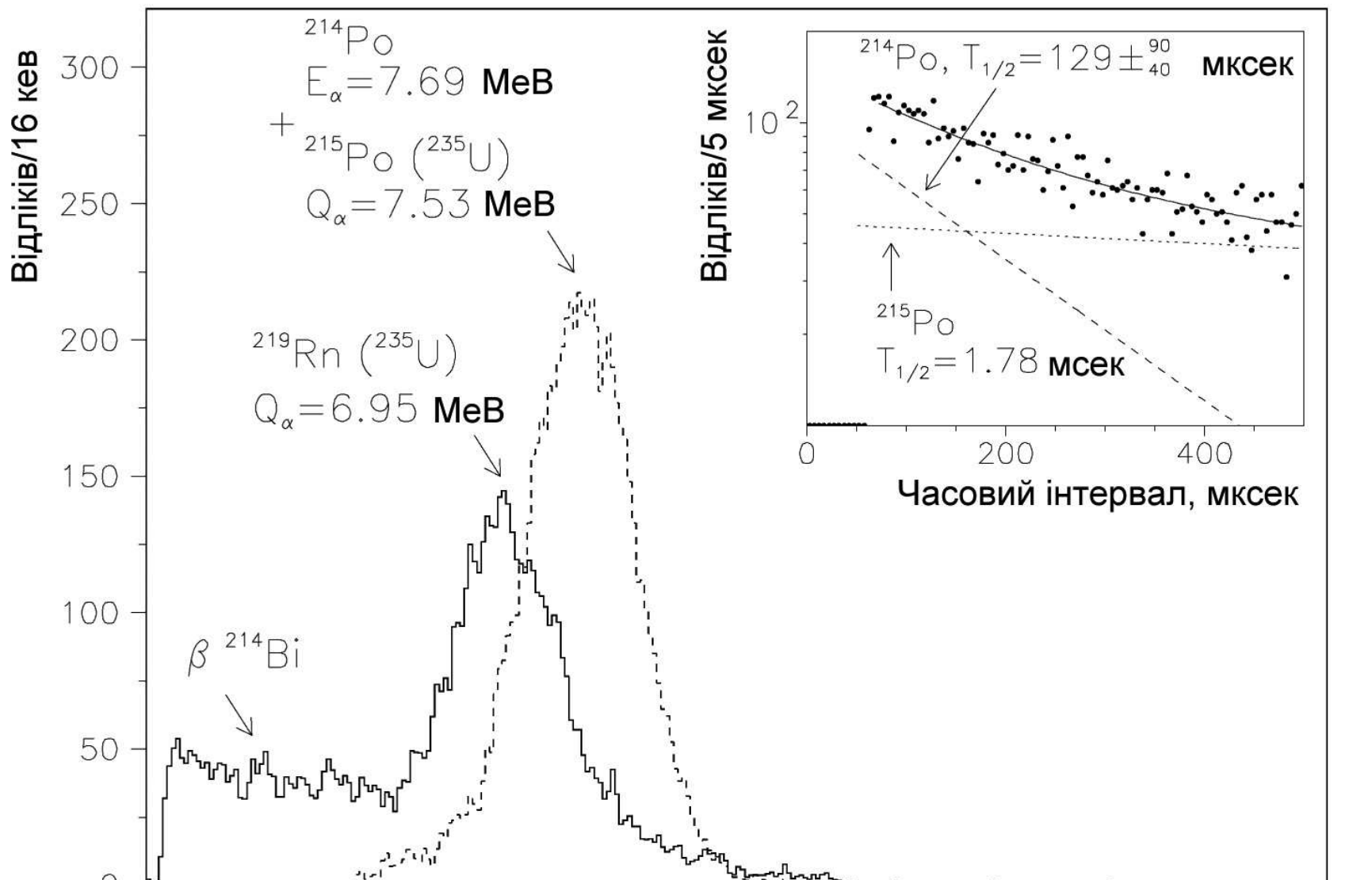

Рис. 4.3. Енергетичні спектри та розподіл часових проміжків між подями β– та α–розпадів в ланцюжку $^{214}$Bi → $^{214}$Po → $^{210}$Pb, відібраними за допомогою часово–амплітудного аналізу даних вимірювань з детектором GSO.

Енергетична залежність α/β–співвідношення була визначена по виділеним за допомогою часово–амплітудного аналізу α–пікам $^{214}$Po, $^{215}$Po, $^{216}$Po та $^{220}$Rn від внутрішнього забруднення кристала: $\alpha/\beta = 0.152 + 0.01765 \times E_\alpha$, де $E_\alpha$ – енергія альфа–частинок в МеВ.

Крім визначення компонент фону, часово–амплітудний аналіз було застосовано для відкидання фонових подій. Завдяки часово–амплітудному аналізу фон в околі енергії 0$\nu$2β–розпаду $^{160}$Gd (1648 – 1856 кеВ) було зменшено в 2.3 рази, до величини 1.01(1) відліки на добу/(кеВ×кг).

### 4.1.3. Інтерпретація фону, радіоактивна забрудненість сцинтиляторів $Gd_2SiO_5$(Ce).

З метою оцінки обмеження на період напіврозпаду ядра $^{160}$Gd відносно 2β–розпаду, фоновий спектр (після відкидання корельованих в часі подій) був промодельований за допомогою програми GEANT3.21. Генератор подій DECAY4 використовувався для опису початкової кінематики розпадів. Він враховував розпади на основний та збуджені стани дочірнього ядра, а також комплексні процеси зняття їх збудження. Можливість випромінення конверсійних електронів або $e^+e^-$ пар замість γ–квантів та кутова кореляція теж були прийняті до уваги. Експериментальний спектр підганявся сумою модельних функцій. Коефіцієнти останніх були визначені на підставі наступних даних.

(i) Значення активності $^{228}$Th, $^{227}$Ac та $^{226}$Ra (та їх короткоживучих дочірніх ізотопів), що присутні в якості забруднення в кристалах GSO. Їх активності були визначені досить точно (з точністю, кращою за 1%) за допомогою часово–амплітудного аналізу. Частину розпадів $^{224}$Ra, $^{220}$Rn, $^{216}$Po; $^{223}$Ra, $^{219}$Rn, $^{215}$Po; $^{214}$Bi та $^{214}$Po було видалено з фонового спектру, а та, що залишилась, була обчислена з великою точністю. Таким чином можна було описати спектр, спричинений розпадами, наприклад $^{228}$Th та його дочірніх, сумою модельних спектрів з відомою площею. Якщо припустити вікову рівновагу для визначених ядер, можна було б визначити активність довгоживучих членів сімейств $^{232}$Th, $^{235}$U та $^{238}$U. Однак відомо, що в ході виробництва кристалів вікова рівновага природних радіоактивних сімейств порушується внаслідок різних хімічних властивостей різних елементів – членів радіоактивних рядів. Для врахування цієї можливості, при підгонці активності довгоживучих нуклідів (а саме $^{232}$Th, $^{228}$Ra, $^{235}$U+$^{238}$U+$^{234}$U, $^{231}$Pa, $^{230}$Th та $^{210}$Pb) були прийняті в якості вільних параметрів підгонки. Три ізотопи урану $^{235}$U, $^{238}$U та $^{234}$U є хімічно подібними, тому в якості незмінного параметру було прийнято співвідношення їх активностей (0.046:1:1).

(ii) Головним джерелом зовнішнього фону являється радіоактивна забрудненість ФЕП. Ці значення для ФЕП–110 були виміряні раніше [263]: 3.0(3) Бк ($^{40}$K), 0.8(2) Бк ($^{226}$Ra) та 0.17(7) Бк ($^{228}$Th). При підгонці ці активності були прийняті в якості вільних параметрів і варіювалися в межах означених похибок, взятих з множником ×3.

Крім того, до процедури підгонки були включені спектри $^{40}$K та $^{138}$La, природних ізотопів, які можуть бути присутні в кристалі GSO. Останньою компонентою фону є експонента, яка описується двома вільними параметрами і стосується залишкового зовнішнього фону (багатократне розсіяння γ–квантів, фон від нейтронів, тощо). Експоненційна поведінка цієї компоненти була підтверджена у вимірюваннях з кристалом $CdWO_4$ (454 г) в тій самій установці [259].

Підгонка експериментального спектра в районі 0.1–3.0 МеВ дала наступні значення активностей радіонуклідів в кристалі GSO: $^{40}$K ≤ 14 мБк/кг, $^{138}$La ≤ 55 мБк/кг (це обмеження відповідає можливій концентрації лантану в кристалі GSO на рівні ≤ 67 ppm, яке не протирічить даним хімічного аналізу ), $^{232}$Th ≤ 6.5 мБк/кг, $^{228}$Ra ≤ 9 мБк/кг, $^{238}$U ≤ 2 мБк/кг, $^{231}$Pa ≤ 0.08 мБк/кг, $^{230}$Th ≤ 9 мБк/кг, $^{210}$Pb ≤ 0.8 мБк/кг. Вклад експоненти, знайдений за допомогою процедури підгонки, був малий (≤ 2% в енергетичному інтервалі 1–2 МеВ). Крива підгонки експериментальних даних показана на вкладці рис. 4.1.

Дані про активність радіонуклідів в сцинтиляторів $Gd_2SiO_5$(Ce) зведені в табл. 4.1.

Таблиця 4.1. Активності радіоактивних домішок в кристалі $Gd_2SiO_5$(Ce).

| Сімейство | Нуклід | Активність (мБк/кг) |
|---|---|---|
| $^{232}$Th | $^{232}$Th | ≤ 6.5 |
| | $^{228}$Ra | ≤ 9 |
| | $^{228}$Th | 2.287(13) |
| $^{238}$U | $^{238}$U | ≤ 2 |
| | $^{230}$Th | ≤ 9 |
| | $^{231}$Pa | ≤0.08 |
| | $^{226}$Ra | 0.271(4) |
| | $^{210}$Pb | ≤ 0.8 |
| $^{235}$U | $^{227}$Ac | 0.948(9) |
| | $^{40}$K | ≤ 14 |
| | $^{138}$La | ≤ 55 |

4.1.4. Обмеження на процеси 2β–розпаду ядра $^{160}$Gd.

Оскільки у виміряному спектрі відсутній пік від можливого 0ν2β–розпаду $^{160}$Gd, можна встановити лише обмеження на його імовірність. Для оцінки періоду напіврозпаду була використана формула 3.1. Для обчислення значень ефективності реєстрації ($\eta$) функція відгуку GSO детектора до шуканого ефекту була промодельована за допомогою програм GEANT3.21 та DECAY4. Було прийнято, що для 0ν2β–розпаду функція відгуку являє собою функцію Гауса з максимумом при енергії 1730 кеВ, та шириною на половині висоти FWHM = 176 кеВ. Крайові ефекти (виліт з кристалу одного чи двох електронів, гальмівних гамма–квантів) видаляють ≈ 5% подій з піку, отже ефективність реєстрації становить $\eta$ = 0.95.

Число подій ефекту, які можуть бути відкинуті з заданим рівем довірчої імовірності (lim*S*), визначалось двома шляхами. В першому підході кількість подій, що можуть бути непоміченими в спектрі, оцінюється, як корінь квадратний з числа подій у вибраному інтервалі енергій. Не дивлячись на простоту цього методу, він дає надійну оцінку чутливості досліду. Наприклад, фоновий спектр в інтервалі 1648 – 1856 кеВ (містить 82% площі очікуваного піку) налічує 74 500 відліків; оцінка цим методом дає lim*S* = 273 подій. Використовуючи це значення lim*S* і враховуючи експозицію експерименту ($6.04 \cdot 10^{23}$ ядер $^{160}$Gd × рік) та ефективність реєстрації ($\eta$ = 0.78), отримуємо $T_{1/2} \geq 1.2 \cdot 10^{21}$ років (з 68% C.L.).

Більш точний спосіб оцінки значення lim*S* враховує форму функції відгуку детектора та модель фонового спектра. Значення lim*S* було отримане в результаті підгонки експериментального спектра методом найменших квадратів. Слід наголосити, що найбільша частина фону (73%) в області енергій 1600–1900 кеВ визначається розпадами $^{226}$Ra, $^{227}$Ac і $^{228}$Th (та їх короткоживучими дочірніми) в кристалі GSO. Значення активностей цих радіонуклідів були обчислені дуже точно за допомогою часово–амплітудного аналізу даних, що дає змогу коректно побудувати модель фону. В результаті підгонки експериментального спектра в області 1.3–2.1 МеВ, отримана площа піку 0ν2β–розпаду склала 160±233 відліки ($\chi^2$ = 0.85), що вказує на відсутність ефекту. Різниця виміряного та модельного спектрів в області 0ν2β–розпаду $^{160}$Gd показана на рис. 4.4, де зображено й виключений ефект (суцільна лінія). Максимальне число подій, яке виключається з 90% (68%) C.L., було обчислене з урахуванням рекомендацій [280] і становить 298±169. Це дає наступне обмеження на період напіврозпаду ядра $^{160}$Gd відносно 0ν2β–розпаду на основний стан $^{160}$Dy:

$$T_{1/2}^{0\nu2\beta}\,(\text{o.c.} - \text{o.c.}) \geq 1.3\ (2.3) \times 10^{21} \text{ років} \qquad \text{з } 90\%\ (68\%)\ \text{C.L.}$$

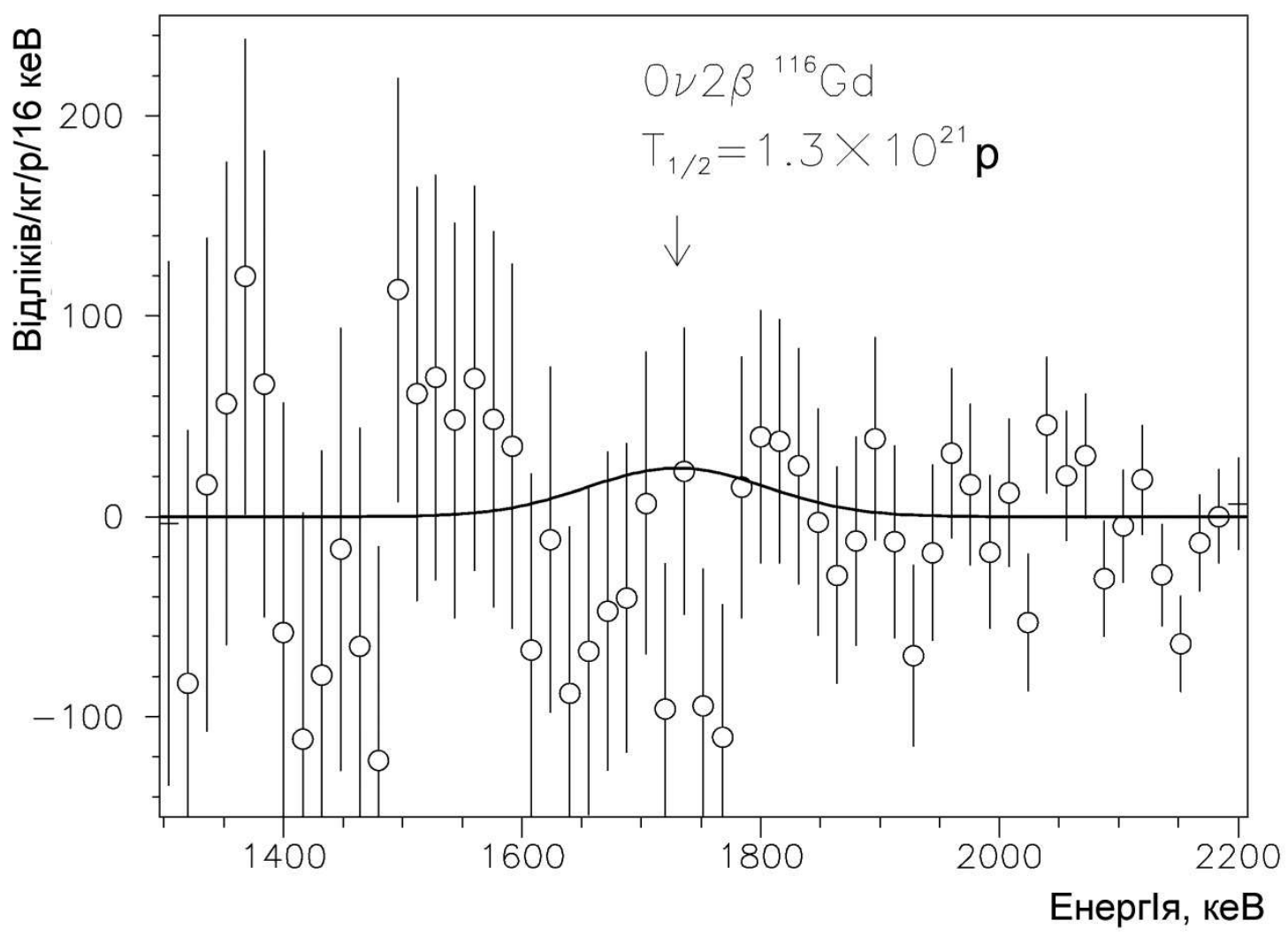

Рис. 4.4. Різниця між експериментальним та модельним спектрами в області 0ν2β–розпаду $^{160}$Gd. Суцільною лінією показано виключений з довірчою імовірністю 90% 0ν2β–пік з $T^{0\nu}_{1/2} = 1.3\times10^{21}$ років.

Порівнюючи цей результат з теоретичними підрахунками [198], можна вивести обмеження на масу нейтрино $\langle m_\nu \rangle \leq 26$ (19) еВ з 90% (68%) C.L.

Обмеження $T_{1/2}$ було також встановлене на 0ν2β–перехід на перший збуджений стан $^{160}$Dy (2+, 87 кеВ). Оскільки γ–квант з енергією 87 кеВ майже повністю поглинається в кристалі, ефективність реєстрації ефекту практично не відрізняється від ефективності реєстрації 0ν2β–розпаду на основний стан дочірнього ядра:

$$T_{1/2}^{0\nu2\beta}(\text{o.c.} - 2^{+}) \geq 1.3\ (2.3)\times10^{21}\ \text{років} \qquad \text{з } 90\%\ (68\%)\ \text{C.L.}$$

Отримані обмеження набагато вищі, ніж опубліковані раніше [281, 282].

Обмеження на період напіврозпаду ядра $^{160}$Gd відносно двохнейтринної моди 2β–розпаду на основний стан $^{160}$Dy було отримане двома методами. Консервативна оцінка була одержана з аналізу експериментального спектру, від якого був віднятий модельний спектр точно визначених забрудненостей кристалу та γ–квантів від ФЕП. Залишок у вибраному енергетичному діапазоні був прирівняний до розподілу від 2ν2β–розпаду, і це значення ($\lim S = 7.6\times10^{5}$ подій) було прийняте, як виключений ефект. Це дало обмеження на період напіврозпаду $T^{2\nu}_{1/2}$(о.с. – о.с.) ≥ 5.5 × $10^{17}$ років для 2ν2β–розпаду $^{160}$Gd з 99% CL. Однак в такому спрощеному підході не враховується інформація про форму виміряного спектру фону, тому для дослідження 2ν–моди 2β–розпаду була застосована процедура моделювання фону та його підгонки, описана раніше. Серія підгонок була повторена з різними значеннями параметрів в області енергій від (100–760) кеВ до (2400–3000) кеВ. Найбільше значення lim$S$ = 2.2 (1.3) × $10^4$ подій з 90% (68%) C.L., отримане в результаті підгонки в інтервалі 760–2600 кеВ, було взяте для оцінки обмеження на період напіврозпаду $^{160}$Gd відносно двохнейтринної моди 2β–розпаду на основний стан $^{160}$Dy:

$T^{2\nu}_{1/2}$(о.с. – о.с.) ≥ 1.9 (3.1)×$10^{19}$ років з 90% (68%) C.L.

Модельний спектр 2ν2β–розпаду $^{160}$Gd на основний стан $^{160}$Dy показано на рис.4.5 разом з основними компонентами фону. Як видно з малюнка, модель добре відображає фон навіть за межами вибраного інтервалу підгонки.

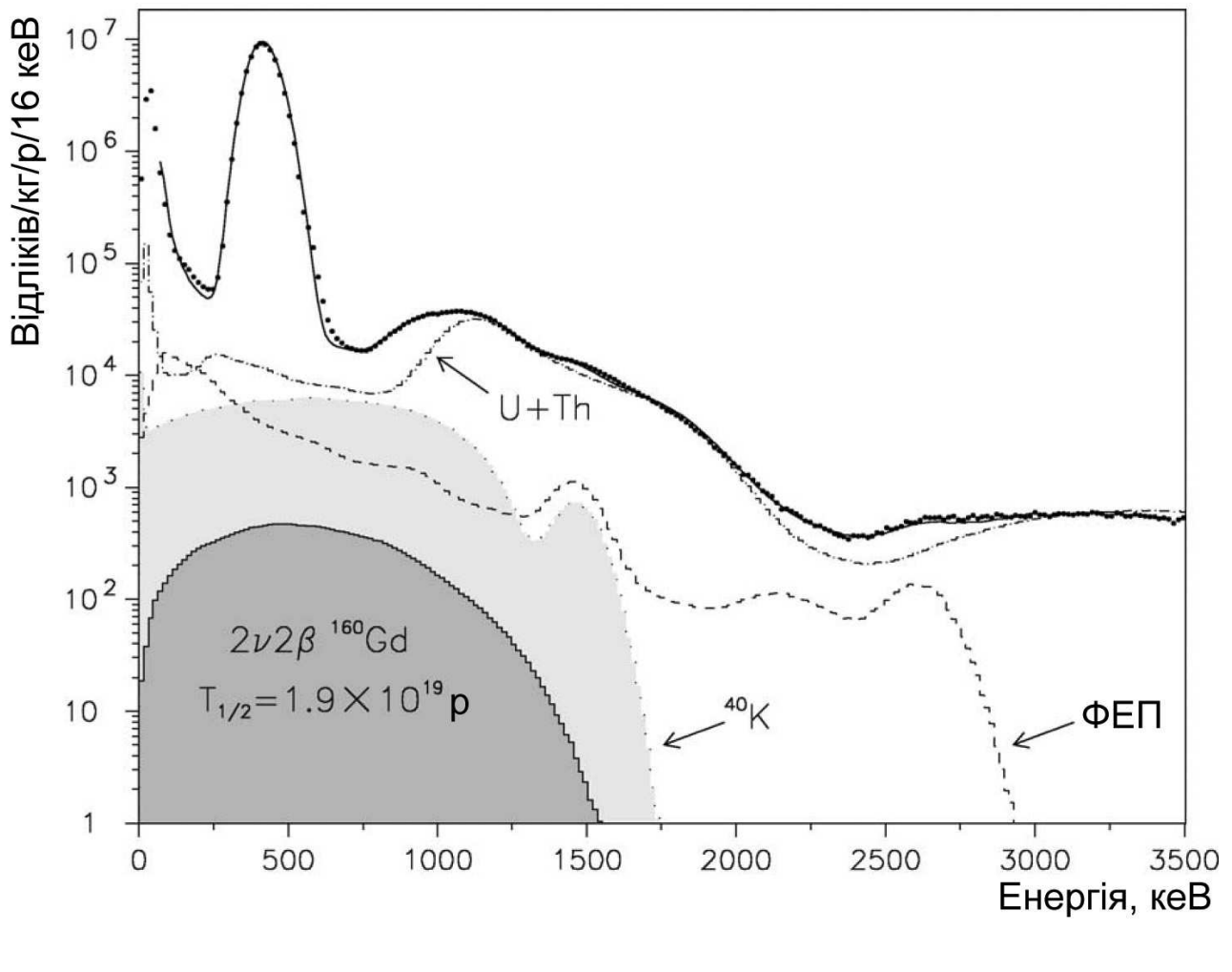

Рис. 4.5. Фон детектора GSO (експозиція 0.969 років × кг, точки) та модель фону (суцільна лінія) отримані підгонкою спектрі в області 760–2600 кеВ. Показані найбільш важливі внутрішні ($^{40}$K і сума $^{238}$U, $^{235}$U, $^{232}$Th) та зовнішні (γ–випромінювання від ФЕП) компоненти фону. Показано також виключений з 90% C.L. спектр 2ν2β–розпаду $^{160}$Gd на основний стан $^{160}$Dy з $T^{2\nu}_{1/2}$(о.с. – о.с.) = 1.9×$10^{19}$ р.

Таким же шляхом було отримано обмеження на 2ν2β–перехід до першого збудженого стану ($2^+$) $^{160}$Dy:

$$T^{2\nu}_{1/2}\,(\text{o.c.} - 2+) \geq 2.1\ (3.4)\times10^{19}\ \text{років} \qquad \text{з } 90\%\ (68\%)\ \text{C.L.}$$

Існує всього три теоретичні роботи, в яких розглянуто 2ν2β–розпад $^{160}$Gd [198,279,283]. Ядерні матричні елементи для 2ν2β–розпаду $^{160}$Gd були обчислені методом QRPA у роботі [198] та з використанням методу Operator Extension Method в роботі [283]. Перший метод дав величину $T^{2\nu}_{1/2} = 5\times10^{18}$ р, другий $T^{2\nu}_{1/2} = 3\times10^{21}$ р. В роботі [279] вказано, що завдяки великій деформації $^{160}$Gd двохнейтринний 2β–розпад просто заборонений ($T^{2\nu}_{1/2} = \infty$). Отже, передбачення, зроблене методом QRPA для $^{160}$Gd, виключене за допомогою нашого експериментального обмеження.

Подібним методом здійснювались пошуки безнейтринного подвійного бета–розпаду з випроміненням майоронів. Отримано нижні границі для періодів напіврозпаду $^{160}$Gd відносно 0ν2β–розпаду з випроміненням майоронів:

$$T^{0\nu2\beta M}_{1/2} \geq 3.5\ (5.3)\times10^{18}\ \text{р з } 90\%\ (68\%)\ \text{C.L.}$$

$$T^{0\nu2\beta MM}_{1/2} \geq 1.3\ (2.0)\times10^{19}\ \text{р з } 90\%\ (68\%)\ \text{C.L.}$$

### 4.1.5. Пошук 2ε, $\varepsilon\beta^+$, $2\beta^+$ та $2\beta^-$ процесів в ядрах $^{136}$Ce, $^{138}$Ce та $^{140}$Ce.

Концентрація церію в сцинтиляторі GSO(Ce) (0.8%) відома з умов вирощування кристалу та з результатів хімічного аналізу. Це дозволяє провести пошук 2β–процесів для трьох ізотопів церію: подвійного позитронного розпаду ($2\beta^+$), захвату електрону з вильотом позитрона ($\varepsilon\beta^+$), та подвійного електронного захвату (2ε) в $^{136}$Ce (різниця мас атомів $\Delta M_A = 2397(48)$ кеВ; ізотопна розповсюдженість $\delta = 0.185\%$); подвійного електронного захвату в $^{138}$Ce ($\Delta M_A = 693(11)$ кеВ, $\delta = 0.251\%$); та 2β–розпаду в $^{142}$Ce ($\Delta M_A = 1417(2)$ кеВ, $\delta = 11.114\%$). Загальна кількість ядер $^{136}$Ce, $^{138}$Ce та $^{142}$Ce в кристалі становить $4.1\cdot10^{19}$, $5.4\cdot10^{19}$ та $2.4\cdot10^{21}$, відповідно. Функція відгуку детектора по відношенню до різних мод 2β–процесів цих ізотопів церію моделювалась за допомогою програм GEANT3.21 та DECAY4.

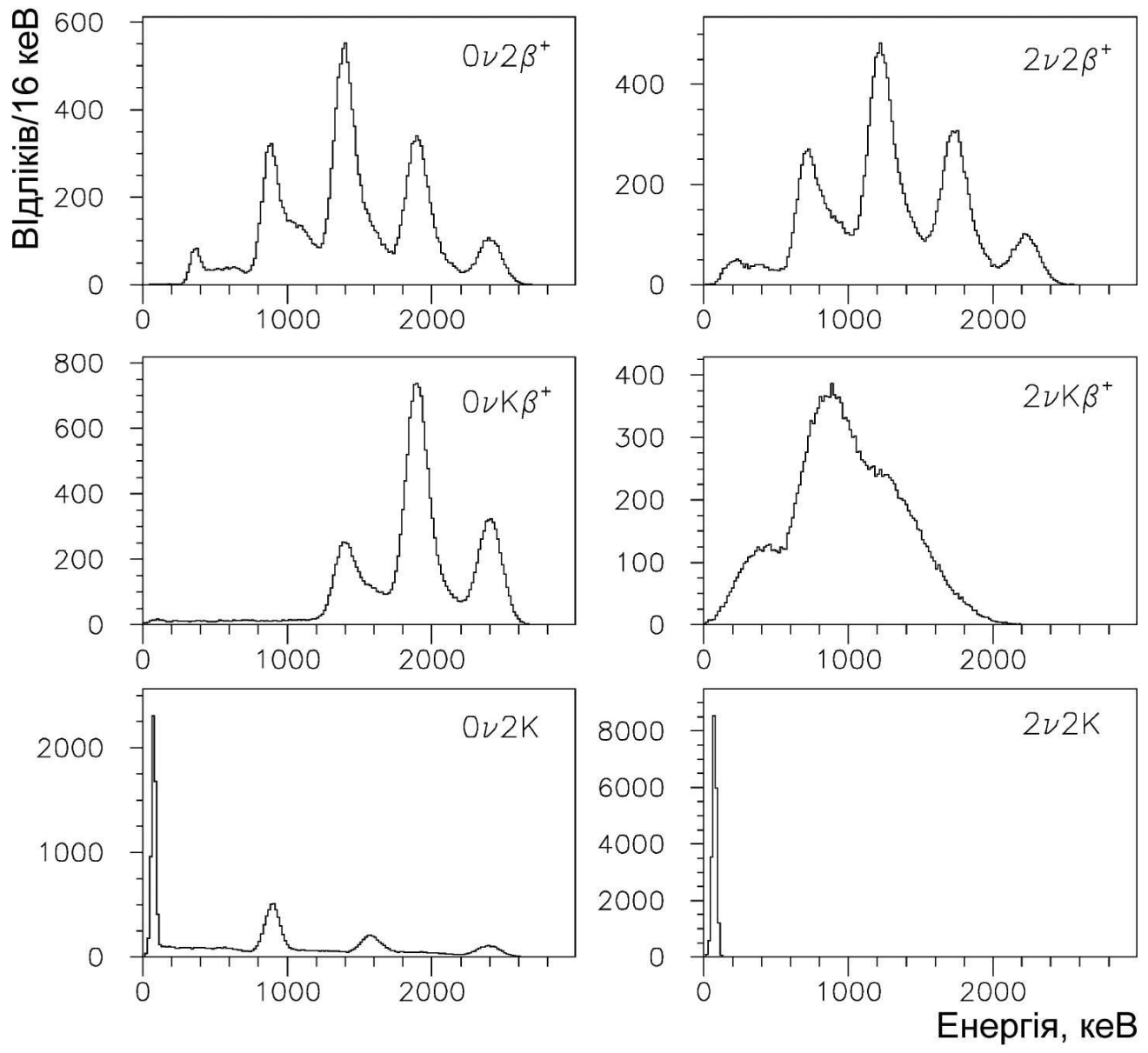

Рис. 4.6. Функції відгуку детектора GSO(Ce) до безнейтринного та двохнейтринного $2\beta^+$ ($\varepsilon\beta^+$ та $2\varepsilon$) розпадів $^{136}$Ce.

На рис. 4.6 показані функції відгуку детектора GSO(Ce) до безнейтринного та двохнейтринного $2\beta^+$ ($\varepsilon\beta^+$ та $2\varepsilon$) розпадів $^{136}$Ce. Обмеження на різні моди 2β–процесів в ізотопах церію були обчислені по аналогії з $^{160}$Gd і наведені в таблиці 4.2. Крім двох обмежень на $0\nu2\beta^+$ розпад $^{136}$Ce та 0ν2β–розпад $^{142}$Ce (встановлені за допомогою сцинтиляторів $CeF_3$ [284]), інші результати отримано в експерименті вперше.

Таблиця 4.2. Обмеження на періоди напіврозпадів ядер $^{160}Gd$, $^{136}Ce$, $^{138}Ce$ та $^{142}Ce$ відносно процесів 2β–розпаду.

| Нуклід | Канал та мода розпаду | | | Обмеження на $T_{1/2}$ (р) | |
|---|---|---|---|---|---|
| | | | | дана робота 90% (68)% C.L. | інші роботи (C.L.) |
| $^{160}Gd$ | $2\beta^-$ | $0\nu$ | o.c. – o.c., $2^+$ | $1.3\ (2.3)\times10^{21}$ | $1.4\times10^{19}$ (90%) [278,281]<br>$3.0\times10^{20}$ (68%) [282]<br>$8.2\times10^{20}$ (90%) [285] |
| | | $2\nu$ | o.c. – o.c. | $1.9\ (3.1)\times10^{19}$ | $1.3\times10^{17}$ (99%) [278,282] |
| | | | o.c. – $2^+$ | $2.1\ (3.4)\times10^{19}$ | |
| | | $0\nu M$ | o.c. – o.c. | $3.5\ (5.3)\times10^{18}$ | $2.7\times10^{17}$ (99%) [278,282] |
| | | $0\nu MM$ | o.c. – o.c. | $1.3\ (2.0)\times10^{19}$ | |
| $^{136}Ce$ | $2\beta^+$ | $0\nu$ | o.c. – o.c. | $1.9\ (3.2)\times10^{16}$ | $6.9\times10^{17}$ (68%) [284] |
| | | $2\nu$ | o.c. – o.c. | $1.8\ (3.8)\times10^{16}$ | |
| | $K\beta^+$ | $0\nu$ | o.c. – o.c. | $3.8\ (6.0)\times10^{16}$ | |
| | | $2\nu$ | o.c. – o.c. | $1.8\ (3.0)\times10^{15}$ | |
| | 2K | $0\nu$ | o.c. – o.c. | $6.0\ (8.0)\times10^{15}$ | |
| | | $2\nu$ | o.c. – o.c. | $0.7\ (1.1)\times10^{14}$ | |
| $^{138}Ce$ | 2K | $0\nu$ | o.c. – o.c. | $1.8\ (1.9)\times10^{15}$ | |
| | | $2\nu$ | o.c. – o.c. | $0.9\ (1.5)\times10^{14}$ | |

| Нуклід | Канал та мода розпаду | | | Обмеження на $T_{1/2}$ (р) | |
|---|---|---|---|---|---|
| | | | | дана робота 90% (68)% C.L. | інші роботи (C.L.) |
| $^{142}$Ce | 2β$^{-}$ | 0ν | o.c. – o.c. | 2.0 (3.3)×10$^{18}$ | 1.5×10$^{19}$ (68%) [284] |
| | | 2ν | o.c. – o.c. | 1.6 (2.6)×10$^{17}$ | |

Ядро $^{160}$Gd, враховуючи високу ізотопну розповсюдженість (22%), є одним з найбільш перспективних об‘єктів для крупномасштабного 2β–експерименту. Дійсно, високий вміст $^{160}$Gd дозволяє вирощувати кристали GSO з незбагаченого гадолінію, таким чином значно зменшується вартість досліду у порівнянні з більшістю ізотопів кандидатів на 2β–розпад. Тобто, цілком реальною задачею є створення установки з кристалами GSO з загальною масою $^{160}$Gd кілька тон.

Чутливість досліду обмежена радіоактивною забрудненістю кристалу GSO (головним чином, $^{232}$Th, $^{235}$U та $^{238}$U). Тому виробництво радіочистих кристалів GSO має важливе значення. Зменшення забрудненості кристалів GSO до рівня кількох мкБк/кг дозволить в кілька десятків разів підвищити чутливість експерименту. Така чистота вже досягнута в кристалах $CdWO_4$ [159,261,286].

Фон сцинтиляційного детектора може бути зменшений при розміщенні його в надчистому рідкому середовищі (воді або сцинтиляторі), який буде служити світлопроводом та захистом одночасно, як це було запропоновано у проекті CAMEO [324]. Існуючі установки (SNO [287], BOREXINO [229], KamLand [288]) можуть бути використані для здійснення експерименту з кількома тонами кристалів GSO. Було показано також можливість досягнення просторового розділення 1–5 мм в кристалах $CdWO_4$, розміщених в рідкому сцинтиляторі CTF, який проглядається 200 ФЕП [289]. Із детекторами GSO воно може становити 4–10 мм (це пов‘язано з меншим коефіцієнтом заломлення кристалів GSO у порівнянні з $CdWO_4$: 1.7 і 2.3, відповідно). Ми оцінили чутливість досліду з двома тонами кристалів GSO (розміщеними наприклад, в установках SNO, KamLand чи BOREXINO). Для експозиції 5–10 р вона склала $T^{0\nu}_{1/2} \geq 2\times10^{26}$ р, що відповідає обмеження на масу нейтрино $m_{\nu} \geq 0.07$ еВ. Це порівняне з чутливістю сучасних проектів по вивченню 2β–розпаду.

Більше того, експерименти з надчистими кристалами GSO становлять значний інтерес для спектроскопії сонячних нейтрино за допомогою $^{160}$Gd. Це викликане тим, що захват сонячних нейтрино $^{160}$Gd (з низьким енергетичним порогом) може бути відокремлений від фону завдяки дуже вираженій ознаці сигналу: швидкій послідовності двох подій [290].

## 4.2. Дослідження можливостей застосування сцинтиляторів $CeF_3$ для пошуку 2β–процесів в ізотопах церію

### 4.2.1. Сцинтиляційні властивості кристалів $CeF_3$.

Протягом останніх десятиріч кристали $CeF_3$ вважаються одними з найбільш перспективних для наступних експериментів в галузі фізики високих енергій завдяки їх високій питомій вазі, малому часу висвітлювання та високій радіаційній стійкості. Ці сцинтилятори відносно нові, їхні характеристики були досліджені для зразків малого об'єму (декілька кубічних сантиметрів) лише в кінці 80–х років [291,292]. Наступні роки проводились інтенсивні дослідження властивостей $CeF_3$, наприклад колаборацією Crystal Clear Collaboration [293,294]. Основні фізичні характеристики $CeF_3$:

(i) питома вага до 6.16 г/см$^3$;

(ii) світловихід становить 4–5% у порівнянні з NaI(Tl);

(iii) коефіцієнт заломлення світла становить 1.62 на довжині хвилі 400 нм;

(iv) піки в спектрі емісії: 286, 300 та 340 нм.

У сцинтилятора $CeF_3$ було спостережено декілька часових компонентів висвітлювання, серед яких найбільш важливі мають час розпаду в інтервалі від 5 до 30 нс. Матеріал негігроскопічний, температура плавлення складає 1443$^{o}$C; світловихід слабко залежить від температури (≈ 0.05% па градус С).

Кристали $CeF_3$ являють інтерес не тільки для фізики високих енергій, а й для інших завдань, наприклад, для пошуку подвійного бета–розпаду та інших рідкісних процесів ядерного розпаду. Так, перше застосування сцинтиляторів $CeF_3$ (який містить три ядра–кандидати на подвійний бета–розпад: $^{136}$Ce, $^{138}$Ce та $^{142}$Ce) для вивчення подвійного бета–розпаду описано в роботі [284]. Для досліду використовувались два кристали $CeF_3$ (один вагою 345 г, вироблений чеською компанією Preciosa–Crytur, а інший масою 74.5 г, виробництва китайської фірми).

Сцинтиляційні властивості цих кристалів, такі як світловихід, механізм люмінесценції та радіаційна стійкість, важливі для застосування для фізики високих енергій, були вивчені в багатьох роботах. В той же час, властивості кристалів $CeF_3$, важливі для низькофонових досліджень не були досліджені. Зокрема, вперше було досліджено відносний світловихід при реєстрації $\alpha$–частинок відносно $\beta$–частинок та $\gamma$–квантів ($\alpha/\beta$–співвідношення), можливість дискримінації між $\alpha$– та $\beta$–частинками ($\gamma$–квантами) за формою імпульсу, радіочистота кристалів.

В даній роботі було вивчено спектрометричні властивості та можливість розділення частинок по формі імпульсу з новим кристалом $CeF_3$ вагою 49,3 г (Китай). Вперше було також кількісно досліджено радіоактивну забрудненість сцинтиляторів $CeF_3$. Результати низькофонових вимірювань було використано для пошуку процесів двохнейтринної електронної конверсії ($2\nu 2\varepsilon$) в ядрах $^{136}$Ce та $^{138}$Ce, а також для пошуку $\alpha$–активності $^{142}$Ce.

### 4.2.2. $\alpha/\beta$–Співвідношення.

Відносний світловихід при реєстрації $\alpha$–частинок у порівнянні з $\beta$–частинками (або $\gamma$–променями), так зване $\alpha/\beta$ відношення, було виміряне з колімованим джерелом $\alpha$–частинок $^{241}$Am. Енергетична шкала детектора була визначена за допомогою $\gamma$–джерел $^{137}Cs$, $^{22}Na$, $^{133}Ba$. Енергія $\alpha$–частинок при проходженні пучку променів через тонкі поглинаючі шари повітря та майлару була виміряна поверхнево–бар'єрним напівпровідниковим кремнієвим детектором. Енергетична роздільна здатність детектора становила FWHM = 0.6% при енергії 5.25 МеВ. Розміри коліматора становили $\varnothing 0.75 \times 2$ мм, а товщина одного шару поглинача з майлару 0.65 мг/см$^2$. Завдяки використанню різних наборів таких поглиначів були отримані $\alpha$–частинки з енергіями 2.07, 3.04, 3.88, 4.58 та 5.25 МеВ. Крім того, були використані $\alpha$–частинки від внутрішнього забруднення кристалів $^{220}Rn$, $^{216}Po$ та $^{212}Po$. Це дозволило перевірити точність вимірювання $\alpha/\beta$–співвідношення та розширити енергетичний діапазон $\alpha$–частинок до 8.8 МеВ. Визначена залежність $\alpha/\beta$–співвідношення показана на 4.7 і може бути описана лінійною функцією $\alpha/\beta = 0.084 + 1.09 \times 10^{-5} E_\alpha$, де $E_\alpha$ – енергія $\alpha$–частинок в кеВ. Піки від $^{241}Am$, так само як і від $^{220}Rn$, $^{216}Po$ та $^{212}Po$, були використані для визначення енергетичної роздільної здатності детектора для $\alpha$–частинок ($FWHM_\alpha$), яке добре описується лінійною функцією: $FWHM_\alpha$(кеВ) $= 25 + 0.022 \times E_\alpha$. Була також вивчена залежність світловиходу кристалу від напрямку опромінення відносно кристалічних осей. Для цього кристал $CeF_3$ опромінювався $\alpha$–частинками в трьох напрямках перпендикулярно до поверхні кристалу. В межах експериментальної похибки такої залежності $\alpha/\beta$–співвідношення в кристалі $CeF_3$ не було виявлено.

### 4.2.3. Низькофоновий експеримент з кристалом $CeF_3$.

Вимірювання проводились на низькофоновому обладнанні в лабораторії Gran Sasso. Сцинтилятор $CeF_3$ ($2 \times 2 \times 2$ см$^3$, масою 49.3 г) був оптично приєднаний до двох низькорадіоактивних фотопомножувачів EMI9265FLB53 з діаметром вхідного вікна 3 дюйми. ФЕП були включені в режим збігів для зменшення кількості шумових подій в енергетичному спектрі. Детектуюча система була розміщена всередині герметичного контейнера з низькорадіоактивної міді. Для запобігання проникненню всередину радіоактивного радону з оточуючого середовища, об'єм всередині контейнера постійно продувався азотом високої чистоти, який перед тим зберігався під землею тривалий час. Мідний контейнер був оточений пасивним захистом з високочистої міді товщиною 10 см, низькорадіоактивного свинцю (15 см), 1.5–міліметровим шаром кадмію та шаром поліетилену (парафіну) завтовшки 4 – 10 см для зменшення зовнішнього фону. Вся установка розміщувалась всередині герметичної оболонки з органічного скла, яка теж продувалась азотом.

Система реєстрації записувала для кожної події амплітуду сигналу, час приходу та форму імпульсу (200 каналів, шириною 6.25 нс кожний). Енергетична шкали детектора та залежність роздільної здатності від енергії $\gamma$–квантів вимірювалась в енергетичному діапазоні 60–1275 кеВ та контролювалась за допомогою джерел $\gamma$–випромінювання ($^{22}$Na, $^{57}$Co, $^{137}$Cs, $^{133}$Ba та $^{241}$Am). Виміряна залежність роздільної здатності може бути виражена як $FWHM_{\gamma}$(кеВ) = 5 + (20 ×

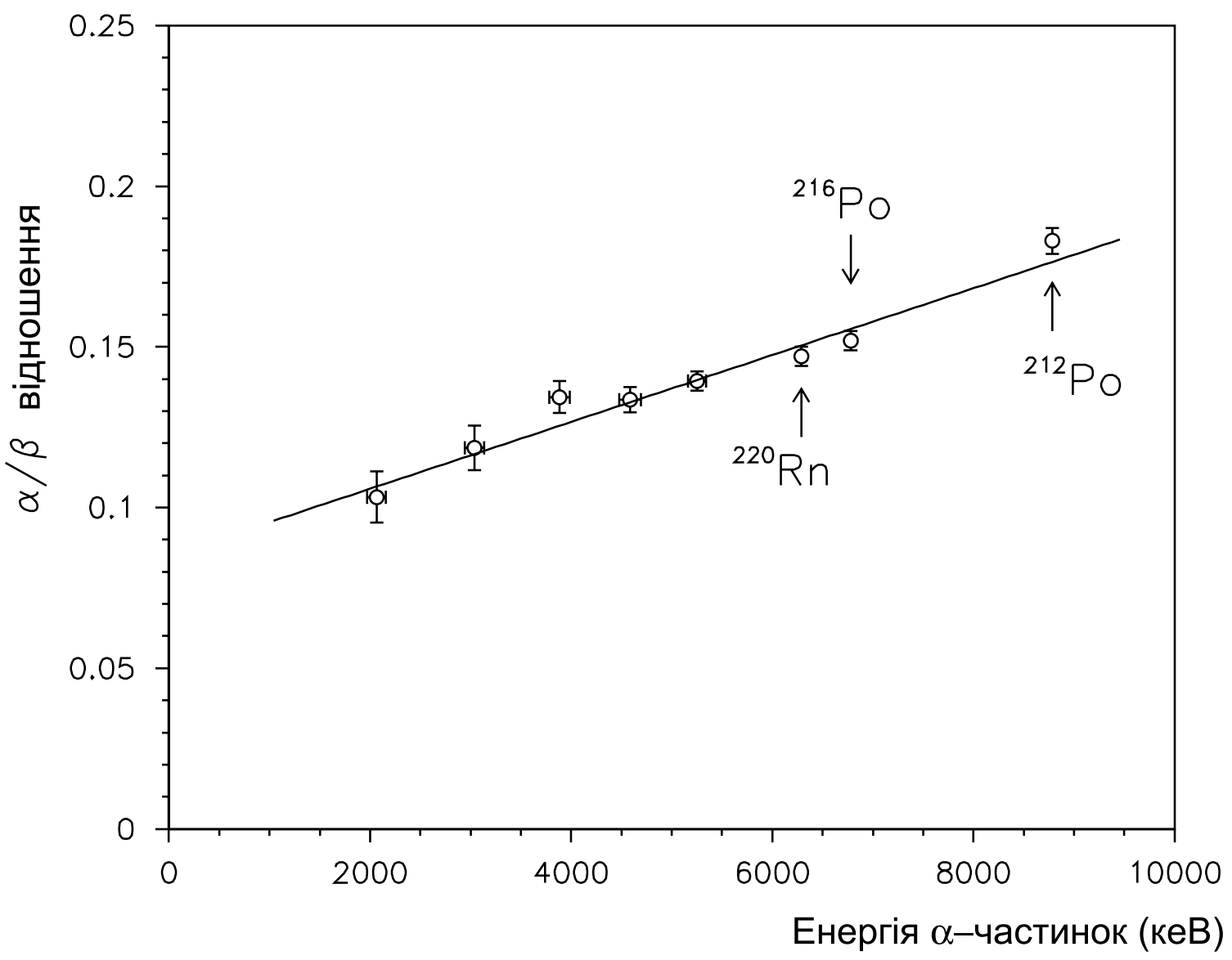

Рис. 4.7. Енергетична залежність $\alpha/\beta$–співвідношення для сцинтилятора $CeF_3$, виміряна з джерелом альфа–частинок $^{241}$Am та по $\alpha$–пікам від радіонуклідів торієвого та уранового сімейств, що присутні в кристалі в слідових кількостях.

$E_\gamma)^{1/2}$, де $E_\gamma$ – енергія γ–квантів в кеВ. Енергетичний поріг спектрометра перевірявся за допомогою джерела γ–випромінювання $^{241}$Am (було визначено, що детектор має ≈ 100% ефективність реєстрації при енергії 60 кеВ та більш ніж 95% при 45 кеВ).

4.2.4. Дискримінація частинок за формою імпульсів.

Сцинтиляційні імпульси, викликані α–частинками в $CeF_3$ помітно коротші за ті, що пов'язані з реєстрацією γ–квантів (β–частинок), що дозволяє розділяти події від α–частинок та γ–квантів (β–частинок) за допомогою техніки аналізу форми імпульсу. Для цього ми використали метод оптимального цифрового фільтру [246].

Для отримання числової характеристики форми сцинтиляційного імпульсу $CeF_3$ (SI, shape indicator – індикатор форми), було застосовано наступну формулу для кожного сигналу $SI = \Sigma f(t_k) \times P(t_k)$, де сумування по k часовим каналам починається з початку сигналу і до 125 нс, а $f(t_k)$ – оцифрована амплітуда імпульсу в даному каналі, нормалізована до загальної площі імпульсу. Вагова функція P(t) описується виразом $P(t) = f_\alpha(t) - f_\gamma(t)$; де $f_\alpha(t)$ та $f_\gamma(t)$ – форми сцинтиляційних імпульсів для α–частинок та γ–квантів, визначені по великій кількості подій. Форма імпульсів сцинтиляторів $CeF_3$ для γ–квантів була визначена в енергетичному діапазоні 0.06–1.33 МеВ (з джерелами $^{137}$Cs та $^{60}$Co) та для α–частинок в діапазоні 5.5 – 6.8 МеВ (з використанням α–піків від радіонуклідів внутрішнього α–активного забруднення кристалу).

Виміряна залежність SI для γ–квантів слабко залежить від енергії: $SI_\gamma = 26.8 - 0.16 \times 10^{-3} E_\gamma$ (в цій формулі величина SI є безрозмірною, а енергія гамма–квантів $E_\gamma$ виражена в кеВ). Залежність індикатору форми від енергії для даних 20.9–годинних вимірювань фону вимірювань з кристалами $CeF_3$ показана на рис. 4.8. Розподіл значень індикатора форми для α–подій трохи зміщений відносно значень для γ(β)–подій. Як можна бачити на рис. 4.8, розподіли індикатора форми імпульсів для γ(β) та α–подій добре описується функціями Гауса.

Енергетична залежність стандартного відхилення розподілу Гауса для $\gamma$–квантів ($\sigma_\gamma$) була отримана з підгонки дослідних даних: $\sigma_\gamma = 0.7 + 122 / E_\gamma^{1/2}$. Інтервал значень SI ± $\sigma_\gamma$ також показаний на рис. 4.8. Відповідні параметри для $\alpha$–частинок в інтервалі енергій 5.5 – 6.8 МеВ становлять $SI_\alpha = 30.2$ та $\sigma_\alpha = 4.8$.

Крім того, завдяки помітній часовій різниці між сигналами сцинтилятору $CeF_3$ (≈30 нс) та шумами ФЕП (≈5 нс), яка призводить до значної різниці у значеннях індикатора форми, метод оптимального фільтра, розроблений для розділення сигналів від $\alpha$–частинок від $\gamma$–квантів та $\beta$–частинок, був використаний для видалення з спектра подій від шумів ФЕП.

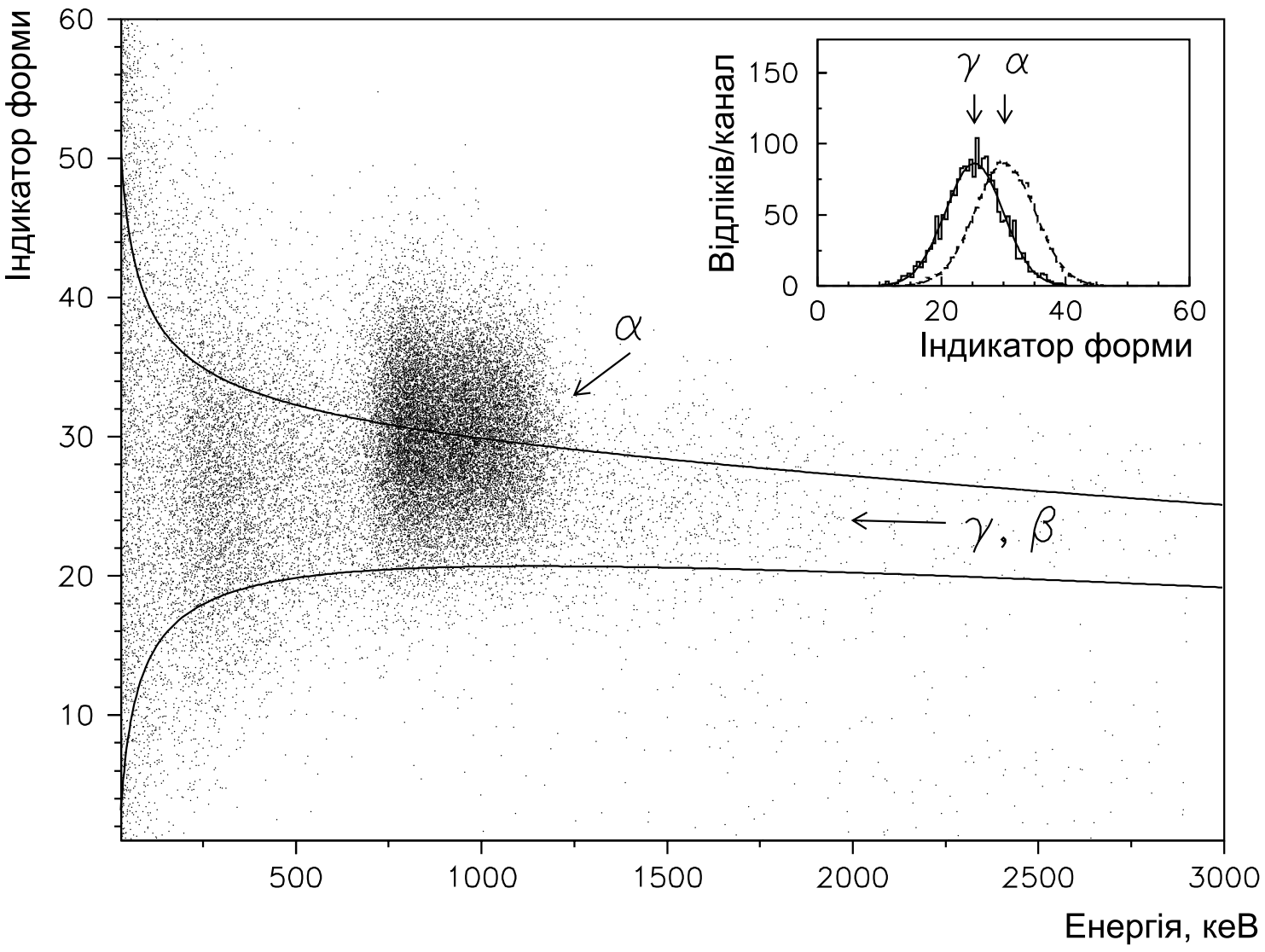

Рис. 4.8. Залежність індикатора форми від енергії $\gamma$–квантів для експозиції 20.9 г фонових вимірювань з детектором $CeF_3$. Зображено також інтервал $\pm\sigma_\gamma$ для значень індикатора форми, які відносяться до $\gamma$–квантів ($\beta$–частинок). Точки із значенням SI ≤ 15 – 20 викликані подіями розпадів $^{212}$Bi – $^{212}$Po – $^{208}$Pb. На вкладці показано розподіли індикатору форми для $\gamma$–квантів та $\alpha$–подій.

Для відкидання фонових подій від послідовності розпадів нуклідів з сімейства $^{232}$Th: $^{212}$Bi ($Q_\beta$ = 2254 кеВ) → $^{212}$Po ($E_\alpha$ = 8784 кеВ; $T_{1/2}$ = 0.299 мкс) → $^{208}$Pb було застосовано іншу техніку. Типовий приклад такого аналізу наведено на рис. 4.9, де зображені β–спектр $^{212}$Bi, α–пік $^{212}$Po та розподіл часових інтервалів між першою та другою подіями. Енергетичні та часові спектри знаходяться у відповідності з очікуваними для β–розпаду $^{212}$Bi та α–розпаду $^{212}$Po. Всі подібні подвійні імпульси (рис. 4.9) в інтервалі затримки Δt = (0.11 – 0.65) мкс були відкинуті з фонових даних. Слід відзначити, що у випадку менших Δt ці події успішно відділяються за допомогою аналізу по формі імпульсу.

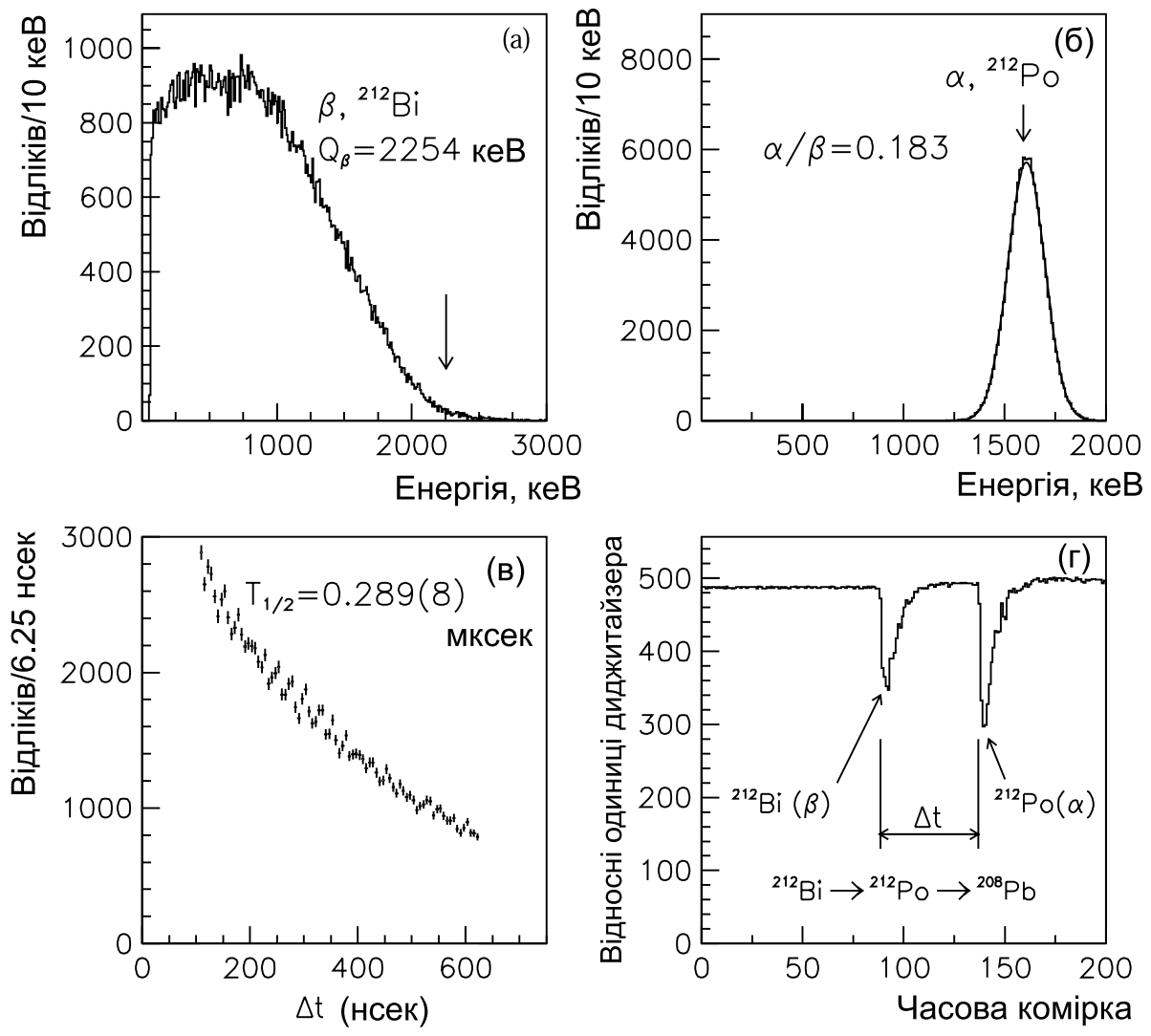

Рис. 4.9. Енергетичний (а,б) та часовий (в) розподіли β–розпадів $^{212}$Bi ($Q_\beta$ = 2254 кеВ) та α–розпадів $^{212}$Po ($E_\alpha$ = 8784 кеВ, $T_{1/2}$ = 0.299 мкс), відібрані за допомогою аналізу форми імпульсу форми сигналів, накопичених за 2142 годин вимірювань. Аналіз часового розподілу дає значення періоду напіврозпаду $T_{1/2}$ = 0.289(8) мкс, яке знаходиться у добрій відповідності з літературними даними для $^{212}$Po. (г) приклад подібної події в сцинтиляторі $CeF_3$.

4.2.5. Радіоактивна забрудненість кристалів $CeF_3$.

Фоновий спектр кристалів $CeF_3$, виміряний протягом 2142 г в низькофоновій установці, приведено на рис. 4.10. Приймаючи до уваги α/β–співвідношення, широкий пік в інтервалі енергій 0.6–1.3 МеВ може бути пояснений α–розпадами $^{232}$Th, $^{235}$U, $^{232}$U та їх дочірніх α–активних ядер. Природа подій вказаних піків, а саме той факт, що це α–частинки, була підтверджена за допомогою аналізу по формі імпульсу. Відмінності між спектрами, показаними на рис.10, пояснюються вкладом шумів ФЕП в низькоенергетичній частині спектру та подіями розпадів в ланцюжку $^{212}$Bi – $^{212}$Po – $^{208}$Pb (для енергій вище 1.2 МеВ). Пік з енергією ≈1.55 МеВ належить α–розпадам $^{212}$Po. Пік при енергії 482±12 кеВ пояснюється активністю $^{232}$Th в кристалі на рівні 37 (16) мБк/кг.

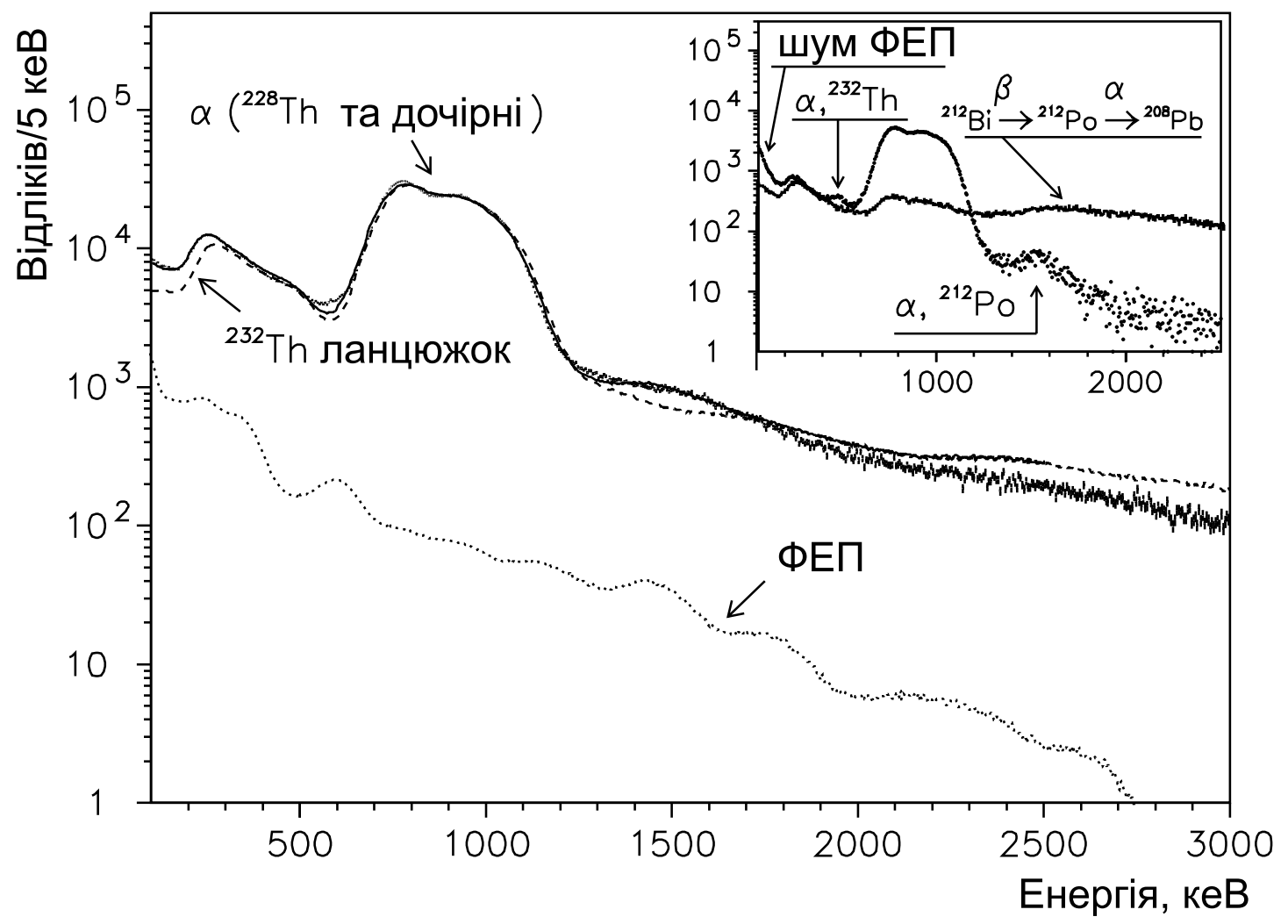

Рис. 4.10. Енергетичний спектр сцинтилятора $CeF_3$ масою 49.3 г виміряний за 2142 годин в низькофоновій установці. Показаний результат підгонки спектра (суцільна лінія) та найбільш важливі компоненти фону (внутрішня забрудненість дочірніми ядрами $^{232}$Th та зовнішнє γ–випромінювання від двох ФЕП). На вкладці: енергетичний спектр, відібраний за допомогою аналізу по формі імпульсу для γ(β)–подій (суцільна лінія) та α–подій (крапки).

Наявність в кристалі дочірніх торію була підтверджено часово–амплітудним аналізом записаних подій. Наприклад, піки від α–розпадів $^{220}$Rn та $^{216}$Po (дочірні $^{232}$Th), отримані за допомогою аналізу амплітуди та часу подій, так само як розподіл часових інтервалів між першою та другою подією, представлені на рис 4.11. Період напіврозпаду $^{216}$Po, отриманий за допомогою підгонки цього розподілу експоненційною функцією ($T_{1/2}$ = 0.144 с), знаходиться у відповідності з табличними даними. Активність $^{228}$Th у кристалах $CeF_3$, визначена за допомогою даного методу аналізу, є досить значною і дорівнює 1.01(1) Бк/кг.

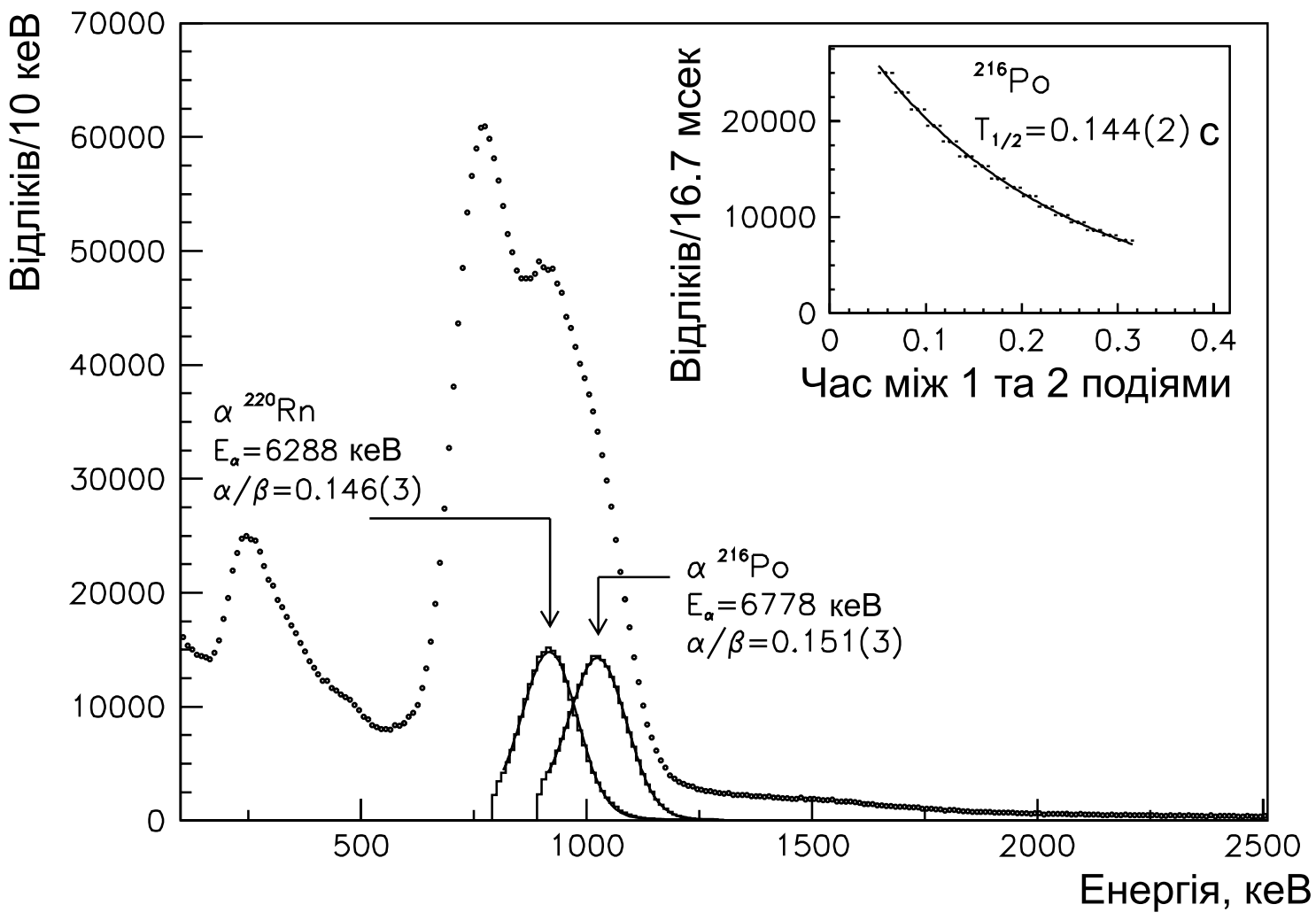

Рис. 4.11. Фоновий енергетичний спектр, набраний за 2142 г (крапки) та спектри першої та другої α–частинок з ланцюжка розпаду $^{220}$Rn – $^{216}$Po – $^{212}$Pb, відібрані за допомогою часово–амплітудного аналізу (гістограми). На вкладці показаний розподіл часових інтервалів між першою та другою подіями з експоненціальною підгонкою.

Так само було визначено, що широкий розподіл в енергетичному інтервалі 0.1–0.6 МеВ пояснюється, в основному, забрудненням кристалів радіонуклідами, що належать до сімейства $^{232}$Th. Бета– та альфа–випромінювання, які являються наслідком розпаду нуклідів в ланцюжку $^{212}$Bi – $^{212}$Po – $^{208}$Pb, дають події в інтервалі енергій 1.2 – 4 МеВ, тоді як розпади $^{208}$Tl ($Q_\beta$ = 5001 кеВ) дають вклад до 5 МеВ. Експериментальний спектр був промодельований за допомогою програм GEANT3.21 та генератора подій DECAY4. Окрім описаних вище компонентів, радіоактивна забрудненість ФЕП (виміряна попередньо [295]) та залишкові шуми ФЕП (експоненційна функція) були також включені в підгонку. Щоб врахувати цілком можливе порушення вікової рівноваги в процесі виробництва кристалів, окремі фрагменти ланцюжків розпадів в сімействах урану та торію були взяті з незалежними коефіцієнтами. Підгонка по методу найменших квадратів експериментального спектру в інтервалі енергій 0.1 – 2.5 МеВ сумою перерахованих вище компонентів дала значення активностей радіонуклідів внутрішнього забруднення кристалів $CeF_3$ $^{232}$Th, $^{228}$Ra, $^{228}$Th, $^{238}$U/$^{234}$U та $^{226}$Ra; які приведені в таблиці 4.3. Основна частина фону викликана розпадами дочірніх ізотопів $^{232}$Th та γ–випромінюванням від радіоактивних домішок в ФЕП (95 та 2%, відповідно).

Вищеописану процедуру було повторено з включенням до моделі фону наступних радіоактивних домішок: $^{230}$Th; $^{210}$Pb; $^{235}$U; $^{231}$Pa; $^{227}$Ac (та їх короткоживучих дочірніх), а також $^{40}$K; $^{138}$La та $^{176}$Lu. Отримані обмеження на їх активності приведені в таблиці 4.3.

На відміну від $^{228}$Ra та $^{228}$Th, активність яких досить висока, активність материнського ізотопу $^{232}$Th в сцинтиляторі порівняно низька: 55(30) мБк/кг. Однак, присутність α–піку $^{232}$Th все–таки була підтверджена за допомогою аналізу форми імпульсу. Результати показані на рис. 4.10, де зображені енергетичні спектри γ(β)– та α–подій. Перший було відібрано за умови: $SI \leq SI_\gamma - 1.6\sigma_\gamma$. Він містить ≈5.5% γ(β)–подій, в той час як α–події в цьому спектрі практично відсутні. Другий спектр було відібрано за умови $SI > SI_\gamma + 1.6\sigma_\gamma$ і він знову–таки містить ≈5.5% γ(β)–подій та близько (6–9)% α–частинок. Окрім видимої різниці біля енергій 5.5 – 6.8 МеВ, ці спектри відрізняються також і в інших областях. Так, зростання першого спектру γ–квантів та β–частинок при енергіях більших 1.2 МеВ викликане подіями з послідовності $^{212}Bi - {}^{212}Po - {}^{208}Pb$. Надлишок другого розподілу в низькоенергетичній частині може бути пояснений шумами ФЕП, в той час як пік при енергії ≈1.55 МеВ викликаний α–розпадом $^{212}Po$. У випадку, коли енергія випроміненої $^{212}Bi$ α–частинки знаходилась нижче порогу реєстрації детектора, подія α–розпаду $^{212}Po$ реєструється як окрема і не може бути відібрана за допомогою аналізу форми.

Таблиця 4.3. Радіоактивна забрудненість кристалів $CeF_3$.

| Родина | Джерело | Активність, мБк/кг |
|---|---|---|
| $^{232}Th$ | $^{232}Th$ | 55(30) |
| | $^{228}Ra$ | 890(270) |
| | $^{228}Th$ | 1010(10) |
| $^{238}U$ | $^{238}U$ | $\leq 70$ |
| | $^{234}U$ | $\leq 60$ |
| | $^{230}Th$ | $\leq 60$ |
| | $^{226}Ra$ | $\leq 60$ |
| | $^{210}Pb$ | $\leq 280$ |
| $^{235}U$ | $^{235}U$ | $\leq 40$ |
| | $^{231}Pa$ | $\leq 50$ |
| | $^{227}Ac$ | $\leq 20$ |
| | $^{40}K$ | $\leq 330$ |
| | $^{138}La$ | $\leq 60$ |
| | $^{176}Lu$ | $\leq 20$ |
| | $^{147}Sm$ | $\leq 80$ |

Чіткий пік невеликої інтенсивності є в $\alpha$–спектрі при енергії 482 ± 12 кеВ. Енергія піку знаходиться у добрій відповідності з очікуваним положенням альфа–піку $^{232}$Th: 509 ± 25 кеВ. Відповідна активність $^{232}$Th, визначена за допомогою аналізу форми імпульсу, дорівнює 37(16) мБк/кг. Така розбіжність між активностями $^{232}$Th та $^{228}$Th (1.010 Бк/кг) може бути пояснена порушенням вікової рівноваги сімейства торію під час вирощування кристалів. Отже, до кристалу попадав головним чином $^{228}$Ra, який призводив до утворення $^{228}$Th. Цей факт треба приймати до уваги при вирощуванні радіоактивно чистих кристалів $CeF_3$.

Результати вимірювань із сцинтиляторами $CeF_3$ були використані для визначення можливостей застосування сцинтиляторів для пошуку рідкісних розпадів ізотопів церію. Для двох ізотопів церію, $^{136}$Ce та $^{138}$Ce, енергетично дозволений захват двох K–електронів з атомної оболонки. В обох описаних процесах повна енергія, вивільнювана в детекторі $CeF_3$ становить $2E_K$ (де $E_K$=37.4 кеВ – енергія зв'язку електронів на K–оболонці атомів барію), в той час як нейтрино виносить іншу частину доступної енергії.

Отримані обмеження на періоду напіврозпаду [284] знаходяться значно нижче сучасних теоретичних прогнозів (див. роботу [76]) для 2ν2K захвату в $^{136}$Ce ($10^{18}$ – $10^{22}$ р). Приблизно таких же періодів напіврозпадів можна очікувати і для 2ν2K захвату в $^{138}$Ce, для якого теоретичні розрахунки відсутні. Тим не менше, ці експериментальні обмеження на два порядки вищі за ті, що були отримані в попередньому досліді з використанням активованих церієм кристалів ортосилікату гадолінію.

**Висновки розділу.** Досліджені сцинтиляційні властивості та радіочистота сцинтиляційних кристалів ортосилікату гадолінію (GSO). Вперше сцинтилятори GSO були застосовані для пошуку 2$\beta$–розпаду. З даних довгострокових вимірювань фону в Солотвинській підземній лабораторії отримані обмеження на процесу $2\beta$–розпаду $^{160}$Gd та $2\beta$–процесів в $^{136}$Ce, $^{138}$Ce та $^{142}$Ce. Встановлені найбільш жорсткі обмеження на періоди напіврозпаду ядра $^{160}$Gd відносно $0\nu2\beta$–розпаду на основний та перший збуджений стан ядра $^{160}$Dy (з 90% CL): $T_{1/2}(0^+ \to 0^+) \geq 1.3 \times 10^{21}$ років та $T_{1/2}(0^+ \to 2^+) \geq 1.3 \times 10^{21}$ років, а також на період напіврозпаду ядра $^{160}$Gd відносно $2\nu2\beta$–розпаду на основний та перший збуджений стан $^{160}$Dy (з 90% CL): $T_{1/2}(0^+ \to 0^+) \geq 1.9 \times 10^{19}$ років та $T_{1/2}(0^+ \to 2^+) \geq 2.1 \times 10^{19}$ років. Вперше була вивчена функція відгуку кристалічного сцинтилятору $CeF_3$ по відношенню до реєстрації $\alpha$–частинок у широкому енергетичному діапазоні від 2 до 8.8 МеВ та продемонстрована можливість дискримінації по формі імпульсу між $\alpha$–частинками та $\gamma$–квантами. Було проведено вивчення радіоактивної забрудненості сцинтилятора $CeF_3$. Досліджені можливості застосування детекторів $CeF_3$ для пошуку $2\beta$–розпаду ізотопів Ce.

Результати, викладені в цьому розділі, опубліковані в роботах:

# РОЗДІЛ 5

# ПОШУКИ 2β–РОЗПАДУ ЦИНКУ ЗА ДОПОМОГОЮ СЦИНТИЛЯТОРІВ $ZnWO_4$

## 5.1. Сцинтиляційні властивості та рівень радіочистоти кристалів $ZnWO_4$

### 5.1.1 Енергетична роздільна здатність.

В дослідженнях були використані три кристали вольфрамату цинку ($ZnWO_4$), вирощені методом Чохральського. Розміри кристалів ∅14×4 мм, ∅14×7 мм і ∅13×28 мм. Кристали мають слабке рожеве забарвлення. Сцинтиляційні властивості кристалів $ZnWO_4$ наведені в табл. 5.1. Кристали $ZnWO_4$ вивчались в роботах [255,296,297,298]. Світловихід сцинтилятора $ZnWO_4$ вивчався в роботі [297] за допомогою кремнієвого фотодіода. Автори дали значення 9300 фотонів на

Таблиця 5.1. Властивості сцинтиляторів $ZnWO_4$.

| Характеристика | |
|---|---|
| Густина (г/см$^3$) | 7.8 |
| Точка плавлення (°С) | 1200 |
| Структурний тип | Вольфраміт |
| Площина спайності | (010) |
| Твердість за Моосом | 4 – 4,5 |
| Довжина хвилі максиму спектру емісії (нм) | 480 |
| Показник заломлення | 2.1 – 2.2 |
| Ефективний час світіння (мкс) | 24 |
| Вихід фотоелектронів (відносно NaI(Tl)) | 13% |

1 МеВ поглинутої в кристалі енергії гамма–квантів, що становить близько 23% від світловиходу NaI(Tl). Нами був виміряний вихід фотоелектронів з кристалом $ZnWO_4$ ∅14 × 7 мм, оптично приєднаним до ФЕП EMI D724KFLB з рубідій–цезієвим фотокатодом. Вимірювання проводилися відносно сцинтилятора NaI(Tl) ∅40 × 40 мм і було одержано значення 13%.

В попередніх роботах з кристалами вольфрамату цинку для гамма–квантів $^{137}Cs$ з енергією 662 кеВ була отримана енергетична роздільна здатність 13% [295] і 11.5% [298]. В роботі [299] з кристалом $ZnWO_4$ ∅14×7 мм, оптично приєднаним до ФЕП Philips XP2412, для γ–випромінювання $^{137}Cs$ нами було отримано енергетичну роздільну здатність 11.0%. Враховуючи час висвічування сцинтиляторів $ZnWO_4$, час формування спектрометричного підсилювача було вибрано рівним 24 мкс.

Суттєвого покращення світлозбору (40%) та енергетичної роздільної здатності було досягнуто завдяки розміщенню кристала $ZnWO_4$ ∅14 × 7 мм у фторопластовий контейнер розмірами ∅70 × 90 мм, заповнений силіконовою олією. Кристал був зафіксований в центрі контейнера і проглядався двома ФЕП XP2412. Вперше з детектором $ZnWO_4$ було отримано роздільну здатність 9.1% і 6% для гамма–ліній з енергіями 662 кеВ ($^{137}Cs$) і 1333 ($^{60}Co$). Відповідні енергетичні спектри показані на рис. 5.1 (а). З більшим кристалом в тих самих умовах для гамма–квантів $^{137}Cs$ була отримана роздільна здатність 11.3%.

Здатність детектора з сцинтилятором $ZnWO_4$ реєструвати низькоенергетичні події була продемонстрована у вимірюваннях з джерелами $^{55}Fe$ (6 кеВ) $^{241}Am$ (≈18 кеВ і 59.5 кеВ). З рис. 5.1 (б) видно, що навіть рентгенівське випромінювання з енергією 6 кеВ все ще може бути виділене на фоні шумів ФЕП.

### 5.1.6. Альфа/бета–співвідношення.

Альфа/бета–співвідношення з кристалами $ZnWO_4$ було виміряне за допомогою колімованого джерела $^{241}Am$ і тонких майларових плівок (0.65 мг/см$^2$) та шару повітря для отримання енергій α–частинок в діапазоні енергій 1 – 5.3 МеВ.

Кристал $ZnWO_4$ ∅14 × 7 мм опромінювався в трьох напрямках перпендикулярних до його основних кристалографічних осей. Було знайдено залежність α/β–співвідношення від енергії і напрямку опромінення (див. рис. 5.2), подібну до тієї, що була виміряна із сцинтиляторами $CdWO_4$.

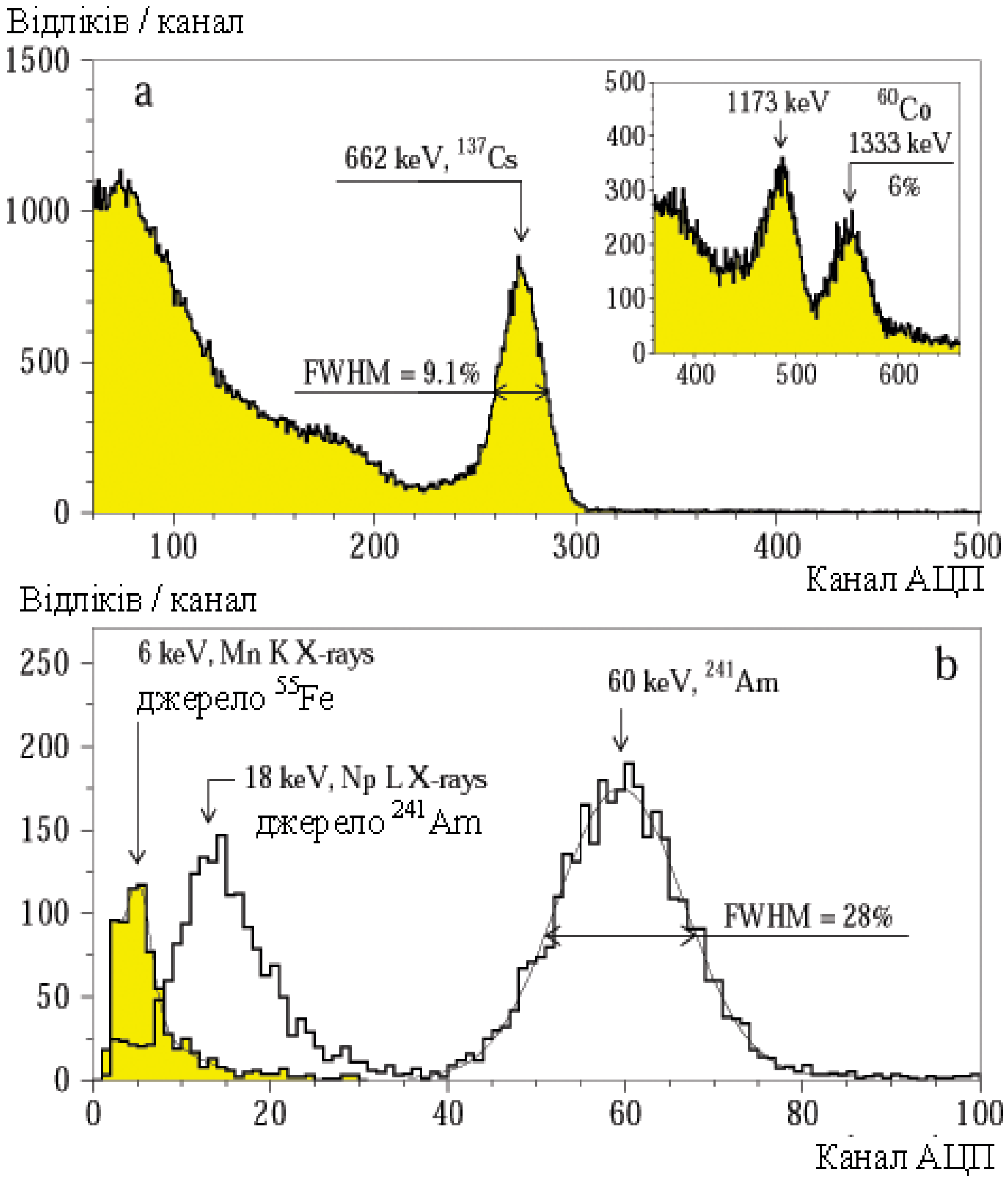

Рис 5.1. Енергетичні спектри γ–джерел $^{137}Cs$, $^{60}Co$, $^{55}Fe$, $^{241}Am$, виміряні сцинтиляційним детектором з кристалом $ZnWO_4$.

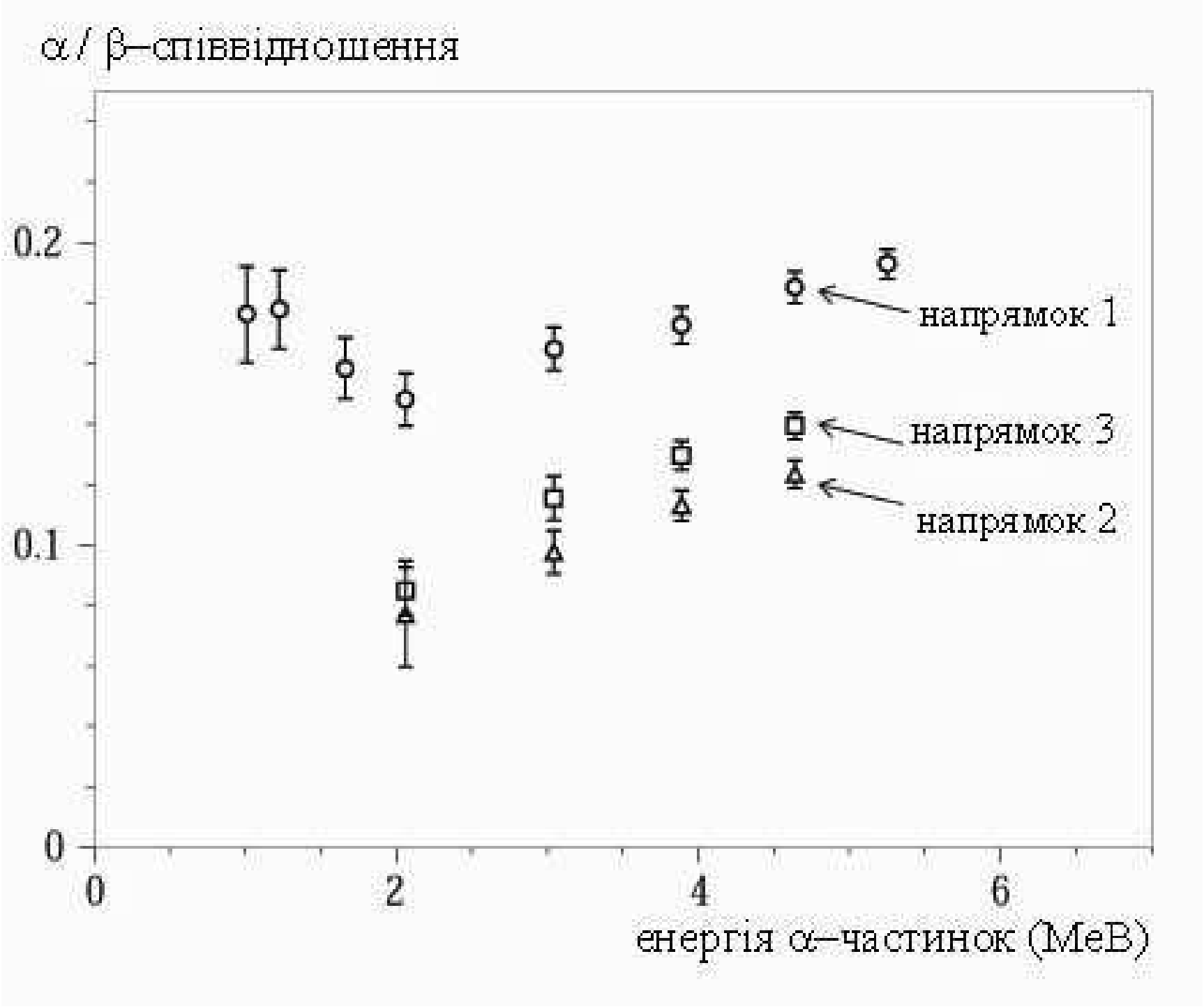

Рис. 5.2. Залежність α/β–співвідношення в детекторі $ZnWO_4$ від енергії та напрямку опромінення α–частинок.

### 5.1.3. Форма сцинтиляційного спалаху і дискримінація подій від α–частинок та γ–квантів.

Часові характеристики сцинтиляторів вольфрамату цинку були вивчені за допомогою 12–розрядного оцифровщика форми сигналів з частотою 20 МГц. Сцинтилятор опромінювався α–частинками з енергією 4.6 МеВ в трьох напрямках, перпендикулярних до основних кристалографічних осей. Форма сигналів при збудженні гамма–квантами вивчалась за допомогою опромінення сцинтилятора гамма–квантами від джерела $^{60}$Co. Форми сцинтиляційних сигналів для α–частинок та γ–квантів, отримані усередненням 5 тисяч окремих сцинтиляційних спалахів, показані на рис. 5.3. Підгонка експериментально виміряних форм трьома експоненційними функціями дає значення ≈0.7, ≈7 і ≈25 мкс для часів окремих компонент сцинтиляційного спалаху. При цьому інтенсивності різних компонент сигналів, викликаних α–частинками та γ–квантами, відрізняються, як це можна бачити з таблиці 5.2. Отримане значення повільної компоненти (≈25 мкс) узгоджується з результатом роботи [296], в той час як компоненти ≈0.7 і ≈7 мкс були спостережені вперше. В той же час, в зв'язку з недостатньо високою частотою оцифровки (20 МГц), компонента 0.1 мкс, що була спостережена авторами роботи [296], в даних вимірюваннях не була зафіксована.

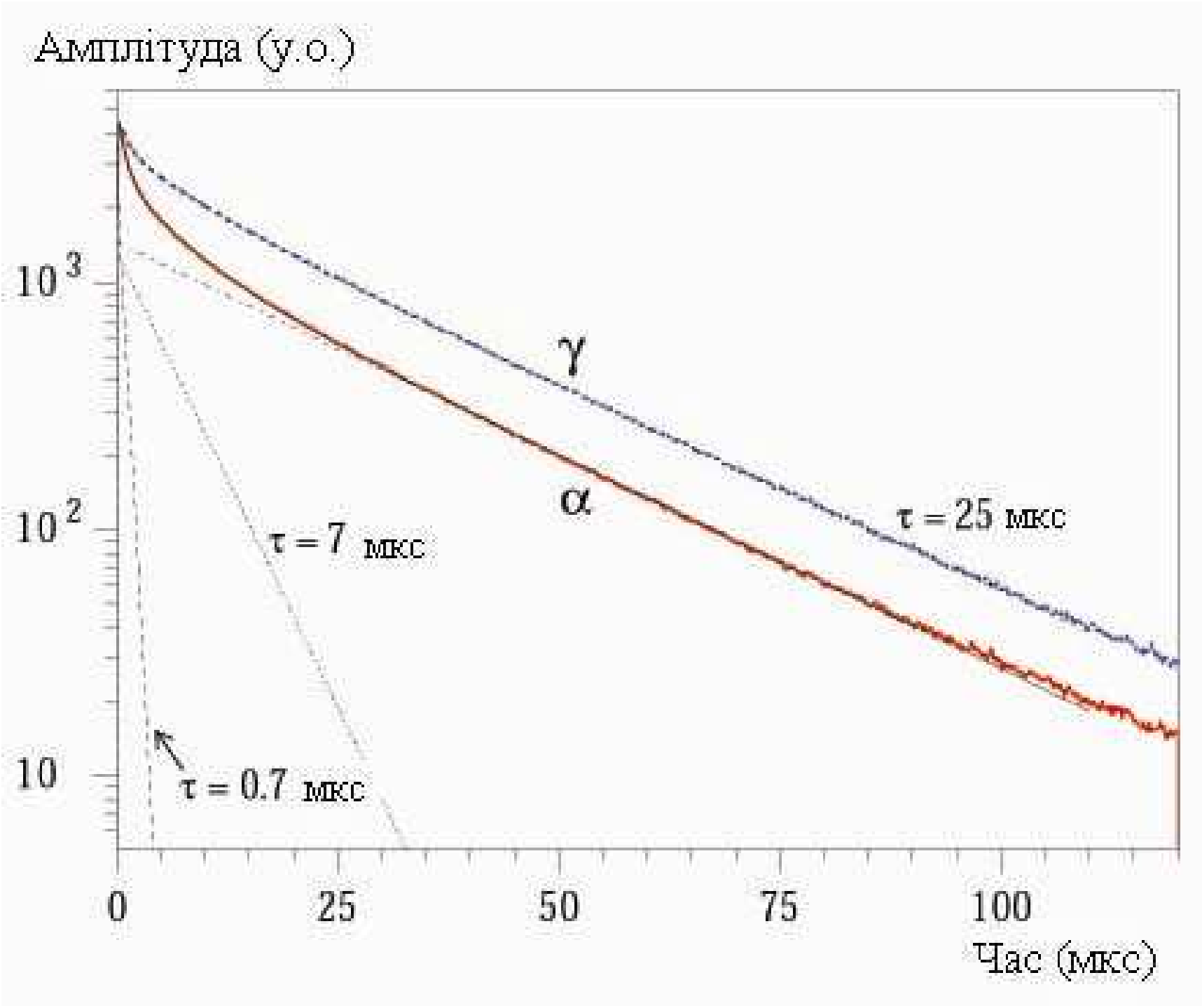

Рис. 5.3. Форми сцинтиляційних сигналів сцинтилятора $ZnWO_4$ виміряні з α–частинками та γ–квантами, та їх підгонка сумою трьох експоненційних функцій.

Різниця у формі сцинтиляційних сигналів дозволила ефективно розділяти події від гамма–квантів та альфа–частинок. При цьому, так як і для сцинтиляторів $CdWO_4$, була спостережена залежність форми від енергії та напрямку опромінення альфа–частинками (рис. 5.4).

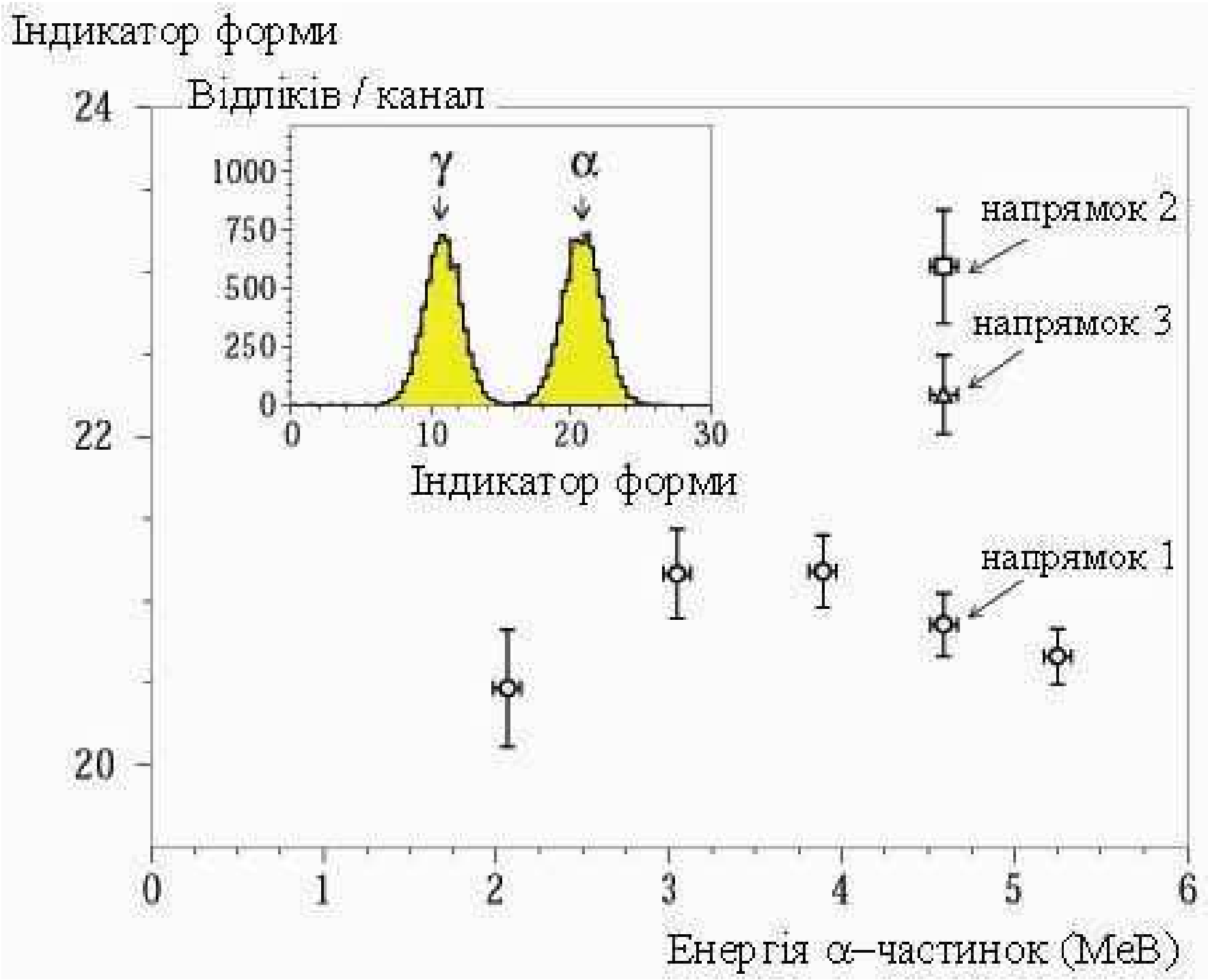

Рис. 5.4. Залежність індикатора форми в детекторі $ZnWO_4$ від енергії та напрямку опромінення α–частинок.

Таблиця 5.2. Постійні (мкс) та інтенсивності сцинтиляційного спалаху сцинтиляторів $ZnWO_4$.

| Вид опромінення | Постійні сцинтиляційного спалаху (мкс) та інтенсивності | | |
|---|---|---|---|
| | $\tau_1$ ($A_1$) | $\tau_1$ ($A_1$) | $\tau_3$ ($A_3$) |
| γ–кванти | 0.7 (2%) | 7.5 (9%) | 25.9 (89%) |
| α–частинки | 0.7 (4%) | 5.6 (16%) | 24.8 (80%) |

5.1.4. Низькофонові вимірювання із сцинтиляторами вольфрамату цинку.

Радіочистота кристалу вольфрамату цинку була виміряна із зразком розмірами $\varnothing 14 \times 4$ мм (масою 4.5 г) в низькофоновій установці в Солотвинській підземній лабораторії. Кристал проглядався ФЕП ФЭУ–137 через кварцовий світловод $\varnothing 49 \times 250$ мм з надчистого кварцу. Детектор був захищений шаром поліметилметакрилату (органічного скла) товщиною 6 – 13 см, електролітичної міді (3 – 6 см) та свинцю. Система реєстрації записувала для кожної події амплітуду та час. Енергетична роздільна здатність спектрометра була виміряна з гамма–джерелами $^{241}$Am (60 кеВ) та $^{207}$Bi (570 і 1064 кеВ) і становила 37%, 15% і 10%, відповідно.

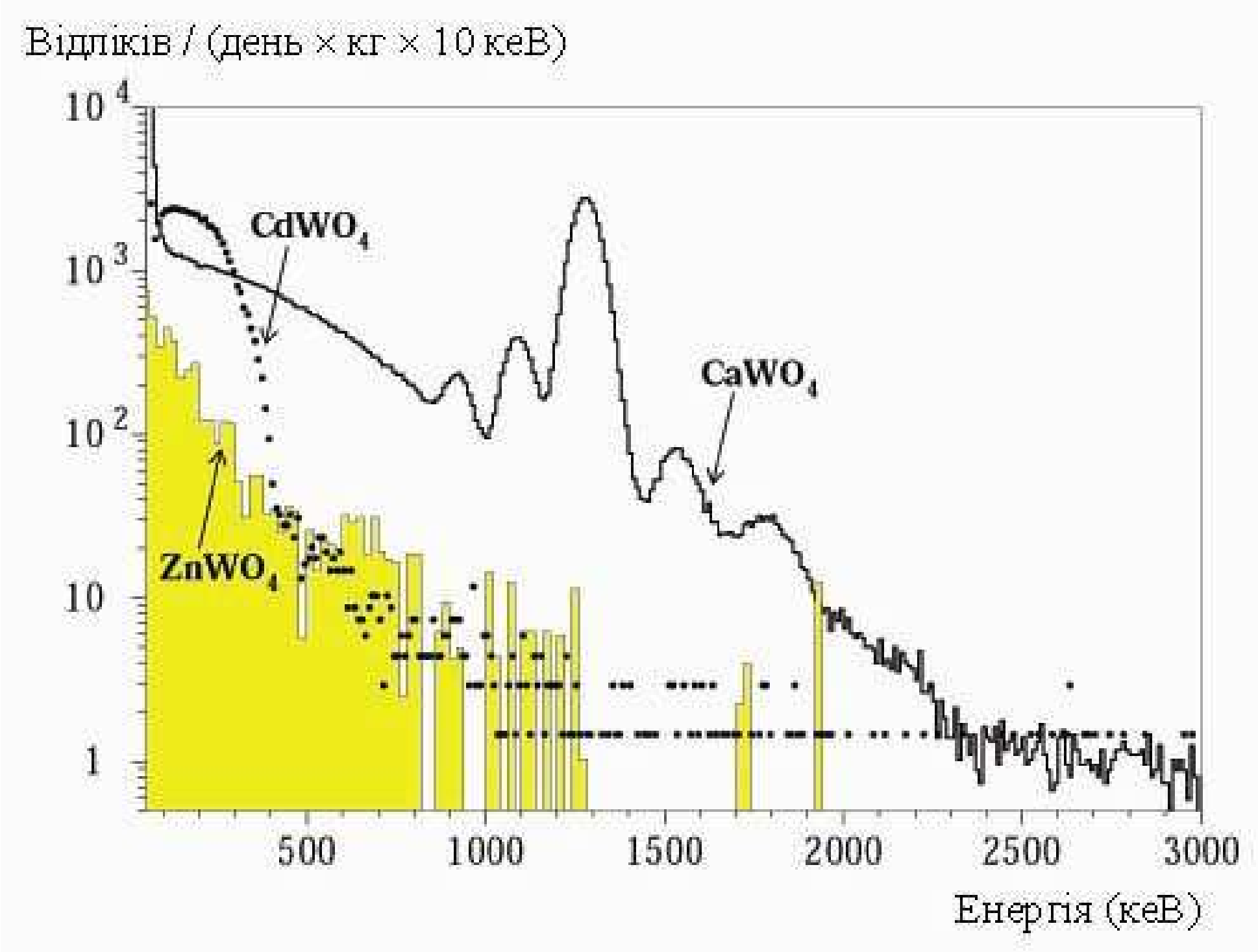

Рис. 5.5. Спектри фону виміряні з детекторами $ZnWO_4$, $CaWO_4$, $CdWO_4$ в низькофоновій установці в Солотвинській підземній лабораторії. Спектри нормовані на час вимірювання та масу сцинтиляційних кристалів.

5.1.5. Оцінка рівня радіоактивної чистоти кристалів $ZnWO_4$.

Енергетичний спектр, виміряний з детектором вольфрамату цинку, показаний на рис. 5.5. Спектри кристалів двох інших вольфраматів, $CdWO_4$ та $CaWO_4$, нормовані на масу та час вимірювань, також приведені для порівняння. Видно, що фон кристалів $ZnWO_4$ значно нижчий за $CaWO_4$ і порівняний з $CdWO_4$. При енергіях менше 0.5 МеВ фон кристала вольфрамату кадмію значно вищий за фон вольфрамату цинку, що пояснюється β–розпадом ізотопу кадмію–113, котрий визначає фон детектора $CdWO_4$ при низьких енергіях. Фон $ZnWO_4$, що має експоненційний характер, обумовлюється, головним чином, зовнішніми гамма–квантами. Ніяких особливостей, які б могли бути спричинені радіоактивними домішками в кристалах, в спектрі не видно. Аналіз форми енергетичного спектру фону дозволив одержати обмеження на активності домішок урану, торію, $^{40}$K в $ZnWO_4$, які приведені в таблиці 5.3.

Більш чутливі оцінки вмісту $^{228}$Th та $^{226}$Ra були одержані за допомогою часово–амплітудного аналізу подій (п. 2.3.1). Не було знайдено жодної пари подій обумовленої розпадами $^{220}Rn \to {}^{216}Po \to {}^{212}Pb$ та $^{214}Bi \to {}^{214}Po \to {}^{210}Pb$. Тому, враховуючи ефективності відбору цих подій, були встановлені обмеження на активності $^{228}$Th та $^{226}$Ra в $ZnWO_4$, що також наведені в таблиці 5.3.

Таблиця 5.3. Обмеження на активності можливих радіоактивних домішок в кристалі $ZnWO_4$

| Сімейство | Нуклід | Активність (мБк/кг) |
|---|---|---|
| $^{232}$Th | $^{232}$Th | ≤ 3.3 |
| | $^{228}$Th | ≤ 0.2 |
| $^{238}$U | $^{238}$U | ≤ 3.2 |
| | $^{230}$Th | ≤ 4.5 |
| | $^{226}$Ra | ≤ 0.4 |
| | $^{210}$Pb | ≤ 1 |
| $^{235}$U | $^{227}$Ac | ≤ 0.2 |
| | $^{40}$K | ≤ 12 |
| | $^{147}$Sm | ≤ 1.8 |
| | $^{137}$Cs | ≤ 20 |

## 5.3. Пошуки процесів подвійного бета–розпаду ізотопів цинку

Два ізотопи цинку потенційно є 2β–активними: $^{64}$Zn та $^{70}$Zn. Енергії переходів, ізотопні розповсюдженості цих ізотопів, можливі канали та моди розпаду приведені в таблиці 5.4.

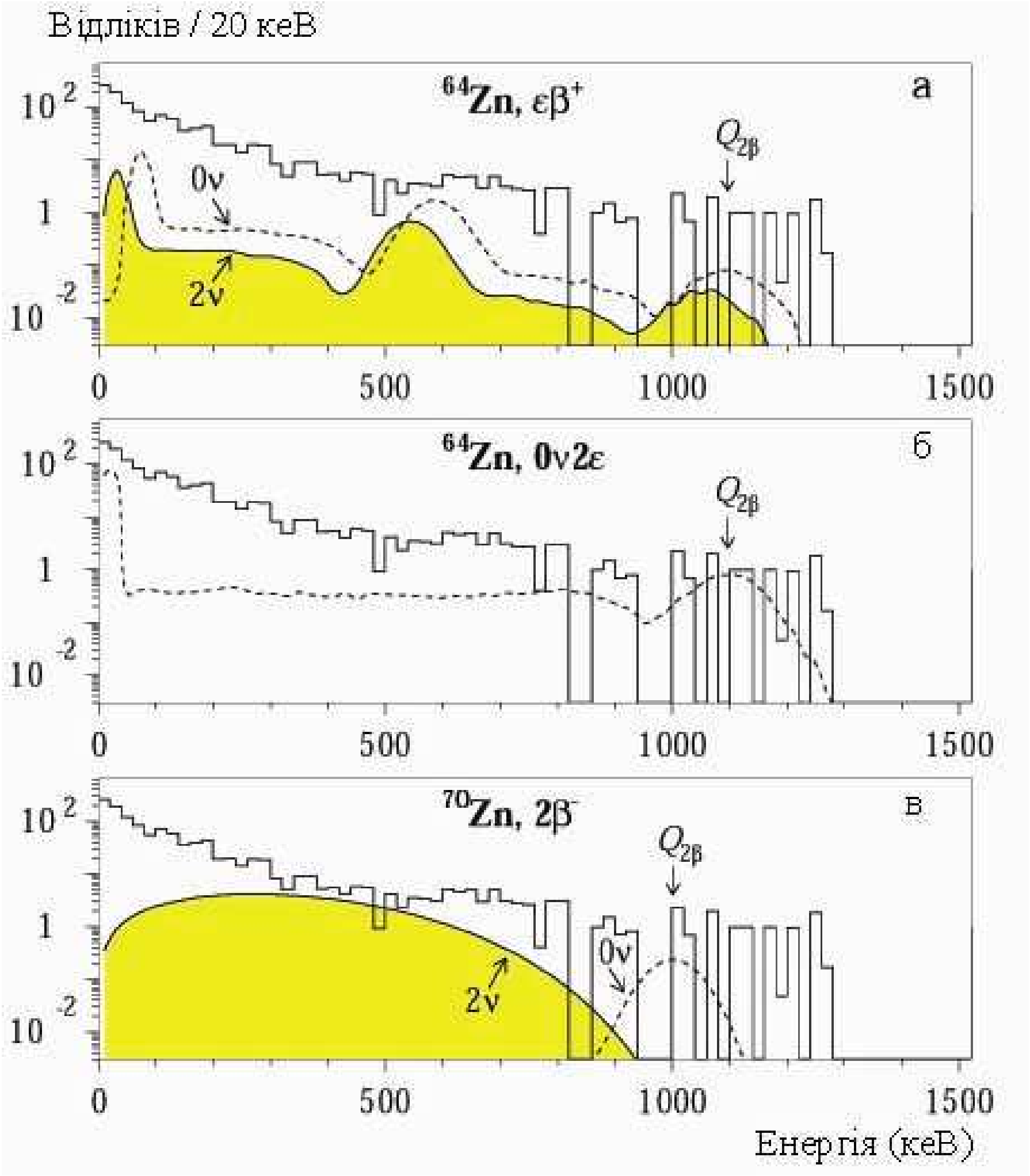

Рис. 5.5. Спектр фону виміряний протягом 429 годин з детектором $ZnWO_4$ (масою 4.5 г) в низькофоновій установці в Солотвинській підземній лабораторії. Показані виключені з довірчою імовірністю 68% енергетичні спектри різних мод і каналів очікуваних 2β–процесів в ядрах цинку.

Таблиця 5.4. Властивості потенційно 2β–активних ізотопів цинку та одержані обмеження на періоди напіврозпаду відносно різних каналів та мод 2β–розпаду.

| Перехід, енергія переходу (кеВ), розповсюдженість материнського ізотопу (%) | Канал Розпаду | Мода розпаду | Обмеження на період напіврозпаду (років), одержане в даній роботі 90% (68%) CL | Обмеження на період напіврозпаду (років), попередні роботи |
|---|---|---|---|---|
| $^{64}Zn \to {}^{64}Ni$ 1096.4(0.9) 48.63(0.60) | 2ε | 0ν | $0.7\ (1.0)\times10^{18}$ | $8(1.0)\times10^{15}$ [300] |
| | $\varepsilon\beta^+$ | 2ν | $4.3\ (8.9)\times10^{18}$ | $2.3\times10^{18}$ (68% CL) [301] |
| | | 0ν | $2.4\ (3.6)\times10^{18}$ | $2.3\times10^{18}$ (68% CL) [300] $T_{1/2} = (1.1\pm0.9)\times10^{19}$ [302] |
| $^{70}Zn \to {}^{70}Ge$ 1000.9(3.4) 0.62(0.03) | $2\beta^-$ | 2ν | $1.3\ (2.1)\times10^{16}$ | – |
| | | 0ν | $0.7\ (1.4)\times10^{18}$ | $1.3\times10^{16}$ (90% CL) [303] |

5.2.1. Пошук 2β–розпаду $^{70}Zn$.

В енергетичному спектрі фону детектора з кристалом $ZnWO_4$ розмірами $\varnothing 14\times 4$ мм немає особливостей, які можна було б інтерпретувати як процеси подвійного бета–розпаду ізотопів цинку. Була побудована модель фону, що включала в себе експоненційну функцію (фон, обумовлений розсіянням гамма–квантів від оточуючих сцинтиляційний кристал матеріалів) та шуканий ефект. Функції відгуку детектора до різних процесів 2β–розпаду всередині кристала були обраховані методом Монте–Карло за допомогою пакету програм GEANT4 та генератора подій DECAY0 (ці розрахунки були здійснені співробітником відділу фізики лептонів Кобичевим В.В., генератор подій DECAY0 розроблений співробітником відділу фізики лептонів Третяком В.І.). Для оцінки кількості подій ефекту, які можуть бути присутніми в спектрі фону, показаному на рис. 5.5, були здійснені підгонки спектру в різних інтервалах енергій. Так, наприклад, підгонка спектру фону методом максимуму правдоподібності в інтервалі енергій 0.7 – 1.3 МеВ дає –0.6±1.9 відліків для площі шуканого ефекту, що відповідає (приймаючи до уваги рекомендації Particle Data Group [304]) кількості відліків ефекту, які можуть бути виключені з заданим рівнем довірчої імовірності, lim*S* = 2.6 (1.3) 90% (68%). Враховуючи кількість ядер $^{70}Zn$ в кристалі ($5.4\times10^{19}$) та практично 100% ефективність реєстрації ефекту, одержано обмеження на період напіврозпаду ядра $^{70}Zn$ відносно $0\nu2\beta$–розпаду:

$$T_{1/2}(0\nu2\beta) \geq 0.9\ (1.2)\times10^{18} \text{ років при 90\% (68\%) CL.}$$

Таким же чином було одержане обмеження на двохнейтринну моду $0\nu2\beta$–розпаду $^{70}Zn$:

$$T_{1/2}(2\nu2\beta) \geq 1.3\ (2.1)\times10^{16} \text{ років при 90\% (68\%) CL.}$$

На рис 5.6 показано розподіли, що відповідають двохнейтринному та безнейтринному подвійному бета–розпаду ядра $^{70}Zn$, які можуть бути виключені з 68% довірчої імовірності.

5.2.2. Пошук $2\varepsilon$ та $\varepsilon\beta^+$ процесів в ядрі $^{64}Zn$.

Ядро $^{64}Zn$ цікаве значним вмістом (48.6%) в природній суміші ізотопів цинку, що дозволяє планувати крупномасштабний експеримент без використання дорогих збагачених ізотопів. Щоб оцінити значення lim*S* для двохнейтринної моди $\varepsilon\beta^+$–розпаду $^{64}Zn$, експериментальний спектр підганявся в енергетичному інтервалі 380 – 800 кеВ сумою експоненти та ефекту. Підгонка методом найменших квадратів дала значення для ефекту – 15 ± 29 відліків, що дає площу ефекту, яку можна виключити, $\lim S = 34\,(16)$ відліків при 90% (68%) CL. Це дозволяє, враховуючи кількість ядер $^{64}$Zn в кристалі: $4.21 \times 10^{21}$, обмежити період напіврозпаду ядра $^{64}$Zn відносно $2\nu\varepsilon\beta^+$–розпаду:

$$T_{1/2}\,(2\nu\varepsilon\beta^+) \geq 4.2\ (8.9)\times10^{18} \text{ років при 90\% (68\%) CL.}$$

Таким же чином було одержане обмеження на безнейтринну моду $\varepsilon\beta^+$–розпаду $^{64}$Zn:

$$T_{1/2}(0\nu\varepsilon\beta^+) \geq 2.4\ (3.6)\times10^{18} \text{ років при 90\% (68\%) CL,}$$

та процес безнейтринного подвійного електронного захвату:

$$T_{1/2}(0\nu2\varepsilon) \geq 0.7\ (1.8)\times10^{18} \text{ років при 90\% (68\%) CL,}$$

Розподіли, що відповідають виключеним з 68% CL процесам подвійного бета–розпаду $^{64}$Zn, показані на рис. 5.6 (а, б). Одержані обмеження на процеси подвійного бета–розпаду ядер цинку приведені в табл. 5.4.

В роботі [301] повідомляється про спостереження процесу електронного захвату з випроміненням позитрона з періодом напіврозпаду $T_{1/2} = (1.1\pm0.9)\times10^{19}$ років. Автори протягом 392 годин вимірювали гамма–випромінювання із зразка цинку масою 350 г. Було використано сцинтиляційний детектор NaI(Tl) розмірами ∅7.6×7.6 см у режимі збігів з HPGe детектором (з ефективністю реєстрації 25%). В спектрі збігів спостерігається пік з енергією 511 кеВ, в той час як у вимірюваннях із зразком заліза пік не спостерігався. Тому автори інтерпретують пік як анігіляційний від позитронів, що випромінювались в процесі $\varepsilon\beta^+$–розпаду $^{64}$Zn. Цей результат викликає сумнів, перш за все тому, що автори не аналізували радіоактивну забрудненість зразків цинку та заліза, що були використані в експерименті. Тим не менше, цей результат необхідно перевірити в незалежному досліді, який ми плануємо здійснити з детектором на основі сцинтиляційних кристалів $ZnWO_4$.

Було оцінено чутливість досліду по пошуку подвійних бета–процесів в цинку з кристалом масою 1 кг. Як показали розрахунки, ефективність реєстрації ефекту значно більша в детекторі такого розміру. Крім того, фон детектора очікується меншим в кілька разів за рахунок використання більшого детектора, застосування дискримінації частинок за формою сцинтиляційних кристалів, вдосконалення захисту. За рік вимірювань з таким детектором можна досягнути чутливості $T_{1/2}$ ~ $5 \times 10^{21}$ для обох мод (0ν і 2ν) $\varepsilon\beta^{+}$–розпаду $^{64}$Zn.

### 5.2.3. Оцінки можливості застосування кристалів $ZnWO_4$ для пошуків процесів подвійного бета–розпаду ядер вольфраму.

Кристали $ZnWO_4$ можуть бути використані також для пошуків процесів подвійного бета–розпаду ізотопів вольфраму: $^{180}$W та $^{186}$W, дослідження яких було здійснене за допомогою сцинтиляторів $^{116}CdWO_4$ значно більшої маси (330 г) і за більший час вимірювань для $^{180}$W (692 годин) та $^{186}$W (13 316 годин). Але очевидною перевагою детектора $ZnWO_4$ є відсутність бета–активного ізотопу $^{113}$Cd. Наявність цього β–активного ізотопу в кристалах $CdWO_4$ (енергія β–переходу становить 316 кеВ) призводить до фону в області енергій аж до 450–500 кеВ, де якраз і очікуються ефекти від 2β–процесів в ядрах вольфраму (див. табл. 5.5). З аналізу даних вимірювань з кристалом $ZnWO_4$ були отримані попередні оцінки чутливості експерименту, які наведені в табл. 5.5. Можна очікувати суттєвого підвищення чутливості експерименту з використанням кристалів $ZnWO_4$ масою близько 1 кг. Зокрема, для ядра $^{180}$W за рік вимірювань чутливість досліду може досягати рівня $\lim T_{1/2} \sim 10^{18}$ років. Подальше підвищення чутливості може бути досягнуте завдяки використанню кристалів $ZnWO_4$ в якості болометричного детектора із значно кращою, ніж у сцинтилятора, енергетичною роздільною здатністю.

Таблиця 5.5. Властивості потенційно 2β–активних ізотопів вольфраму та одержані обмеження на періоди напіврозпаду відносно різних каналів та мод 2β–розпаду.

| Перехід, енергія переходу (кеВ), розповсюдженість материнського ізотопу (%) | Канал розпаду | Мода розпаду | Обмеження на період напіврозпаду (років), одержане в даній роботі 90% (68%) CL | Обмеження на період напіврозпаду (років), попередня робота [161] |
|---|---|---|---|---|
| $^{180}W \rightarrow ^{180}Hf$ 145(5) 0.12(0.01) | 2K | 2ν | $0.7\ (0.8)\times10^{16}$ | $0.7\ (0.8)\times10^{17}$ 90% (68%) CL |
| | 2ε | 0ν | $0.9\ (1.1)\times10^{16}$ | $0.9\ (1.3)\times10^{17}$ 90% (68%) CL |
| $^{186}W \rightarrow ^{186}W$ 488.0(1.7) 28.43(0.19) | $2\beta^-$ | 2ν | $1.4\ (2.5)\times10^{18}$ | $3.7\ (5.3)\times10^{18}$ 90% (68%) CL |
| | | 0ν | $1.1\ (1.7)\times10^{19}$ | $1.1\ (2.1)\times10^{21}$ 90% (68%) CL |

**Висновки розділу.** Були досліджені сцинтиляційні властивості, форми сигналів та здатність до ідентифікації γ–квантів та α–частинок, радіочистота кристалів вольфрамату цинку. З даних вимірювань фону в Солотвинській підземній лабораторії отримані обмеження на процесу 2β–розпаду цинку, що на один–два порядки краще за попередні результати. Оцінені можливості підвищення чутливості експерименту для пошуку подвійних бета–процесів в ядрах цинку та вольфраму.

Результати, викладені в цьому розділі, опубліковані в роботі:

F.A.Danevich, V.V.Kobychev, S.S.Nagorny, D.V.Poda, V.I.Tretyak, S.S.Yurchenko, Yu.G.Zdesenko, $ZnWO_4$ crystals as detectors for 2β decay and dark matter experiments, Nucl. Instr. Meth. Phys. Res. A. 544 (2005) 553–564.

РОЗДІЛ 6

# ДОСЛІДЖЕННЯ СЦИНТИЛЯТОРІВ $CaWO_4$ ТА YAG:Nd ЯК МОЖЛИВИХ ДЕТЕКТОРІВ 2β–РОЗПАДУ ЯДЕР $^{48}Ca$ ТА $^{150}Nd$

## 6.1. Сцинтиляційні властивості та рівень радіочистоти сцинтиляторів вольфрамату кальцію

Ядро $^{48}Ca$ має найбільшу енергію переходу серед усіх потенційних 2β–активних ядер. Велика енергія переходу, по–перше, значно спрощує проблему боротьби з фоном, а по–друге, забезпечує більше значення фазового інтегралу, а отже, збільшує ймовірність процесу 2β–розпаду. Тому важливо знайти детектор, який містив би в своєму складі ядра кальцію. Вже один з ранніх дослідів по пошуку 2β–розпаду $^{48}Ca$ був здійснений з використанням сцинтиляторів фториду кальцію, активованого європієм ($CaF_2$(Eu)) [184]. Дослідження велись також, з використанням сцинтиляторів $CaF_2$ без активатора [185]. Нами було вивчено можливість застосування сцинтиляційних кристалів вольфрамату кальцію ($CaWO_4$) для експерименту по пошуку подвійного бета–розпаду $^{48}Ca$.

Кристали вольфрамату кальцію були одними з перших, що їх застосували в якості сцинтиляторів [305,306]. Beard та Kelly використали кристал $CaWO_4$ в низькофоновому експерименті для пошуків альфа–активності природного вольфраму [307]. Властивості сцинтиляторів $CaWO_4$ приведені в табл. 6.1. Кристали негігроскопічні і дуже стійкі до впливу кислот, лугів та органічних розчинників.

Таблиця 6.1. Властивості сцинтиляторів $CaWO_4$.

| Характеристика | |
|---|---|
| Густина (г/см$^3$) | 6.1 |
| Точка плавлення (°C) | 1570–1650 |
| Структурний тип | Шеєліт |
| Площина спайності | Слабка (101) |
| Твердість за Моосом | 4.5 – 5 |
| Довжина хвилі максимуму спектру емісії (нм) | 420–425 |
| Показник заломлення | 1.94 |
| Ефективний час світіння (мкс)* | 8 |
| Вихід фотоелектронів по відношенню до сцинтилятора NaI(Tl)* | 18% |

* При кімнатній температурі та при опроміненні γ–квантами

6.1.1. Спектрометричні властивості сцинтиляторів $CaWO_4$.

В дослідженнях були використані чотири кристали $CaWO_4$ розмірами $20\times20\times10$ мм, $40\times34\times23$ мм, $\varnothing40\times39$ мм і $\varnothing60\times120$ мм, вирощені методом Чохральського. Лінійність енергетичної шкали та енергетична роздільна здатність були виміряні з γ–джерелами $^{60}$Co, $^{137}$Cs, $^{207}$Bi, $^{232}$Th та $^{241}$Am в діапазоні енергій 60 – 2615 кеВ. Сцинтилятори буди обгорнуті плівкою PTFE (Bikron) і оптично з’єднувалися з фотопомножувачем XP2412 (Philips). Сигнал з фотопомножувача через попередній підсилювач подавався на вхід спектрометричного підсилювача з постійною формування 12 мкс, а потім на вхід аналогово–цифрового (чутливого до максимуму імпульсу напруги) перетворювача. На рис. 6.1 (а) показані енергетичні спектри гамма–квантів $^{60}$Co та $^{137}$Cs, виміряні при опроміненні кристала розмірами $\varnothing40\times39$ мм. Енергетична роздільна здатність становить 7.8% по γ–лінії 662 кеВ $^{137}$Cs.

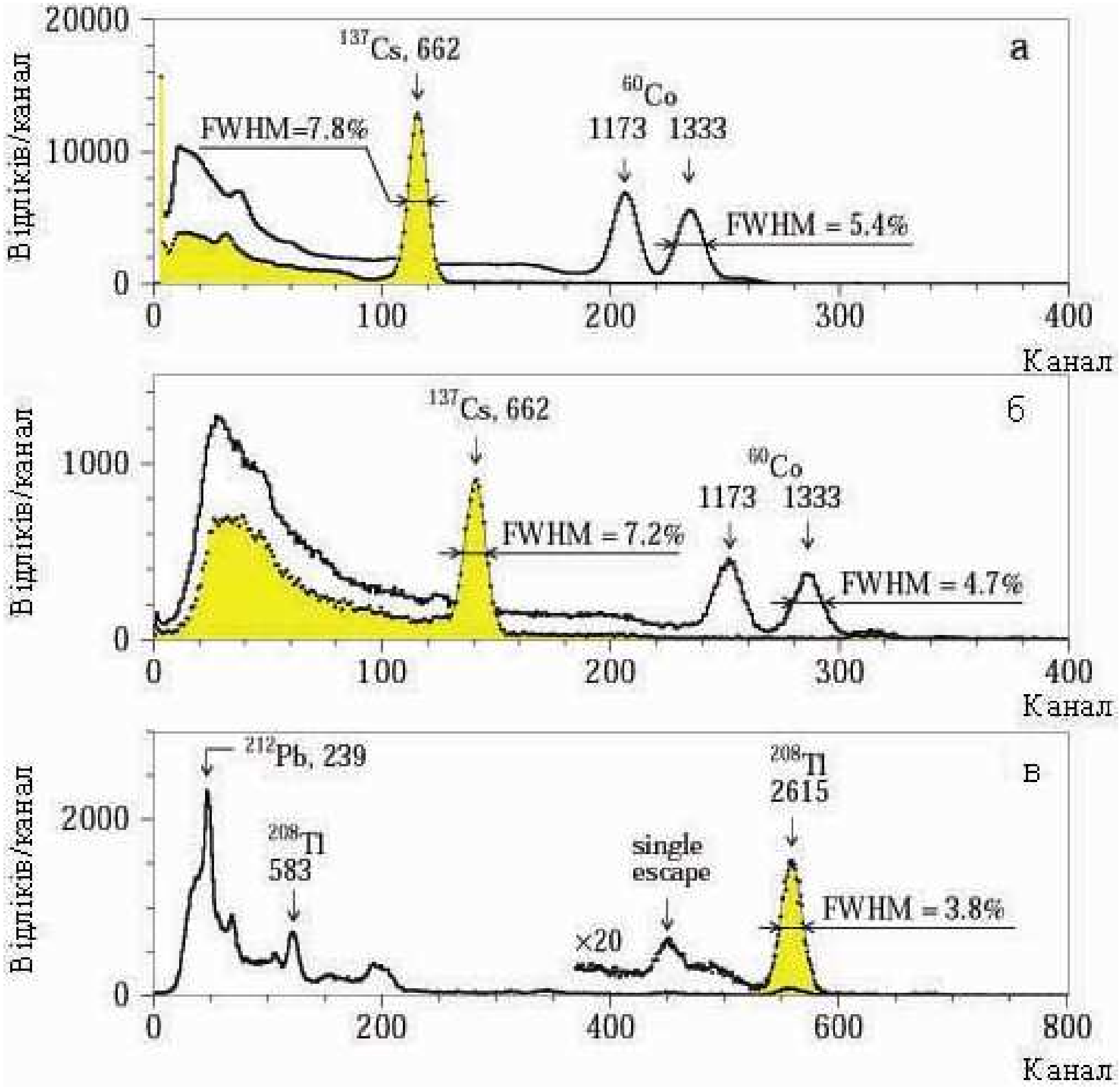

Рис. 6.1. Енергетичні спектри $^{137}Cs$ та $^{60}Co$, $^{232}Th$, виміряні із сцинтиляційним кристалом $CaWO_4$ розмірами ∅40×39 мм в двох геометріях збору світла. Геометрія (а): кристал встановлено на ФЕП; (б, в) – кристал розміщений всередині фторопластового контейнера, наповненого силіконовою олією, і проглядається двома ФЕП.

Значного покращення світловиходу та роздільної здатності було досягнуто завдяки розміщенню цього кристала в рідину (силіконова олія з коефіцієнтом заломлення 1.5). Кристал був зафіксований всередині фторопластового контейнера розмірами ∅70×90 мм і проглядався з двох сторін двома ФЕП XP2412. Як видно з рис. 6.1 (б), світловихід зріс на 20%, а енергетична роздільна здатність для $^{137}Cs$ становила 7.2%. Енергетичні спектри, виміряні з джерелами $^{60}Co$, $^{137}Cs$ та $^{232}Th$, показані на рис. 6.1 (б, в). Зокрема, для γ–квантів $^{232}$Th з енергією 2615 кеВ отримана роздільна здатність 3.8%.

6.1.2. Відгук до α–частинок.

Відгук сцинтиляційних детекторів $CaWO_4$ до α–частинок був вивчений із джерелом $^{241}$Am. Крім того, для визначення α/β–співвідношення були використані виділені за допомогою аналізу форми сцинтиляційних сигналів α–піки нуклідів, що присутні в слідових кількостях у кристалі $CaWO_4$ розмірами 40×34×23 мм. Кристал

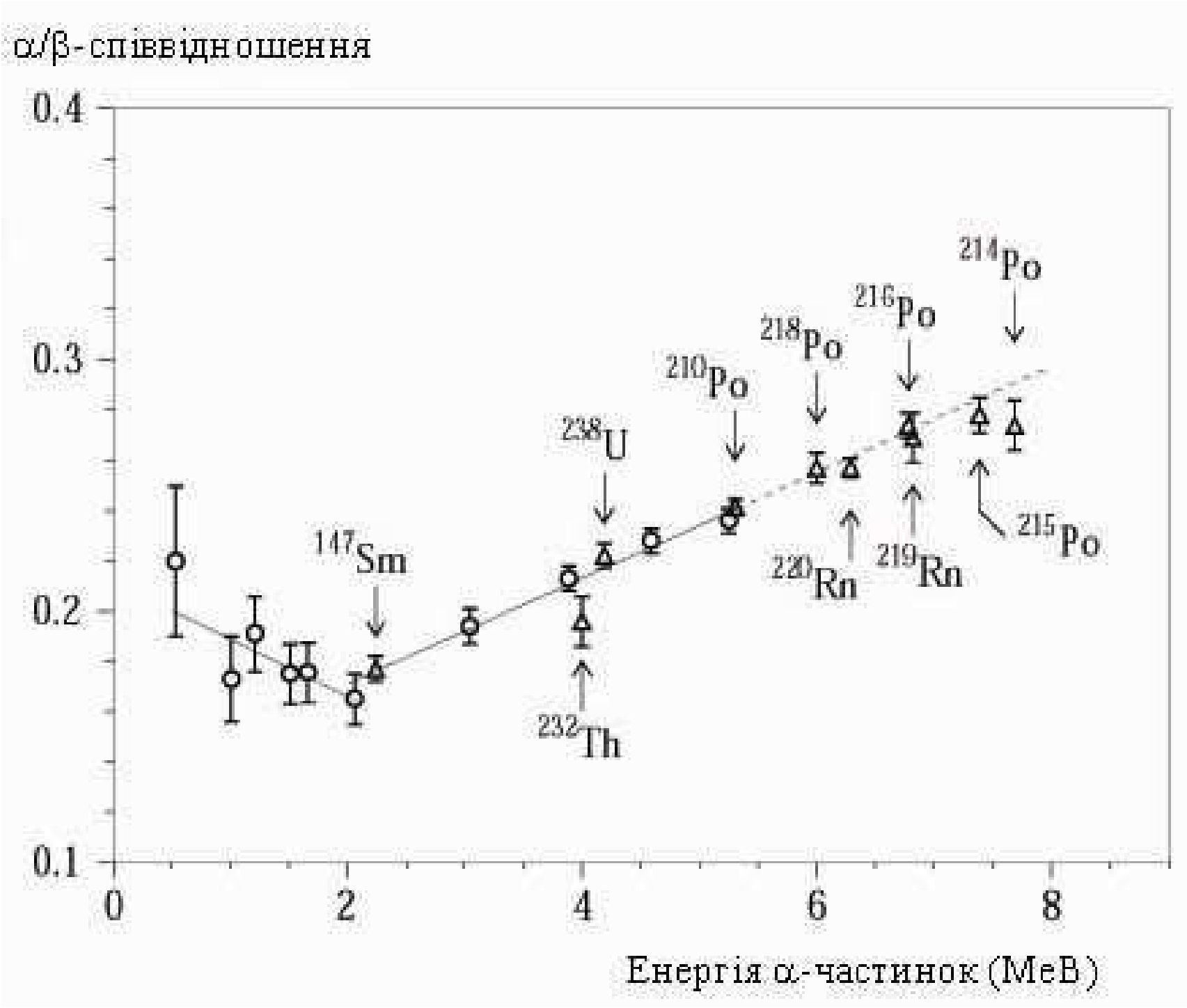

Рис. 6.2. Енергетична залежність α/β–співвідношення, виміряна з кристалом $CaWO_4$ розмірами 40 × 34 × 23 мм. Кружочками показані точки отримані у вимірюваннях з α–джерелом $^{241}$Am, трикутниками – від слідових забрудненостей кристала α–активними радіонуклідами рядів урану, торію та $^{147}$Sm.

опромінювався α–частинками в трьох напрямках, перпендикулярних до головних кристалічних площин. На відміну від сцинтиляторів $CdWO_4$ та $ZnWO_4$, у вольфрамату кальцію не було виявлено залежності α/β–співвідношення від напрямку опромінення. Виміряна з детектором $CaWO_4$ розмірами 40×34×23 мм залежність α/β–співвідношення від енергії α–частинок показана на рис. 6.2. В інтервалі енергій 0.5 – 2 МеВ α/β–співвідношення зменшується з ростом енергії:

$$\alpha/\beta = 0.21(2) - 0.023(14)E_\alpha,$$

тут $E_\alpha$ – енергія α–частинок в МеВ. При енергіях $E_\alpha > 2$ МеВ α/β–співвідношення зростає з ростом енергії:

$$\alpha/\beta = 0.129(12) + 0.021(3)E_\alpha.$$

### 6.1.3. Дослідження форми сцинтиляційних сигналів та ідентифікація α–частинок та γ–квантів за формою сцинтиляційного спалаху.

Форма сцинтиляційних сигналів $CaWO_4$ досліджувалась за допомогою оцифровщика форми з частотою 20 МГц та розрядністю 12 біт. Кристал опромінювався α–частинками від джерела $^{241}$Am та γ–квантами від різних джерел. На рис. 6.3 показані форми сигналів від α–частинок та γ–квантів, отримані завдяки сумуванню 4 тисяч окремих імпульсів. Підгонка цих розподілів сумою трьох експонент дає значення постійних спаду та інтенсивностей компонент сцинтиляційних сигналів $CaWO_4$, що наведені в табл. 6.2.

Таблиця 6.2. Постійні ($\tau_i$, мкс) та інтенсивності ($A_i$) сцинтиляційного спалаху сцинтиляторів $CaWO_4$.

| Вид опромінення | Постійні сцинтиляційного спалаху (мкс) та інтенсивності | | |
|---|---|---|---|
| | $\tau_1$ ($A_1$) | $\tau_1$ ($A_1$) | $\tau_3$ ($A_3$) |
| γ–кванти | 0.3 (3%) | 4.4 (15%) | 9.0 (82%) |
| α–частинки | 0.3 (6%) | 3.2 (18%) | 8.8 (76%) |

Різниця в формі сцинтиляційних сигналів дозволяє розділяти події від

альфа–частинок та гамма–квантів, що дуже важливо для інтерпретації та зменшення фону в експериментах по пошуку рідкісних процесів, зокрема, подвійного β–розпаду.

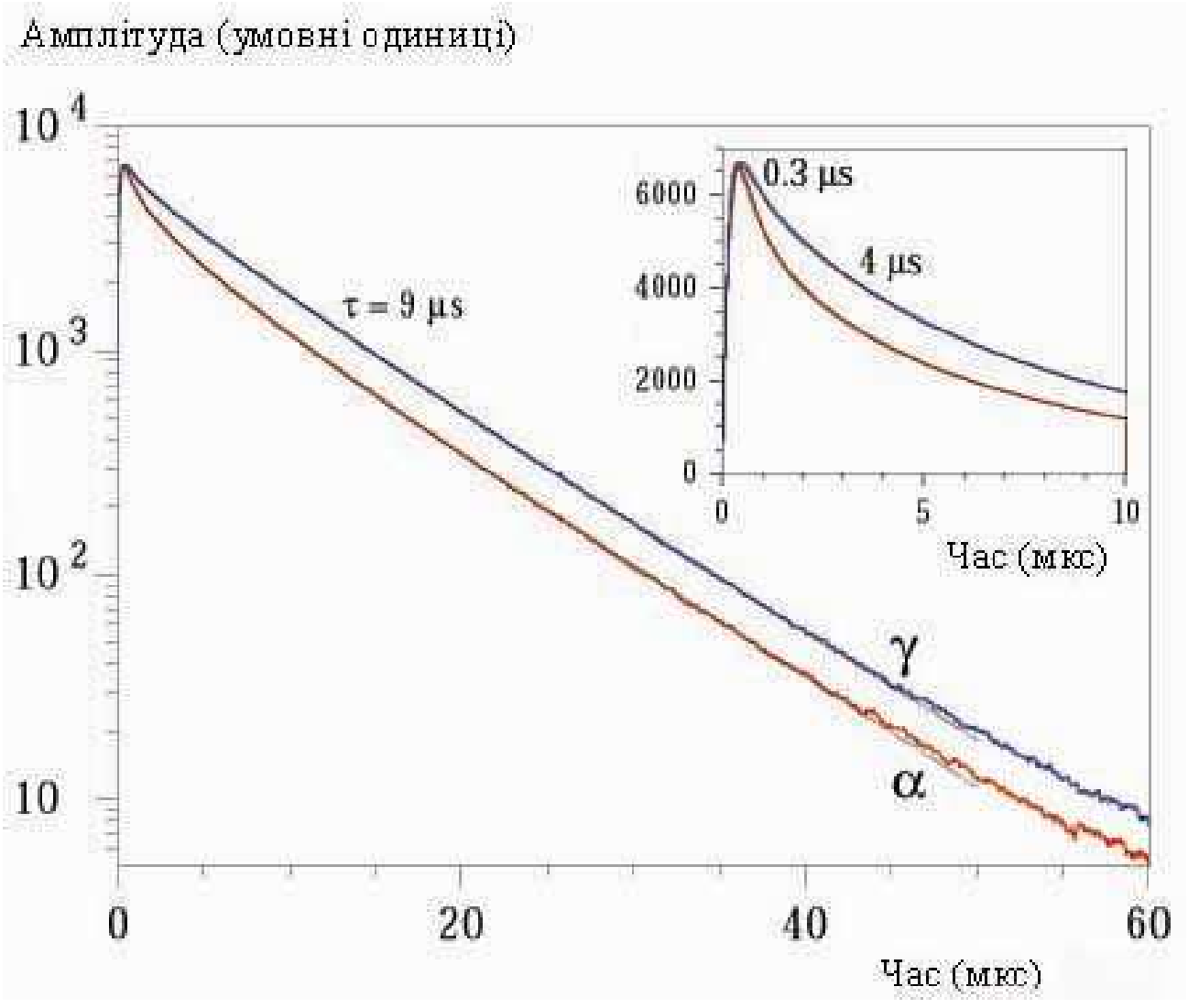

Рис. 6.3. Форма сцинтиляційних сигналів сцинтиляційних сигналів $CaWO_4$ при опроміненні α–частинками та γ–квантами.

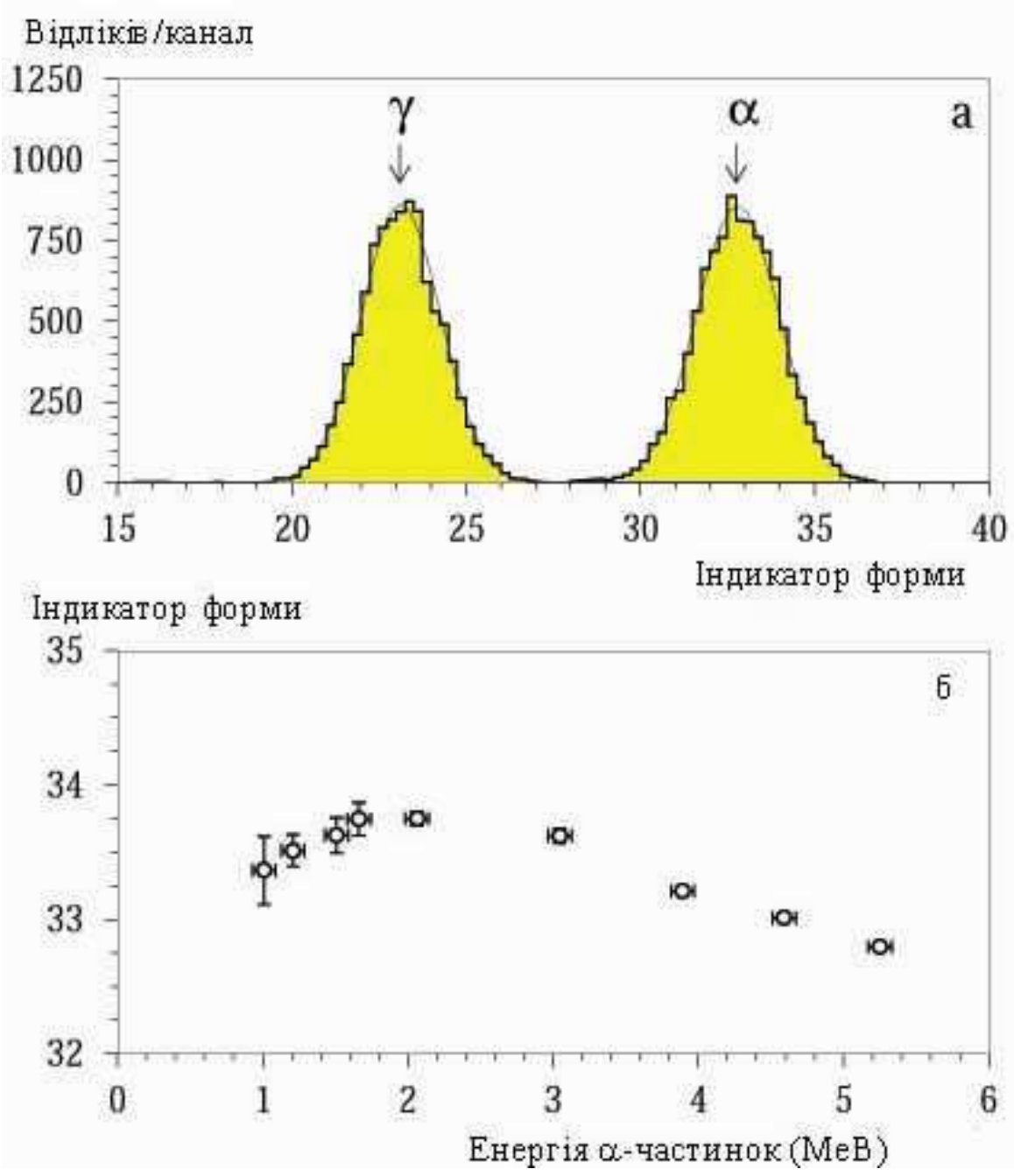

Рис. 6.4. (а) Розподіли індикатора форми для γ–квантів та α–частинок та (б) залежність індикатора форми від енергії α–частинок, виміряні сцинтиляційним детектором з кристалом $CaWO_4$.

Для дискримінації сигналів був застосований метод оптимального фільтра, що вже був описаний в розділі 2. Було отримано практично повне розділення сигналів від $\alpha$–частинок та $\gamma$–квантів (рис. 6.4 (а)).

Була також виміряна залежність форми сцинтиляційних сигналів (точніше, індикатора форми) для $\alpha$–частинок від їх енергії та напрямку опромінення відносно головних кристалічних площин. В той час, як енергетична залежність якісно така ж як (див. рис. 6.4.(б)) для сцинтиляторів $CdWO_4$ і $ZnWO_4$, залежності форми сцинтиляцій від напрямку у сцинтиляторів вольфрамату кальцію не виявлено.

### 6.1.4. Низькофонові вимірювання з кристалом $CaWO_4$ в Солотвинській підземній лабораторії.

Вимірювання з метою вивчення рівня радіоактивної забрудненості сцинтилятора $CaWO_4$ та оцінки можливостей їх застосування для пошуків подвійного бета–розпаду були проведені в Солотвинській підземній лабораторії. Фон детектора з кристалом $CaWO_4$ розмірами 40×34×23 мм вимірювався в низькофоновій установці. Сцинтилятор проглядався через кварцовий світловод ∅100×330 мм у формі логарифмічної спіралі низькофоновим фотопомножувачем (EMI D724KFLB). Детектор був оточений захистом із фторопласту (3–5 см), органічного скла (6–13 см), міді (3–6 см), свинцю (15 см) та поліетилену (16 см). Два детектори з пластиковими сцинтиляторами розмірами 120×130×3 см, встановлені над установкою, виробляли сигнал заборони реєстрації подій, пов’язаних з проходженням космічних мюонів. Для кожної події в детекторі записувалась енергія, час, факт збігу з сигналами, що генерувалися антимюонним захистом та форма сцинтиляційних сигналів (з розрядністю 12 біт та частотою 20 МГц). Енергетична калібровка та роздільна здатність спектрометра була виміряна з $\gamma$–джерелами $^{60}Co$, $^{137}Cs$, $^{207}Bi$ та $^{232}Th$ в інтервалі енергій 60 – 2615 кеВ. Залежність роздільної здатності від енергії може бути описана функцією

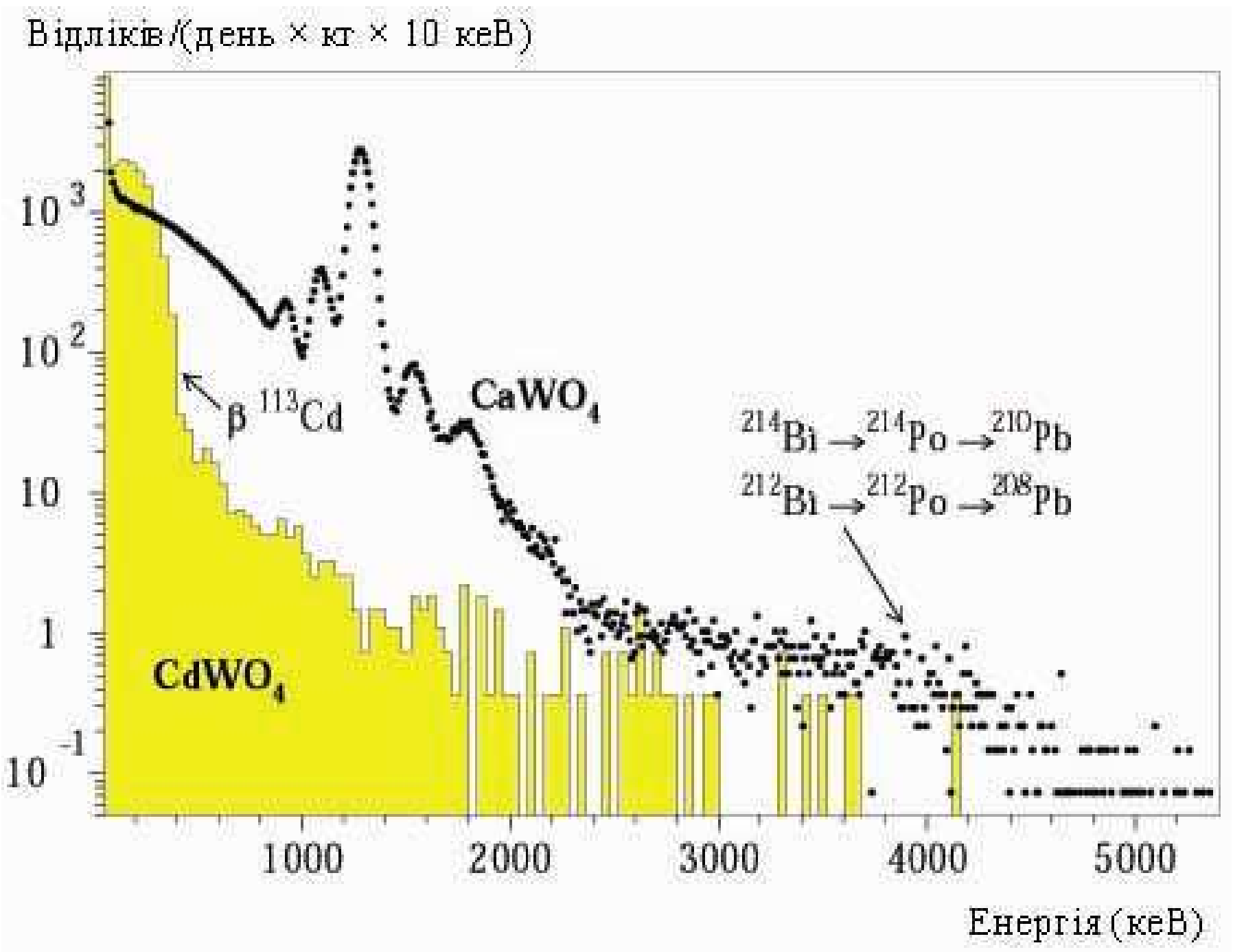

Рис. 6.5. Енергетичні спектри, виміряні сцинтиляційним детектором з кристалом $CaWO_4$ в низькофоновій установці в Солотвинській підземній лабораторії. Для порівняння зображено також спектр, виміряний з кристалом $CdWO_4$ в тих же умовах. Спектри нормовані на масу сцинтиляторів та час вимірювання.

$FWHM_\gamma$(кеВ) = –3 + $\sqrt{6.9E\gamma}$, де $E_\gamma$ – енергія γ–квантів в кеВ. Періодичні (раз на тиждень) калібровки проводились за допомогою γ–джерел $^{207}$Bi та $^{232}$Th.

### 6.1.5. Аналіз фону та визначення рівня радіоактивної чистоти кристалів $CaWO_4$.

Енергетичний спектр фону, виміряний детектором з кристалом $CaWO_4$ протягом 1734 годин, зображений на рис. 6.5. Спектр детектора $CdWO_4$, виміряний в таких же умовах, нормований на масу кристала та час вимірювання, також приведено для порівняння. Видно, що радіоактивна забрудненість сцинтиляторів $CaWO_4$ значно вища. З метою визначення радіонуклідів, що спричинюють фон детектора, та їх активностей, було проведено часово–амплітудний аналіз подій, аналіз форми сигналів та форми енергетичних спектрів фонових подій.

6.1.5.1. Часово–амплітудний аналіз.

З метою пошуку та виділення подій ланцюжка $^{220}$Rn $\rightarrow$ $^{216}$Po $\rightarrow$ $^{212}$Pb (сімейство $^{232}$Th) всі події в інтервалі енергій 1.4 – 2.2 MeB були використані в якості тригера. В цей енергетичний діапазон, враховуючи виміряне значення $\alpha/\beta$–співвідношення, потрапляють практично всі $\alpha$–розпади $^{220}$Rn. Був здійснений

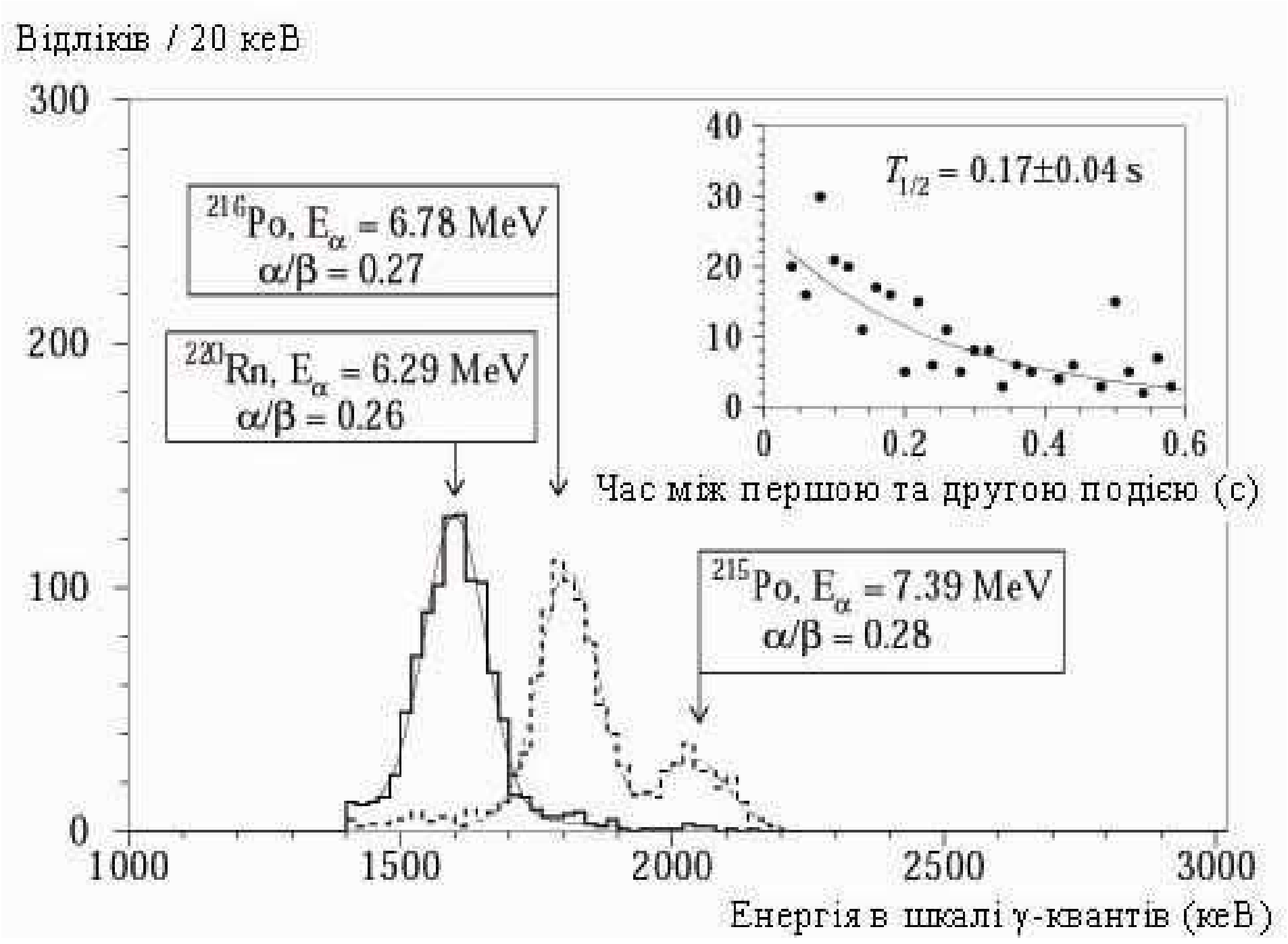

Рис. 6.6. Енергетичні спектри $\alpha$–частинок $^{220}$Rn та $^{216}$Po та розподіл часових інтервалів між подіями, відібраними за допомогою часово–амплітудного аналізу з даних вимірювань фону сцинтиляційним детектором з кристалом $CaWO_4$. $\alpha$–Пік $^{215}$Po належить до іншого ланцюжка розпадів (сімейство $^{235}$U).

пошук всіх подій (з тими ж енергіями), які слідують за тригером впродовж 20 – 600 мкс. Враховуючи період напіврозпаду ядра $^{216}$Po, таким чином можуть бути відібрані 85.2% розпадів $^{216}$Po. Енергія отриманих піків (з урахуванням $\alpha/\beta$–співвідношення) $^{220}$Rn та $^{216}$Po, а також період напіврозпаду $^{216}$Po, отриманий в результаті підгонки часового розподілу подій, відповідає табличним значенням (див. рис. 6.6). Активність $^{228}$Th (сімейство $^{232}$Th), визначена таким методом, складає 0.6(2) мБк/кг.

За допомогою часово–амплітудного аналізу даних було також визначено активності $^{227}$Ac (з ряду $^{235}$U) та $^{226}$Ra (з ряду $^{238}$U). Активність $^{227}$Ac була оцінена за допомогою відбору подій розпадів в ланцюжку $^{219}$Rn $\rightarrow$ $^{215}$Po $\rightarrow$ $^{211}$Pb, а $^{226}$Ra – з аналізу подій розпадів $^{214}$Bi $\rightarrow$ $^{214}$Po $\rightarrow$ $^{210}$Pb. Дані про виміряні значення або обмеження на активності радіонуклідів, що у слідових кількостях присутні в кристалі $CaWO_4$ розмірами $40 \times 34 \times 23$ мм, наведені в таблиці 6.3.

Таблиця 6.3. Активності радіоактивних домішок в кристалі $CaWO_4$

| Сімейство | Нуклід | Активність (мБк/кг) |
|---|---|---|
| $^{232}$Th | $^{232}$Th | 0.69(10) |
| | $^{228}$Th | 0.6(2) |
| $^{238}$U | $^{238}$U | 14.0(5) |
| | $^{226}$Ra | 5.6(5) |
| | $^{210}$Pb | $\leq$ 430 |
| | $^{210}$Po | 291(5) |
| $^{235}$U | $^{227}$Ac | 1.6(3) |
| | $^{40}$K | $\leq$ 12 |
| | $^{147}$Sm | $\leq$ 1.8 |
| | $^{137}$Cs | $\leq$ 20 |

6.1.5.2. Аналіз фонового $\alpha$–спектру.

Енергетичний спектр $\alpha$–подій, відібраних за допомогою аналізу форми сцинтиляційних сигналів, показано на рис. 6.7. Сумарна $\alpha$–активність в кристалі $CaWO_4$ становить $\approx 0.4$ мБк/кг. Інтенсивний пік в спектрі з енергією $\approx$1.28 МеВ належить ізотопу $^{210}$Po. Ізотоп $^{210}$Po є дочірнім радіоактивного свинцю $^{210}$Pb з сімейства $^{238}$U. Очевидно, що вікова рівновага в ряду ізотопів $^{238}$U сильно порушена в кристалі $CaWO_4$, оскільки активність $^{238}$U, оцінена по $\alpha$–піку, становить лише 14.0(5) мБк/кг.

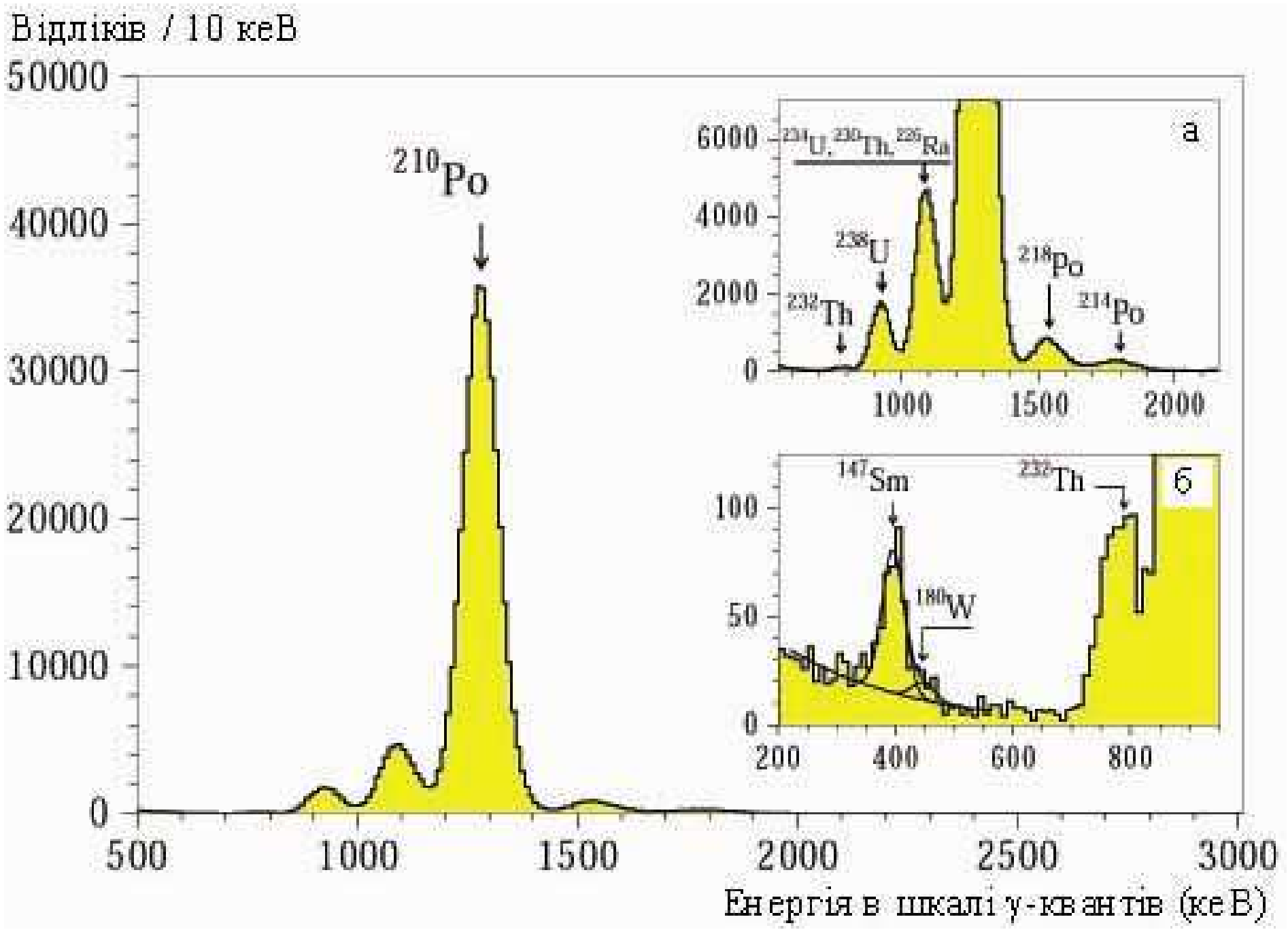

Рис. 6.7. Енергетичний спектр $\alpha$–частинок виділений за допомогою аналізу форми сцинтиляційних сигналів з даних вимірювань з кристалом $CaWO_4$. $\alpha$–Спектр може бути пояснений розпадами радіонуклідів сімейств урану і торію. В низькоенергетичній частині спектру (б) присутні $\alpha$–піки $^{147}$Sm та $^{180}$W.

Альфа–піки дочірніх радіонуклідів урану: $^{234}$U, $^{230}$Th, $^{226}$Ra не розділяються в енергетичному спектрі, оскільки їхні енергії альфа–розпадів дуже близькі. Повна площа піку (з енергією $\approx$ 1.1 МеВ) погоджується з активністю $^{238}$U, визначеною з площі піку в α–спектрі, та $^{226}Ra$, активність якого була визначена за допомогою часово–амплітудного аналізу. Ще один член уранового ряду, ізотоп $^{222}Rn$, не розділяється від α–піку $^{210}Po$, в той час як α–пік $^{218}Po$ спостерігається в спектрі. Активність $^{226}Ra$, обрахована за цим піком, становить 5.9(8) мБк/кг і узгоджується з активністю, розрахованою за допомогою часово–амплітудного аналізу.

В низькоенергетичній частині спектру α–пік з енергією ≈ 0.8 МеВ може бути віднесений до $^{232}Th$ з активністю 0.69(10) мБк/кг. Пік з енергією в шкалі γ–квантів 395(2) кеВ (що відповідає енергії α–частинок 2243(9) кеВ) може бути пояснений α–розпадами ізотопу самарію $^{147}Sm$ ($E_{\alpha} = 2247$ кеВ, $T_{1/2} = 1.06\times10^{11}$ років) з активністю в сцинтиляторі 0.49(4) мБк/кг. Присутність самарію в кристалах вольфрамату кальцію на рівні ≈ 6 мБк/кг була також спостережена в досліді [308], в якому цей кристал працював в режимі кріогенного болометричного детектора. Крім того, в спектрі α–частинок спостерігається особливість на енергії 447(8) кеВ, що відповідає енергії α–частинок 2471(30) кеВ. Ці α–події можуть бути пояснені α–активністю природного вольфраму, а саме α–розпадом ізотопу $^{180}W$, що був вперше спостережений у вимірюваннях з детектором $^{116}CdWO_4$ [257]. Кількість ядер $^{180}W$ в детекторі $CaWO_4$ становить $4.7 \times 10^{20}$. Враховуючи площу α–піку (38±16 відліків), отриману в результаті підгонки форми α–спектра та ефективність відбору α–подій (59%) за допомогою аналізу форми сцинтиляційних сигналів, отримаємо період напіврозпаду ядра $^{180}W$ відносно α–розпаду: $T_{1/2} = 1.0^{+0.7}_{-0.3}\times10^{18}$ років. Це значення, в межах похибок, співпадає з нашим попереднім результатом: $T_{1/2} = 1.1^{+0.8}_{-0.4}\times10^{18}$ років [257], а також з даними вимірювань періоду напіврозпаду ядра $^{180}W$ відносно α–розпаду за допомогою кріогенного болометричного детектора з високою енергетичною роздільною здатністю [307].

6.1.5.3. Інтерпретація спектру подій від γ–квантів та β–частинок.

Енергетичний спектр γ(β)–подій в детекторі $CaWO_4$, відібраних за допомогою аналізу форми сигналів, показаний на рис. 6.8. Фон в низькоенергетичній частині спектру, найбільш ймовірно, пов’язаний з β–розпадами радіоактивного свинцю $^{210}Pb$, що має енергію β–розпаду 64 кеВ. Значення активності $^{210}Pb$, оцінене за площею β–спектра, не протирічить результатам оцінки активності в кристалі α–активного $^{210}Po$, дочірнього ізотопу $^{210}Pb$.

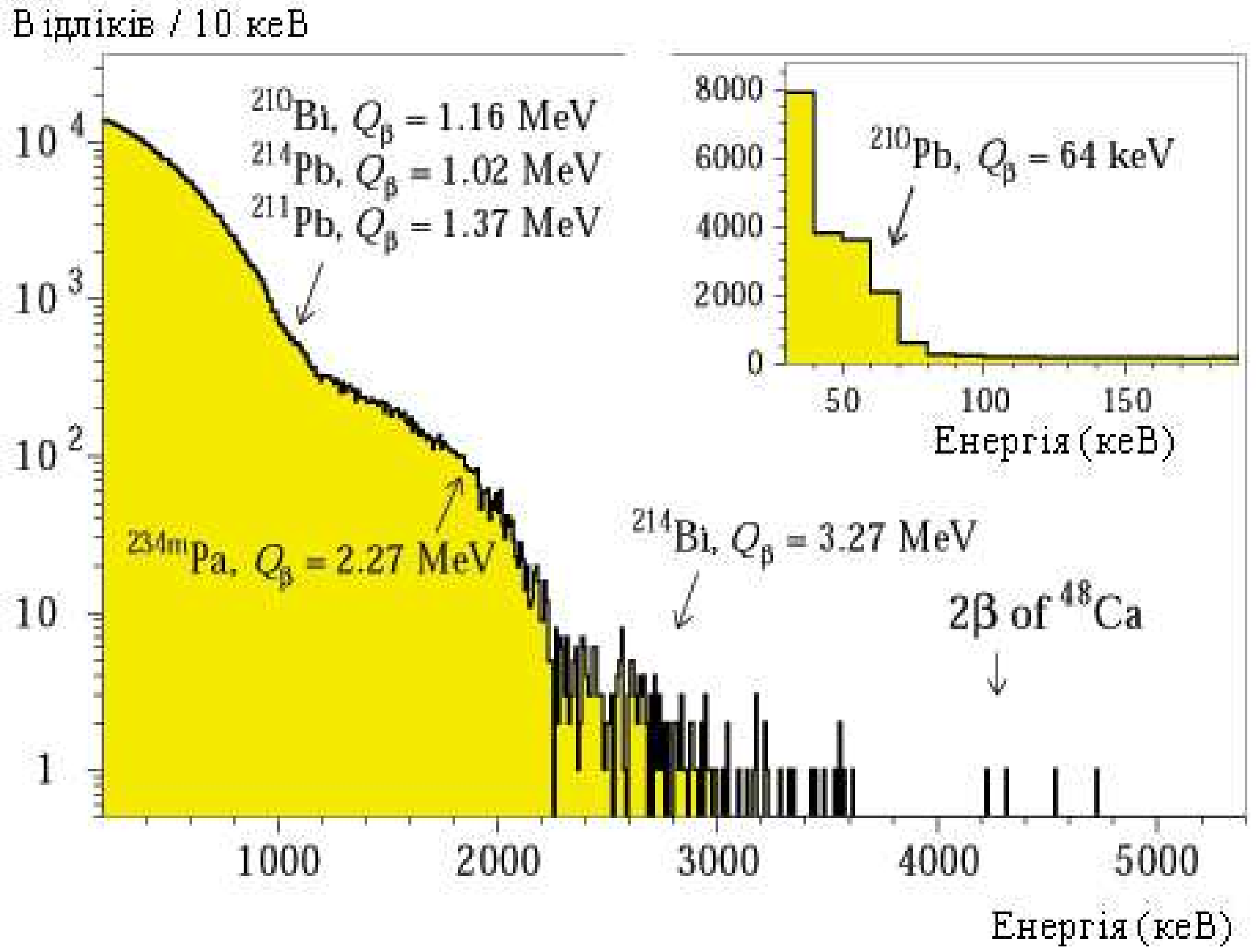

Рис. 6.8. Енергетичний спектр β–частинок та γ–квантів, виділений за допомогою аналізу форми сцинтиляційних сигналів з даних вимірювань з кристалом $CaWO_4$.

## 6.2. Розробка чутливого експерименту для пошуку 2β–розпаду ядра $^{48}Ca$

### 6.2.1. Обмеження на період напіврозпаду ядра $^{48}Ca$ відносно 0ν2β–розпаду.

Кристали $CaWO_4$ містять кілька ізотопів, які можуть розпадатися з випроміненням двох β–частинок або шляхом захвату двох електронів. Властивості цих потенційних 2β–нестабільних ядер представлені в табл. 6.4.

Таблиця 6.4. Властивості потенційно 2β–активних ізотопів кальцію та вольфраму, що містяться в кристалах $CaWO_4$.

| Перехід | Енергія переходу, кеВ | Розповсюдженість материнського ізотопу, % | Канал розпаду |
|---|---|---|---|
| $^{40}Ca \rightarrow {}^{40}Ar$ | 193.78(0.29) | 96.941(0.156) | 2ε |
| $^{46}Ca \rightarrow {}^{46}Ti$ | 990.4(2.4) | 0.004(0.003) | $2\beta^-$ |
| $^{48}Ca \rightarrow {}^{48}Ti$ | 4272(4) | 0.187(0.021) | $2\beta^-$ |
| $^{180}W \rightarrow {}^{180}Hf$ | 146(5) | 0.12(0.01) | 2ε |
| $^{186}W \rightarrow {}^{186}Os$ | 488.0(1.7) | 28.43(0.19) | $2\beta^-$ |

Ізотоп $^{48}Ca$ має найбільшу енергію β–розпаду серед усіх потенційно 2β–активних ядер. Сцинтиляційний детектор на основі кристалів $CaWO_4$ має сприятливі для високочутливого 2β–експерименту властивості: високі енергетичну здатність та ефективність дискримінації за формою сигналів. Фон детектора може бути значно знижений завдяки аналізу форми сигналів. Така можливість була перевірена у вимірюваннях з кристалом $CaWO_4$ розмірами 40 × 34 × 23 мм. Фон в околі енергії 2β–розпаду $^{48}Ca$ (3.6 – 5.4 МеВ) дуже низький: він становить усього 0.07 відліків/(рік×кг×кеВ), незважаючи на досить високу забрудненість кристалу.

Енергетична роздільна здатність детектора була виміряна в інтервалі енергій 60 – 2615 кеВ. Зокрема, для лінії 2615 кеВ енергетична здатність становила FWHM = 132 кеВ (5.1%). Екстраполюючи визначену залежність до енергії 2β–розпаду $^{48}Ca$, отримаємо, що у випадку безнейтринного 2β–розпаду $^{48}$Ca в спектрі мав би спостерігатися пік на енергії 4.27 МеВ з шириною FWHM = 170 кеВ (4%). Такого піку в спектрі не спостерігається, а отже, на основі експериментальних даних можна встановити обмеження на період напіврозпаду $^{48}$Ca відносно 0ν2β–розпаду за формулою 3.1. Величину ефекту 0ν2β–розпаду $^{48}$Ca, яку можна відкинути з довірчою імовірністю 68% (lim*S*), було оцінено згідно рекомендацій [304]. В околі очікуваного піка 0ν2β–розпаду $^{48}$Ca є лише два відліки, в той час як оцінки дають 2.9 відліки, що слідує з виміряної активності $^{228}$Th та моделі спектра фону від розпадів його дочірніх (практично, лише $^{208}$Tl). Таким чином отримаємо lim*S* = 1.5 відліків. Враховуючи, що кількість ядер в кристалі становить $7.4 \times 10^{20}$, ефективність реєстрації процесу 0ν2β–розпаду, розрахована методом Монте–Карло, дорівнює 87%, одержимо обмеження на період напіврозпаду ядра $^{48}$Ca відносно 0ν2β–розпаду на основний стан $^{48}$Ti:

$$T_{1/2}(0\nu2\beta) \geq 6 \times 10^{19} \text{ років} \qquad \text{з 68\% CL.}$$

Це обмеження, отримане з невеликим за масою кристалом (причому, виготовленим з незбагаченого кальцію) поступається результатам робіт [162,163,188,189], де використовувалися або збагачений ізотоп, або детектори значно більшої маси. Поряд з тим, цей попередній експеримент демонструє перспективність використання саме сцинтиляційних детекторів з кристалами вольфрамату кальцію для крупномасштабного досліду по пошуку подвійного бета–розпаду $^{48}$Ca (див. Розділ 7).

## 6.3. Дослідження сцинтиляційних властивостей та рівня радіочистоти кристала алюмо–ітрієвого граната, активованого неодимом (YAG:Nd), як можливого детектора для пошуку 2β–розпаду $^{150}$Nd.

Ізотоп $^{150}$Nd є одним з найперспективніших для пошуків безнейтринного

$2\beta$–розпаду. По–перше, завдяки другій по величині (після $^{48}Ca$) енергії переходу ($Q_{2\beta}$ = 3368 кеВ), теоретично обраховане значення фазового інтегралу для $^{150}Nd$ є найбільшим серед усіх потенційно $2\beta$–нестабільних ядер [14]. Теоретичне передбачення для добутку $T_{1/2} \times \langle m_\nu \rangle^2$ знаходиться в межах $3.4 \times 10^{22} - 3.4 \times 10^{24}$ років × еВ$^2$ [20,22]. Крім того, велика енергія розпаду важлива з огляду на можливості зменшення фону в досліді по пошуку $2\beta$–розпаду $^{150}Nd$. Адже фон від природної радіоактивності різко спадає після енергії 2615 кеВ ($\gamma$–лінія $^{208}Tl$ з найбільшою енергією, серед порівняно інтенсивних $\gamma$–ліній радіонуклідів уранових та торієвих сімейств). До того ж, як показують розрахунки, вплив космогенних радіонуклідів практично не буде давати фонових подій при такій енергії, чого не можна сказати про більшість інших ядер, з якими плануються чутливі експерименти по пошуку $0\nu2\beta$–розпаду.

Не існувало жодного підходящого детектора, який містив би ядра неодиму в своєму складі, що дозволило б провести чутливий „калориметричний” дослід. А враховуючи той факт, що існуючі методи збагачення неодиму дуже дорогі, якраз такий підхід, котрий дозволяє забезпечити максимальну ефективність реєстрації ефекту, варто використати. Адже ефективність в експерименті з використання досліджуваного ізотопу у вигляді фольги є значно (приблизно на порядок величини) нижчою. Тому метою роботи було дослідження можливості використання кристалів алюмо–ітрієвого гранату, активованого неодимом (YAG:Nd, $Y_3Al_5O_{12}$:Nd), в якості сцинтиляційного детектора для пошуку $2\beta$–розпаду $^{150}Nd$.

### 6.3.1. Сцинтиляційні властивості кристалів YAG:Nd.

Властивості кристалів YAG:Nd приведені в таблиці 6.5. Ці кристали широко використовуються в лазерній техніці і тому їх виробництво добре налагоджено. Свого часу були досліджені сцинтиляційні властивості алюмо–ітрієвих гранатів активованих церієм (YAG:Ce) [309, 310] та ітербієм (YAG:Yb) [311]. Сцинтиляцій-Таблиця 6.5. Властивості кристалів YAG:Nd ($Y_3Al_5O_{12}$:Nd).

| Характеристика | |
|---|---|
| Густина (г/см$^3$) | 4.56 |
| Температура плавлення (°C) | 1970 |
| Структурний тип | Кубічний гранат |
| Твердість за Моосом | 8.5 |
| Показник заломлення | 1.82 |
| Ефективний час світіння (мкс)* | 4 |
| Вихід фотоелектронів по відношенню до сцинтилятора NaI(Tl)* | 8% |

* При кімнатній температурі та при опроміненні γ–квантами

ні властивості кристалів YAG:Nd, наскільки нам відомо, ніколи не були вивчені.

6.3.1.1. Енергетична роздільна здатність.

Всі вимірювання були проведені з кристалом YAG:Nd розмірами ∅17 × 6 мм (масою 7.16 г) з концентрацією неодиму на рівні 2 мольних %. Вихід фотоелектронів був виміряний за допомогою ФЕП (Philips, XP2412) відносно виходу стандартного сцинтилятора NaI(Tl) розмірами ∅40 × 40 мм. Кристал YAG:Nd був обгорнутий трьома шарами тефлонової плівки. Обидва детектори опромінювались γ–квантами $^{137}$Cs. Амплітуда сигналів від детектора з кристалом YAG:Nd становила 8% від NaI(Tl). Енергетична роздільна здатність, виміряна детектором з кристалом YAG:Nd для γ–лінії 662 кеВ $^{137}$Cs, склала 13.6%. Суттєвого покращення світловиходу та роздільної здатності вдалося досягти завдяки розміщенню кристала YAG:Nd у фторопластовому контейнері розмірами ∅70×90 мм, заповненому силіконовою олією. З двох боків кристал проглядався ФЕП (Philips, XP2412). На рис. 6.9 показані енергетичні спектри, виміряні цим детектором з γ–квантами від джерел $^{137}Cs$ та $^{207}Bi$. Видно, що піки повного поглинання мають дуже малу інтенсивність у порівнянні з розподілом від розсіяних квантів. Це пов'язано з низьким ефективним атомним номером кристалу і є сприятливим фактором для проведення низькофонового досліду. Для γ–лінії з енергією 662 кеВ від джерела $^{137}Cs$ отримано роздільну здатність 9.3%.

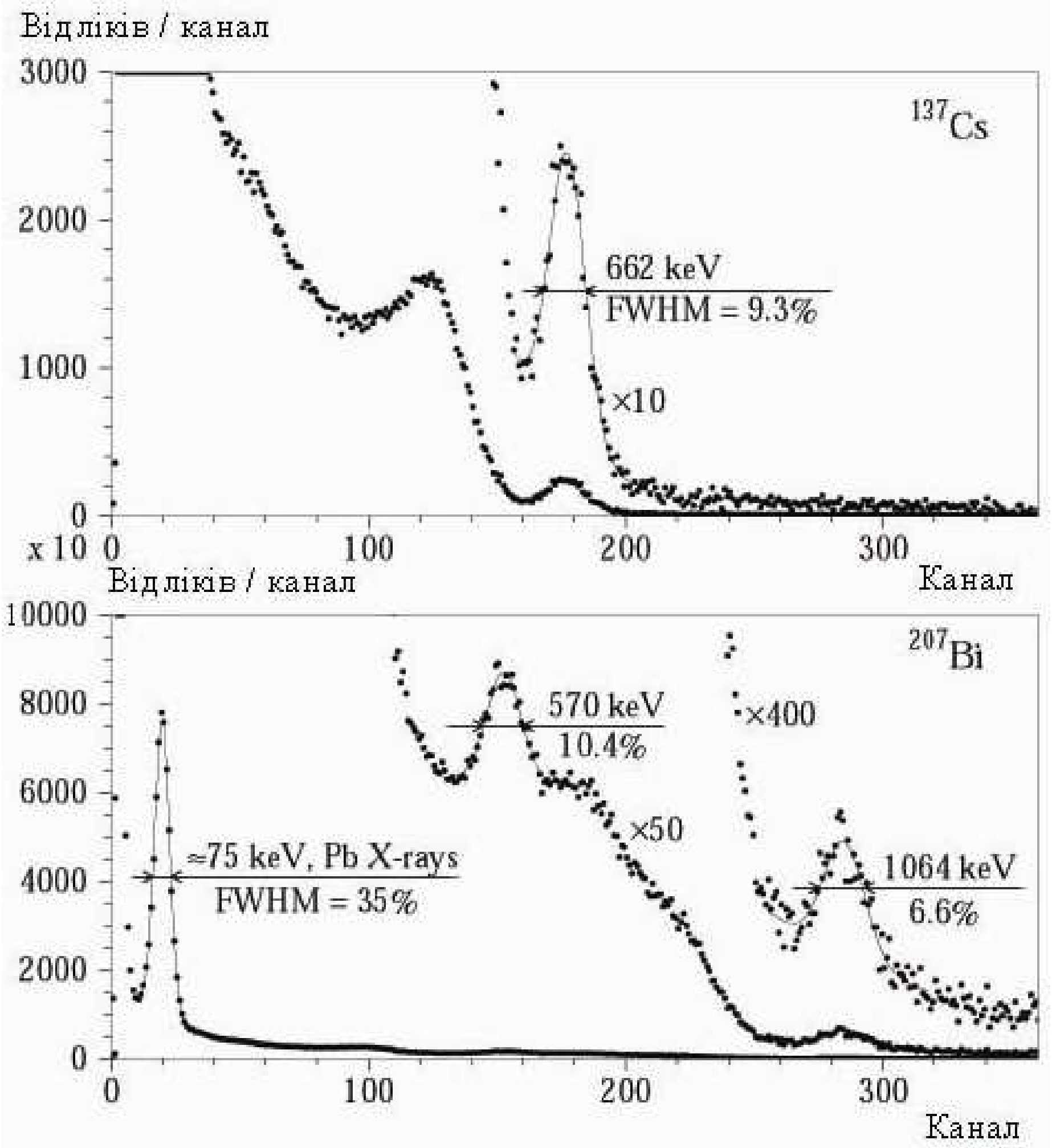

Рис. 6.9. Енергетичні спектри $^{137}Cs$ та $^{207}Bi$, виміряні з кристалом YAG:Nd в двох геометріях збору світла. Кристал встановлено на ФЕП (верхній рис.); кристал був розміщений всередині фторопластового контейнера, наповненого силіконовою олією і проглядався двома ФЕП (нижній рис.).

Енергетичну роздільну здатність сцинтиляційного детектора з кристалом

YAG:Nd на енергії 2β–розпаду $^{150}$Nd можна оцінити за допомогою апроксимації даних вимірювань з джерелами $^{137}$Cs та $^{207}$Bi, результати яких зображені на рис. 6.9. Для підгонки даних була взята корінна залежність FWHM = $a + (b \times E_\gamma)^{1/2}$. Коефіцієнти, отримані в ході підгонки, дорівнюють: $a$ = 2 кеВ, $b$ = 5.2 кеВ. Енергія γ–квантів $E_\gamma$ та роздільна здатність FWHM тут виражені в кеВ. За цією формулою енергетична роздільна здатність детектора з кристалом YAG:Nd на енергії 2β–розпаду $^{150}$Nd становить ≈ 4%.

6.3.1.2. α/β–Співвідношення.

Відгук детектора з кристалом YAG:Nd до α–частинок був виміряний з джерелом $^{241}$Am. Спектр α–частинок від колімованого джерела $^{241}$Am показано на рис. 6.10. Виявилося, що α/β–співвідношення не залежить від напрямку опромінення і становить 0.33 для енергії альфа–частинок 5.25 МеВ.

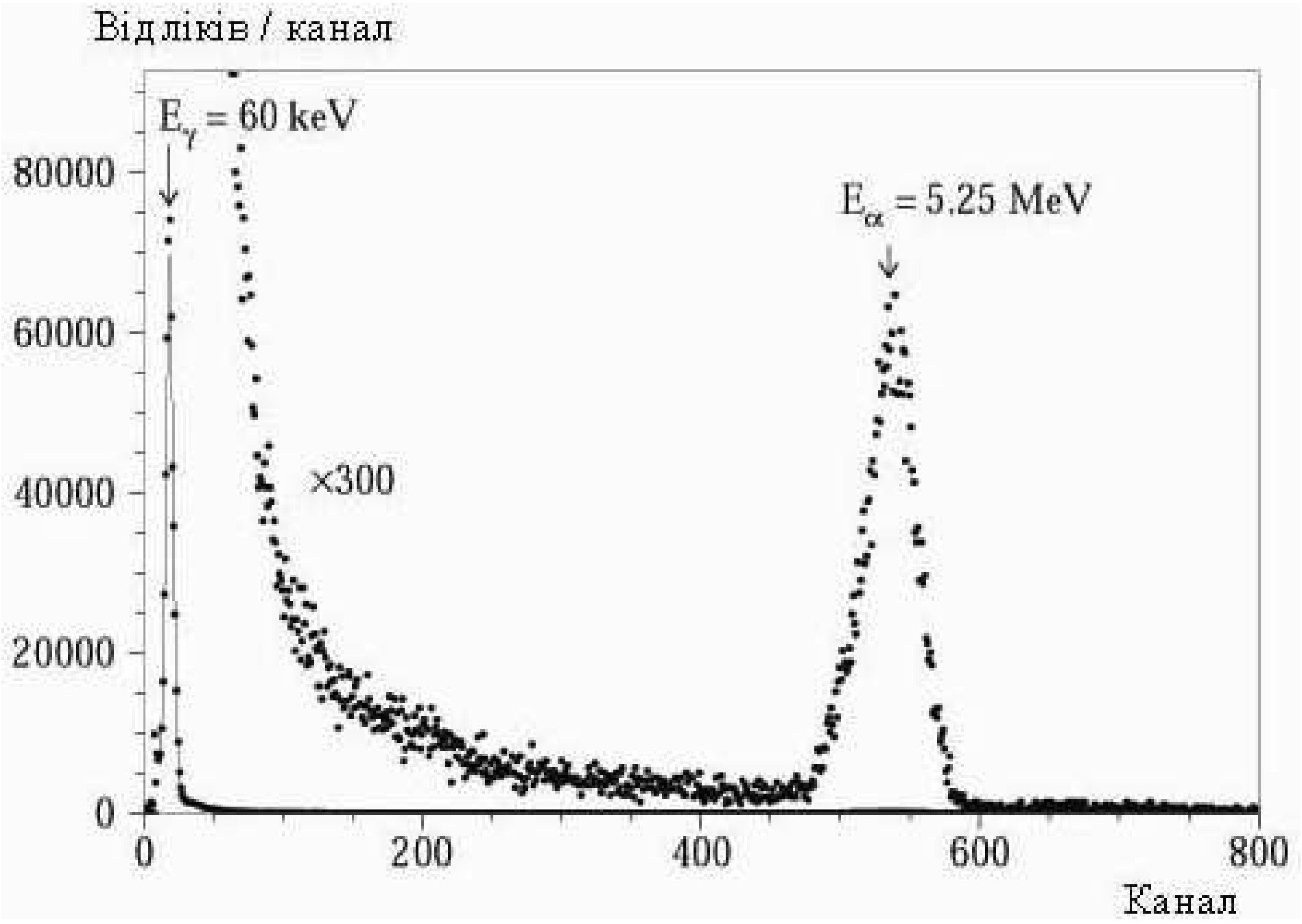

Рис. 6.10. Енергетичний спектр α–частинок від колімованого джерела $^{241}$Am.

6.3.1.3. Форма сцинтиляційного спалаху.

Форми сцинтиляційного спалаху в кристалі YAG:Nd було вивчено з α–частинками (з енергією ≈ 5.25 МеВ) та γ–квантами за допомогою 12–розрядного оцифровщика форми з частотою 20 МГц. Форма сцинтиляційних спалахів (*f(t)*) може бути описана сумою експоненційних функцій:

$$f(t) = \Sigma A_i / (\tau_i - \tau_0)\,(e^{-t/\tau_i} - e^{-t/\tau_0}), \qquad t > 0,$$

де $A_i$ – інтенсивності компонент сцинтиляційного спалаху, $\tau_i$ – постійні затухання сцинтиляційного спалаху, $\tau_0$ – постійна інтегрування сигналу (≈ 0.2 мкс). Виявилося, що форма сцинтиляційних спалахів може бути добре описана трьома компонентами (рис. 6.11). Значення $A_i$ та $\tau_i$, отримані в результаті підгонки сумарних сцинтиляційних сигналів від α–частинок та γ–квантів в інтервалі від 0 до 60 мкс (були просумовані по три тисячі сигналів), приведені в таблиці 6.6.

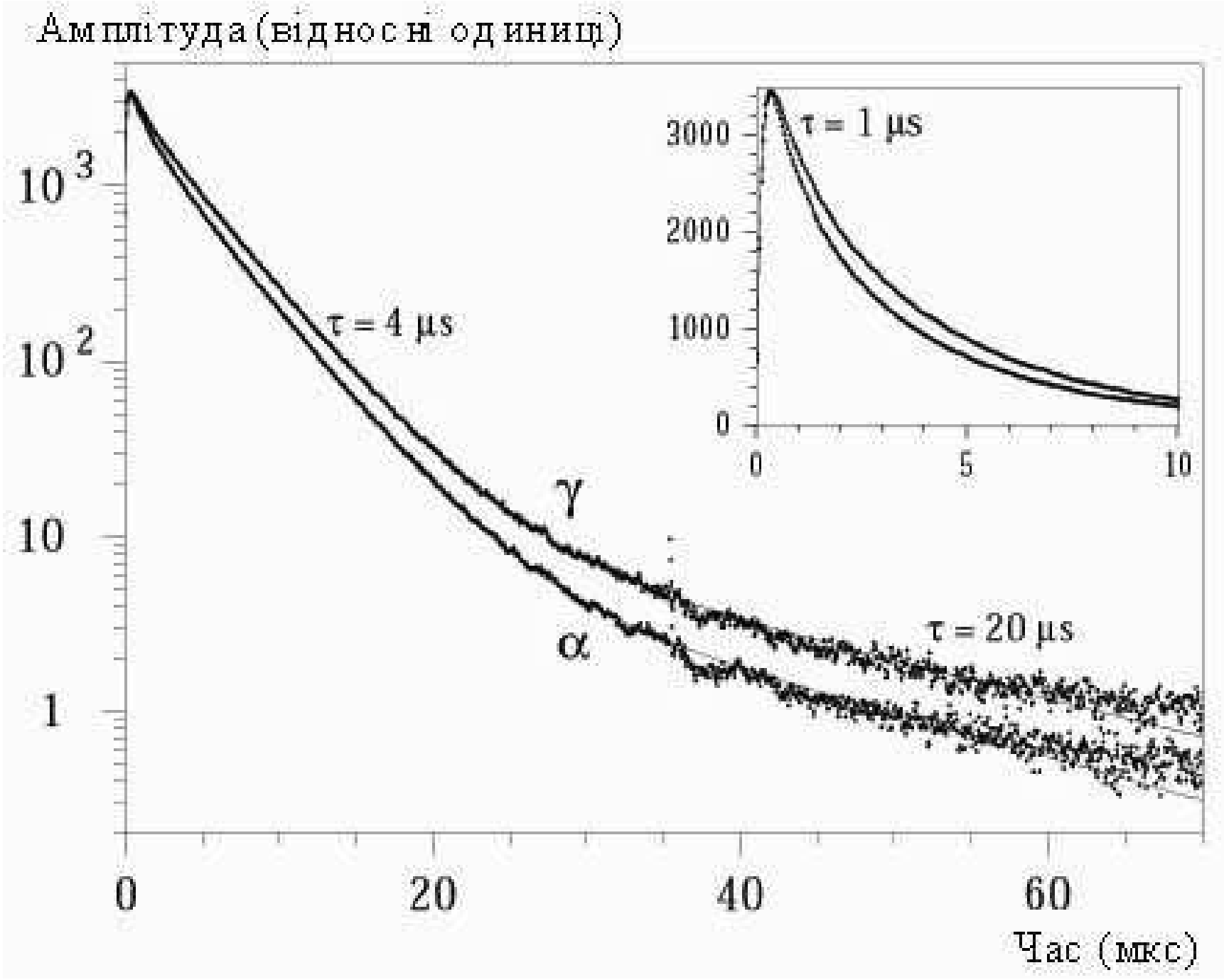

Рис. 6.11. Форма сцинтиляційних сигналів YAG:Nd при опроміненні α–частинками та γ–квантами.

Таблиця 6.6. Постійні ($\tau_i$, мкс) та інтенсивності ($A_i$) сцинтиляційного спалаху сцинтиляторів YAG:Nd.

| Вид опромінення | Постійні сцинтиляційного спалаху (мкс) та інтенсивності | | |
|---|---|---|---|
| | $\tau_1$ ($A_1$) | $\tau_1$ ($A_1$) | $\tau_3$ ($A_3$) |
| γ–кванти | 1.1 (10%) | 4.1 (87%) | 20 (3%) |
| α–частинки | 1.0 (15%) | 3.9 (83%) | 18 (2%) |

6.3.1.4. Дискримінація подій від α–частинок та γ–квантів за формою.

Різниця у формах сцинтиляційних сигналів дозволяє відділяти події від α–частинок та γ–квантів. Для цього був застосований метод оптимального цифрового фільтру (формули 3.2 та 3.3). Для отримання індикатора форми сумування проводилося по 1000 часових каналах оцифровщика, починаючи від 0 до 50 мкс. Ефективність розділення проілюстрована на рис. 6.12, де показані розподіли індикаторів форми для α–частинок з енергією ≈5.3 МеВ та γ–квантів з енергією ≈1.5 – 1.8 МеВ.

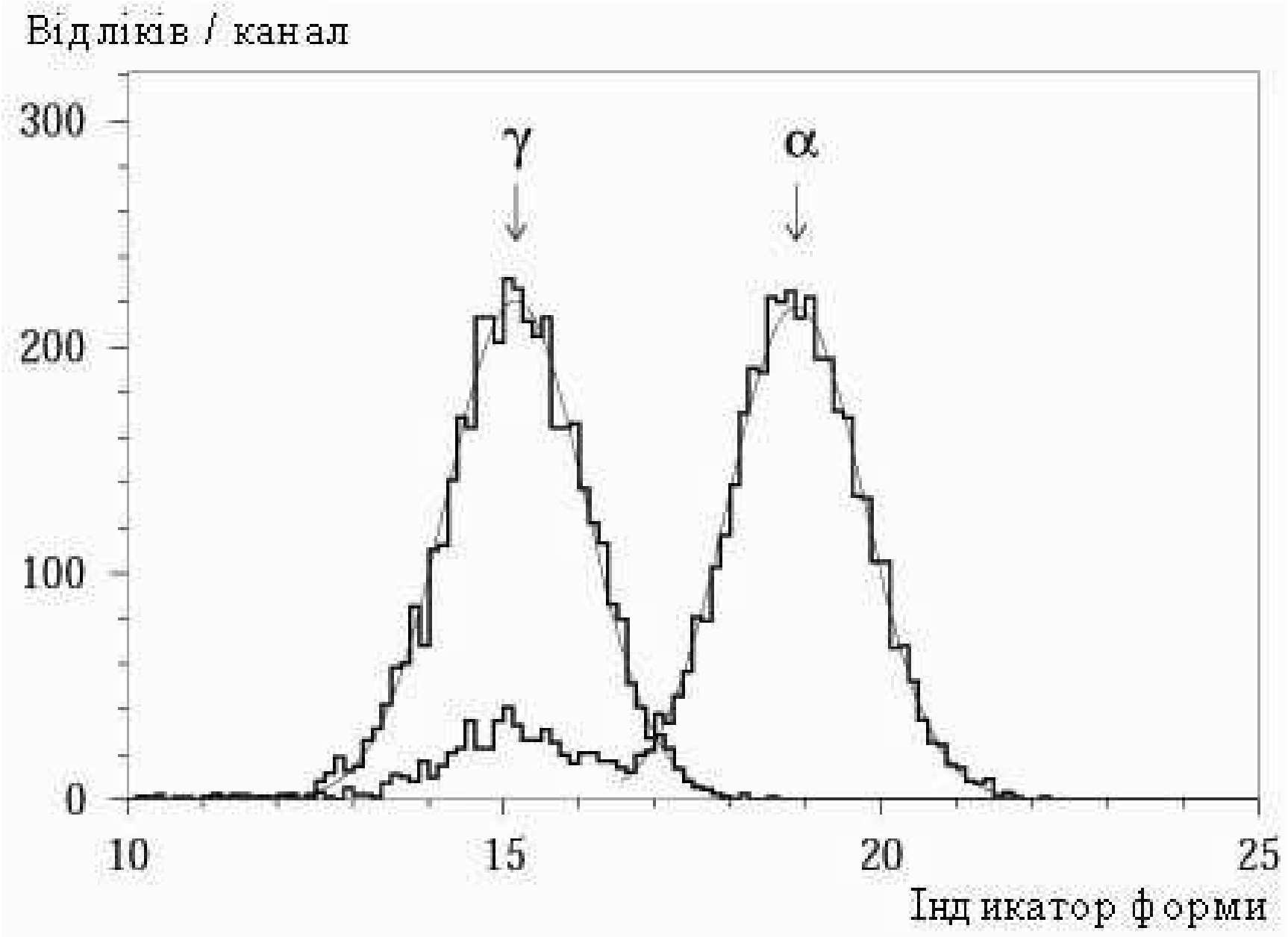

Рис. 6.12. Розподіл індикаторів форми для сцинтиляційних сигналів YAG:Nd при опроміненні α–частинками та γ–квантами.

### 6.3.2. Рівень радіочистоти кристалів YAG:Nd.

Оцінка активності радіоактивних домішок в кристалі YAG:Nd була зроблена за допомогою низькофонових вимірювань в надземній лабораторії в Києві. Кристал YAG:Nd проглядався низькофоновим ФЕП (EMI D724KFLB) через світловод з пластикового сцинтилятора ∅10 × 4 см. Сигнали від пластикового світловода, так само як і накладені сигнали YAG:Nd + пластик, відділялись за допомогою аналізу форми сигналів. Це дозволило придушити, зокрема, фон від космічних мюонів. Детектор був захищений шаром міді (2 см) та свинцю (10 см). У вимірюваннях протягом 14.8 годин не було спостережено яких–небудь особливостей, що могли бути викликані радіоактивними розпадами всередині кристалу, і дані цих вимірювань були використані для встановлення обмежень на активності радіонуклідів. Було встановлено, що сумарна $\alpha$–активність радіонуклідів урану і торію в кристалі YAG:Nd не перевищує 20 мБк/кг.

### 6.3.3. Оцінка чутливості експерименту по пошуку 2$\beta$–розпаду $^{150}$Nd з використанням сцинтиляторів YAG:Nd.

Серед ізотопів неодиму є три потенційно 2$\beta$–активних. Властивості цих ядер приведені в таблиці 6.7. Крім подвійного бета–розпаду $^{150}$Nd, детектор з кристалами YAG:Nd може бути застосований для пошуків подвійних $\beta$–процесів ще двох ізотопів: $^{146}$Nd та $^{148}$Nd. Як уже зазначалося, енергія 2$\beta$–переходу $^{150}$Nd є другою за своєю величиною серед усіх ядер – кандидатів на 2$\beta$–розпад.

Двонейтринна мода з періодом напіврозпаду $T_{1/2} = 1.9^{+0.7}_{-0.4}\times10^{19}$ років була спостережена в експерименті [177] з використанням часово–проекційних камер і зразків як збагаченого $^{150}$Nd неодиму, так і незбагаченого. В роботі [223] процес вивчався за допомогою схожої методики, але з використанням лише збагаченого ізотопу. Була виміряна 2$\nu$2$\beta$–активність ядра $^{150}$Nd з періодом напіврозпаду $T_{1/2} = (6.8\pm0.8)\times10^{18}$ років. Нещодавно було заявлено про спостереження 2β–розпаду $^{150}$Nd на другий ($0^+_1$) збуджений рівень $^{150}$Sm з періодом напіврозпаду $T_{1/2} = 1.4^{+0.5}_{-0.4}\times10^{20}$ років [ 312 ], в той час як в роботі [ 313 ] процес не був спостережений і було встановлене лише обмеження $T_{1/2} \geq 1.5\times10^{20}$ років. Найбільш жорстке обмеження на період напіврозпаду відносно безнейтринного процесу було встановлене в роботі [223]: $T_{1/2} \geq 1.2\times10^{21}$ років.

Таблиця 6.7. Властивості потенційно 2β–нестабільних ізотопів неодиму.

| Перехід | Енергія переходу, кеВ | Розповсюдженість материнського ізотопу, % | Канал розпаду |
|---|---|---|---|
| $^{146}$Nd → $^{146}$Sm | 70.2(2.9) | 17.2(0.3) | $2\beta^-$ |
| $^{148}$Nd → $^{148}$Sm | 1928.8(1.9) | 5.7(0.1) | $2\beta^-$ |
| $^{140}$Nd → $^{150}$Sm | 3367.5(2.2) | 5.6(0.2) | $2\beta^-$ |

Сцинтиляційні кристали YAG:Nd дозволяють здійснити експеримент по пошуку подвійного бета–розпаду $^{150}$Nd. Важливо, що із сцинтиляційним детектором YAG:Nd можна ідентифікувати α– та β–частинки (γ–кванти), а отже придушувати фон від внутрішніх забрудненостей в кристалі. Звичайно, низький вміст неодиму є суттєвим недоліком детектора YAG:Nd. В кристалі, що використовувався в наших дослідженнях, концентрація ізотопу неодиму–150 становить всього 0.03% від маси кристалу. Треба відмітити, що кристали YAG:Nd можуть бути вирощені із значно більшою концентрацією неодиму, аж до 8 мольних %, що відповідає концентрації $^{150}$Nd ≈ 0.1%. Це все ще досить низька концентрація досліджуваного ізотопу, але, наприклад, в проекті CANDLES [314] пропонується застосувати сцинтилятори $CaF_2$ для пошуку 2β–розпаду $^{48}$Ca. В той же час, концентрація цього ізотопу в кристалах фториду кальцію є такою ж: ≈ 0.1%.

Як показано в роботі [233], отримана оцінка енергетичної роздільної здатності, FWHM ≈ 4%, дозволяє планувати дослід по пошуку 0ν2β–розпаду $^{150}$Nd з можливістю *спостереження* процесу з періодом напіврозпаду $T_{1/2} \geq 1.2\times10^{21}$ років, що відповідає масі нейтрино на рівні ≈0.06 – 0.6 еВ (приймаючи до уваги всі існуючі розрахунки матричних елементів). *Чутливість* експерименту може бути значно вищою. Експеримент з використанням ≈20 тон незбагачених кристалів YAG:Nd з концентрацією 8 мольних % може досягти чутливості до 0ν2β–розпаду $^{150}$Nd на рівні $T_{1/2} \approx 3\times10^{26}$ років (припускаючи нульовий фон протягом 10 років експозиції), що відповідає масі нейтрино майоранівської природи ~ 0.01 – 0.1 еВ.

**Висновки розділу.** Досліджені сцинтиляційні властивості кристалів вольфрамату кальцію ($CaWO_4$). Отримана висока, порівняна із сцинтиляторами NaI(Tl), енергетична роздільна здатність. Виміряні α/β–співвідношення та форма сцинтиляційних сигналів для α–частинок та γ–квантів. Розроблено метод розділення подій від α–частинок та γ–квантів за формою імпульсів. Проведені вимірювання фону сцинтилятора $CaWO_4$ в Солотвинській підземній лабораторії, які дозволили визначити радіочистоту $CaWO_4$. Вперше досліджено сцинтиляційні властивості, відгук до α–частинок, форму сцинтиляційних сигналів та здатність до дискримінації частинок за формою з кристалом алюмо–ітрієвого гранату, активованого неодимом (YAG:Nd). Було оцінено радіоактивну забрудненість кристалу та запропоновано використати цей детектор для пошуку 2β–розпаду $^{150}$Nd.

Результати, описані в цьому розділі були, опубліковані в статтях:

# РОЗДІЛ 7

# МОЖЛИВОСТІ ПІДВИЩЕННЯ ЧУТЛИВОСТІ ЕКСПЕРИМЕНТІВ ПО ПОШУКУ 2β–РОЗПАДУ

## 7.1. Чутливість 2β–експериментів

### 7.1.1. Чи був зареєстрований безнейтринний 2β–розпад $^{76}$Ge?

В грудні 2001 р. група з Гейдельберга повідомила про спостереження 0ν2β–розпаду $^{76}$Ge [135]. З аналізу даних експерименту, що проводився в підземній лабораторії Гран Сассо колаборацією Heidleberg–Moscow, було отримано значення періоду напіврозпаду $^{76}$Ge $T_{1/2} = 1.5 \times 10^{25}$ років. Це повідомлення одразу викликало критику [201,108,267]. Фактично, ця дискусія продовжується і зараз. Тим більше, що група з Гейдельбергу опублікувала нові дані, набрані з дещо більшою експозицією і після більш ретельної енергетичної калібровки [136,137,203].

Очевидно, проблема полягає в дуже малій кількості відліків, що присутні в спектрі, зокрема, виміряному в експерименті Heidleberg–Moscow. Врешті–решт, це питання постає в усіх наднизькофонових експериментах, а саме: „як відрізнити ефект від фону в умовах дуже низької статистики?” У випадку експериментів, метою яких є пошук 2β–розпаду, ситуація ускладнюється ще й тим, що фон у виміряному спектрі може бути описаний з досить обмеженою точністю. Так, у випадку даних експерименту Heidleberg–Moscow площа піка 0ν2β–розпаду $^{76}$Ge залежить від інтерпретації фону в околі ефекту. Власне, дискусія розгорнулась навколо питання про коректність моделі фону в експерименті.

Самим простим, а разом з тим досить надійним шляхом оцінки фону в 2β–експериментах, є аналіз енергетичних діапазонів зліва і справа від очікуваного піку. При цьому, з одного боку, обидва діапазони мають бути якомога більшими, щоб забезпечити якнайменші статистичні флуктуації. Але з іншого боку, надмірне збільшення інтервалів оцінки фону призводить до того, що для його інтерпретації необхідно використовувати все більш складну модель. Найпростішою є ситуація, коли фон можна описати простою функцією, наприклад, постійною величиною.

На рис. 7.1 показано можливі моделі фонів отриманих в експерименті Heidleberg–Moscow. Один з них (верхній спектр) відповідає даним без застосування дискримінації за формою імпульсів, а інший (нижній) отримано в результаті досить складної, і почасти сумнівної процедури відкидання подій, що відбулись в декількох (принаймні двох) точках детектора за допомогою аналізу форми імпульсів. Таким чином події від багаторазового розсіяння γ–квантів можна відрізнити від подій 2β–розпаду, що мають відбуватися в одному місці в детекторі.

В обох випадках була вибрана ширина області ефекту 6 кеВ, оскільки енергетична роздільна здатність детектора становить 4 кеВ. Інтервал 6 кеВ містить 97% піка 0ν2β–розпаду $^{76}$Ge. В спектрі без застосування аналізу форми сигналів є дві особливості, які можна розглядати як вказівки на наявність піків з енергіями 2010 і 2054 кеВ. Щоб уникнути включення цих піків в модель фону, ми вибрали такі енергетичні інтервали для оцінки фону: справа від очікуваного піку від 2042 до 2051.5 кеВ ($SB_R$ = 9.5 кеВ) і зліва – від 2021.5 до 2036 кеВ ($SB_L$ = 14.5 кеВ). Відповідно сумарний енергетичний діапазон, в якому обраховувалась швидкість набору фону становив 24 кеВ. Повна кількість подій в цьому інтервалі становить 198, а отже, очікуваний фон в інтервалі очікуваного піку становить 49.5±3.5 відліків. В спектрі, отриманому після застосування дискримінації сигналів за формою $SB_R$ = 38 кеВ і $SB_L$ = 36 кеВ. В цих інтервалах міститься 126.7 відліків, що дає можливість оцінити очікуваний фон в області 0ν2β–розпаду: 10.3±0.9 відліків.

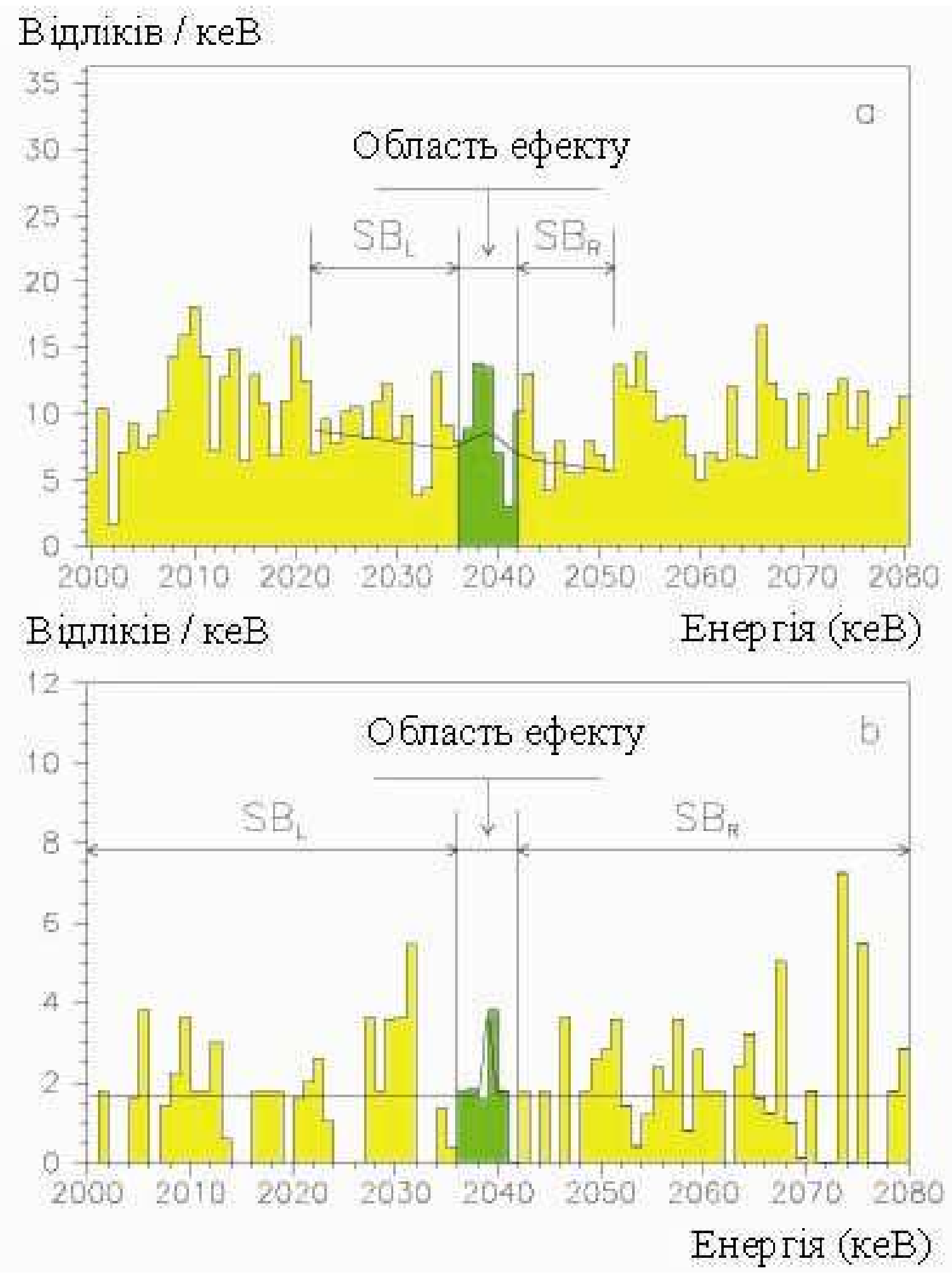

Рис. 7.1. Спектри фону, набрані в експерименті Heidleberg–Moscow за допомогою HPGe детекторів, збагачених $^{76}$Ge. (а) Сумарний спектр, що відповідає експозиції 54.98 кг×років; (б) сумарний спектр детекторів (експозиція 28.05 кг×років) після відбору одиночних подій за допомогою аналізу форми сигналів. Показано вибір інтервалів справа ($SB_R$) і зліва ($SB_L$) від очікуваного піку $0\nu2\beta$–розпаду $^{76}$Ge, які використовувались в нашій оцінці фону. Показані підгонки спектрів простою моделлю, яка включає шуканий пік від $0\nu2\beta$–розпаду $^{76}$Ge і фон.

Ефект $0\nu2\beta$–розпаду може бути розпізнаний, якщо кількість відліків в інтервалі очікуваного піку перевищує фон з достатнім рівнем статистики. Важливо відмітити, що за умови малої статистики лише статистичне перевищення кількості відліків в інтервалі очікуваного ефекту є єдиним критерієм наявності ефекту, оскільки за умови малої статистики особливості форми шуканого ефекту (у випадку обговорюваних даних – піка гаусової форми) не можуть бути використані [315].

Таким чином, єдиним шляхом з’ясувати, чи є в спектрі ефект безнейтринного подвійного бета–розпаду $^{76}$Ge, є застосування простих статистичних оцінок. В спектрі, отриманому без застосування аналізу сигналів за формою, в інтервалі від 2036 до 2042 кеВ (область ефекту) є 55.2±7.4 відліки. Порівнюючи це значення з отриманою вище оцінкою очікуваного фону (49.5±3.5 відліків), ми можемо бачити, що різниця між ефектом і фоном, а саме, 5.7±8.2 відліки, не дає підстав стверджувати, що в спектрі спостерігається пік 0ν2β–розпаду $^{76}$Ge.

Так само, аналіз спектра отриманого після застосування дискримінації подій за формою імпульсів дає 10.8 відліків в області ефекту. Враховуючи раніше оцінений рівень фону, отримаємо площу ефекту 0.5±3.4 відліки, що, знову таки, не дає підстав стверджувати наявність якого–небудь статистичного відхилення від фону.

Для повноти аналізу можна оцінити межу періоду напіврозпаду $^{76}$Ge відносно 0ν2β–розпаду за формулою 3.1. Спочатку розглянемо спектр без застосування аналізу сигналів за формою. Виходячи з отриманої оцінки площі ефекту (яка, повторюємо, не дає ніяких підстав стверджувати, що ефект спостерігається), отримаємо, що lim$S$ = 19.2 (13.9) відліків з довірчою імовірністю 90% (68%). Це дозволяє обмежити період напіврозпаду $^{76}$Ge відносно 0ν2β–розпаду на рівні lim$T_{1/2}$ = 1.2 (1.7) × $10^{25}$ років. Крім того, значення lim$S$ було оцінене за допомогою стандартної процедури методу найменших квадратів. Для цього експериментальний спектр, показаний на рис. 7.1, підганявся в інтервалі 2022 – 2051 кеВ сумою двох функцій: постійним фоном і піком очікуваного 0ν2β–розпаду. В результаті підгонки було одержане значення площі піка 4.8 ± 8.1 відліків, що відповідає величині lim$S$ = 18.1 (12.9) відліків з довірчою імовірністю 90% (68%).

Це значення близьке до отриманого перед цим за допомогою лише аналізу кількості відліків в спектрі.

Подібним чином було проаналізовано спектр детекторів після обробки сигналів за формою. Порівняння кількості відліків в області ефекту та інтервалах справа і зліва (рис. 7.1 (б)) дало значення площі очікуваного піку 0.5±3.4 відліки, що дає обмеження на площу ефекту lim$S$ = 6.1 (3.9) відліків і обмеження на період напіврозпаду lim$T_{1/2}$ = 2.0 (3.1) $\times 10^{25}$ років з довірчою імовірністю 90% (68%).

Таким чином, різні підходи дають близькі значення границі періоду напіврозпаду і не дають підстав стверджувати про наявність піка 0ν2β–розпаду $^{76}$Ge у виміряних спектрах. Важливо відмітити, що отримані нами обмеження на періоди напіврозпаду близькі до значень, що були свого часу опубліковані колаборацією Heidelberg–Moscow [195,116]. Але на відміну від роботи [135], в якій стверджується спостереження піку 0ν2β–розпаду $^{76}$Ge на рівні 2.2σ та 3.1σ , ми, виходячи з проробленого простого аналізу енергетичного спектру, не бачимо підстав стверджувати, що в даних експерименту присутній пік від безнейтринного подвійного бета–розпаду $^{76}$Ge. Єдиним поясненням „спостереження” піку в спектрах авторами роботи [135] є вибір ними занадто вузького енергетичного діапазону для оцінки наявності ефекту. Автори обрали інтервали справа і зліва від ефекту шириною лише ≈6 кеВ, що і привело до „появи” піку.

Оскільки є результати ще одного експерименту, в якому пошук 2β–розпаду ядра $^{76}$Ge здійснювався також за допомогою HPGe детекторів, збагачених ізотопом германію–76, ми можемо спробувати «збільшити» статистику даних, об’єднавши результати двох дуже схожих вимірювань. Дійсно, в експерименті IGEX [166,167] як енергетична роздільна здатність, так і досягнутий рівень фону дуже близькі до відповідних параметрів експерименту Heidelberg–Moscow. Спектри фону, отримані в ході експерименту IGEX, показані на рис. 7.2.

По–перше, нами був проведений аналіз спектру за допомогою простих статистичних оцінок. З аналізу спектру (рис. 7.2 (а)), отриманого після відбору подій за формою, виявилося, що період напіврозпаду $^{76}Ge$ може бути обмежений на рівні $\lim T_{1/2}$ = (1.1 – 1.6) × $10^{25}$ років, що узгоджується з оцінками авторів [167].

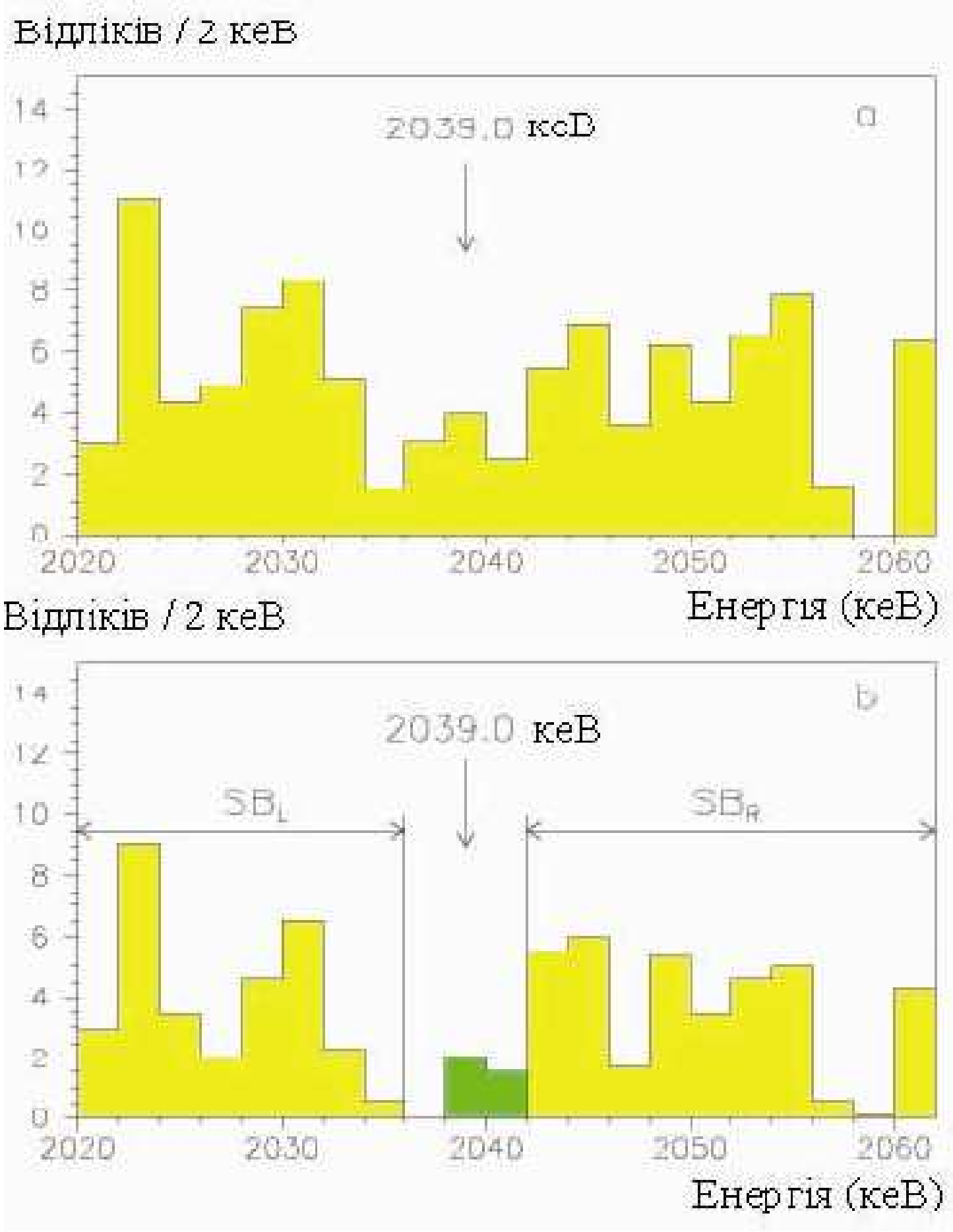

Рис. 7.2. Фонові спектри набрані в експерименті IGEX за допомогою збагачених ізотопом германію–76 напівпровідникових HPGe детекторів. Експозиція вимірювань становить 8.87 кг × років в перерахунку на $^{76}Ge$. (а) Сумарний спектр без застосування аналізу сигналів за формою; (б) спектр з використанням аналізу форми сигналів.

Нами були побудовані спектри отримані шляхом об'єднання даних двох експериментів, метою яких був пошук 0ν2β–розпаду $^{76}Ge$. Це дало змогу збільшити статистику даних, а отже підвищити чутливість досліду. Сумарні спектри показані на рис. 7.3.

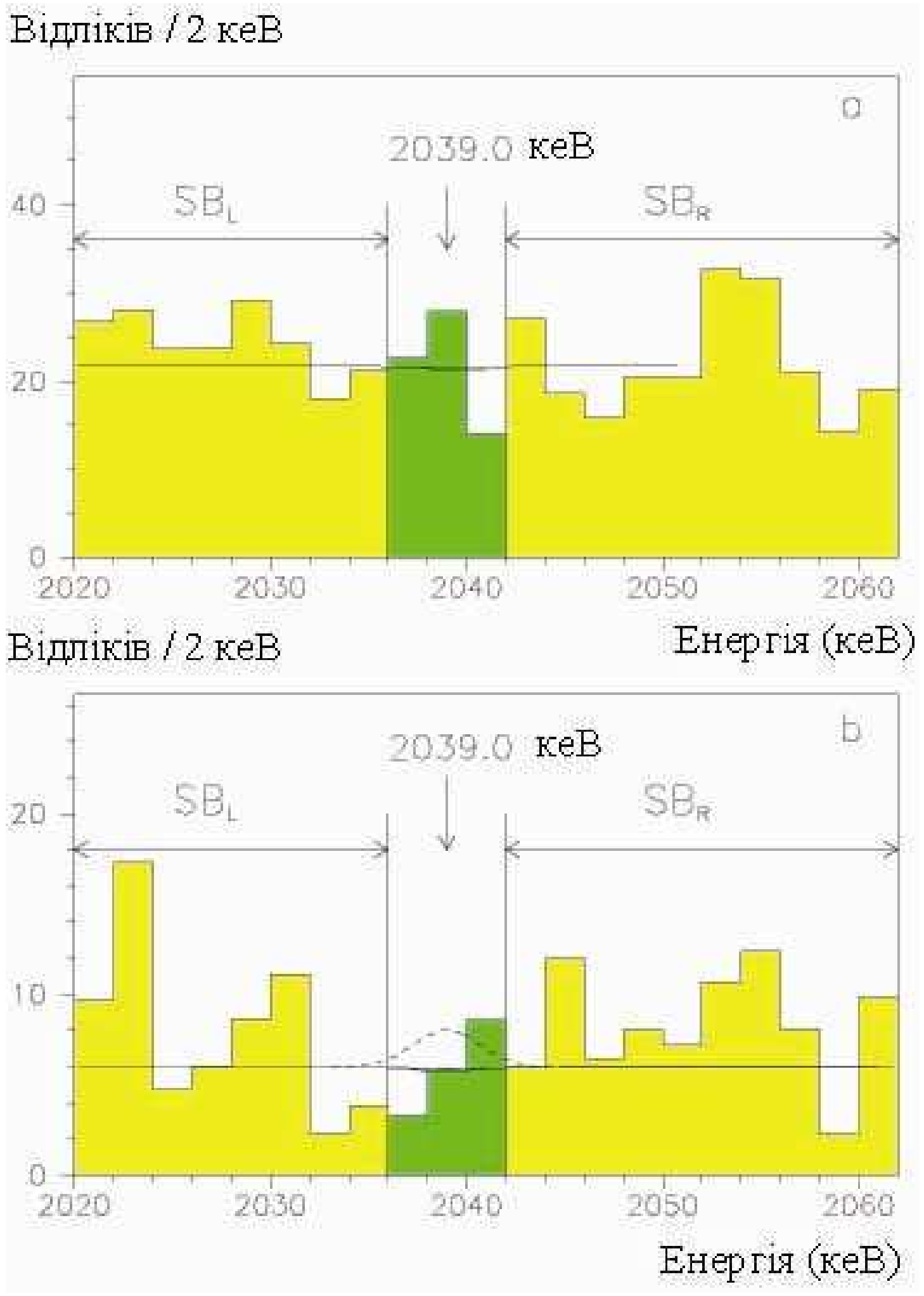

Рис. 7.3. Фонові спектри, отримані об'єднанням даних експериментів Heidelberg–Moscow та IGEX. Експозиція вимірювань становить 56.16 кг × років для спектру без відбору подій за формою (а) і 39.4 кг × років для спектру отриманого за допомогою дискримінації подій за формою (б). Експозиція розрахована для маси $^{76}$Ge. На рисунках показані криві, отримані в результаті підгонки фону. Пунктирна крива на нижньому рисунку відповідає піку від 0ν2β–розпаду $^{76}$Ge (період напіврозпаду $T_{1/2}^{0\nu} = 2.5 \times 10^{25}$ років), виключеному з довірчою імовірністю 90%.

Можна повторити для сумарних спектрів простий аналіз, що був застосований окремо для даних обох експериментів. В спектрі, показаному на рис. 7.3 (а) немає, окрім можливого піку з енергією ≈ 2054 кеВ і площею 44 ± 8 відліків, ніяких особливостей. Це дає змогу описати фон просто постійною, незалежною від енергії, величиною. В інтервалі ефекту (з 2036 до 2042 кеВ) присутні $64.6 \pm 8.0$ відліків. В інтервалах оцінки фону (величина взятого нами лівого інтервалу $SB_L$ = 16 кеВ, правого – $SB_R$ = 20 кеВ) присутні $417 \pm 20$ відліків, якщо вважати, що в спектрі немає піка з енергією 2054 кеВ. Припускаючи, що такий пік присутній, фон становить $373 \pm 22$ відліків. Враховуючи сумарну ширину інтервалів оцінки фону (36 кеВ), отримаємо оцінку фону в інтервалі очікуваного ефекту: $69.5 \pm 3.4$ відліків ($62.2 \pm 3.7$ відліків з урахуванням можливого піка 2054 кеВ). Різниця між швидкістю набору подій в інтервалі ефекту та фону становить $-4.9 \pm 8.7$ відліків ($2.4 \pm 8.8$). Так само підгонка даних за допомогою простої моделі, що складається з постійного фону та піка $0\nu2\beta$–розпаду, дає площу ефекту $-1.5 \pm 8.2$ відліків. Таким чином, використовуючи різні підходи, ми бачимо, що немає підстав стверджувати наявність ефекту $0\nu2\beta$–розпаду $^{76}Ge$ у сумарному спектрі. З отриманих значень площі ефекту та помилок її визначення можуть бути розраховані лише обмеження на період напіврозпаду: $T_{1/2}^{0\nu} = (2.0 - 2.3) \times 10^{25}$ років з довірчою імовірністю 90% і $T_{1/2}^{0\nu} = (3.2 - 3.8) \times 10^{25}$ років з довірчою імовірністю 68%.

Так само був проаналізований сумарний спектр, отриманий з даних, до яких була застосована дискримінація за формою сигналів. (рис. 7.3. (б)). В цьому спектрі в обох інтервалах оцінки фону (шириною $SB_L$ = 16 кеВ, лівий і $SB_R$ = 20 кеВ, правий) присутні 147 відліків. Таким чином, очікуваний в інтервалі оцінки ефекту (шириною 6 кеВ) фон становить $24.5 \pm 2$ відліки. В той же час, в цьому інтервалі є $17.7 \pm 4.2$ відліки, а отже можна дати таку оцінку різниці, тобто шуканого ефекту $\Delta = -6.8 \pm 4.7$ відліків. Процедура підгонки цього спектру (за модель фону було взято постійний фон і очікуваний пік $0\nu2\beta$–розпаду) методом найменших квадратів дала оцінку ефекту $-0.7 \pm 5.2$ відліки. Крива підгонки показана на рис. 7.3. (б).

Таким чином, і в сумарних даних двох експериментів по пошуку безнейтринного подвійного бета–розпаду ядра $^{76}Ge$, не дивлячись на збільшення статистики в 1.2 рази для даних без використання аналізу за формою сигналів і в 1.6 рази для даних, отриманих в результаті застосування аналізу форми, також не спостерігається пік з енергією ≈ 2039 кеВ, який можна приписати процесу $0\nu2\beta$–розпаду $^{76}$Ge. Отже, ці дані можна використати, щоб оцінити обмеження на період напіврозпаду з чутливістю кращою, ніж в обох, окремо взятих, експериментах.

Спробуємо оцінити обмеження на період напіврозпаду $^{76}$Ge відносно $0\nu2\beta$–розпаду. З отриманої нами різниці між кількістю відліків в інтервалах оцінки ефекту та фону ($\Delta = -6.8 \pm 4.7$) слідує величина ефекту, гіпотезу про наявність якого можна відкинути з заданою довірчою імовірністю: lim$S$ = 7.5 (4.7) відліків з 90% (68%) C.L. Це дає обмеження на період напіврозпаду $T_{1/2}^{0\nu} \geq 2.6$ (4.2) $\times 10^{25}$ років. Крім того, величина lim$S$ була оцінена за допомогою підгонки даних методом найменших квадратів. Підгонка дала площу ефекту $-0.7 \pm 5.2$ відліки, звідки слідує значення lim$S$ = 8.7 (5.2) відліків з 90% (68%) C.L. (тут ми консервативно відкинули від'ємну площу піка і взяли її рівною 0). Це відповідає обмеженню на період напіврозпаду ядра $^{76}$Ge відносно $0\nu2\beta$–розпаду:

$$T_{1/2}^{0\nu} \geq 2.5\ (4.2) \times 10^{25}\ \text{років} \quad \text{з } 90\%\ (68\%)\ \text{C.L.}$$

Як бачимо, обидва підходи дають досить близькі оцінки обмеження на ефект, що не спостерігається в сумарному спектрі, який був отриманий сумуванням даних двох експериментів після відкидання подій кількаразової взаємодії за формою імпульсів. Треба відмітити, що отримане нами обмеження є найкращим серед отриманих в прямих експериментах по пошуку подвійного бета–розпаду. Порівнюючи цей результат з теоретичними розрахунками [198], отримаємо таке обмеження на ефективну масу електронного нейтрино майоранівської природи та параметри домішок правих струмів в слабкій взаємодії (з довірчою імовірністю 90%):

$$\langle m_\nu \rangle \leq 0.36\ \text{еВ},\ \eta \leq 3.5 \times 10^{-9},\ \lambda \leq 5.5 \times 10^{-7}.$$

Нехтуючи внеском правих токів, отримаємо обмеження на масу нейтрино:

$$\langle m_\nu \rangle \leq 0.31\ (0.24)\ \text{еВ з довірчою імовірністю 90\% (68\%)}.$$

У відповідності з розрахунками [67] з отриманого обмеження на період напіврозпаду можна встановити також обмеження на параметр мінімальної суперсиметричної моделі з порушенням R–парності:

$$\varepsilon \leq 2.7 \times 10^{-4} \qquad \text{з 90\% C.L.}$$

Використовуючи розрахунки [316], отримаємо:

$$\varepsilon \leq 1.1 \times 10^{-4} \qquad \text{з 90\% C.L.}$$

Таким чином, статистичний аналіз даних двох найбільш чутливих експериментів по пошуку подвійного безнейтринного бета–розпаду $^{76}Ge$ дозволяє стверджувати, що повідомлення про спостереження цього процесу в експерименті Heidelberg–Moscow є принаймні передчасним.

Нещодавно група з Гейдельбергу опублікувала дані експерименту Heidelberg–Moscow після повторної обробки та з більшою на 30 % статистикою і продовжує наполягати на спостереженні $0\nu2\beta$–розпаду $^{76}Ge$ з $T_{1/2} \approx 1.2 \cdot 10^{25}$ років. З порівняння з теоретичними розрахунками слідує ефективна маса електронного нейтрино $m_\nu \sim 0.4$ еВ. Цей результат знову був підданий критиці [4]. Зрозуміло, що лише експеримент з більшою чутливістю може його спростувати або підтвердити. Можна очікувати, що в двох експериментах, метою яких є пошук $0\nu2\beta$–розпаду ядра $^{76}Ge$ на більш високому рівні чутливості [231,317], результат [135,166,167,203] може бути спростований або підтверджений. Ці експерименти, що зараз готуються, розглянуті нижче (див. 7.3).

### 7.1.2. Чутливість та здатність детектора зареєструвати $0\nu2\beta$–розпад.

Результати експериментів, в яких спостерігається явище осциляцій нейтрино вимагають значного підвищення чутливості експериментів по пошуку $0\nu2\beta$–розпаду ядер до рівня чутливості $T_{1/2} > 10^{26}$ років, що дозволить перевірити шкалу майоранівської маси нейтрино на рівні $m_\nu < 0.05$ еВ. Очевидною мірою має бути збільшення кількості досліджуваних потенційно 2β–активних ядер. Зараз обговорюються експерименти з масою 2β–активних ізотопів в десятки і сотні кілограмів, а отже вартість таких дослідів значно зростає. В цій ситуації дуже важливо оцінити чутливість експериментів, що плануються. При цьому важливо розрізняти, з однієї сторони, чутливість експерименту до *реєстрації* 0ν2β–розпаду, і, з іншого боку, граничну *чутливість* досліду, тобто здатність встановити обмеження на період напіврозпаду, коли ефект не спостерігається.

Вже зараз суттєвий внесок в фон детекторів в околі піку 0ν2β–розпаду дають події від двохнейтринної моди. Проблему фону від 2ν2β–моди розпаду в експерименті по пошуку 0ν2β–розпаду $^{116}$Cd проілюстровано на рис. 7.4. При заданих значеннях енергетичної роздільної здатності та періоду напіврозпаду відносно 2ν2β–розпаду, 2ν–події починають відігравати роль фону для спостереження піку 0ν2β–розпаду. При цьому, чим більшим буде період напіврозпаду відносно 0ν2β–розпаду, тим важче буде стверджувати, що пік спостерігається. У випадку (а), коли розподіл від 2ν–моди перетинається з піком від 0ν–розпаду на 1/10 висоти останнього, пік 0ν2β–розпаду (з $T_{1/2}^{0\nu} = 6.7\times10^{23}$ років) чітко відділяється від подій 2ν2β–моди. Якщо при $T_{1/2}^{0\nu} = 1.6\times10^{25}$ років (б) ще можна говорити про спостереження 0ν2β–розпаду, то у випадку (в) ($T_{1/2}^{0\nu} = 3.8\times10^{26}$ років) це буде практично неможливо.

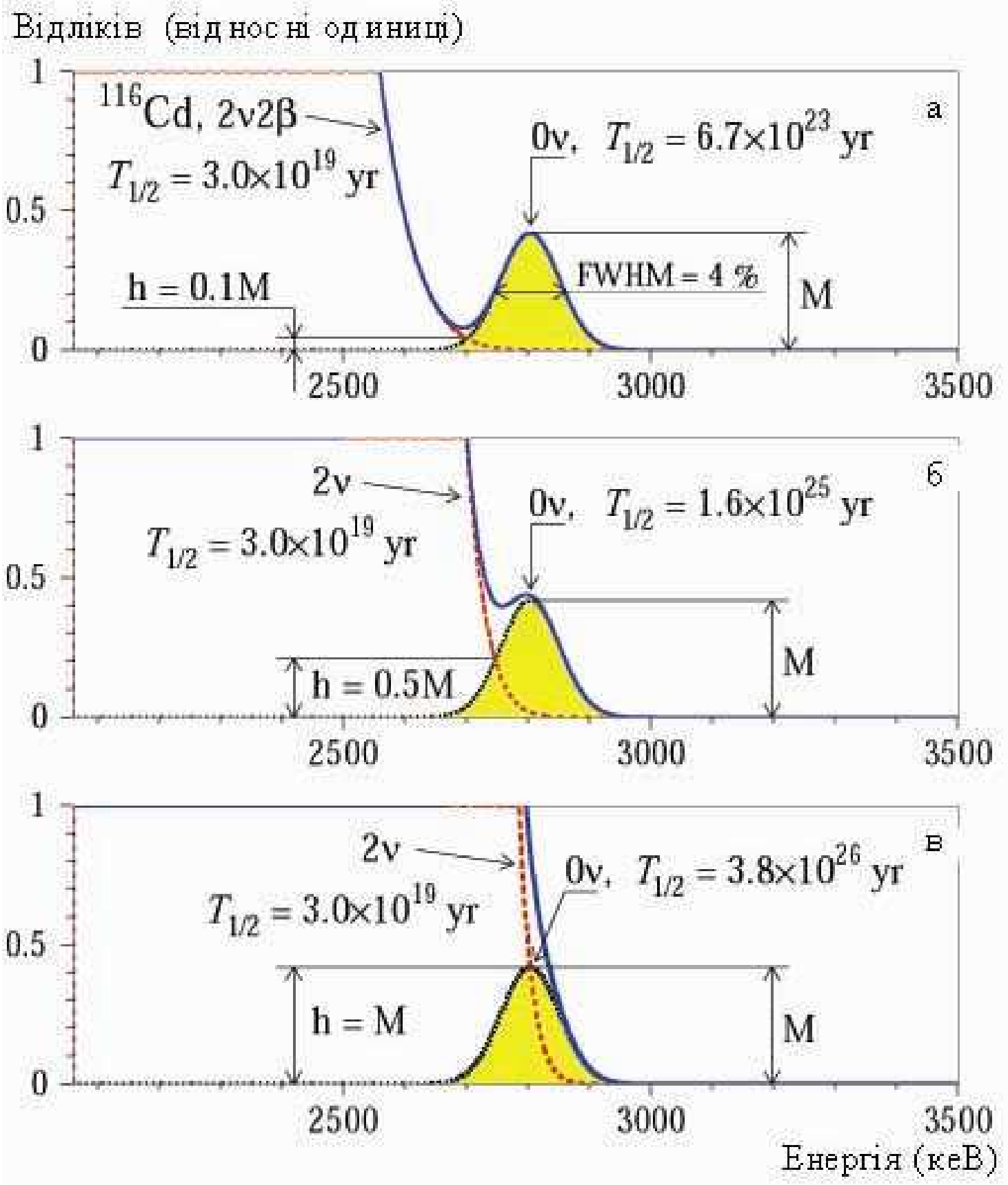

Рис. 7.4. Визначення потенціальної можливості експерименту зареєструвати 0ν2β–розпад. Розподіл двохнейтринної моди 2β–розпаду ядра $^{116}$Cd з $T_{1/2}^{2\nu} = 3\times10^{19}$ років починає відігравати роль фону в околі піка безнейтринного 2β–розпаду $^{116}$Cd. Якщо у випадку (а) ($T_{1/2}^{0\nu} = 6.7\times10^{23}$ років) пік 0ν2β–розпаду чітко відділяється від подій 2ν2β–моди, при $T_{1/2}^{0\nu} = 1.6\times10^{25}$ років (б) ще можна говорити про спостереження 0ν2β–розпаду, то у випадку (в) ($T_{1/2}^{0\nu} = 3.8\times10^{26}$ років) це буде практично неможливо.

Очевидно, що покращення енергетичної роздільної здатності детектора відіграє ключову роль для підвищення чутливості досліду до реєстрації ефекту 0ν2β–розпаду. На рис. 7.5 показано залежність чутливості експерименту до спостереження процесу 0ν2β–розпаду від роздільної здатності детектора. Видно, що для спостереження безнейтринного процесу з періодом напіврозпаду $T_{1/2}^{0\nu} \sim 10^{26}$ років, роздільна здатність має бути кращою, ніж кілька відсотків на енергії 0ν2β–розпаду.

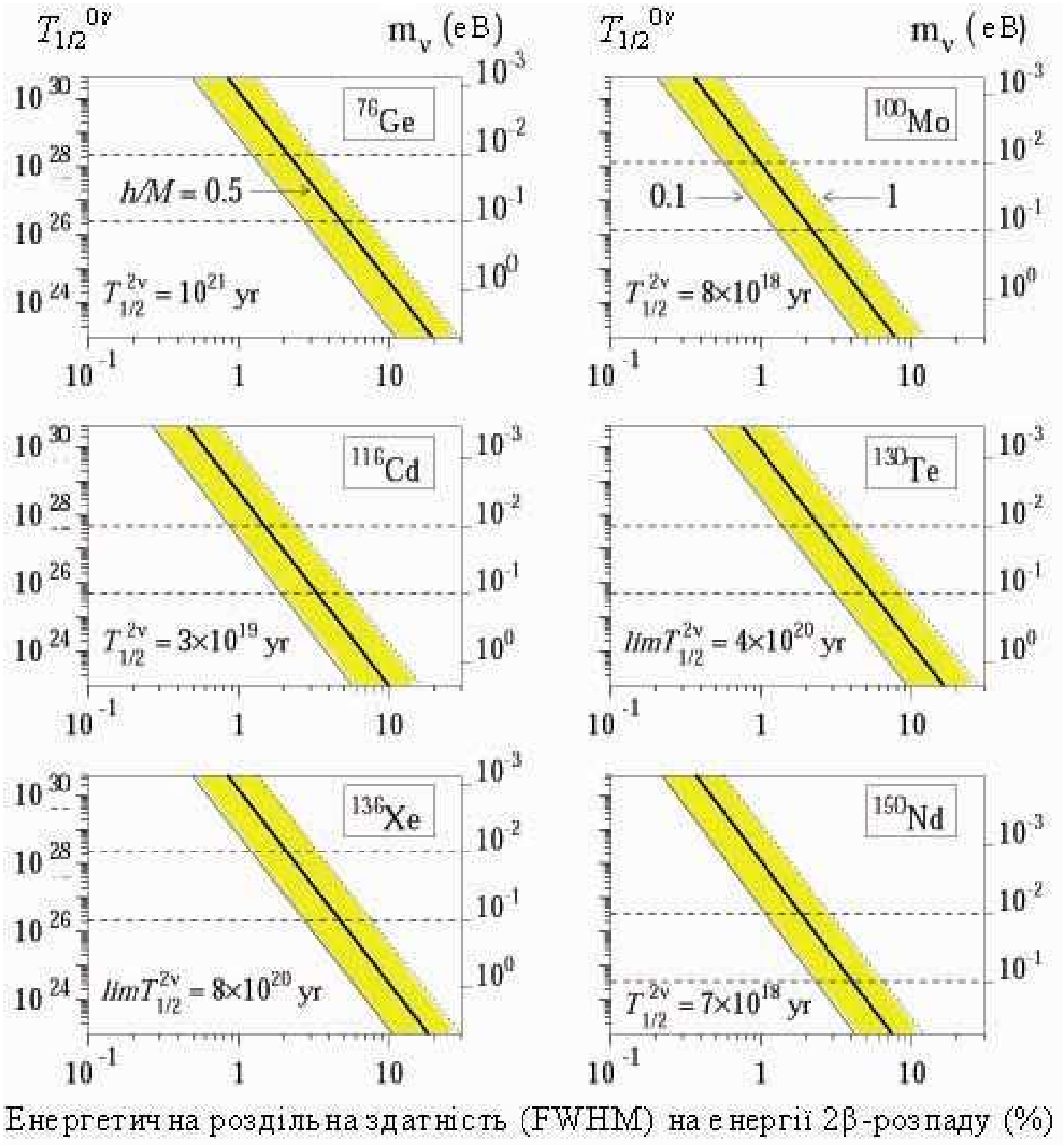

Рис. 7.5. Залежність чутливості до спостереження 0ν2β–розпаду від роздільної здатності детектора для різних ядер в припущенні, що фон спричинюється лише 2ν2β–розпадом досліджуваного ядра.

Здавалося б, при достатньо високій статистиці в піку 0ν2β–розпаду, можна, після віднімання розподілу для 0ν2β–розпаду від експериментальних даних, отримати пік 0ν2β–розпаду. Але треба пам‘ятати, що кількість відліків шуканого 0ν2β–розпаду практично не буде достатньо високою. На рис. 7.6. показано залежність необхідної експозиції в експерименті по пошуку 0ν2β–розпаду від періоду напіврозпаду ядра за умови, що в досліді буде зареєстровано лише 10 відліків. Таку кількість подій ефекту не можна назвати статистично високою, але навіть для спостереження 0ν2β–розпаду з періодом напіврозпаду $T_{1/2}^{0\nu} \sim 10^{26}$ років необхідно створити наднизькофоновий детектор з масою досліджуваного ізотопу більше ніж 100 кг.

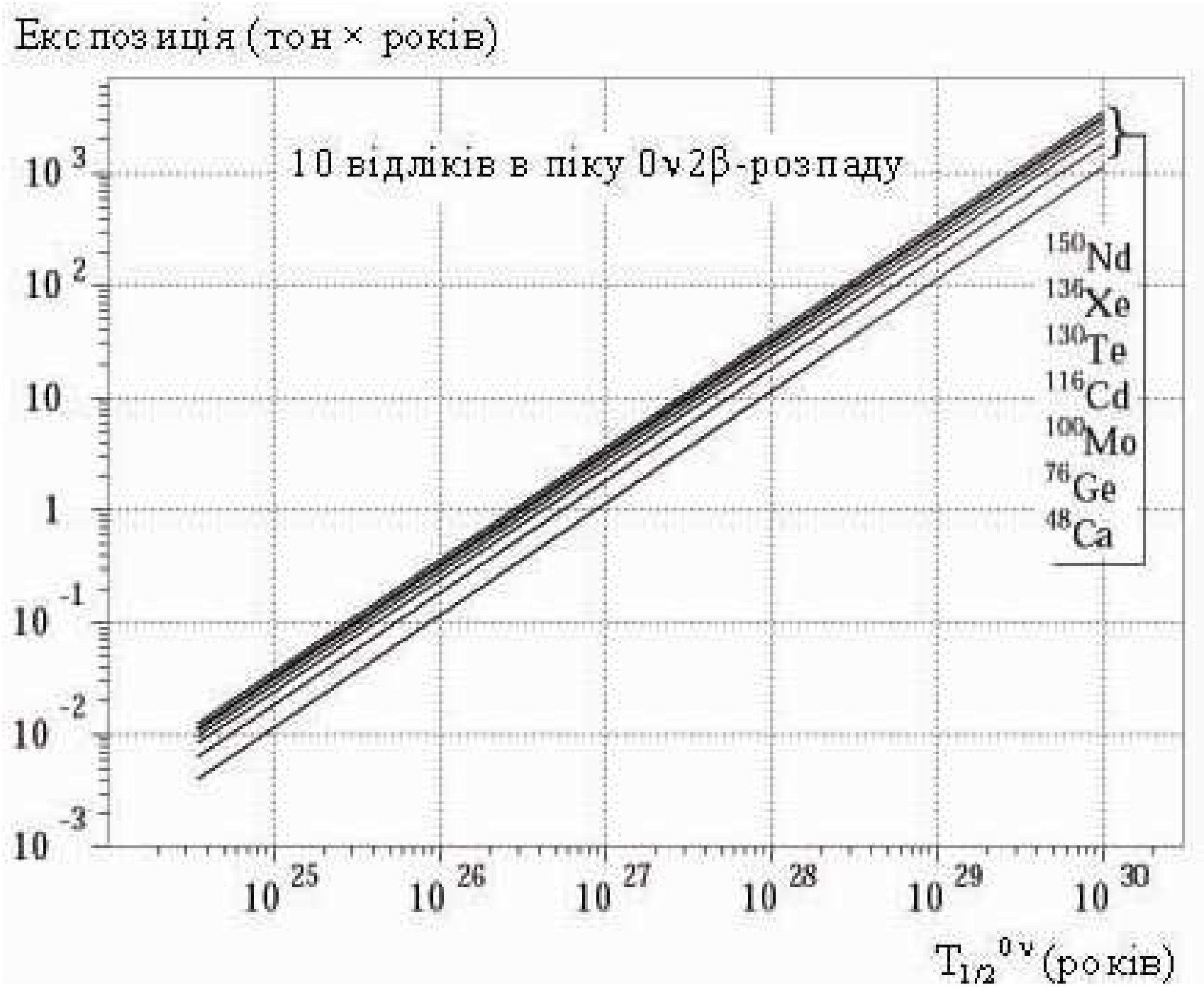

Рис. 7.6. Залежність експозиції, необхідної для спостереження 10 подій 0ν2β–розпаду (при 100% ефективності реєстрації), від періоду напіврозпаду різних ядер відносно 0ν2β–розпаду.

Щодо граничної чутливості досліду, то вона обмежена лише статистикою подій в околі 0ν2β–піку. Тому гранична чутливість 2β–досліду можу бути оцінена за формулою:

$$T_{1/2}^{0\nu} \sim \varepsilon \cdot \delta \cdot [(m \cdot t)/(R \cdot B)]^{1/2} \quad (7.1),$$

де $\varepsilon$ – ефективність реєстрації ефекту 0ν2β–розпаду, $\delta$ – збагачення досліджуваного ізотопу, $m$ та $R$ – маса та роздільна здатність детектора, $t$ – час вимірювань, $B$ – рівень фону детектора. В формулі 7.1 приймається, що гранична кількість подій ефекту, гіпотеза про присутність якого відкидається з 68% довірчою імовірністю, прирівнюється до кореня квадратного від кількості фонових подій в енергетичному інтервалі оцінки.

## 7.2. Проекти експериментів для пошуку 2β–розпаду

За останніх кілька років запропоновано більше як півтора десятка проектів для дослідження 0ν2β–розпаду різних ядер на рівні чутливості до періоду напіврозпаду більш як $T_{1/2}^{0\nu} \sim 10^{26}$ років. Тут ми коротко розглянемо ті, які, на наш погляд, є найбільш перспективними.

Проекти MAJORANA [231] та GERDA [317] націлені на пошуки 0ν2β–розпаду ядра $^{76}$Ge ($Q_{\beta\beta}$=2039 кеВ) за допомогою напівпровідникових германієвих детекторів. Проекти є логічним продовженням вже закінчених дослідів, здійснених колабораціями Heidelberg–Moscow та IGEX, котрі дали найкращі обмеження на масу нейтрино. Метою обох запланованих експериментів є також перевірка повідомлення про спостереження безнейтринного 2β–розпаду ядра $^{76}$Ge.

В проекті MAJORANA планується використати 210 напівпровідникових HPGe детекторів (збагачення $^{76}$Ge до $\approx 86\%$, маса одного кристала $\approx 2.4$ кг) загальною масою 0.5 тон, які планується помістити в наднизькофоновий кріостат (21 кристал в один кріостат) [231]. Для виготовлення кріостатів планується використовувати спеціальну технологію виробництва міді, яка має забезпечити принаймні на порядок величини більш чисту, по відношенню до радіоактивної забрудненості, мідь, ніж та, що була застосована в попередніх експериментах. Для запобігання космогенній активації детекторів та кріостатів, виробництво детекторів очікується здійснити в підземних умовах. Навколо детекторів буде пасивний захист з надчистого свинцю чи міді. Кожен кристал устатковується 6–ма азимутальними і 2–ма аксіальними контактами, що дозволить отримувати просторову інформацію про подію. Автори оцінюють, що така сегментація кристалів і аналіз експериментальних даних за формою імпульсів, дозволять знизити додатково рівень фону детекторів до ≈ 0.01 відліків/(кг × кеВ × рік) на енергії 2 МеВ, що в шість разів менше, ніж в сучасних найбільш чутливих експериментах із $^{76}$Ge. Автори вважають, що основний внесок в фон даватимуть космогенні довгоживучі ізотопи, зокрема $^{60}$Co. Після 10 років вимірювань чутливість цього проекту оцінюється на рівні $T_{1/2}^{0\nu} \approx 8\times10^{27}$ років, а отже чутливість до маси нейтрино на рівні: $m_\nu \leq 0.02$ еВ [318].

На першому етапі проекту GERDA пропонується використовувати збагачені $^{76}$Ge HPGe детектори, які раніше були застосовані в експериментах Heidelbrg–Moscow та IGEX. Очевидною перевагою проекту є якраз те, що ці детектори довгий час зберігалися під землею, а отже космогенні радіонукліди значною мірою вже розпалися. В подальшому планується збільшення маси детекторів до рівня кількох сотень кг. Детектори будуть розміщені у великому (4 м в діаметрі і висотою майже 9 м) мідному кріостаті, наповненому рідким азотом або аргоном, які одночасно будуть виконувати функції охолоджуючого середовища і пасивного захисту від зовнішніх γ–квантів. Таким чином автори проекту сподіваються знизити вплив навіть тієї дуже малої кількості радіонуклідів, що містяться в міді. Мідний кріостат, в свою чергу, буде занурений у бак з водою розмірами близько 10–12 метрів для захисту від γ–променів з оточуючого середовища. В цьому експериментів також планується застосовувати дискримінацію сигналів за формою. Цей проект значною мірою використовує ідеї робіт [319] та [320][1]. Фон в області очікуваного піку від 0ν2β–розпаду може бути знижений до ≈ 0.001 відліків/(кг × кеВ × рік) [317]. Для другої фази експерименту (35 кг $^{76}$Ge, 3 роки вимірювань) оцінюється чутливість на рівні $T_{1/2}^{0\nu} \approx 2 \times 10^{26}$ років, яка відповідає масі нейтрино: $m_{\nu} \approx 0.1$ еВ.

[1] Треба зауважити, наукова співдружність вже визнала, що ідея помістити германієвий детектор в рідкий азот належить проф. Герду Хойсеру з Інституту Макса Планка в Гейдельберзі (Німеччина).

В проекті CUORE (Cryogenic Underground Observatory for Rare Events) [220], який є розвитком діючого експерименту CUORICINO, пропонується використовувати близько тисячі охолоджених до ≈10 мК кристалів $TeO_2$ загальною масою 760 кг. Висока енергетична роздільна здатність кріогенних детекторів з кристалами $TeO_2$ (≈ 5 кеВ на енергії 2.5 МеВ) – потужний засіб для виділення сигналу 0ν2β–розпаду від фонових подій, зокрема для запобігання впливу 2ν2β–розпаду $^{130}$Te. Чутливість проекту “CUORE”, за умови зниження фону до рівня 0.1 – 0.01 відліків/(кг×кеВ×рік), оцінюється авторами на рівні $T_{1/2}^{0\nu} \sim (0.3 - 4)\times10^{26}$ років, що відповідає масі нейтрино $m_{\nu} \sim 0.1 - 0.04$ кеВ. Треба зауважити, що проблемою болометричних детекторів, і зокрема, детекторів з кристалами $TeO_2$, є порівняно високий рівень фону. По–перше, комплекс кріогенної техніки використовує велику кількість різних конструкційних матеріалів, що розташовані безпосередньо коло кристалів. Крім того, суттєвим є фон від α–частинок, що випромінюються з поверхні кристалів $TeO_2$, або з поверхні мідних деталей кріостата. Енергія таких подій, в зв’язку з частковим поглинанням α–частинок в тонких шарах матеріалів, може бути рівною енергії 2β–розпаду $^{130}$Te.

В проекті EXO (Enriched Xenon Observatory) пропонується використовувати близько тони збагаченого до 60 – 80% $^{136}$Xe в якості джерела і детектора одночасно у вигляді газу під тиском 5 – 10 атм для часово–проекційних камер. Автори розглядають також можливість використання рідкого ксенону в якості сцинтиляційного детектора. Одним з основних недоліків обох типів детекторів є недостатньо висока енергетична роздільна здатність (у випадку використання часово–проекційної камери енергетична роздільна здатність очікується на рівні 5% на енергії 2β–переходу $^{136}$Xe), а отже і значний вплив фонових 2ν2β–подій. Для зниження фону запропоновано також застосувати детектування іонів $^{136}Ba^{2+}$, що виникатимуть в процесі 2β–розпаду $^{136}$Xe. На жаль, ця досить складна методика не дозволить відділяти події 0ν2β–розпаду від подій 2ν–моди. В той же час, установка буде досить складною. Чутливість проекту EXO до маси нейтрино оцінена авторами на рівні ≈ 0.05 еВ [321]. Прототип детектора з ≈200 кг збагаченого $^{136}$Xe зараз споруджується, і можна очікувати, що цей експеримент вже в найближчі роки дасть цікаві результати, зокрема, може бути перевірена шкала маси нейтрино ≈0.1 еВ. Це дозволить перевірити повідомлення [135,136] про спостереження 0ν2β–розпаду в досліді з іншим ядром.

Проект Super–NEMO являє собою розвиток експерименту NEMO–3. В експерименті планується шукати 0ν2β–розпад $^{82}$Se. Близько ≈100 кг збагаченого ізотопу $^{82}$Se будуть розміщені у вигляді тонких фольг. Детектор вимірюватиме треки, енергію, час прольоту електронів. Період напіврозпаду $^{82}$Se порівняно великий: $T_{1/2}^{2\nu} = 9.5\times10^{19}$ років, а отже фон від 2ν2β–подій в околі піку 0ν2β–розпаду буде меншим. Але досягнення якнайкращої роздільної здатності залишається однією з найважливіших задач проекту.

Проект MOON має на меті вивчення 0ν2β–розпаду $^{100}$Mo ($Q_{\beta\beta}$=3034 кеВ) та дослідження в реальному часу сонячних нейтрино [322]. Детектор має складатися з 60 000 пластикових сцинтиляторів (6 × 0.2 × 0.25 см), світловий сигнал з яких буде збиратися через 866 000 світлопроводів зі зміщувачами спектру (∅1.2 мм × 6 м), на 6800 16–анодних ФЕП. Планується використання 34 тон натурального молібдену (тобто 3.3 тони $^{100}$Mo) на модуль у вигляді фольги товщиною ≈ 50 мг/см$^2$. Чутливість експерименту до маси нейтрино оцінюється авторами на рівні ≈ 0.05 еВ.

## 7.3. Проект експерименту по пошуку 0ν2β–розпаду $^{116}Cd$ (проект CAMEO)

Солотвинський експеримент по пошуку 2β–розпаду $^{116}Cd$ за допомогою сцинтиляційних кристалів $^{116}CdWO_4$ продемонстрував перспективність обраного методу. Справді, усього із ≈100 г збагаченого ізотопу було здійснено один з найбільш чутливих 2β–експериментів. Високий рівень чистоти кристалів вольфрамату кадмію відносно радіонуклідів дозволяє планувати проведення крупномасштабного експерименту з використанням цих сцинтиляторів. Можливості підвищення чутливості експерименту по пошуку 2β–розпаду ядра $^{116}Cd$ було продемонстровано в роботах [323,324]. Ідея полягає в тому, щоб розмістити сцинтиляційні кристали в надчистому рідкому сцинтиляторі і реєструвати світло віднесеними на певну відстань фотопомножувачами. Це дозволить практично повністю придушити зовнішній γ–фон, а також на кілька порядків зменшити вплив внутрішньої радіоактивності кристалів завдяки можливості визначати координату сцинтиляційних спалахів всередині кристалів.

В експерименті CAMEO для дослідження 2β–розпаду $^{116}Cd$ планується використання вже існуючої дослідницької установки BOREXINO Counting Test Facility (CTF), що розміщена в підземній лабораторії Гран Сассо. Експериментальна установка CTF описана в [237,325]. Установка розміщена в підземній лабораторії Гран Сассо і складається з зовнішнього бака (∅11×10 м), наповненого ≈1000 т води, яка є пасивним захистом для рідкого сцинтилятора об'ємом 4.8 м$^3$, розміщеного в сферичній нейлоновій оболонці ∅2.1 м. В установці використовується надчиста вода (рівень радіоактивної забрудненості ≈$10^{-14}$ г/г для U/Th, ≈$10^{-10}$ г/г для K). Активність радону у воді не перевищує 5 мкБк/літр. Рідкий сцинтилятор являє собою розчин 1.5 г/л PPO (дифенілоксазолу) у псевдокумолі. Максимум спектру емісії рідкого сцинтилятора знаходиться при 365 нм, світловихід становить близько $10^4$ фотонів на МеВ, прозорість досить висока:

постійна затухання перевищує 5 м для довжини хвилі більшої за 380 нм. Час затухання сцинтиляційного імпульсу складає 3.5 нс в малому об'ємі (4.5–5.0 нс у всьому об'ємі з джерелом, розміщеним в центрі CTF). Рідкий сцинтилятор має надзвичайно низький рівень забрудненості радіонуклідами уранових та торієвого рядів та калію. Це досягнуто завдяки високій початковій радіочистоті псевдокумолу, яка притаманна взагалі більшості органічних рідин, а також завдяки багатоступінчастій додатковій очистці сцинтилятора. Забруднення $^{232}$Th і $^{238}$U не перебільшує $(2–5)\times 10^{-16}$ г/г. Внутрішній резервуар для рідкого сцинтилятора виготовлений з прозорої нейлонової плівки, товщиною 500 мкм, яка дозволяє збирати сцинтиляційне світло за допомогою 100 фотоелектронних помножувачів (ФЕП), розміщених на допоміжній конструкції діаметром 7 м всередині водного баку. ФЕП Thorn EMI935 виготовлені з низькорадіоактивного скла Schott 8246, і характеризуються високою квантовою ефективністю (≈ 26% при 420 нм), гарними часовими характеристиками (постійна спаду сцинтиляційного спалаху ≈ 5 нс) та відмінним енергетичним розділенням для одноелектронних імпульсів (пік/долина ≈ 2.5). Для покращення світлозбору в експерименті CAMEO планується збільшити кількість фотопомножувачів в кілька разів. Установка CTF на сьогодні є однією з найкращих низькофонових установок великого об'єму. Насправді, загальна інтенсивність фону в області енергій 250 – 2500 кеВ (важливій для вивчення 2β–розпаду) становить близько 0.03 відліки/(рік × кеВ × кг), а в діапазоні 250 – 800 кеВ (так званому "енергетичному вікні сонячних нейтрино") ≈ 0.3 відліки/(рік × кеВ × кг) і визначається випромінюванням радону з води (≈ 5 мкБк/м$^3$).

7.3.1. Енергетична роздільна здатність.

Оцінку енергетичної роздільної здатності, яку можна досягнути з кристалами $CdWO_4$, можна оцінити за формулою 2.2. Незважаючи на те, що формула є наближеною, вона дає тим точніші значення роздільної здатності, чим більша енергія γ–квантів (електронів), які реєструються детектором. Для енергії 0ν2β–розпаду $^{116}Cd$ компонента $R_i$ буде визначальною, якщо детектор оптимізований з огляду на неоднорідність збору фотонів на фотокатод ФЕП. Отже, за формулою 2.2, щоб досягнути роздільної здатності 4%, потрібно забезпечити умову: $N_{phe} \geq 3500$. Кількість зібраних на першому диноді фотоелектронів $N_{phe}$ при поглинанні в сцинтиляторі $CdWO_4$ β–частинок або γ–квантів з енергією $E$ можна записати у вигляді добутку:

$$N_{phe} = N_{ph} \times E \times QE \times LC \tag{7.2}$$

де $N_{ph}$ – середня кількість утворених сцинтиляційних фотонів на один МеВ поглинутої енергії, $QE$ –квантова ефективність фотокатода ФЕП, $LC$ – коефіцієнт збирання фотонів на ФЕП.

Величина $QE$ обраховується як інтегральна згортка спектру висвічування кристала $CdWO_4$ та квантової чутливості ФЕП (див. рис. 7.7). Вона виявилася рівною 0.17 для найкращого з трьох ФЕП типу D724KFL з рубідій–цезієвим (RbCs) фотокатодом, які були виготовлені компанією THORN EMI спеціально для Солотвинського експерименту. Для ФЕП з мультилужним фотокатодом ми отримали $QE = 0.13$, що співпадає із значенням, яке використовувалося в роботі [326].

Для величини кількості фотонів $N_{phe}$, що випромінюються в сцинтиляторах вольфрамату кадмію, в літературі зустрічаються значення в діапазоні $N_{ph} = (15 - 28) \times 10^3$ фотонів / МеВ [224, 327 ]. З кристалом $CdWO_4$ розміром ∅40×30 мм, встановленим на ФЕП з мультилужним фотокатодом (XP2412, Philips), була отримана роздільна здатність FWHM = 6.7 % для γ–лінії 1.064 МеВ $^{207}Bi$. Коефіцієнт збирання світла в цій геометрії було розраховано методом Монте–Карло за допомогою пакета GEANT4. Було одержано, що на фотокатод

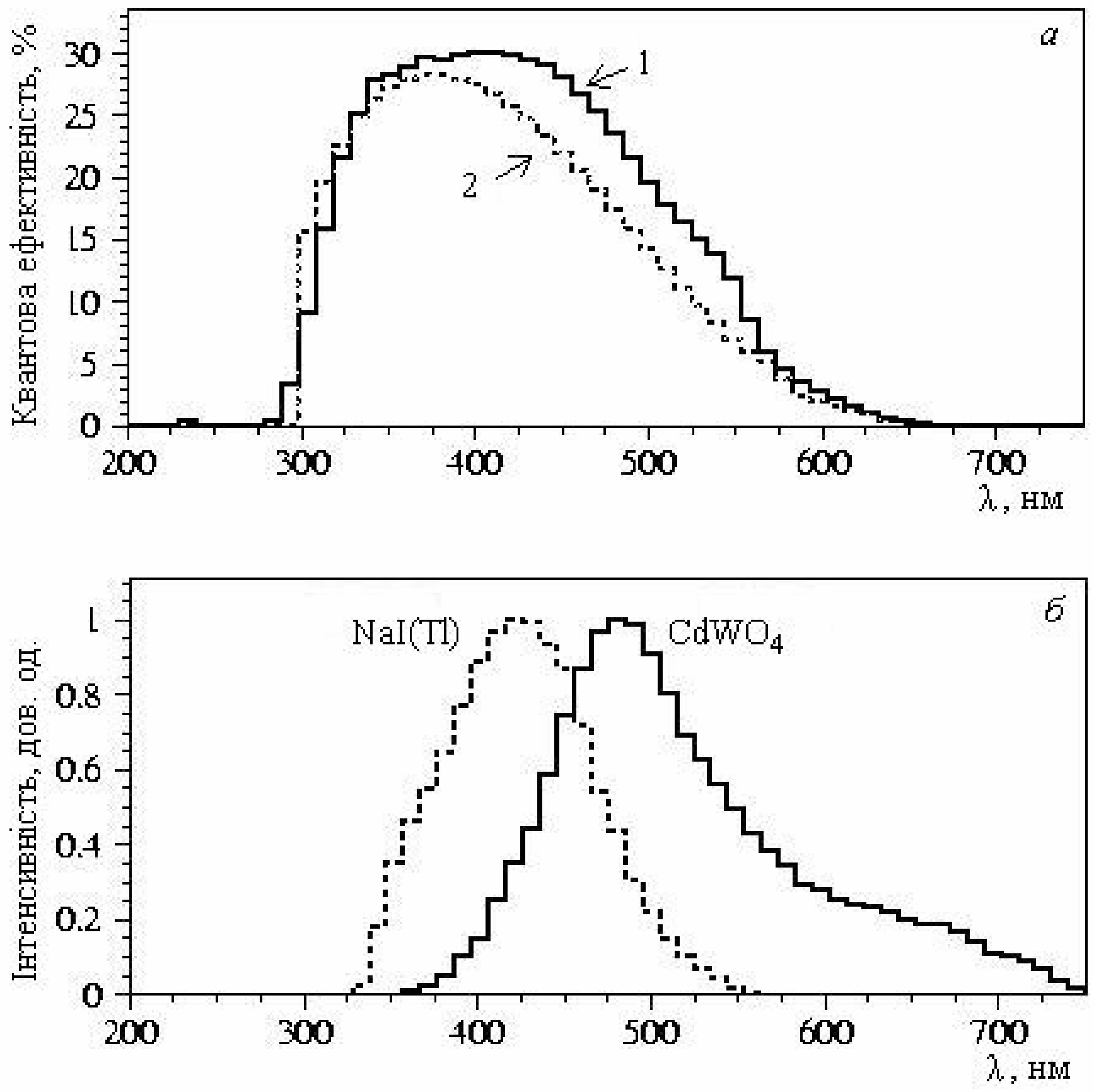

Рис. 7.7. Спектральна залежність квантової ефективності фотопомножувачів з різними фотокатодами (1 – RbCs, 2 – мультилужний) (а) та спектр висвічування сцинтиляторів NaI(Tl) [328] і $CdWO_4$ [329] (б).

ФЕП потрапляє 27 % від загальної кількості фотонів, що випромінювалися в об'ємі сцинтилятора. Таким чином, за формулами 2.2 і 7.2 ми оцінили, що в кристалі $CdWO_4$ утворюється не менше як $30 \times 10^3$ фотонів/МеВ.

Підставляючи е формулу 7.2 значення $\overline{N}_{phe}$, маємо, що для отримання *FWHM* = 4 % на енергії $Q_{2\beta}$ $^{116}$Cd світлозбір повинен становити не менше як 25 %, тобто 90 % порівняні зі стандартною конструкцією (кристал оптично приєднаний до ФЕП) за умови використання ФЕП з *QE* = 0.17. Якщо дещо послабити вимоги до якості ФЕП і припустити використання RbCs ФЕП з $QE \approx 0.15$ або ФЕП з мультилужним фотокатодом ($QE = 0.13$), одержимо $LC = (28 - 32)$ %. Тобто світлозбір у детекторі потрібно забезпечити таким же, як у випадку, коли кристал установлено безпосередньо на ФЕП, або навіть більшим у ≈ 1.2 рази. Крім того, оскільки розрахунки проведені для ідеального випадку, оптимізація детектора передбачає не лише можливість передачі необхідної кількості фотонів на фотокатод ФЕП, а також і компенсацію неоднорідностей світлозбору.

В експерименті CAMEO планується розмістити ≈ 100 кг збагачених сцинтиляційних кристалів $^{116}CdWO_4$ (40 кристалів ∅7×8 см кожен) на сфері діаметром 0.8 м всередині нейлонового контейнера з рідким сцинтилятором. Моделювання поширення світла методом Монте–Карло в розглянутій геометрії за допомогою пакета GEANT3.21 дало $\approx 4\times10^3$ фотоелектронів на 2.8 MeB енергетичних втрат. Таким чином, пік $0\nu2\beta$–розпаду $^{116}Cd$ може бути виміряний з енергетичною роздільною здатністю FWHM = 4%.

Важливим є те, що показник заломлення кристалів $CdWO_4$ значно більший за показник заломлення рідкого сцинтилятора, в якому поміщені кристали. В результаті кількість світла, що виходить з кристала в різних напрямках, залежить від координати сцинтиляційного спалаху. Це дозволяє визначати координату сцинтиляційного спалаху по кутовому розподілу світла, яке планується реєструвати кількома сотнями фотопомножувачів. Точність визначення положення точки випромінювання сцинтиляційних фотонів залежить від їх кількості (тобто від енергії поглинутої частинки) і від місця сцинтиляції в кристалі. Інформація про координату сцинтиляційного спалаху дозволить, по–перше, покращити енергетичну роздільну здатність спектрометра, а по–друге, відкидати фонові події різної природи: як події від зовнішніх γ–квантів, так і події розпадів радіонуклідів сімейств урану і торію. Останнє можливе завдяки визначенню положення в кристалі місця, де знаходяться радіоактивні ядра (за допомогою

часово–амплітудного аналізу та аналізу форми сигналів) і відслідковування подальших розпадів радіонуклідів, що можуть давати події в околі шуканого 0ν2β–піку $^{116}Cd$.

Принципова можливість отримання енергетичної роздільної здатності ≈4% на енергії 2.8 МеВ з кристалом $CdWO_4$, який поміщується в рідину, була продемонстрована у вимірюваннях. Циліндричний кристал $CdWO_4$ (40 мм в діаметрі та висотою 30 мм) закріплювався в центрі тефлонового контейнера з внутрішнім діаметром 70 мм. До нього з протилежних боків приєднувалися два ФЕП Philips XP2412. Таким чином, відстань від кожної поверхні кристала в напрямку ФЕП становила 30 мм, а проміжок між бічною поверхнею кристала і

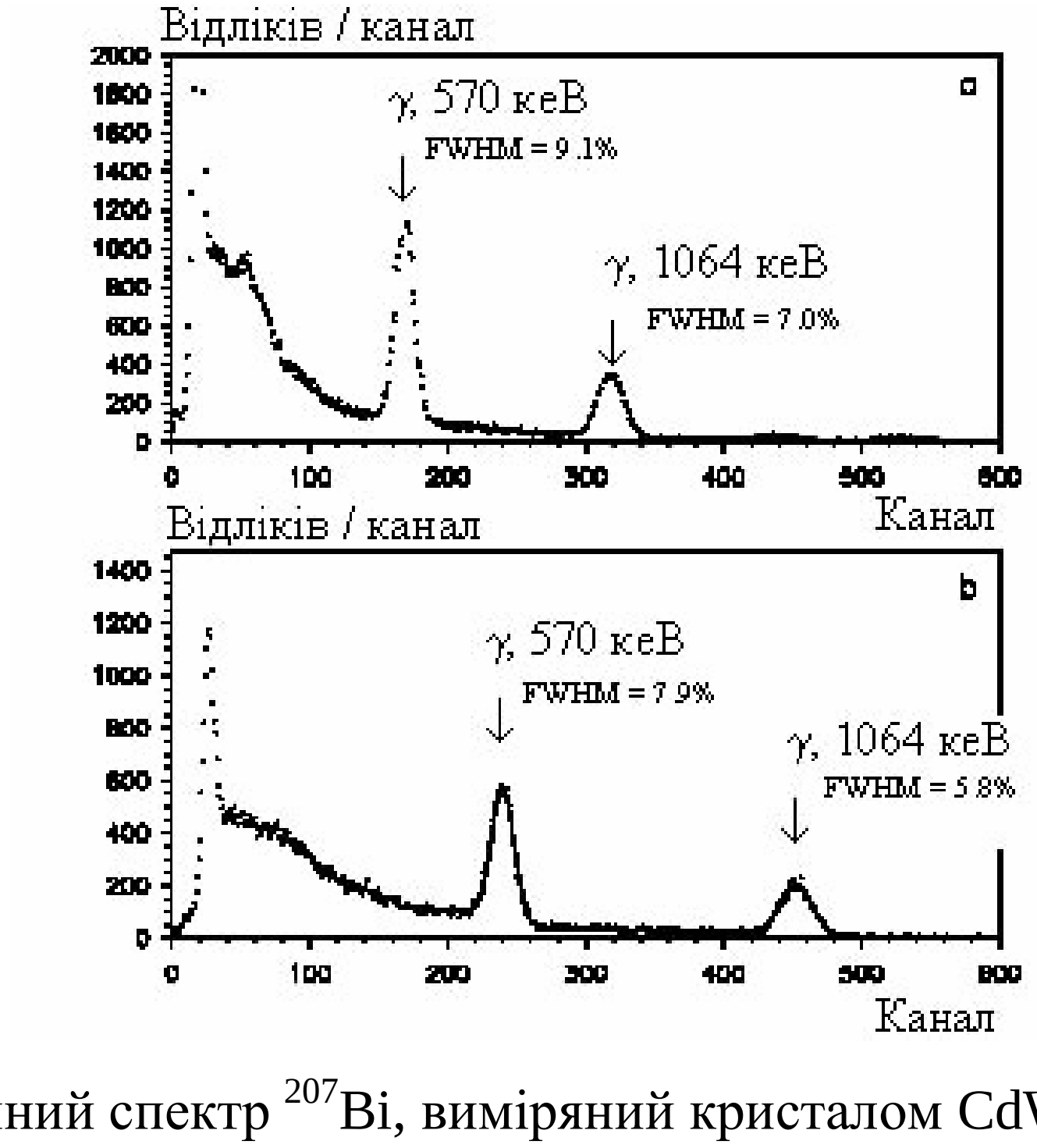

Рис. 7.8. Енергетичний спектр $^{207}Bi$, виміряний кристалом $CdWO_4$ (∅40x30 мм) для двох випадків: (а) стандартна, коли кристал $CdWO_4$, обгорнутий дифузним тефлоновим рефлектором, був оптично з'єднаний з фотокатодом ФЕП; (б) кристал $CdWO_4$ знаходився в рідині, і світло з нього збиралось двома ФЕП.

внутрішньою поверхнею контейнера дорівнював 15 мм. Контейнер заповнювався парафіновою олією (з показником заломлення ≈ 1.5). Два ФЕП працювали в режимі збігів. Результат вимірювань з джерелом $^{207}$Bi наведено на рис. 7.8, де для порівняння показаний також спектр, отриманий за стандартною методикою (кристал $CdWO_4$ обгорнутий тефлоновою плівкою і оптично приєднаний до ФЕП). Як видно з рис. 7.8, суттєве (≈42%) покращення світлозбору отримане з $CdWO_4$ в рідині. В результаті такого способу збору світла отримано також поліпшення енергетичної роздільної здатності у всьому енергетичному діапазоні 0.6–2.6 МеВ (див. рис.7.9, де зображені виміряні спектри $^{137}$Cs, $^{60}$Co та $^{232}$Th). Варто відмітити, що значення роздільних здатностей (7.4% при 662 кеВ; 5.8% при 1064 кеВ; 5.4% при 1173 кеВ та 4.3% при 2615 кеВ) подібні до значень, які можуть бути виміряні із сцинтиляторами NaI(Tl).

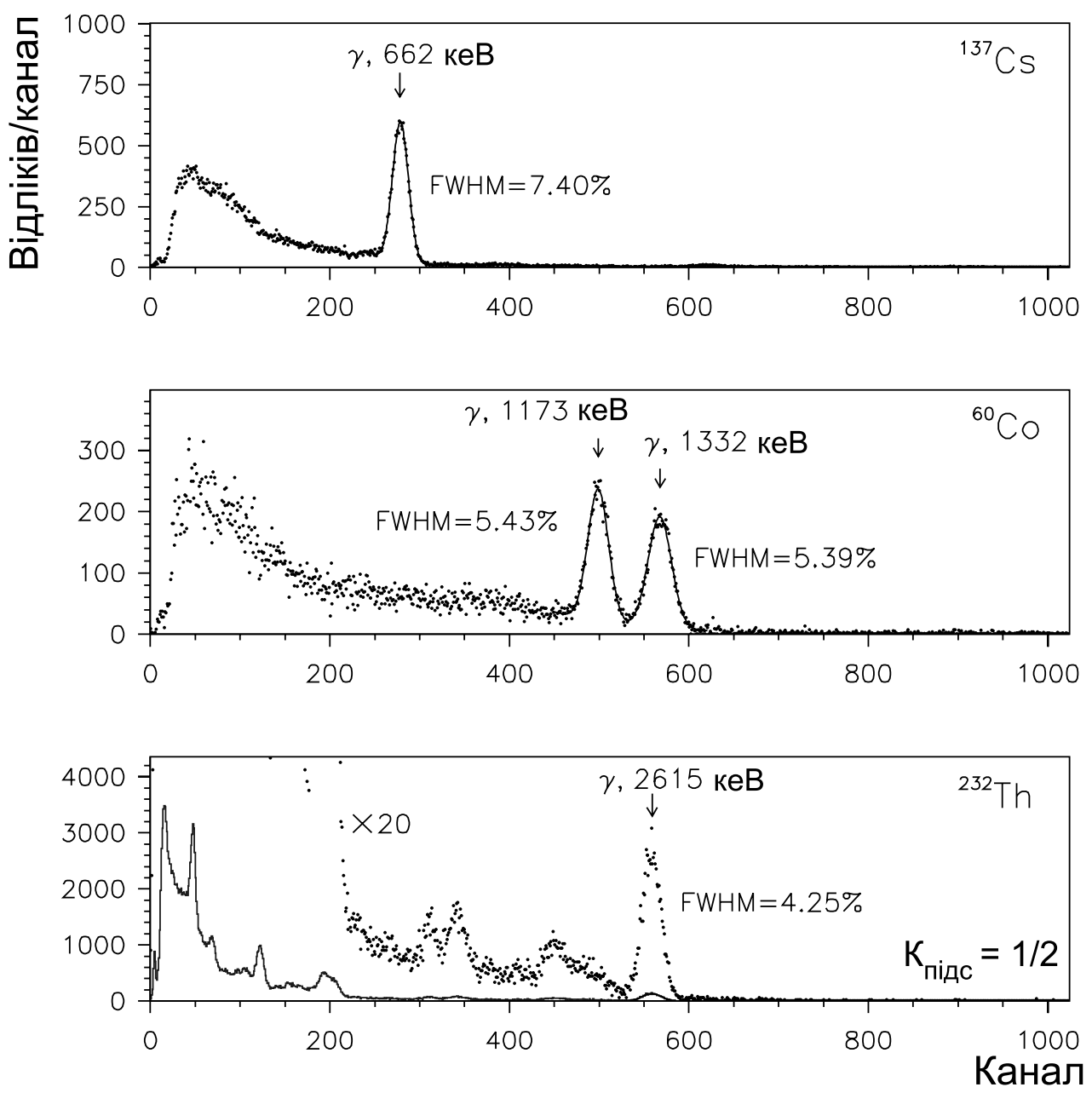

Рис. 7.9. Енергетичні спектри $^{137}$Cs, $^{60}$Co та $^{232}$Th, виміряні детектором з кристалом $CdWO_4$ (∅40x30 мм), розміщеним в рідині. Світло із сцинтилятора збиралося двома ФЕП.

### 7.3.2. Оцінки фонових умов експерименту та чутливість до безнейтринного 2β–розпаду $^{116}Cd$.

Фон детектора був промодельований О.А. Понкратенком методом Монте–Карло за допомогою пакета програм GEANT3.21 та DECAY4. Було показано, що навіть за умови використання кристалів з рівнем радіоактивних домішок на досягнутому зараз рівні, фон може бути знижений до $\approx 10^{-3} - 10^{-4}$ відліків/(рік × кеВ × кг). Важливою проблемою, яку треба буде подолати, є космогенна активація кристалів. Для зменшення впливу радіонуклідів космогенного походження кристали необхідно якнайшвидше опустити під землю і витримати там близько року. Чутливість експерименту оцінена на рівні $T_{1/2}^{0\nu} \geq 10^{26}$ років, що відповідає верхній межі для маси майоранівського нейтрино $m_\nu \leq 0.06$ еВ. Подальше підвищення чутливості до $m_\nu \sim 0.02$ еВ можливе завдяки збільшенню маси кристалів $^{116}CdWO_4$ до ≈ 1 тони і розміщення їх в одному з детекторів, таких як Borexino, KamLAND або SNO.

### 7.3.3. Можливості використання кристалів $CdWO_4$ в якості кріогенного детектора для високочутливого експерименту по пошуку 0ν2β–розпаду $^{116}Cd$.

Як уже зазначалося, енергетична роздільна здатність детектора має надзвичайно велике значення для постановки високочутливого експерименту по пошуку 0ν2β–розпаду. В той час як сцинтиляційний детектор з кристалом вольфрамату кадмію має роздільну здатність ≈ 4% на енергії 2β–переходу $^{116}Cd$, кріогенні детектори зараз здатні забезпечити роздільну здатність на рівні ≈ 0.2%.

У співробітництві з групою Міланського університету та університету Комо (Італія) була перевірено можливість використання кристалу вольфрамату кадмію в якості болометричного детектора.

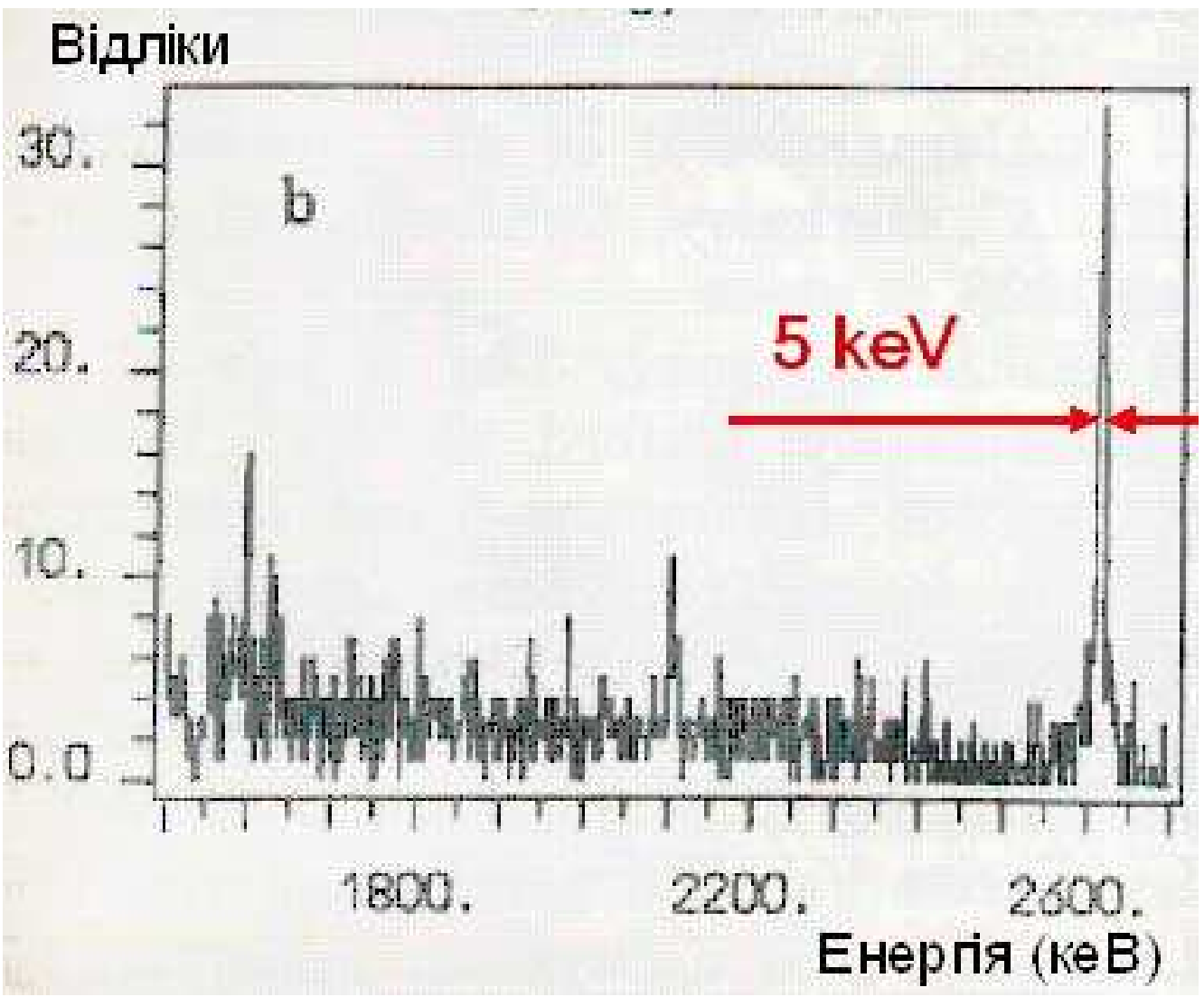

Рис. 7.10. Спектр γ–квантів $^{232}$Th виміряний з кристалом $CdWO_4$, охолодженим до 20 мК. Енергетична роздільна здатність по γ–лінії 2615 кеВ становить 5 кеВ.

Кристал $CdWO_4$ циліндричної форми ∅25×15 мм масою 58 г був охолоджений до температури ≈20 мК в спеціальній кріогенній камері, яка була розміщена в підземній лабораторії Гран Сассо. Для зменшення фону від зовнішніх γ–квантів, деталі кріостату виготовлені з безкислородної міді. Шар мідного захисту становив 10 см. Всі компоненти установки були відібрані за рівнем радіоактивних домішок. Кріостат оточений шаром міді 5 см та 10 см низькорадіоактивного свинцю. Енергетична роздільна здатність детектора була перевірена з γ–джерелами $^{60}$Co та $^{232}$Th і становила ≈ 5 кеВ в широкому енергетичному інтервалі після корекції нестабільності детектора. Високоенергетична частина спектру, виміряного з γ–джерелом $^{232}$Th, показана на рис. 7.10.

Вимірювання фону були проведені впродовж 340 годин. Фон детектора в енергетичному інтервалі, де очікується пік безнейтринного 2β–розпаду $^{116}$Cd, становить ≈ 9 відліків/(кеВ × рік × кг) [ 330 ]. Цей рівень фону, отриманий в попередньому досліді, є значно вищим, аніж той, що досягнутий в Солотвинській підземній лабораторії з кристалами $^{116}CdWO_4$ в якості сцинтиляційного детектора. Треба також зауважити, що рівень фону, який досягнуто в експерименті по пошуку 2β–розпаду $^{130}Te$, де використовуються кристали оксиду телуру, також поступається рівню фону солотвинського експерименту. Як уже зазначалося в розділі 1, це пояснюється реєстрацією частини енергії альфа–частинок, що потрапляють в чутливий об’єм детектора з поверхні кристалів або оточуючих матеріалів. Важливою перевагою кристалів $CdWO_4$ є можливість реєстрації не лише температурного, а й світлового сигналу, що дозволить ефективно відкидати події від α–частинок і таким чином зменшити фон у порівнянні з досягнутим в експерименті [216,217,219,243]. Можливість такої дискримінації подій в кристалах $CaF_2$, $CaWO_4$, BGO була продемонстрована в роботах [331,332,333].

## 7.4. Дослідження можливості застосування сцинтиляційних кристалів $PbWO_4$ в експерименті по пошуку 2β–розпаду $^{116}Cd$

Однією з найважливіших характеристик планованих зараз експериментів, метою яких є пошук 0ν2β–розпаду ядер з чутливістю до маси нейтрино $m_\nu \sim 0.1 – 0.05$ еВ (а отже чутливість до періоду напіврозпаду має бути на рівні $T_{1/2}^{0\nu} \sim 10^{26}$ і більше років), є значне, у порівнянні з досягнутим, зниження рівня фону детектора. Справді, якщо в кращих сучасних експериментах рівень фону в області піка 0ν2β–розпаду (як, наприклад, в експерименті по пошуку 2β–розпаду $^{116}Cd$ за допомогою сцинтиляторів $CdWO_4$, що проводився в Солотвинській підземній лабораторії) не перевищує сотих відліку за рік вимірювань в енергетичному інтервалі 1 кеВ в кілограмі детектора, реалізація сучасних проектів вимагає, щоб швидкість надходження фонових подій не перевищувала такої величини, але вже на тону маси детектора. Зниження фону на ≈ 3 порядки величини вимагає суттєвого розвитку технологій отримання радіочистих матеріалів, нових підходів до конструкції детектора та аналізу даних. В більшості проектів планується значно збільшити розміри установки, що пов'язано з використанням для пасивного захисту рідин, для яких можна досягнути розумного співвідношення між радіочистотою і вартістю. Адже маса захисту в запланованих проектах сягає кілька десятків і навіть сотень тон. Так в проектах GERDA [317], GENIUS [319], CAMEO [324], GEM [320] запропоновано використовувати надчисту воду, органічну рідину, рідкий азот. В результаті розміри цих установок сягають кількох метрів. В той же час, площі і об'єми низькофонових лабораторій, як правило, обмежені. Крім того, привабливою є така конструкція детектора, коли він оточений ефективним до $\gamma$–квантів детектором активного захисту. Виходячи з цих міркувань, нами були проведення дослідження можливості використання сцинтиляційних кристалів вольфрамату свинцю ($PbWO_4$) в якості, по–перше, світловодів, а по–друге, активного захисту для кристалів вольфрамату кадмію.

7.4.1. Сцинтиляційні властивості кристалів $PbWO_4$.

Властивості сцинтиляторів $PbWO_4$ приведені в таблиці 7.1. Цей сцинтиляційний матеріал набув протягом останнього десятиріччя широкого використання у фізиці високих енергій завдяки таким якостям як високі густина і ефективний атомний номер, дуже малий час світіння, висока радіаційна стійкість. Сцинтиляційні властивості $PbWO_4$ вивчались багатьма групами авторів [334,335,336,337,338,339] і ця робота продовжується (див. огляд [340] і список літератури в ньому). Крім застосування в фізиці високих енергій, сцинтилятори $PbWO_4$ виглядають перспективним матеріалом для використання в багатьох областях як фундаментальних, так і прикладних досліджень, а також в різних галузях техніки і медицині. Виробництво вольфрамату свинцю добре розвинуте і сягає кількох тон на рік (див., наприклад [341]).

Таблиця 7.1. Властивості сцинтиляційних кристалів $PbWO_4$.

| Характеристика | |
|---|---|
| Густина (г/см$^3$) | 8.28 |
| Температура плавлення (°C) | 1123 |
| Структурний тип | Шеєліт |
| Твердість за Моосом | 3 |
| Показник заломлення | 2.2 |
| Ефективний час світіння (мкс) | 0.01 |

В наших вимірюваннях були використані два кристали $PbWO_4$, вирощені методом Чохральського. Кристал розмірами $45 \times 22 \times 22$ мм (далі цей кристал будемо називати (PWO–1) був вироблений в Інституті сцинтиляційних матеріалів НАНУ (Харків), другий кристал розмірами $32 \times 32 \times 10$ мм (PWO–2) – на Богородицькому технологічному хімічному комбінаті (Богородицьк, Росія).

Енергетична роздільна здатність була виміряна з кристалом PWO–1, який був обгорнутий тефлоновою плівкою PTFE і оптично з’єднаний з ФЕП XP2412 (Philips) за допомогою оптичного з’єднувача Q2–3067 (Dow Corning), при температурах від кімнатної (+24°C) до –18°C. Кристал опромінювався $\gamma$–квантами від джерел $^{137}$Cs та $^{207}$Bi. Спектри, виміряні при +24°C та –18°C, показані на рис. 7.11. При охолодженні детектора світловихід збільшився в 3 рази, енергетична роздільна здатність при температурі –18°C становила 45% і 36% для $\gamma$–ліній з енергіями 570 кеВ ($^{207}$Bi) і 662 кеВ ($^{137}$Cs), відповідно.

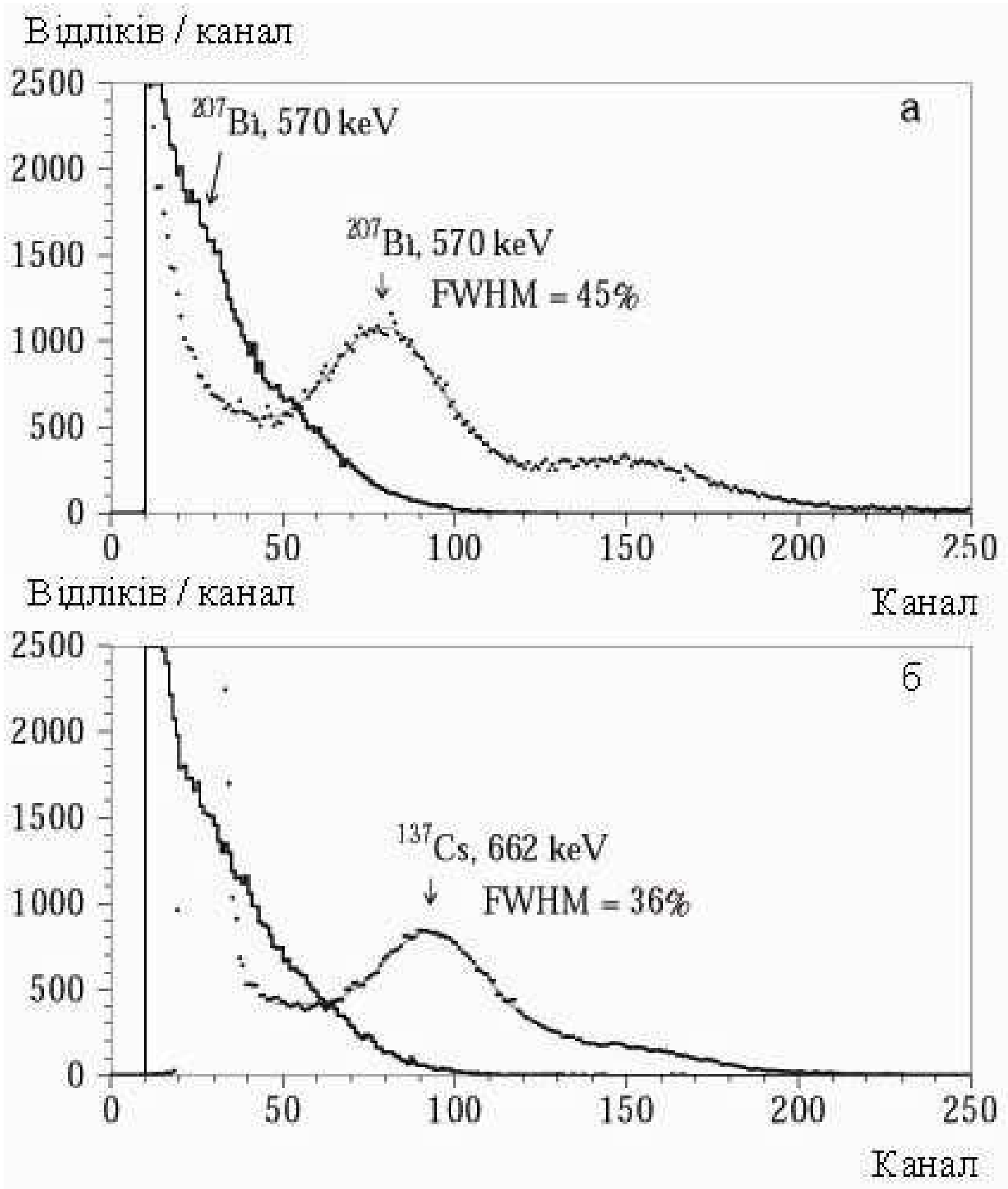

Рис. 7.11. Енергетичні спектри, виміряні із сцинтиляційним кристалом вольфрамату свинцю розмірами 45 × 22 × 22 мм з джерелами γ–квантів $^{207}$Bi (а) та $^{137}$Cs (б) при температурах +24°C (суцільні лінії) та –18°C (точки).

Відгук сцинтиляційного детектора $PbWO_4$ до α–частинок вивчався з кристалом PWO–1 при температурі –18°C. Кристал опромінювався колімованими α–частинками від джерела $^{241}$Am. За допомогою поглиначів з майлару товщиною ≈ 0.65 мг/см$^2$ були отримані α–частинки в інтервалі енергій 2.1 – 4.6 МеВ. Енергії α–частинок були виміряні за допомогою поверхнево–бар’єрного детектора. Крім того, для визначення α/β–співвідношення був використаний α–пік $^{210}$Po (енергія α–частинок 5.3 МеВ), що є дочірнім радіоактивного свинцю $^{210}$Pb (сімейство $^{238}$U), який присутній в кристалі в досить значній кількості (див. нижче). Виміряна залежність α/β–співвідношення від енергії α–частинок показана на рис 7.11.

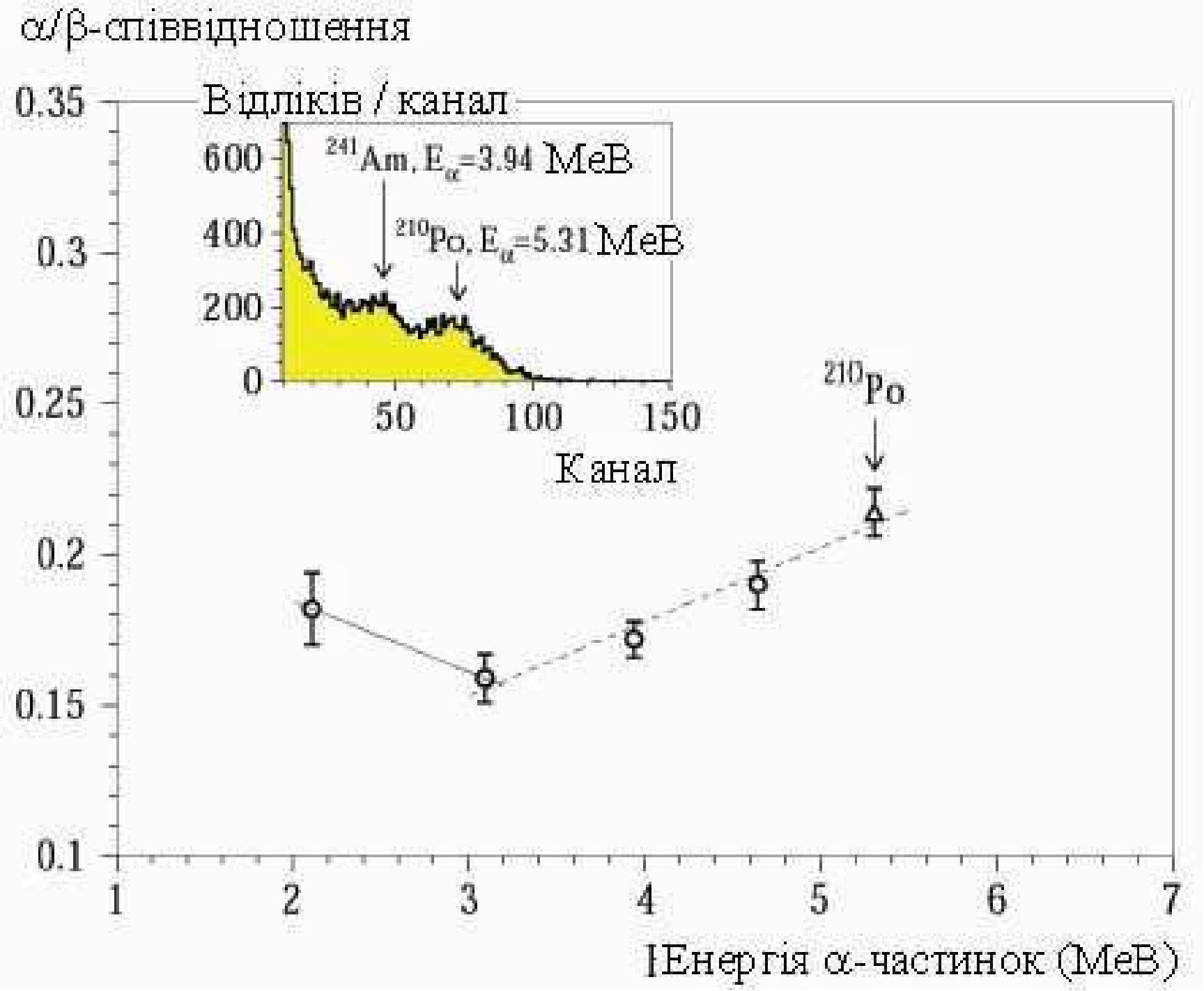

Рис. 7.12. Залежність α/β–співвідношення від енергії α–частинок, виміряна із сцинтиляційним кристалом вольфрамату свинцю розмірами $45 \times 22 \times 22$ мм. Кристал опромінювався α–частинками від джерела $^{241}$Am (кружки). α–Пік $^{210}$Po від внутрішнього забруднення кристалу радіоактивним свинцем показаний трикутником. На вставці показано енергетичний спектр α–частинок.

При енергіях, менших 3 МеВ, α/β–співвідношення спадає з ростом енергії як $\alpha/\beta = 0.23(4) - 0.024(14)\times E_\alpha$, в той час як при енергіях більше 3 МеВ спостерігається зростання α/β–співвідношення з енергією: $\alpha/\beta = 0.08(2) + 0.025(5)\times E_\alpha$, де $E_\alpha$ – енергія α–частинок в МеВ. Схожа поведінка α/β–співвідношення була виміряна для сцинтиляторів вольфраматів кальцію [191], цинку [299], кадмію [257]. Енергетична роздільна здатність, виміряна по α–частинкам $^{210}$Po, становить 39%.

7.4.2. Радіоактивна забрудненість вольфрамату свинцю.

Щоб оцінити рівень радіоактивної забрудненості, фон кристалу PWO–1 був виміряний в Солотвинській підземній лабораторії, кристалу PWO–2 – в низькофоновій установці в Києві. Кристал PWO–1 був оптично з'єднаний з ФЕП XP2412 і охолоджений в холодильній камері до –18°C. Постійна формування спектрометричного підсилювача була взята рівною 0.8 мкс. Система реєстрації записувала амплітуду і час приходу сигналів. Калібровка енергетичної шкали спектрометра була проведена з $\gamma$–джерелом $^{207}$Bi.

Енергетичний спектр фону детектора показаний на рис. 7.13. Інтенсивний пік з енергією близько 1.2 МеВ може бути пояснений $\alpha$–розпадами всередині сцинтилятора ізотопу $^{210}$Po, який утворюється в результату розпаду радіоактивного свинцю $^{210}$Pb (з сімейства $^{238}$U), присутнього в свинці, виплавленому порівняно недавно (період напіврозпаду $^{210}$Pb становить ≈22 роки). Отже, більша частина подій до енергії ≈ 1 МеВ спричинена розпадами іншого дочірнього ізотопу свинцю–210, а саме $\beta$–активного $^{210}$Bi. Активність $^{210}$Po в кристалі PWO–1 становить 53(1) Бк/кг, а кристалу PWO–2 – 79(3) Бк/кг.

Дані вимірювань з кристалом PWO–1 були проаналізовані за допомогою часово–амплітудного аналізу. Був здійснений пошук розпадів $^{214}$Bi ($Q_\beta$ = 3.27 МеВ) → $^{214}$Po ($E_\alpha$ = 7.69 МеВ, $T_{1/2}$ = 164 мкс) → $^{210}$Pb (сімейство $^{238}$U). Щоб відібрати такі розпади, події з енергіями більшими за 0.3 МеВ були тригерами (при цьому за допомогою розрахунків методом Монте–Карло було отримано, що таким чином відбираються близько 76% розпадів $^{214}$Bi). Другу подію шукали в часовому інтервалі 90 – 1000 мкс та енергетичному 1.6 – 2.6 МеВ. Таким умовам задовольняють ≈67% розпадів $^{214}$Po. В отриманих енергетичних спектрах не було особливостей, які можна було б інтерпретувати як розпади в ланцюжку $^{214}$Bi → $^{214}$Po → $^{210}$Pb, що дозволило встановити верхню межу на активність $^{226}$Ra в кристалі на рівні ≤ 10 мБк/кг. Очевидно, що рівновага радіонуклідів уранового ряду в кристалі дуже сильно порушена.

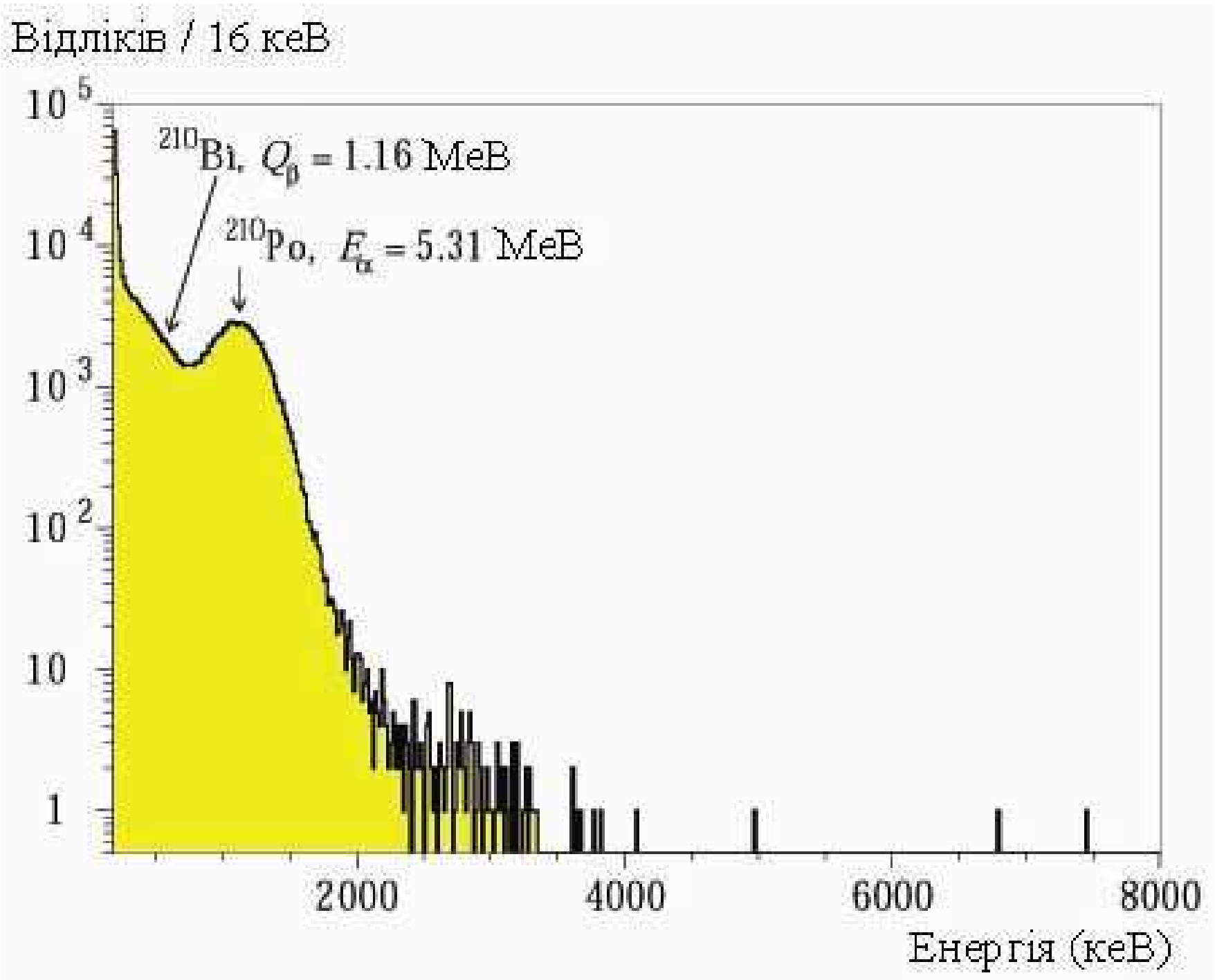

Рис. 7.13. Енергетичний спектр фону детектора з кристалом $PbWO_4$ (45 × 22 × 22 мм), виміряний протягом 2.15 годин в Солотвинській підземній лабораторії.

Активність в кристалі $^{228}$Th (з сімейства $^{232}$Th) була оцінена за допомогою аналізу енергетичного спектру, зображеного на рис. 7.13. Оскільки час формування спектрометричного підсилювача становив 0.8 мкс, події з ланцюжка розпадів $^{212}$Bi ($Q_\beta$ = 2.25 МеВ) → $^{212}$Po ($E_\alpha$ = 8.78 МеВ, $T_{1/2}$ = 0.3 мкс) → $^{208}$Pb (сімейство $^{232}$Th) будуть реєструватися як одна подія з енергією в інтервалі 2.5 – 5 МеВ. Оскільки в діапазоні енергій 3.4 – 5 МеВ є всього 7 подій, походження яких невідоме (найбільш ймовірно, що це події від γ–квантів), можна оцінити верхню межу на активність $^{228}$Th в сцинтиляторі. Враховуючи частину розпадів, які потрапляють в цей енергетичний діапазон (≈60%), отримаємо, що активність $^{228}$Th не перевищує 13 мБк/кг.

Порівняно слабкі обмеження на ступінь забрудненості $PbWO_4$ пов'язані з коротким часом вимірювання і значною активністю $^{210}$Po в кристалі, яка заважає дослідити радіочистоту цього сцинтилятора на більш високому рівні. Очевидною мірою може бути виготовлення кристалів $PbWO_4$ з археологічного свинцю, виплавленого кілька сот років тому. Відомо, що якщо в свинці, виплавленому в наш час, активність сягає сотень Бк/кг, то в свинці, виплавленому в античні часи, вона не виявлена. Наприклад, в роботі [342] досліджувався фон кристала $PbWO_4$ в якості низькотемпературного детектора з високою (кілька кеВ) енергетичною роздільною здатністю. Оскільки $\alpha$–пік $^{210}$Po не спостерігався, автори отримали обмеження на активність $^{210}$Pb: $\leq$ 4 мБк/кг. В той же час, свинець є надзвичайно чистим по відношенню до забрудненості торієм та ураном матеріалом. В роботі [234]] у вимірюваннях на наднизькофоновому напівпровідниковому германієвому детекторі встановлені лише верхні обмеження на активність цих радіонуклідів на рівні $\leq$ 0.3 мБк/кг ($^{232}$Th) та $\leq$ 0.6 мБк/кг ($^{238}$U), відповідно. Виключно високий рівень радіочистоти свинцю був продемонстрований в експерименті IGEX [166,167], де свинець використовувався для захисту наднизькофонових германієвих детекторів. Отже, свинець є одним з найбільш чистих металів по відношенню до радіоактивних домішок. Таким чином, можна очікувати, що сцинтиляційні кристали вольфрамату свинцю виявляться навіть більш чистими по відношенню до забруднень торієм та ураном, ніж сцинтилятори вольфрамату кадмію.

### 7.4.3. Використання вольфрамату свинцю в якості світловода для сцинтиляторів $CdWO_4$.

Можливість використання кристалу $PbWO_4$ в якості світловода для сцинтилятора $CdWO_4$ була перевірена у вимірюваннях з сцинтиляційним кристалом $CdWO_4$ розмірами 10 × 10 × 10 мм. Спочатку були проведені вимірювання з цим кристалом, оптично з'єднаним з фотопомножувачем. Енергетичний спектр, отриманий з $\gamma$–джерелом $^{232}$Th, показаний на рис. 7.14 (а). Енергетична роздільна здатність по $\gamma$–піку з енергією 2615 кеВ становить 3.7%.

Потім кристал $CdWO_4$ проглядався фотопомножувачем через кристал $PbWO_4$ розмірами $45 \times 22 \times 22$ мм. Відповідний спектр показаний на рис. 7.14 (б). Енергетична роздільна здатність в цьому випадку становить 4.1%, амплітуда сигналів 86% у порівнянні із звичайною геометрією вимірювань. Таким чином, можна зробити висновок, що кристали вольфрамату свинцю можуть бути успішно застосовані в якості світловодів для сцинтиляційних детекторів з кристалами $CdWO_4$.

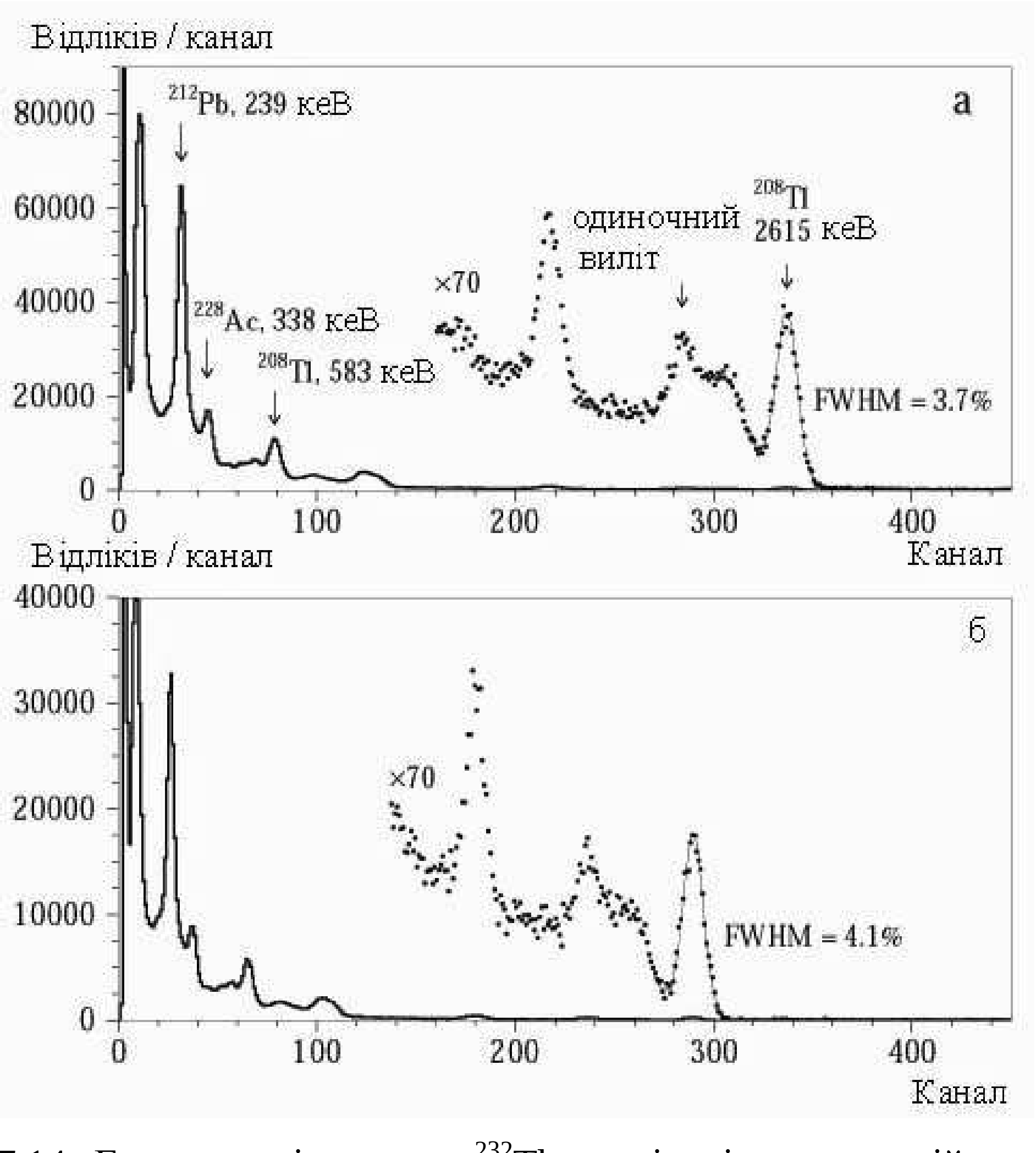

Рис. 7.14. Енергетичні спектри $^{232}Th$, виміряні сцинтиляційним кристалом $CdWO_4$ розмірами $10 \times 10 \times 10$ мм, що був встановлений безпосередньо на ФЕП (а) і проглядався через кристал $PbWO_4$ розмірами $45 \times 22 \times 22$ мм (б).

7.4.4. Оцінки фонових умов експерименту.

Були розглянуті такі основні джерела фону детектора з кристалами $CdWO_4$, оточеними сцинтиляційними кристалами $PbWO_4$, які працюють як детектори активного захисту і світловоди: забрудненість кристалів $PbWO_4$ радіонуклідами $^{228}$Th і $^{226}$Ra разом з продуктами їх розпадів; фон від урану і торію, що міститься в ФЕП, та міді, що оточує детектор, космогенна активація кристалів $PbWO_4$. Розрахунки за допомогою пакета GEANT4 були проведені В.В. Кобичевим, розрахунки космогенної активації $PbWO_4$ – В.І. Третяком.

Щоб оцінити вплив радіоактивних домішок в $PbWO_4$ була вибрана наступна геометрія детектора. Сцинтиляційний кристал $CdWO_4$ розмірами $\varnothing 5 \times 5$ см розміщений всередині циліндра із $PbWO_4$ розмірами $\varnothing 45 \times 45$ см. Енергетична роздільна здатність детектора $CdWO_4$ була взята рівною 4% на енергії 2β–розпаду $^{116}$Cd (2.8 МеВ). Розрахований енергетичний спектр фону детектора $CdWO_4$ від розпадів $^{228}$Th і його дочірніх показано на рис. 7.15 (а). Спектр відповідає експозиції вимірювань з детектором $CdWO_4$ 250 кг $\times$ років при умові, що вміст торію знаходиться на рівні $10^{-12}$ г/г. На рис 7.15 (б) показано спектр фону детектора $CdWO_4$ у збігах з детектором $PbWO_4$. Поріг останнього був взятий рівним 0.5 МеВ. З рис. 7.15 видно, що навіть при такому, досить високому порозі детектора активного захисту, фон від $^{232}$Th дуже ефективно може бути придушений. В інтервал енергій 2.7 – 2.9 МеВ потрапили лише дві події. Внесок від забрудненості кристалів $PbWO_4$ радієм ще менший. Ні одна подія від розпадів $^{226}$Ra всередині захисного детектора $PbWO_4$ з енергією більшою за 2 МеВ не потрапила в спектр детектора $CdWO_4$.

Вплив забрудненості торієм мідного захисту та ФЕП був розрахований у припущенні про наступну конструкцію детектора. 32 детектори $CdWO_4$ були розміщені всередині $PbWO_4$ розмірами $\varnothing 70 \times 70$ см. Мідний захист товщиною 5 см оточує $PbWO_4$. Забрудненість міді торієм була взята з роботи [343] і становила $10^{-11}$ г/г. Кристали $CdWO_4$ проглядаються ФЕП через світловоди із вольфрамату свинцю довжиною 33 см. Забрудненість ФЕП торієм взята на рівні $2.5 \times 10^{-7}$ г/г, що відповідає даним компанії Electron Tubes Limited [344] для низькофонових ФЕП. Як показали розрахунки, в пік з енергією 2615 кеВ потрапляє лише 13 відліків від міді і 3 відліки від забрудненості ФЕП. В інтервал 2.7 – 2.9 МеВ, де очікується пік 0ν2β–розпаду ядра $^{116}$Cd, потрапляє лише ≈ 0.5 відліків.

Космогенна активація кристалів $PbWO_4$ була розрахована за допомогою програми COSMO [345]. Розрахунки були проведені у припущенні, що кристали виготовляються на поверхні землі протягом місяця, а потім рік знаходяться під землею. Лише п'ять ізотопів з 175 можуть давати фонові відліки в інтервал піку 0ν2β–розпаду $^{116}$Cd: $^{68}$Ge, $^{88}$Y, $^{88}$Zr, $^{106}$Ru, $^{110m}$Ag. Фон, пов'язаний з розпадами цих радіонуклідів в кристалах $PbWO_4$ був промодельований за допомогою методу Монте–Карло. Як показали розрахунки, лише 0.9 відліків за 10 років вимірювань в 25 кг детекторів $CdWO_4$ потрапить в інтервал енергії 2.7 – 2.9 МеВ.

Таким чином, фон детектора $CdWO_4$, може бути ефективно придушений активним захистом із сцинтиляційних кристалів вольфрамату свинцю. Ці кристали також можуть бути застосовані в якості світловодів для детекторів з $CdWO_4$. Це дозволить значно зменшити розміри захисту. Справді, розміри детектора з 32 кристалами $CdWO_4$, оточеними захистом із кристалів $PbWO_4$, не перевищують $70 \times 70 \times 70$ см. Пасивний захист із міді (5 см), свинцю (50 см) і поліетилену (50 см) матиме розміри $2.8 \times 2.8 \times 3.1$ м, що значно менше, аніж, наприклад, розміри установки Borexino Counting Test Facility ($\approx \varnothing 11 \times 10$ м) [323] в якій міг би розміститись детектор CAMEO [324].

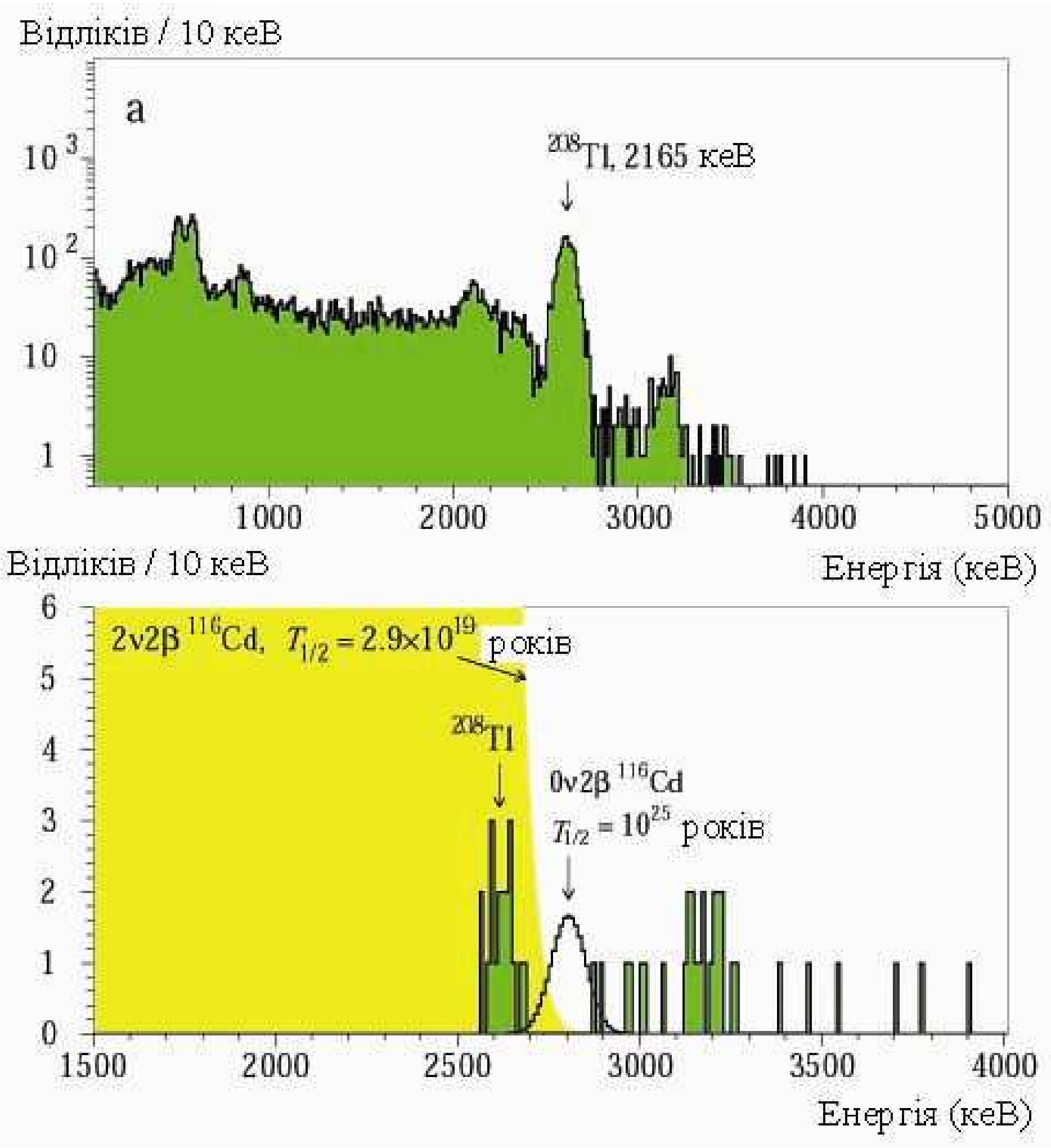

Рис. 7.15. Промодельовані методом Монте–Карло енергетичні спектри фону детектора із сцинтиляційними кристалами $^{116}CdWO_4$, оточеного кристалами $PbWO_4$. Модель відповідає експозиції 250 років × кг. (а) Фон, спричинений розпадами $^{228}$Th і його дочірніх, за умови використання $PbWO_4$ лише як пасивного захисту; (б) те ж саме, що на рис. (а), але в режимі антизбігів з детектором $PbWO_4$, що спрацьовує при енерговиділенні, більшому за 0.5 МеВ. Показані розподіли 2ν2β–розпаду $^{116}$Cd $T_{1/2} = 2.9 \times 10^{19}$ років) та пік 0ν2β–розпаду $^{116}$Cd з $T_{1/2} = 10^{25}$ років.

## 7.5. Оцінки чутливості експерименту пошуку 2β–розпаду ядра $^{48}Ca$ за допомогою сцинтиляторів $CaWO_4$ (проект CARVEL).

На основі вимірювань з кристалом $CaWO_4$ розмірами 40×34×23 мм (див. Розділ 6) було запропоновано проект експерименту по пошуку подвійного бета–розпаду ядра $^{48}Ca$ під назвою CARVEL (CAlcium Research for VEry Low neutrino mass). Пропонується використати близько 100 кг кристалів $CaWO_4$, виготовлених із збагаченого ізотопу $^{48}Ca$. Будова установки значною мірою подібна до успішно використаної в Солотвинському експерименті по пошуку 2β–розпаду $^{116}Cd$ за допомогою сцинтиляційних кристалів $^{116}CdWO_4$ [161]. Планується використати 50 кристалів $^{48}CaWO_4$ розмірами ∅7.5×7.5 см. Кожний сцинтилятор проглядається з двох протилежних сторін через кварцові світловоди діаметром 10 і довжиною 70 см двома наднизькофоновими фотопомножувачами діаметром 125 мм. Кожний світловод склеєний з пластикового сцинтилятора (30 см) і кварцу (40 см). Всі 50 детекторів оточені пластиковими сцинтиляторами що дозволяє, разом з активними світловодами, реалізувати 4π–геометрію активного захисту. Пасивний захист складається з міді (5 см), свинцю (30 см) та поліетилену (16 см).

Чутливість експерименту залежить від маси досліджуваного ізотопу (≈16 кг $^{48}Ca$, що становить $\approx 2\times10^{26}$ ядер) та рівня фону, який може бути досягнутий на енергії 2β–переходу $^{48}Ca$ (4274 кеВ).

### 7.5.1. Оцінки фону в експерименті зі сцинтиляційними кристалами $CaWO_4$.

Очікувані джерела фону спектрометра такі: 1) радіоактивна забрудненість кристалів $^{48}CaWO_4$ радіонуклідами $^{228}Th$ і $^{226}Ra$; 2) космогенна активація сцинтиляторів $^{48}CaWO_4$; 3) зовнішній фон; 4) фон від радону, що присутній в повітрі; 5) β– та 2ν2β–розпад $^{48}Ca$. Радіоактивна забрудненість кристалів $^{48}CaWO_4$ радіонуклідами ланцюжка $^{228}$Th була прийнятою такою, що вже досягнута в деяких кристалах вольфрамату кадмію: ≈ 5 мкБк/кг. Активність $^{226}$Ra, як показали розрахунки, може бути більшою: ≈20 мкБк/кг. Якщо вдасться виготовити сцинтилятори з такою радіочистотою, фон від розпадів $^{208}$Tl з ланцюжка $^{228}$Th може бути відкинутий шляхом аналізу попередніх α–розпадів $^{212}$Bi. Справді, при наявності $^{228}$Th в кристалах на вищевказаному рівні за 10 років вимірювань в детекторі буде зареєстровано ≈$1.6\times10^{5}$ розпадів $^{228}$Th. Як показали розрахунки методом Монте–Карло, в енергетичне вікно, де очікується 0ν2β–розпад $^{48}$Ca (4.17 – 4.38 МеВ), потрапить ≈700 подій. Однак, використовуючи часово–амплітудний аналіз та аналіз форми сцинтиляційних спалахів, фон може бути зменшений до ≈1.7 відліків за 10 років вимірювань. Ці розрахунки показують, наскільки важливими є вимоги, щоб детектор мав якнайкращу енергетичну здатність до α–частинок (це дозволить виділити саме α–розпади $^{212}$Bi від розпадів інших дочірніх α–активних ізотопів урану та торію), а також якнайкращу здатність до розділення сигналів від α–частинок та γ–квантів. Важливо відмітити, що вимоги до радіочистоти сцинтиляторів $^{48}CaWO_4$ є значно м’якшими, ніж в усіх інших проектах по пошуку 2β–розпаду. Це обумовлене великою енергією розпаду ядра $^{48}$Ca. Велика кількість фонових подій матиме меншу енергію і не буде потрапляти в область енергії, де очікується пік 0ν2β–розпаду, оскільки, як правило, фон детектора має чітку тенденцію до зменшення по мірі збільшення енергії.

В той час як фон, спричинений проходженням через установку космічних променів, може бути зменшений до необхідного рівня шляхом розміщення установки глибоко під землею та за допомогою детекторів активного захисту, активація матеріалів детектора може викликати фон, який практично неможливо усунути. Були проведені розрахунки активації кристалів $CaWO_4$ за допомогою програми COSMO [345]. Для розрахунків було припущено, що кристали з моменту свого виготовлення перебуватимуть протягом місяця на поверхні, а потім будуть знаходитись рік в підземній лабораторії. Було показано, що серед радіонуклідів, що утворюються в кристалах $CaWO_4$ під дією космічних променів (а всього утворюється 158 радіонуклідів з періодом напіврозпаду більше 25 днів) немає таких, що могли б давати фонові події в околі піку 0ν2β–розпаду $^{48}$Ca. Очевидно, що ця перевага експерименту по пошуку 0ν2β–розпаду також завдячує високій енергії 2β–розпаду $^{48}$Ca. В той же час, космогенна активація буде головною проблемою багатьох інших експериментів по пошуку 0ν2β–розпаду. Наприклад, активація космічними променями напівпровідникових германієвих детекторів, які планується застосувати в експерименті по пошуку 0ν2β–розпаду $^{76}$Ge, призвела до утворення ізотопів $^{22}$Na, $^{60}$Co, $^{68}$Ga із загальною активністю на рівні 10 – 100 мкБк на тону детекторів. Ця дуже мала активність, тим не менше, буде давати суттєвий фон при енергії 2β–розпаду $^{76}$Ge (2039 кеВ) після того, як інші компоненти фону будуть придушені.

Велика енергія 2β–переходу $^{48}$Ca має ще одну, дуже суттєву перевагу в порівнянні з іншими ядрами, а саме, можливість використовувати для пасивного захисту матеріали з поточним рівнем забрудненості. Тобто, можуть бути застосовані звичайні мідь, свинець, сталь, поліетилен. В той же час, в більшості проектів експериментів по пошуку 2β–розпаду до радіочистоти конструкційних матеріалів установки ставляться дуже високі вимоги. Більш того, виявляється, що досягти необхідного рівня чистоти в металах неможливо на сучасному рівні технологій. Тому для пасивного захисту пропонується використовувати рідини: воду, рідкі гази, органічні рідини з рівнем забрудненості ураном, торієм, калієм на рівні $\approx 10^{-15}$ г/г. Це призводить, враховуючи низьку густину цих рідин, до значного збільшення установок. Наприклад, розміри установки GERDA [317], в якій для захисту напівпровідникових германієвих детекторів очікується застосувати воду та рідкий азот чи аргон, становлять 10 – 15 метрів. Завдяки можливості використовувати звичайні свинець та мідь (з типовими рівнями забрудненості ураном та торієм $10^{-10} - 10^{-12}$ г/г) для пасивного захисту детектора з кристалами $CaWO_4$, розміри установки не будуть перевищувати 3–4 метрів, що дуже суттєво, враховуючи розташування установок в підземних лабораторіях, де необхідно економити кожний кубічний метр простору. Фон на енергії $\approx$ 4.3 МеВ можуть викликати лише випадкові збіги двох і більше γ–квантів, випромінених протягом часу, який не може бути розрізнений детектором з кристалами $CaWO_4$. Оцінки показують, що таких подій практично не буде. Дійсно, припускаючи забрудненість міді торієм $10^{-10}$ г/г, масу мідного захисту 4 тони, часове вікно спектрометра, в якому дві події будуть розглядатися як одна ($\approx$1 мкс), отримаємо, що за 10 років вимірювань в 100 кг детектора буде зареєстровано $\approx 10^{-3}$ подій, тобто цим внеском в фон можна знехтувати.

Фон від радону є серйозною проблемою експериментів по пошуку 2β–розпаду. Оскільки радон є інертним газом, він дуже легко проникає через різні матеріали, розчиняється в рідинах, не видаляється з рідин хімічними способами очистки, тощо. Найбільш небезпечним є ізотоп радону–222 з сімейства урану. Період напіврозпаду цього ізотопу достатньо високий ($T_{1/2}$ = 3.82 діб), в результаті він встигає проникнути через шари ґрунту чи скельних порід, бетон, захист установки. Але, оскільки найбільша енергія β–розпаду серед дочірніх ізотопів радону–222, а саме $^{214}$Bi ($Q_\beta$ = 3.27 МеВ), менша за енергію 2β–розпаду $^{48}$Ca, радон–222 не буде давати фонових подій на енергії 4.27 МеВ. Збіги подій від β–частинок $^{214}$Bi та α–частинок $^{214}$Po ($T_{1/2}$ = 164 мкс) може викликати енерговиділення в детекторі навіть більше за 4.27 МеВ. Але ця компонента фону ефективно придушується завдяки аналізу форми сцинтиляційних сигналів. Інший ізотоп радону – радон–220 (його ще називають тороном) з сімейства торію має серед своїх дочірніх $^{208}$Tl з енергією ($Q_\beta$ = 5 МеВ), більшою за 4.27 МеВ. Але, оскільки період напіврозпаду $^{220}$Rn порівняно малий ($T_{1/2}$ = 55.4 сек.), він не призводить до появи суттєвого фону, оскільки повністю розпадається, не встигнувши потрапити в чутливий об’єм детектора.

Дозволений законом збереження енергії ($Q_\beta$ = 278 кеВ) процес β–розпаду $^{48}$Ca може викликати фонові події на енергії, де очікується пік 0ν2β–розпаду. Дійсно, енергія β–розпаду $^{48}$Sn дуже висока ($Q_{\beta}$ = 3994 кеВ), а отже збіги подій β–розпаду $^{48}$Ca та $^{48}$Sn можуть призвести до фонових подій на енергії 0ν2β–розпаду $^{48}$Ca. Проте, β–розпад $^{48}$Ca дуже сильно придушений у зв'язку зі значною різницею у спінах материнського та дочірнього ($^{48}$Sn) ядер. Тому β–розпад $^{48}$Ca все ще не спостерігався (експериментально встановлене лише обмеження на період напіврозпаду $T_{1/2} \geq 10^{20}$ років). Оцінки показують, що навіть якщо прийняти обмеження на $T_{1/2}$ за період напіврозпаду, ймовірність появи фонових подій від послідовних β–розпадів $^{48}$Ca – $^{48}$Sc – $^{48}$Ti, які внаслідок збігів за 10 років вимірювань в 100 кг детектора дадуть одну фонову подію з енергією близько 4.27 МеВ, настільки мала, що нею можна знехтувати.

7.5.2. Енергетична роздільна здатність детектора з кристалами $CaWO_4$.

Процес 2ν2β–розпаду $^{48}$Ca може призводити до фонових подій в околі шуканого піку 0ν2β–розпаду цього ядра. Ця компонента фону визначається періодом напіврозпаду відносно двохнейтринної моди і енергетичною роздільною здатністю детектора. Відносна енергетична роздільна здатність ідеального сцинтиляційного детектора (FWHM) визначається за формулою 2.2. В свою чергу, кількість фотонів $N_{pe}$, що випромінюється в сцинтиляторі при енерговиділенні в ньому енергії 1 МеВ (для γ–квантів), може бути оцінена за формулою:

$$N_{pe} = N_{ph} \times E \times LC \times QE \qquad (7.3),$$

де $E$ – енерговиділення в сцинтиляторі, $LC$ – частина фотонів, яка потрапляє на фотокатод ФЕП (цю величину ще називають світлозбором), $QE$ – квантова ефективність ФЕП, в яку вже входить коефіцієнт збору фотоелектронів на перший динод. Величина квантової ефективності залежить від спектру емісії сцинтилятора і спектральної чутливості фотокатода. Величина $N_{pe}$ була оцінена у вимірюваннях з кристалом $CaWO_4$ розмірами ∅40 × 39 мм та сцинтилятором йодистого натрію, активованого талієм, розмірами ∅40 × 40 мм (в стандартній упаковці). Вихід фотонів сцинтилятора NaI(Tl) становить $(3.8 - 4.3)\times10^4$ фотонів на 1 МеВ енерговиділення в кристалі. Обидва сцинтилятори були оптично приєднані до ФЕП Philips XP2412 з мультилужним фотокатодом з відомою залежністю квантової чутливості від довжини хвилі. Згортка цієї залежності зі спектром світіння сцинтиляторів дозволила оцінити величини квантової ефективності ФЕП для обох сцинтиляторів: $QE(CaWO_4) = 0.22$ та $QE$(NaI(Tl)) = 0.24. Відносна амплітуда сигналів від детектора з кристалом $CaWO_4$ була виміряна з γ–квантами $^{137}$Cs та $^{207}$Bi і склала 18% від NaI(Tl). Величини світлозбору для обох сцинтиляторів були розраховані методом Монте–Карло В.В. Кобичевим за допомогою програми GEANT4. Для NaI(Tl) була отримана величина світлозбору $LC \approx 65\%$, а для кристала $CaWO_4$ $LC \approx 52\%$. Таким чином було оцінено, що при поглинання в кристалі $CaWO_4$ γ–квантів з енергією 1 МеВ виникає $(0.8 - 1.2)\times10^4$ фотонів. Отже, на енергії 2β–розпаду $^{48}$Ca величина $N_{pe}$ становить $\approx 4\times10^4$ фотонів. Взявши те саме значення квантової ефективності ФЕП та дещо більше значення величини світлозбору, а саме LC = (65–70)%, (як було показано у вимірюваннях з кристалом $CaWO_4$ зануреним у рідину, світлозбір було збільшено на 20%), ми оцінили ідеальну енергетичну роздільну здатність детектора: FWHM ≈ 2.8%. Таким чином, припустивши деяке покращення сцинтиляційних властивостей та прозорості сцинтиляторів $CaWO_4$, можна сподіватися отримати енергетичну роздільну здатність детектора ≈2.5% на енергії 2β–розпаду $^{48}$Ca.

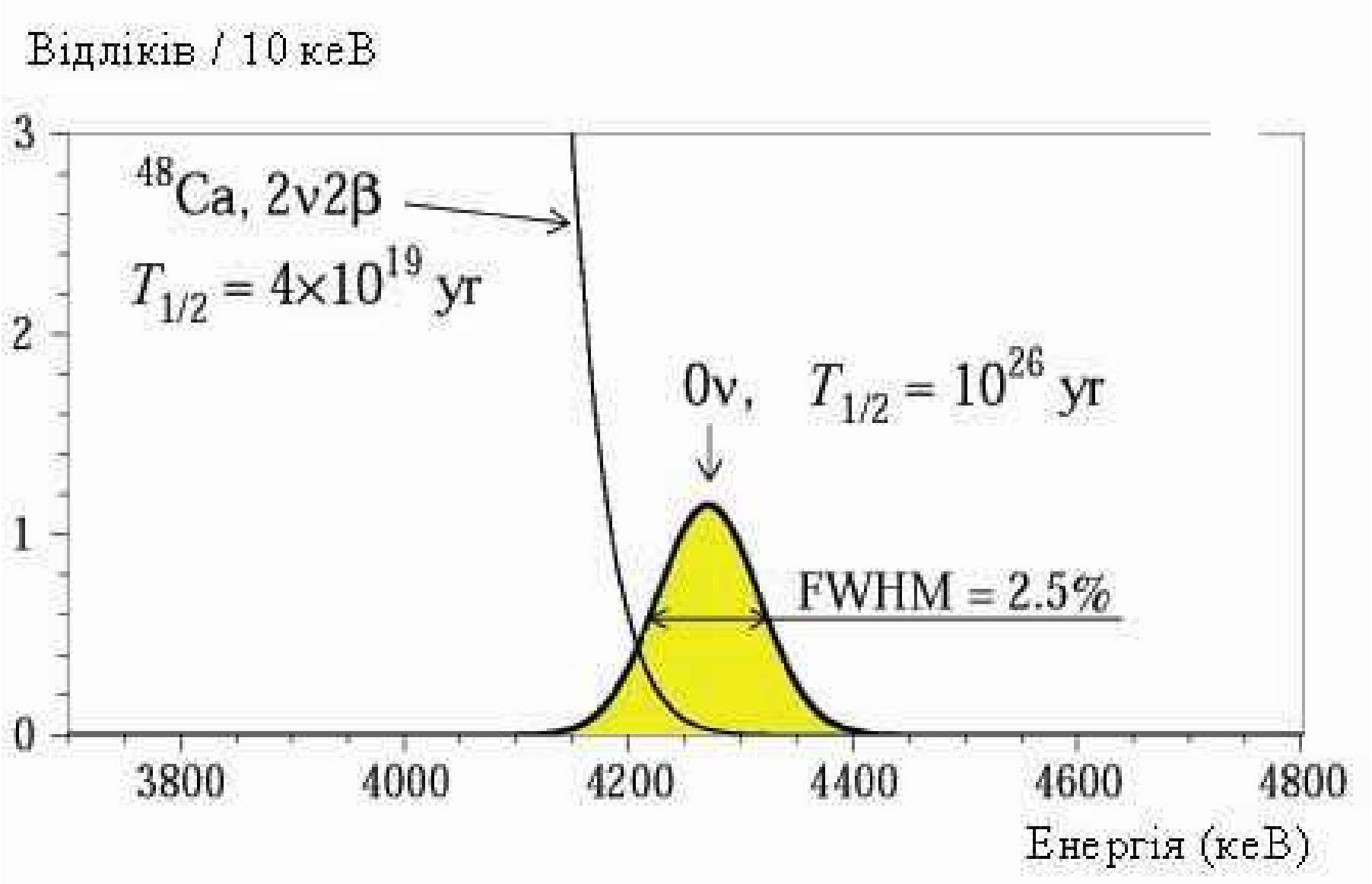

Рис. 7.16. Функція відгуку сцинтиляційного детектора з кристалами $^{48}CaWO_4$ до 2ν та 0ν мод 2β–розпаду $^{48}$Ca з періодами напіврозпаду $T_{1/2} = 4\times10^{19}$ та $T_{1/2} = 10^{26}$ років, відповідно.

7.5.3. Чутливість експерименту CARVEL.

Як видно з рис. 7.16, роздільна енергетична здатність 2.5% на енергії 2β–розпаду $^{48}$Ca дозволить чітко відрізнити пік 0ν2β–розпаду від розподілу 2ν2β–розпаду з періодом напіврозпаду $4 \times 10^{19}$ років. Враховуючи кількість ядер $^{48}$Ca в 100 кг кристалів $^{48}CaWO_4$ ($\approx 2\times10^{26}$), можна оцінити чутливість експерименту після 10 років вимірювань. Вона становитиме $\approx 10^{27}$ років, що відповідає масі нейтрино ≈0.04 – 0.09 еВ, в залежності від теоретично обрахованих значень добутку $T_{1/2}\times\langle m_\nu\rangle^2$. До уваги були взяті розрахунки, здійснені як за допомогою оболонкової моделі, яка дає значення $T_{1/2}^{0\nu} \times \langle m_\nu\rangle^2 = 3.2\times10^{24}$ років×eB$^2$ [13], $1.7 \times 10^{25}$ [346], $5.3 \times 10^{24}$ [347], $(6.4 – 7.8) \times 10^{24}$ [348], $8.8 \times 10^{24}$ [349], так і розрахунки за методом quasi–particle random phase approximation (QRPA). Метод QRPA дає такі значення $T_{1/2}^{0\nu} \times \langle m_\nu\rangle^2 = 1.5 \times 10^{24}$ років × еВ$^2$ [346], $7.1 \times 10^{23}$ [347], $(1.1 – 3.3) \times 10^{24}$ [350], $2.5 \times 10^{24} – 2.8 \times 10^{25}$ [351], $2.3 \times 10^{24}$ [352].

Очевидно, що для постановки такого експерименту необхідно вирішити проблему збагачення ізотопу $^{48}Ca$, вміст якого в природній суміші ізотопів кадмію всього 0.187%. Зараз є кілька проектів установок для великомасштабного виробництва ізотопів, зокрема, $^{48}Ca$. Проект MCIRI [ 353 ], запропонований колаборацією Інституту Курчатова (Москва) та Об'єднаного інституту ядерних досліджень (Дубна), передбачає розділення ізотопів методом іонного циклотронного резонансного розігрівання в плазмі. В США ведеться розробка проекту AVLIS (Atomic Vapor Laser Isotope Separation).

Для пошуку $0\nu2\beta$–розпаду $^{48}Ca$ було також запропоновано проект CANDLES [314]. В цьому проекті пропонується використовувати кристали фториду кальцію ($CaF_2$) виготовлені з незбагаченого кальцію. Кристали планується занурити в контейнери з акрилу, наповнені рідким сцинтилятором, що виконуватиме функції зміщувача спектру, оскільки кристали $CaF_2$ випромінюють в ультрафіолетовій області спектру. В свою чергу, контейнери з кристалами будуть занурені в рідкий сцинтилятор, що буде використовуватись як прозоре середовище для розповсюдження світла та захист від зовнішнього випромінювання. Сцинтиляційне світло буде реєструватись за допомогою фотопомножувачів, занурених в рідкий сцинтилятор. Чутливість експерименту до маси нейтрино оцінена авторами на рівні $m_\nu \leq 0.5$ еВ для 200 кг сцинтиляторів і $m_\nu \leq 0.5$ еВ для 3.2 тон кристалів $CaF_2$. Найбільш суттєвими недоліками проекту CANDLES є складність в здійсненні енергетичної калібровки (у сцинтиляторів $CaF_2$ при реєстрації $\gamma$–квантів практично відсутній пік повного поглинання); велика маса детектора, що призводить до зростання фону, який, як правило, пропорційний масі детектора; поки що недостатня енергетична роздільна здатність. Перевагами проекту CANDLES у порівнянні з запропонованим нами проектом CARVEL, є низька чутливість до гамма–квантів (що зменшує фон від зовнішніх $\gamma$–квантів) та застосування незбагачених кристалів.

Необхідно також відмітити результати вимірювань з кристалами $CaF_2$ в якості болометричних низькотемпературних сцинтилюючих детекторів [ 354 ]. Такі детектори також можуть бути застосовані для пошуку 0ν2β–розпаду $^{48}Ca$ з дуже високою енергетичною роздільною здатністю. Кристали $CaWO_4$ також є болометричними сцинтилюючими детекторами [355,356]. Недоліком кріогенних болометричних детекторів є все ще недостатня стабільність протягом довгострокових вимірювань. В той же час, сцинтиляційні детектори можуть працювати стабільно протягом десятків років, що дуже важливо для проведення експериментів по пошуку 2β–розпаду.

## 7.6. Радіочистота сцинтиляційних детекторів

Рівень забруднень детектора радіонуклідами різного походження відіграє ключову роль в багатьох експериментах по пошуку рідкісних ядерних та суб'ядерних процесів. Після того, як детектор захищений від випромінювання оточуючого середовища потужними екранами та активним захистом, від космічних променів спуском установки глибоко під землю, а матеріали установки ретельно відібрані за рівнем радіочистоти, забрудненість самого детектора, зокрема сцинтилятора, починає давати головний вклад в фон. Оцінки активності радіонуклідів в сцинтиляторах $CaWO_4$, $ZnWO_4$, $CdWO_4$ (як виготовлених із збагаченого ізотопу $^{116}Cd$, так і з незбагаченого кадмію), $CeF_3$, GSO було здійснено в ході аналізу даних експериментів, де ці сцинтилятори були застосовані. При цьому сцинтиляційний детектор реєстрував події розпадів всередині сцинтилятора, а отже сам був одночасно об'єктом вимірювання і вимірювальним приладом. Часово–амплітудний аналіз, ідентифікація за формою сцинтиляційних сигналів та моделювання енергетичних спектрів детекторів методом Монте–Карло використовувались для визначення активностей радіонуклідів в сцинтиляторах.

Радіонукліди уранових та торієвого рядів, а також калій присутні в тих чи інших кількостях в усіх досліджених сцинтиляторах, антропогенні $^{90}Sr$–$^{90}Y$, $^{137}Cs$ є результатом Чорнобильської катастрофи та випробувань ядерної зброї і їх активність залежить від умов, в яких сцинтилятори були виготовлені. Специфічна активність, тобто характерна саме для даного сцинтилятора, спостережена в кристалах вольфрамату кадмію (β–активні $^{113}$Cd та $^{113m}$Cd, 2β–активний $^{116}$Cd), ортосилікату гадолінію (α–активний $^{152}$Gd). Крім того, в експериментах з використанням детекторів вольфраматів кадмію та кальцію вперше було зареєстровано альфа–активність $^{180}$W з періодом напіврозпаду $T_{1/2} \approx 10^{18}$ р. Результати вимірювань радіоактивної забрудненості сцинтиляторів представлені в таблицях 2.2, 4.1, 4.3, 5.3, 6.3. Ці дані зведені в таблицю 7.2, де приведена сумарна α–активність, активності $^{228}$Th, $^{226}$Ra та $^{40}$K, активності специфічних для даного сцинтилятора радіонуклідів. Крім того, в таблиці 7.2 приведені також дані про радіочистоту деяких інших широко використовуваних детекторів.

Таблиця 7.2. Порівняння рівнів радіочистоти різних детекторів (мБк/кг).

| Детектор | Сумарна α–активність U/Th | $^{228}$Th | $^{226}$Ra | $^{40}$K | Специфічна активність | Посилки |
|---|---|---|---|---|---|---|
| $CaWO_4$ | 20 – 400 | 0.6 | 5.6 | <12 | | [191] |
| $ZnWO_4$ | <20 | <0.2 | <0.4 | <12 | | [299] |
| $CdWO_4$ | ≤0.7 – 2.3(3) | <0.003 | <0.004 | 0,3 | 580 ($^{113}$Cd) | [261,357, 286,161] |
| GSO | 40 | 2.3 | 0.3 | <14 | 1200 ($^{152}$Gd) | [230,358] |
| $CeF_3$ | 3400 | 1100 | <60 | <330 | | [359] |
| BGO | <0.4 | <1.2 | | | 700 ($^{207}$Bi) | [360,361] |
| NaI(Tl) | <0.2 | 0.0004 | 0.005 | 1.5 | | [362,363] |
| CsI(Tl) | | <0.3 | <0.3 | 45 | $1.1\times10^{4}$ ($^{87}$Rb) | [364,365] |
| $BaF_2$ | | 400 | 1400 | | | [366] |
| Рідкий сцинтилятор | | $10^{-6}$ | $4\times10^{-7}$ | $3\times10^{-7}$ | | [229] |
| HPGe | | $<2\times10^{-5}$ | $<2\times10^{-5}$ | | | [367,] |

Як видно з таблиці 7.2, рівень радіоактивної забрудненості рідкого сцинтилятора, який планується використовувати в експерименті Borexino, є надзвичайно низьким. Взагалі, рідкі сцинтилятори мають, як правило, дуже низький вміст радіонуклідів. Пояснити такий високий рівень чистоти рідких сцинтиляторів можна тим, що вони виготовлені із органічних сполук, і тому вони є чистими від далеких за фізико–хімічними властивостями сполук металів (калію, а тим більше урану, торію і їх дочірніх продуктів розпаду). Так само технологія виробництва напівпровідникових детекторів з надчистого германію забезпечує надзвичайно високий рівень чистоти (власне, визначені лише верхні межі на вміст $^{232}$Th $^{226}$Ra в цих детекторах). Але при цьому треба пам‘ятати, що рівні фону низькофонових германієвих детекторів визначаються забрудненістю конструкційних матеріалів, в першу чергу, кріостата, виготовленого з електролітичної міді. Видно, що пластмасові сцинтилятори поступаються за рівнем радіочистоти рідким сцинтиляторам. Сцинтиляційні кристали NaI(Tl), вольфрамату кадмію та цинку, CsI(Tl) мають досить низький вміст радіонуклідів, в той час як в детекторах $CaWO_4$, GSO, $BaF_2$, $CeF_3$ питома активність нуклідів рядів урану і торію досить висока.

## 7.7. Про можливість застосування кристалів $ZnWO_4$, $CaWO_4$, YAG:Nd та GSO для пошуків частинок темної матерії

Як було показано в роботах [368,356], кристали $ZnWO_4$ та $CaWO_4$ можуть працювати при низьких температурах як болометричні сцинтиляційні детектори. Важливою перевагою таких детекторів є можливість ефективно розділяти події від

γ–квантів (β–частинок) від α–частинок та ядер віддачі. Остання властивість має особливе значення для експериментів по пошуку гіпотетичних частинок темної матерії, які можуть взаємодіяти з речовиною шляхом вибивання ядер віддачі. Очікується, що кінетична енергія таких ядер має бути досить малою – кілька десятків кеВ і навіть менше. В складі кристалів $ZnWO_4$, $CaWO_4$, YAG:Nd та GSO є ядра з ненульовим спіном (див. табл. 7.3). Згідно з деякими теоретичними моделями [268], частинки темної матерії можуть взаємодіяти з такими ядрами саме завдяки ненульовому спіну. У випадку непружної взаємодії частинок темної матерії з такими ядрами, одночасно може бути зареєстроване ядро віддачі і гамма–квант, що випромінюється при переході ядра в основний стан. Такі події будуть суттєво відрізнятися як від „чистих" γ(β)–подій, так і від подій, пов'язаних з ядрами віддачі за допомогою аналізу співвідношення теплового і світлового сигналів.

Як видно з таблиці 7.3, деякі з цих ядер мають досить значний час життя перших збуджених станів, що може дозволити використати цю властивість для дуже ефективного відбору таких подій від фону. Найбільш цікавим в цьому списку є ядро $^{64}$Zn, оскільки час життя першого збудженого рівня цього ядра достатньо високий для того, щоб спробувати відрізнити дві події, а саме: 1) ядро віддачі від спін–залежної взаємодії частинки темної матерії з ядром $^{64}$Zn; 2) γ–квант з енергією 93.3 кеВ.

Таблиця 7.3. Властивості ядер, присутніх в детекторах $ZnWO_4$, $CaWO_4$, YAG:Nd та GSO, які можуть бути використані для реєстрації спін–залежної взаємодії частинок темної матерії.

| Ядро | Спін, парність | Розповсюдженість ізотопу (%) | Перший збуджений рівень | |
|---|---|---|---|---|
| | | | Енергія (кеВ) | Час життя |
| $^{17}$O | $5/2^+$ | 0.038 | 870.7 | 179 пс |
| $^{27}Al$ | $5/2^+$ | 100 | 843.8 | 35 пс |
| $^{29}Si$ | $1/2^+$ | 7.67 | 1273.4 | 290 фс |
| $^{43}Ca$ | $7/2^-$ | 0.135 | 372.8 | 33 пс |
| $^{67}Zn$ | $5/2^-$ | 4.1 | 93.3 | 9.16 мкс |
| $^{89}Y$ | $1/2^-$ | 100 | 909 | 16.06 с |
| $^{111}Cd$ | $1/2^+$ | 12.49 | 245.4 | 85 нс |
| $^{113}Cd$ | $1/2^+$ | 12.22 | 263.6 | 14.1 років |
| $^{143}Nd$ | $5/2^+$ | 12.18 | 742.0 | 2.8 пс |
| $^{145}Nd$ | $7/2^+$ | 8.30 | 67.2 | 29.4 нс |
| $^{153}Nd$ | $5/2^+$ | 52.2 | 83.4 | 0.79 нс |
| $^{155}Gd$ | $3/2^+$ | 14.80 | 60.0 | 0.193 нс |
| $^{157}Gd$ | $3/2^-$ | 15.65 | 54.5 (63.9) | 130 пс (0.46 мкс) |
| $^{183}W$ | $1/2^-$ | 14.3 | 46.5 | 0.188 нс |

**Висновки розділу.** Було проведено аналіз даних двох експериментів, Heidelberg–Moscow та IGEX, по пошуку 2β–розпаду ядра $^{76}$Ge. Показано, що повідомлення про спостереження 0ν2β–розпаду в експерименті Heidelberg–Moscow зроблені передчасно. Об’єднання даних цих двох експериментів дозволило встановити найбільш високі обмеження на період напіврозпаду $^{76}$Ge і обмежити масу нейтрино на рівні $m_\nu < 0.3$ (0.2) еВ з 90% (68%) рівнем довірчої імовірності. Проаналізовані чутливість експериментів до *реєстрації* 0ν2β–розпаду і гранична *чутливість* досліду, тобто здатність встановити обмеження на період напіврозпаду, коли ефект не спостерігається. Зроблено огляд найбільш перспективних проектів експериментів по пошуку 0ν2β–розпаду ядер. Запрпоновано два проекти експериментів по пошуку подвійного β–розпаду $^{116}Cd$ та $^{48}Ca$ за допомогою сцинтиляційних кристалів вольфраматів кадмію та кальцію. Показано, що кристали $CdWO_4$ можуть бути застосовані в якості болометричних детекторів з високою енергетичною роздільною здатністю. Вивчена радіочистота сцинтиляторів $CaWO_4$, $ZnWO_4$, $CdWO_4$, GSO та $CeF_3$. Запропоновано метод пошуку спін–залежної взаємодії частинок темної матерії за допомогою кристалів вольфраматів кальцію, цинку, а також кристалів алюмо–ітрієвих гранатів, активованих неодимом, та ортосилікату гадолінію.

Результати викладені в цьому розділі опубліковані в роботах:

# ВИСНОВКИ

Основні результати дисертаційної роботи наступні:

1. Розвинутий сцинтиляційний метод для досліджень 2β–розпаду атомних ядер.

1.1. В Солотвинській підземній лабораторії ІЯД НАНУ споруджена низькофонова сцинтиляційна установка для вимірювання наднизьких активностей радіонуклідів в зразках об'ємом до 30 $дм^3$ з чутливістю при реєстрації $^{40}K$ і $^{232}Th$ в зразку масою ≈ 1 кг з точністю ±30% за 24 години вимірювань 0.04 і 0.007 Бк/кг, відповідно, що порівняно з чутливістю низькофонових напівпровідникових детекторів з надчистого германію. На установці проведено відбір конструкційних матеріалів та сцинтиляційних кристалів для наднизькофонових експериментів, спрямованих на пошук 2β–розпаду та інших рідкісних розпадів атомних ядер.

1.2. В Солотвинській підземній лабораторії споруджений наднизькофоновий сцинтиляційний спектрометр з кристалами вольфрамату кадмію, збагаченими ізотопом $^{116}Cd$. Спектрометр включає комплекс пасивного та активного захисту з відібраних за рівнем радіочистоти матеріалів. Рівень фону детектора, 0.04 відліків / (рік кеВ кг) на енергії 2β–розпаду $^{116}Cd$, є одним з кращих серед досягнутих в експериментах по пошуку 2β–розпаду.

1.3. Розроблено багатоканальну систему реєстрації даних для наднизькофонових установок, яка дозволяє записувати енергію та час події, інформацію про збіги подій в основному детекторі та детекторах активного захисту, форму сцинтиляційних сигналів.

1.4. Вперше застосовано аналіз форми сцинтиляційних сигналів для дискримінації подій в сцинтиляційному детекторі з кристалами $^{116}CdWO_4$, що дозволило приблизно в 4 рази підвищити чутливість експерименту по пошуку 2β–розпаду ядра $^{116}$Cd.

1.5. Досліджені сцинтиляційні властивості, відгук до α–частинок, форма сцинтиляційних сигналів та розроблено методику дискримінації частинок за формою в сцинтиляторах $CaWO_4$, $ZnWO_4$, $PbWO_4$, GSO(Ce), $CeF_3$, показані можливості їх застосування в чутливих експериментах з метою пошуку і дослідження 2β–розпаду ізотопів кальцію, вольфраму, цинку, гадолінію, церію, рідкісних α– та β–розпадів, пошуків частинок темної матерії. Запропоновано метод пошуку спін–залежної взаємодії частинок темної матерії за допомогою кристалів вольфраматів кальцію та цинку, а також алюмо–ітрієвих гранатів, активованих неодимом.

1.6. Розроблено методику вимірювань радіочистоти сцинтиляційних кристалів на рівні чутливості $\approx 10^{-6}$ Бк / кг для $^{228}$Th, $^{226}$Ra, $^{227}$Ac, $\approx 10^{-4}$ Бк / кг для α–активних радіонуклідів та $\approx 10^{-2} - 10^{-4}$ Бк / кг для β–активних радіонуклідів. Досліджені активності та особливості (вікова рівновага, просторовий розподіл в об’ємі кристалів) слідових домішок радіонуклідів в сцинтиляторах $CaWO_4$, $ZnWO_4$, $CdWO_4$ (як з природним розподілом ізотопів кадмію, так і збагачених $^{116}$Cd), $PbWO_4$, GSO(Ce), $CeF_3$.

1.7. Вперше досліджені сцинтиляційні властивості, форму сцинтиляційних сигналів, ступінь радіочистоти алюмо–ітрієвого гранату, активованого неодимом (YAG:Nd). Показана можливість використання цього сцинтилятора для пошуку 2β–розпаду ізотопів неодиму.

1.8. Показано, що кристали $CdWO_4$ можуть бути застосовані в якості болометричних детекторів з високою енергетичною роздільною здатністю на рівні 5 кеВ в широкому діапазоні енергій, що робить їх надзвичайно перспективними детекторами для пошуку 2β–розпаду кадмію.

2. Експериментально досліджено різні моди і канали 2β–розпаду ядер $^{48}$Ca, $^{64}$Zn, $^{70}$Zn, $^{106}$Cd, $^{108}$Cd, $^{114}$Cd, $^{116}$Cd, $^{136}$Ce, $^{138}$Ce, $^{142}$Ce, $^{160}$Gd, $^{180}$W, $^{186}$W.

2.1. Виміряний період напіврозпаду ядра $^{116}Cd$ відносно $2\nu2\beta$–розпаду:

$$T_{1/2} = 2.9^{+0.4}_{-0.3} \times 10^{19} \text{ років.}$$

2.2. Встановлене нове обмеження на період напіврозпаду відносно $0\nu2\beta$ розпаду $^{116}Cd$ на основний стан ядра $^{116}Sn$ (з довірчою імовірністю 90%):

$$T_{1/2}\,(0^+ \to 0^+) \geq 1.7 \times 10^{23} \text{ років.}$$

2.3. Встановлені нові обмеження на періоди напіврозпаду $^{116}Cd$ відносно $0\nu2\beta$–розпаду на збуджені стани ядра $^{116}Sn$ (з 90% CL):

$$T_{1/2}(0^+ \to 2^+) \geq 2.9 \times 10^{22} \text{ років,}$$

$$T_{1/2}(0^+ \to 0_1^+) \geq 1.4 \times 10^{22} \text{ років,}$$

$$T_{1/2}(0^+ \to 0_2^+) \geq 0.6 \times 10^{22} \text{ років.}$$

2.4. Встановлені нові обмеження на періоди $0\nu2\beta$–розпаду ядра $^{116}Cd$ з випроміненням майоронів – одного, двох і т.зв. bulk майорону (з 90% CL):

$$T_{1/2}(M1) \geq 0.8 \times 10^{22} \text{ років,}$$

$$T_{1/2}(M2) \geq 0.8 \times 10^{21} \text{ років,}$$

$$T_{1/2}(M^{bulk}) \geq 1.7 \times 10^{21} \text{ років.}$$

2.5. З експериментального обмеження на $0\nu2\beta$–розпад $^{116}Cd$ отримані одні з кращих у світі обмежень на ефективну масу нейтрино майоранівської природи:

$$\langle m_\nu \rangle \leq 1.7 \text{ еВ,}$$

параметри домішок правих токів в слабкій взаємодії:

$$\langle \eta \rangle \leq 2.5 \times 10^{-8},$$

$$\langle \lambda \rangle \leq 2.2 \times 10^{-6},$$

та параметр порушення $R$–парності в мінімальній суперсиметричній СМ з порушенням $R$–парності:

$$\varepsilon \leq 7 \times 10^{-4}.$$

2.6. З експериментального обмеження на $0\nu2\beta$–розпад $^{116}Cd$ з вильотом майорона отримане одне з найбільш жорстких обмежень на константу зв'язку нейтрино з майороном:

$$\langle g_M \rangle \leq 4.6 \times 10^{-5}.$$

2.7. Експериментально досліджені процеси 2β–розпаду ядра $^{160}Gd$. Встановлені обмеження на періоди напіврозпаду ядра $^{160}Gd$ відносно 0ν2β–розпаду на основний та перший збуджений стан ядра $^{160}Dy$ (з 90% CL):

$$T_{1/2}(0^+ \to 0^+) \geq 1.3 \times 10^{21} \text{ років},$$

$$T_{1/2}(0^+ \to 2^+) \geq 1.3 \times 10^{21} \text{ років}.$$

Результати, перераховані в п. 2.1 – 2.7, увійшли до огляду Review of Particle Physics [Phys. Rev. B, Vol. 592 (2004) P. 1] (розділи "Majoron Searches in Neutrinoless Double β Decay" (с. 401 – 402) та "Double–β Decay" (с. 447 – 450).

2.8. Встановлені найбільш жорсткі обмеження період напіврозпаду ядра $^{160}Gd$ відносно 2ν2β–розпаду на основний та перший збуджений стан $^{160}Dy$ (з 90% CL):

$$T_{1/2}(0^+ \to 0^+) \geq 1.9 \times 10^{19} \text{ років},$$

$$T_{1/2}(0^+ \to 2^+) \geq 2.1 \times 10^{19} \text{ років}.$$

2.9. Вперше експериментально досліджені різні моди та канали процесів 2β–розпаду ізотопів вольфраму $^{180}W$ та $^{186}W$. Зокрема, період напіврозпаду ядра $^{180}W$ відносно 0ν2β–розпаду на основний стан ядра $^{180}Os$ (з 90% CL) обмежений на рівні:

$$T_{1/2}(0^+ \to 0^+) \geq 1.1 \times 10^{21} \text{ років}.$$

2.10. Встановлені експериментальні обмеження на різні моди та канали 2β–розпаду ядер $^{48}Ca$, $^{64}Zn$, $^{70}Zn$, $^{106}Cd$, $^{108}Cd$, $^{114}Cd$, $^{136}Ce$, $^{138}Ce$, $^{142}Ce$ (дослідження $^{108}Cd$ та $^{114}Cd$ здійснені вперше) на рівні чутливості $T_{1/2} \approx 10^{17} – 10^{20}$ років.

3. З аналізу даних вимірювань з наднизькофоновими германієвими детекторами (експерименти “Heidelberg–Moscow” та “IGEX”) одержано найкраще в світі обмеження на період піврозпаду ядра $^{76}$Ge відносно безнейтринного подвійного бета-розпаду: $T_{1/2} \geq 4.2\times10^{25}$ р, звідки слідує найбільш жорстке обмеження на масу нейтрино майоранівської природи:

$$\langle m_\nu \rangle \leq 0.24 \text{ еВ} \qquad (68\% \text{ CL}).$$

4. Досліджені можливості підвищення чутливості експериментів по пошуку $0\nu2\beta$–розпаду з використанням сцинтиляційного методу до рівня $T_{1/2} \approx 10^{26} – 10^{27}$ років.

4.1. Показано, яким вимогам мають задовольняти експерименти для реєстрації та встановлення обмеження на процес $0\nu2\beta$–розпаду.

4.2. Запропоновано проект експерименту (CAMEO project) по пошуку подвійного β–розпаду $^{116}$Cd з чутливістю $T_{1/2} \approx 10^{26}$ років. Проаналізовані можливості підвищення чутливості експерименту до рівня $T_{1/2} \approx 10^{27}$ років ($\langle m_\nu \rangle$ ~ 0.02 еВ).

4.3. Показані можливості використання сцинтиляторів вольфрамату свинцю в експерименті по пошуку подвійного β–розпаду $^{116}$Cd з чутливістю $T_{1/2} \approx 10^{26}$ років.

4.4. Запропоновано проект експерименту (CARVEL project) по пошуку подвійного β–розпаду $^{48}$Ca з чутливістю $T_{1/2} \approx 10^{26}$ років, що відповідає масі нейтрино 0.04 – 0.09 еВ.

# ПОДЯКИ

Ця дисертаційна робота була ініційована покійним Юрієм Георгійовичем Здесенком, завідуючим відділом фізики лептонів Інституту ядерних досліджень НАН України, доктором фізико–математичних наук, членом–кореспондентом НАН України, професором, визначним вченим в галузі фізики нейтрино. Юрій Георгійович був моїм керівником, вчителем, співавтором впродовж 26 років. В дуже значній мірі саме завдяки його величезній наполегливості, енергії, вимогливості до себе і підлеглих, таланту вченого і високому професіоналізму були здійснені всі ті дослідження, які описані в цій дисертаційній роботі.

Ця робота була б неможливою без сприяння, уваги і допомоги (особливо це стосується досліджень в Солотвинській підземній лабораторії) директора Інституту ядерних досліджень НАНУ, академіка НАНУ Івана Миколайовича Вишневського. Дякую Івану Миколайовичу також за прочитання тексту дисертації та цінні зауваження щодо її змісту. Я вдячний заступникам директора ІЯД Василю Олександровичу Лавриненку, Володимиру Васильовичу Тришину, Василю Івановичу Слісенку, головному інженеру Володимиру Івановичу Савчуку та багатьом іншим людям з Інституту ядерних досліджень НАНУ за допомогу в забезпеченні функціонування Солотвинської підземної лаюораторії. Автор висловлює щиру подяку академіку–секретарю Відділення ядерної фізики та енергетики академіку НАНУ Неклюдову Івану Матвійовичу за всебічну підтримку досліджень в Солотвинській підземній лабораторії ІЯД НАНУ.

Всі результати, викладені в дисертаційній роботі, були отримані завдяки роботі в тому чудовому колективі, яким є відділ фізики лептонів Інституту ядерних досліджень НАНУ. Автор висловлює особливу подяку Володимиру Іллічу Третяку — талановитому вченому, людині неймовірної наполегливості, працездатності, акуратності. Я щиро вдячний Володимиру за всі роки спільної праці, за плодотворні дискусії, за допомогу і участь у всьому, чим довелось займатись. Дякую Андрію

Степановичу Николайко за ту велику роботу, яка була проведена, зокрема з кристалами вольфрамату кадмію, задовго до того, як автор почав працювати з цими, як потім виявилося, дуже перспективними детекторами. Саме А.С. Ніколайко (за свідченням покійного Юрія Георгійовича Здесенка) належить ідея використання сцинтиляторів $CdWO_4$ для пошуку подвійного бета–розпаду ядра $^{116}Cd$. Як людина дуже скромна, сам Андрій Степанович ніколи не говорив про це. Багато дуже важливої роботи (зокрема, вимірювання в Солотвино, розробка програмного забезпечення для часово–амплітудного аналізу та аналізу за формою сигналів) було зроблено Владиславом Валерійовичем Кобичевим, за що я дуже йому вдячний. Моя щира подяка Борису Миколайовичу Кропив'янському за його вагомий внесок в розробку конструкцій низькофонових установок та проведення вимірювань в Солотвинській лабораторії. Дякую Олега Анатолійовича Понкратенка за багаторічну співпрацю, за розробку і проведення моделювання методом Монте–Карло багатьох описаних в дисертації експериментів. Крім того, саме О.А. Понкратенко запропонував цікаву ідею відтворення координати сцинтиляційного спалаху в зануреному в рідину сцинтиляторі, що описана в проекті експерименту «CAMEO». Дякую Сергію Нагорному за значний внесок у проведення вимірювань сцинтиляційних характеристик, форми сцинтиляційних сигналів, альфа/бета–співвідношення та радіочистоти сцинтиляторів вольфраматів кальцію, кадмію, цинку, свинцю, алюмо–ітрієвого гранату, активованого неодимом. Дякую Анзорію Георгадзе та Сергію Здесенку за їхній внесок в забезпечення наукової роботи відділу та проведення вимірювань в Солотвино. Авторові приємно не лише висловити подяку, а й побажати успіхів молодим співробітникам відділу: Денису Поді, Сергію Юрченко, Юрію Чечеренку, Костянтину Кладько. Щиро дякую Олені Володимирівні Зуєвій за її скромну, але важливу роботу в забезпеченні діяльності відділу фізики лептонів. Не можна не згадати добрим словом працю співробітників Солотвинської лабораторії, які забезпечували і, попри всі негаразди, продовжують забезпечувати діяльність лабораторії, особливо Михайла Степановича Шейчука та Дюло Юліусовича Сєдлака. Дякую всім тим, хто в різні роки працював у відділі фізики лептонів: Василю Миколайовичу Куцу, Віктору Музалевському, Олександру Костежу, Олегу Бондаренку, Валентині Василенко, Олегу Шкворцю за їхню роботу на різних етапах досліджень, що були описані в даному рукописі.

Дуже приємно подякувати чудовим людям, які стали моїми щирими друзями за час нашої співпраці, визначним вченим, професорам фізичного факультету Флорентійського університету та Флорентійського відділення Національного Інституту Ядерної Фізики (Італія) Паоло Маурензігу, П’єру Джорджіо Біззеті, Тіто Фаццині за їх вагомий внесок в Солотвинський експеримент по пошуку β–розпаду ядра $^{116}$Cd, особливо за розробку методів аналізу сцинтиляційних сигналів вольфрамату кадмію за формою. Сердечна подяка професору Риті Бернабей та доктору П´єрлуіджі Беллі (Римський Університет “Tor Vergata”), доктору Антонеллі Інчікітті (Римський Університет “La Sapienza”) за співпрацю в дослідженнях рідкісних процесів за допомогою сцинтиляторів фториду церію. Щиро дякую чудових людей і визначних фізиків Віктора Борисовича Бруданіна та Цвєтана Вилова з Об’єднаного Інституту Ядерних Досліджень (Дубна, Росія), а також професора Університету Південної Кароліни (США) Френка Авіньйона III за співпрацю в дослідженнях кристалів вольфрамату кальцію, як перспективного детектора для пошуку подвійного бета–розпаду ядра $^{48}$Ca. Дякую всім членам колаборації NEMO, особливо Фернанду Шейблінгу, Ізабелль Лінк, Фабрісу Пікмалю, Жану Луї Гійоне, Роджеру Арнольду (Центр Ядерних Досліджень, Страсбург) за співпрацю в експерименті по пошуку подвійного бета–розпаду $^{100}$Mo. Дякую співробітників Інституту Макса Планка в Гейдельберзі, зокрема проф. Ханса Фолкера Клапдора–Кляйнгротхауса, а також Матіаса Фоллінгера, Андреаса Мюллера за співпрацю в експерименті по пошуку подвійних бета–процесів в ядрі $^{106}$Cd. Дякую за співпрацю при роботі над проектом експерименту „CAMEO” всіх учасників колаборації, зокрема, Марко Джіаммаркі, Ліно Мерамонті, Барабару

Каччанігу з Міланського відділення Національного Інституту Ядерної Фізики (Італія), Марка Чена з Королівського Університету м. Кінгстон (Канада), Лотара Оберауера з Технологічного Університету (Мюнхен, Німеччина).

Автору приємно подякувати співробітникам Інституту Монокристалів в Харкові (зараз Інститут сцинтиляційних матеріалів НАНУ) Володимиру Діомидовичу Рижикову, Станіславу Феліксовичу Бурачасу, Людмилі Лаврентіївні Нагорній. Завдяки зусиллям цих вчених був вирощений унікальний кристал вольфрамату кадмію, збагачений ізотопом $^{116}$Cd. Дякую Івану Михайловичу Сольському з Науково–виробничої компанії „Карат” (Львів) не лише за роботу по росту кристалів вольфрамату кадмію та кальцію, які були використані в експериментах, а й за цікаві дискусії, що стосуються росту сцинтиляційних кристалів. Моя щира подяка директору Інституту сцинтиляційних матеріалів НАНУ, член-кореспонденту НАНУ Гриньову Борису Вікторовичу за його інтерес до досліджень подвійного бета–розпаду та співпрацю в дослідженнях сцинтиляторів вольфрамату свинцю та кадмію. Дякую Мирославу Батенчуку (тоді співробітнику Львівського університету) за надані для досліджень сцинтиляційні кристали вольфрамату цинку та Юрію Глінці (тоді співробітнику Інституту хімії поверхні НАНУ) за наданий для досліджень кристал алюмо–ітрієвого гранату, активованого неодимом.

Я дякую всім людям, чиї робота, порада, співчуття сприяли виконанню тих чи інших досліджень, що увійшли до цієї дисертації, і прошу у них вибачення за те, що не висловив їм персональну подяку.

Гадаю, що описані в цій дисертації результати, не варті всіх тих негараздів, що випали на долю моєї дружини Тетяни і наших дітей: Марії, Тетяни і Бориса через мої заняття такою досить малоприбутковою та непрестижною в наш час діяльністю, як наука. Я сердечно вдячний їм, моїм рідним людям, за їхнє терпіння і любов.

# СПИСОК ВИКОРИСТАНИХ ДЖЕРЕЛ